\documentclass[aps,prb,twocolumn,amsmath,amssymb,groupedaddress,dvipdfmx,longbibliography
]{revtex4-2}

\usepackage{graphicx}
\usepackage{dcolumn}
\usepackage{bm}
\usepackage{multirow}

\usepackage{braket}
\usepackage{color}
\usepackage{times}
\usepackage{setspace}

\usepackage{longtable}

\begin{document}
\setcounter{MaxMatrixCols}{16}

\title{Symmetry-adapted modeling for molecules and crystals}

\author{Hiroaki Kusunose$^1$, Rikuto Oiwa$^1$, and Satoru Hayami$^2$}
\affiliation{
$^1$Department of Physics, Meiji University, Kawasaki 214-8571, Japan \\
$^2$Graduate School of Science, Hokkaido University, Sapporo 060-0810, Japan
}

\date{\today}

\begin{abstract}
We have developed a symmetry-adapted modeling procedure for molecules and crystals.
By using the completeness of multipoles to express spatial and time-reversal parity-specific anisotropic distributions, we can generate systematically the complete symmetry-adapted multipole basis set to describe any of electronic degrees of freedom in isolated cluster systems and periodic crystals.
The symmetry-adapted modeling is then achieved by expressing the Hamiltonian in terms of the linear combination of these bases belonging to the identity irreducible representation, and the model parameters (linear coefficients) in the Hamiltonian can be determined so as to reproduce the electronic structures given by the density-functional computation.
We demonstrate our method for the modeling of graphene, and emphasize usefulness of the symmetry-adapted basis to analyze and predict physical phenomena and spontaneous symmetry breaking in a phase transition.
The present method is complementary to \textit{de facto} standard Wannier tight-binding modeling, and it provides us with a fundamental basis to develop a symmetry-based analysis for materials science.
\end{abstract}

\maketitle

\section{Introduction}

Diversity is one of the fascinating aspects of materials science, and the diverse properties of materials are brought about by mutual interplay among electronic degrees of freedom, such as charge, atomic orbital and spin, and underlying molecular or crystal structure.
Moreover, intriguing phenomena emerge by phase transitions with spontaneous symmetry breaking.
In particular, order parameters of spin-orbital-lattice composite objects bring about various off-diagonal responses, and generate nontrivial transport involving atomic internal objects.

In order to analyze proper material properties and predict bright new phenomena quantitatively, one needs microscopic modeling of materials.
For a such purpose, the density-functional (DF) theory and related modelings have been widely used~\cite{Hohenberg_PhysRev.136.B864, Kohn_PhysRev.140.A1133, Parr1994, Koch2001, Martin2004, Marzari_RevModPhys.84.1419}.
Once a tractable model is obtained, one uses it to discuss various response functions, and to give a starting point for taking account of many-body effects such as electron correlations, and electron-phonon interactions.

The Wannier-based tight-binding (Wannier TB) modeling from DF theory is the \textit{de facto} standard, and there are several advantages, such as no need for electronic band fitting to a certain model, capturing covalent-bond feature of wave functions, and so on~\cite{mostofi2008wannier90,mostofi2014updated,pizzi2020wannier90}.
Nevertheless, there are several drawbacks as follows: (1) the obtained Wannier TB model does not satisfy the symmetry of a system rigorously due to a disentangling procedure of bands within a given energy window in addition to simple numerical errors, (2) as the Wannier basis functions differ from atomic orbital ones in general, representation matrices for physical quantities such as the orbital angular-momentum operator become unclear with respect to those bases, and (3) there are considerably small long-range hopping matrices in the Wannier TB model, however, it is quite cumbersome to neglect them without losing the symmetry, which hampers us to compactify the Wannier TB model.

In this paper, we propose a complementary modeling procedure to overcome the above drawbacks, which fully respects the symmetry of a system, and atomic (internal) degrees of freedom, in its construction process.
According to Neumann's principle, any macroscopic responses are characterized by point-group symmetry~\cite{neumann1885vorlesungen}, and in the Landau theory of phase transition, nontrivial irreducible representation determines the fate of an emerging phase in which the order parameter is a macroscopic quantum-mechanical average of a microscopic degree of freedom~\cite{landau1937theory}.
Therefore, a seamless description between macroscopic quantities and microscopic degrees of freedom in accordance with symmetry is indispensable in a promising modeling method.

Realistic materials are characterized by anisotropic distributions in molecular or crystal structure in addition to spatial and time-reversal parities compatible with their symmetry.
The symmetry-adapted multipoles in point group are suitable candidates to describe such parity-specific anisotropic distributions~\cite{hayami2018microscopic, Hayami_PhysRevB.98.165110, Hayami_PhysRevB.102.144441}, as they have the completeness in angular space~\cite{kusunose2020complete}.
By utilizing the completeness of the multipoles, we construct the symmetry-adapted basis to describe any of electronic degrees of freedom in isolated cluster systems (e.g., molecules and quantum dots) and periodic crystals.
Since the present multipole basis can treat the internal atomic degrees of freedom and molecular or crystal structures separately, it is able to bridge explicitly between macroscopic quantities and microscopic degrees of freedom.
Indeed, various physical quantities appearing in ordinary Hamiltonians can be expressed by the multipole basis, whose examples are summarized in Table~\ref{tbl:samb_ham}.

\begin{table}
\caption{\label{tbl:samb_ham}
Correspondence between physical quantities and symmetry-adapted multipole basis (SAMB).
The upper, middle, and lower panels represent one-body, two-body, and hopping terms, respectively.
The site (bond) dependence in the upper (middle) panel is expressed by the site cluster $\mathbb{Q}_{lm}^{\rm (s)}$ (bond cluster $\mathbb{Q}_{lm}^{\rm (b)}$, $\mathbb{T}_{lm}^{\rm (b)}$) SAMBs.
DM int. is the Dzyaloshinsky-Moriya interaction.
The repeated indices are implicitly summed in the expression.
The detailed meaning of these symbols will be explained in the main sections.
}
\begin{spacing}{1.5}
\begin{ruledtabular}
\begin{tabular}{ccc}
Type & Expression & Correspondence \\ \hline
Electric potential & $\phi q$ & $q\to\mathbb{Q}_{0,0}^{\rm (a)}$ \\
Crystal field & $\phi_{lm}Q_{lm}$ & $Q_{lm}\to\mathbb{Q}_{lm}^{\rm (a)}$ \\
Zeeman term & $-h^{a}m^{a}$ & $m^{a}\to\mathbb{M}_{1m}^{\rm (a)}$ \\
Spin-orbit int. & $\zeta l^{a}\sigma^{a}$ & $l^{a},\sigma^{a}\to \mathbb{M}_{1m}^{\rm (a)}$ \\ \hline
Density-density int. & $V_{ij}n_{i}n_{j}$ & $n_{i}n_{j}\to\mathbb{Q}_{0,0}^{\rm (a)}$ \\
Elastic energy & $\epsilon_{ij}^{ab}u_{i}^{a}u_{j}^{b}$ & $u_{i}^{a}u_{j}^{b}\to\mathbb{Q}_{0,0}^{\rm (a)},\mathbb{Q}_{2m}^{\rm (a)}$ \\
Exchange int. & $J_{ij}^{ab}S_{i}^{a}S_{j}^{b}$ & $S_{i}^{a}S_{j}^{b}\to\mathbb{Q}_{0,0}^{\rm (a)},\mathbb{Q}_{2m}^{\rm (a)}$ \\
DM int. & $D_{ij}^{c}\epsilon_{abc}S_{i}^{a}S_{j}^{b}$ & $\epsilon_{abc}S_{i}^{a}S_{j}^{b}\to\mathbb{G}_{lm}^{\rm (a)}$ \\ \hline
Real hopping & $t_{ij}c_{i}^{\dagger}c_{j}+{\rm h.c.}$ & $c_{i}^{\dagger}c_{j}+{\rm h.c.}\to\mathbb{Q}_{lm}^{\rm (b)}$ \\
Imaginary hopping & ${\rm i\,}t_{ij}c_{i}^{\dagger}c_{j}+{\rm h.c.}$ & ${\rm i\,}c_{i}^{\dagger}c_{j}+{\rm h.c.}\to\mathbb{T}_{lm}^{\rm (b)}$ \\
\end{tabular}
\end{ruledtabular}
\end{spacing}
\end{table}

Then symmetry-adapted modeling is achieved by expressing the Hamiltonian in terms of the linear combination of these bases belonging to the identity irreducible representation.
Once a symmetry-adapted model is constructed, one optimizes the model parameters (linear coefficients) to reproduce the electronic states given by DF computation.
The optimization can be carried out by using the machine-learning technique with the deep neural network~\cite{lecun2015deep} having extremely low dependencies of initial guess.
There have been several attempts to construct the TB Hamiltonian~\cite{PhysRevB.87.235109,KORETSUNE2023108645} based on machine-learning technique~\cite{nakhaee_NN_TB_2020, wang_NN_TB_2021}.
The TB models generated by Wang {\it et al.}~\cite{wang_NN_TB_2021} successfully reproduce the DF band dispersions with high accuracy, where the symmetry of the system is not considered and each hopping is regarded as a neuron in their neural network.
On the other hand, in our scheme each symmetry-adapted multipole basis (SAMB) is a neuron in the neural network, and hence the symmetry is always maintained during the optimization process.
We give the prime example of the basis construction and optimization by using graphene, and other examples are given in the supplementary materials~\cite{supplement_SAMB}.

This paper is organized as follows: In Sec.~\ref{sec:samb}, we explain the construction procedure for the complete SAMB set.
We first treat the case of isolated cluster systems by using an example of a fictitious molecule in the C$_{\rm 3v}$ point group, and then the case of periodic crystals is discussed.
After setting up the general basis construction procedure, we give an application of our method to graphene in Sec.~\ref{sec:graphene}.
We construct the symmetry-adapted TB model for graphene up to sixth-neighbor hoppings, and optimize the model parameters to reproduce DF energy dispersion.
The final section summarizes the paper.

\section{Symmetry-Adapted Multipole Basis}\label{sec:samb}

In order to perform symmetry-adapted modeling, we introduce the complete orthonormal basis set that is classified according to the point-group symmetry.
Such a basis set is called SAMB, in which any anisotropy is described by means of multipolar anisotropy~\cite{kusunose2022generalization}.
Let us first discuss the SAMB in the case of isolated cluster systems such as molecules and quantum dots in Sec.~\ref{sec:isolated_cluster}. Then, the case of periodic crystals is explained in Sec.~\ref{sec:crystal}.
The conversions to full matrix form and momentum-space representation for a specified system are discussed in Sec.~\ref{sec:fullmom}.

\subsection{Isolated cluster systems}\label{sec:isolated_cluster}

\begin{figure}
\centering
\includegraphics[width=7.5cm]{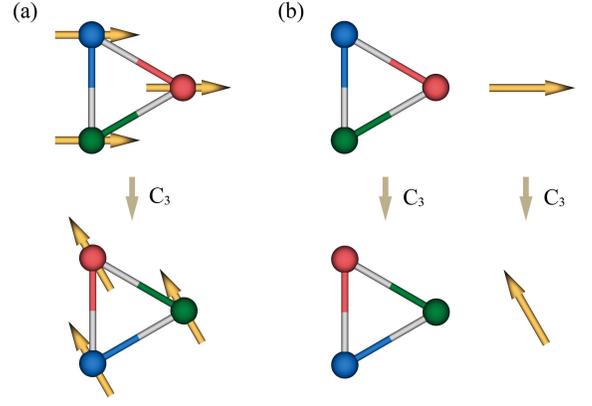}%
\caption{\label{fig:sep_op}
Separation of the symmetry operation, $C_{3}$;
(a) without separation and (b) operations separately to atomic sites/bonds and atomic degrees of freedom, e.g., spins.
}
\end{figure}

We begin with an isolated cluster system which consists of several atoms having internal degrees of freedom, i.e., atomic orbitals and spins, at each atomic site.
Hereafter, we simply call the atomic orbitals including spins as ``atomic orbitals''.
In general, the symmetry operations of the system can be applied separately to positions of atoms and atomic degrees of freedom as shown in Fig.~\ref{fig:sep_op}.
Because of this separable property, we can construct the SAMB separately for atomic sites/bonds, and atomic degrees of freedom as follows.

\subsubsection{Site cluster and bond cluster}\label{sec:sbcluster}

\begin{figure}
\centering
\includegraphics[width=7cm]{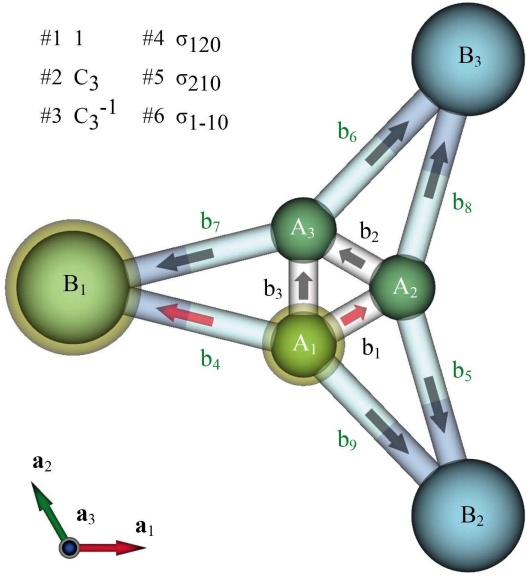}%
\caption{\label{fig:sb_cluster}
Example of site clusters (different colored spheres) and bond clusters (different colored bonds) in the C$_{\rm 3v}$ (31m) point group (vertical mirrors in $xz$ and equivalent planes).
The representative site and bond are denoted by a yellow circle and red arrow, respectively.
The trigonal units are $\bm{a}_{1}=(1,0,0)$, $\bm{a}_{2}=(-1/2,\sqrt{3}/2,0)$, and $\bm{a}_{3}=(0,0,1)$.
}
\end{figure}

First, we explain the SAMB for atomic sites and bonds.
Let us choose a representative atomic site in the isolated cluster, and its position is $\bm{R}_{1}$.
Then, a set of sites $(\bm{R}_{1}',\bm{R}_{2}',\cdots,\bm{R}_{N_{g}}')$ is obtained by applying the symmetry operation $\mathcal{G}_{g}$ in the point group as $\bm{R}_{g}'=\mathcal{G}_{g}\bm{R}_{1}$ ($g=1,2,\cdots,N_{g}$; $\mathcal{G}_{1}$ is assumed to be the identity operation) where $N_{g}$ is the number of symmetry operations.
Since some of the obtained sites are overlapped unless the sites belong to the general point, the total number of cluster sites $N_{s}$ is less than $N_{g}$, and it is equivalent to the number in the Wyckoff symbol.
We call the set of sites without duplication $(\bm{R}_{1},\bm{R}_{2},\cdots,\bm{R}_{N_{s}})$ ``site cluster'', and the one-to-many correspondence is expressed by $g(i)$ so as to satisfy $\bm{R}_{i}=\bm{R}_{g(i)}'$ ($i=1,2,\cdots,N_{s}$).
All the sites in the isolated cluster system can be divided into a set of site clusters.
For example, the division of site clusters in the C$_{\rm 3v}$ (31m) point group with $N_{g}=6$ and $N_s=3$ is shown by different colored spheres in Fig.~\ref{fig:sb_cluster}.

Similarly, the bond cluster is obtained by applying the symmetry operation to a representative bond, $\bm{b}_{1}@\bm{c}_{1}$, where we have introduced the bond-vector $\bm{b}_{1}=\bm{R}_{\rm head}-\bm{R}_{\rm tail}$, and the bond center $\bm{c}_{1}=(\bm{R}_{\rm head}+\bm{R}_{\rm tail})/2$.
The obtained bonds $(\bm{b}_{1}'@\bm{c}_{1}',\bm{b}_{2}'@\bm{c}_{2}',\cdots,\bm{b}_{N_{g}}'@\bm{c}_{N_{g}}')$ are duplicated in general with the equivalent bond centers.
Then, the set of bonds without duplication ($\bm{b}_{1}@\bm{c}_{1},\bm{b}_{2}@\bm{c}_{2},\cdots,\bm{b}_{N_{b}}@\bm{c}_{N_{b}}$) constitutes the ``bond cluster'', where $N_{b}$ is the number of bonds in the bond cluster.
In contrast to the site cluster, some bonds may coincide with other bond with reversed direction.
In this case, we attach a negative sign to the symmetry operation in the correspondence $g(i)$, i.e., the one-to-many correspondence is given by $g(i)$ with $\bm{c}_{i}=\bm{c}_{|g(i)|}'$ and $\bm{b}_{i}={\rm sgn}[g(i)]\bm{b}_{|g(i)|}'$.
All the bonds in the isolated cluster system can be divided into a set of bond clusters, as shown by different colored bonds in Fig.~\ref{fig:sb_cluster}.
The one-to-many correspondences of the site/bond clusters for the example of the fictitious molecule in Fig.~\ref{fig:sb_cluster} are shown in Table~\ref{tbl:mapping}.

\begin{table}
\caption{\label{tbl:mapping}
One-to-many correspondences in the site/bond clusters for the example in Fig.~\ref{fig:sb_cluster}, in which $x=1/6$ and $x'=2/3$ are used.
}
\begin{ruledtabular}
\begin{tabular}{cccccc}
Site A & $\bm{R}_{i}$ & $g(i)$ & Site B & $\bm{R}_{i}$ & $g(i)$ \\ \hline
A$_{1}$ & $[-x,-x,0]$ & $[1,6]$ & B$_{1}$ & $[-x',0,0]$ & $[1,4]$ \\
A$_{2}$ & $[x,0,0]$ & $[2,5]$ & B$_{2}$ & $[0,-x',0]$ & $[2,6]$ \\
A$_{3}$ & $[0,x,0]$ & $[3,4]$ & B$_{3}$ & $[x',x'0]$ & $[3,5]$\\
\end{tabular}
\vspace{1mm}
\begin{spacing}{1.3}
\begin{tabular}{ccc}
Bond A$\to$A & $\bm{b}_{i}@\bm{c}_{i}$ & $g(i)$ \\ \hline
b$_{1}$ & $[\frac{1}{3}, \frac{1}{6}, 0]@[0,-\frac{1}{12}, 0]$ & [1,-5] \\
b$_{2}$ & $[-\frac{1}{6}, \frac{1}{6}, 0]@[\frac{1}{12}, \frac{1}{12}, 0]$ & [2,-4] \\
b$_{3}$ & $[\frac{1}{6}, \frac{1}{3}, 0]@[-\frac{1}{12}, 0, 0]$ & [-3,6] \\
\end{tabular}
\vspace{1mm}
\begin{tabular}{ccc}
Bond A$\to$B & $\bm{b}_{i}@\bm{c}_{i}$ & $g(i)$ \\ \hline
b$_{4}$ & $[-\frac{1}{2}, \frac{1}{6}, 0]@[-\frac{5}{12}, -\frac{1}{12}, 0]$ & [1] \\
b$_{5}$ & $[-\frac{1}{6}, -\frac{2}{3}, 0]@[\frac{1}{12}, -\frac{1}{3}, 0]$ & [2] \\
b$_{6}$ & $[\frac{2}{3}, \frac{1}{2}, 0]@[\frac{1}{3}, \frac{5}{12}, 0]$ & [3] \\
b$_{7}$ & $[-\frac{2}{3}, -\frac{1}{6}, 0]@[-\frac{1}{3}, \frac{1}{12}, 0]$ & [4] \\
b$_{8}$ & $[\frac{1}{2}, \frac{2}{3}, 0]@[\frac{5}{12}, \frac{1}{3}, 0]$ & [5] \\
b$_{9}$ & $[\frac{1}{6}, -\frac{1}{2}, 0]@[-\frac{1}{12}, -\frac{5}{12}, 0]$ & [6] \\
\end{tabular}
\end{spacing}
\end{ruledtabular}
\end{table}

\subsubsection{Symmetry-adapted multipole basis}

Once the site- and bond clusters are introduced, we are ready to construct the SAMB for each cluster.
Let us begin with the SAMB for a site cluster, which enables us to express any site-dependent quantity in a cluster.
To this end, we introduce the normalized spherical harmonics defined by
\begin{align}
O_{lm}(\bm{r})=r^{l}\sqrt{\frac{4\pi}{2l+1}}Y_{lm}(\hat{\bm{r}}),
\end{align}
where $r=|\bm{r}|$, $\hat{\bm{r}}=\bm{r}/r$, and the spherical harmonics, $Y_{lm}(\hat{\bm{r}})$ with the rank $l$~($=0,1,2,\cdots$) and component $m$~($=-l,-l+1,\cdots,l$).
Since the point group is a subgroup of the rotation group supplemented by the inversion operation, we use symmetry-adapted harmonics $O_{l\xi}(\bm{r})$ ($\xi=(\Gamma,n,\gamma)$) instead of $O_{lm}(\bm{r})$ by appropriate linear combination as
\begin{align}
O_{l\xi}(\bm{r})=\sum_{m}U_{m,\xi}^{(l)}O_{lm}(\bm{r}),
\label{eq:rtop}
\end{align}
where $U_{m,\xi}^{(l)}$ is a matrix element of the unitary matrix for basis transformation, and $\Gamma$ and $\gamma$ represent the irreducible representation (irrep.) and its component, respectively.
The label $n$ is the multiplicity to distinguish independent harmonics belonging to the same irrep.
It should be noted that the harmonics in each irrep., especially for two- and three-dimensional ones, must be defined so as to give equivalent representation matrices for all symmetry operations.
For example, the harmonics up to rank 3 in the C$_{\rm 3v}$ (3m1) point group are given in Table~\ref{tbl:harmonics_c3v} (they are also used for D$_{\rm 6h}$ for later purposes).
For point groups with complex characters, i.e., $E_{a}$ ($x+iy$ like) and $E_{b}$ ($x-iy$ like), we treat them together as $E$ irrep. by hermiting as $[O_{l,(E_{a},n)}+O_{l,(E_{b},n)}]/\sqrt{2}$ and $[O_{l,(E_{a},n)}-O_{l,(E_{b},n)}]/\sqrt{2}i$.

\begin{table}
\caption{\label{tbl:harmonics_c3v}
Harmonics up to rank 3 in the C$_{\rm 3v}$ (31m) point group.
The labels in the square bracket represent those for D$_{\rm 6h}$ point group.
The label $g$ and $u$ are exchanged in the irrep. of the axial vector in D$_{\rm 6h}$.
}
\begin{spacing}{1.5}
\begin{ruledtabular}
\begin{tabular}{cccccc}
$l$ & $\Gamma$ & $n$ & $\gamma$ & Form & Axial \\ \hline
0 & $A_{1}$ [$A_{1g}$] & - & - & $1$ \\ \hline
1 & $A_{1}$ [$A_{2u}$] & - & - & $z$ & $Z$ \\
  & $E$ [$E_{1u}$] & - & $u$ & $x$ & $-Y$ \\
  &     &   & $v$ & $y$ & $X$ \\ \hline
2 & $A_{1}$ [$A_{1g}$] & - & - & $\frac{1}{2}(3z^{2}-r^{2}$) \\
  & $E$ [$E_{1g}$] & 1 [-] & $u$ & $\sqrt{3}xz$ \\
  &     &   & $v$ & $\sqrt{3}yz$ \\
  & $E$ [$E_{2g}$]    & 2 [-] & $u$ & $\frac{\sqrt{3}}{2}(x^{2}-y^{2})$ \\
  &     &   & $v$ & $-\sqrt{3}xy$ \\ \hline
3 & $A_{1}$ [$A_{2u}$] & 1 [-] & - & $\frac{1}{2}z(5z^{2}-3r^{2})$ \\
  & $A_{1}$ [$B_{2u}$]        & 2 [-] & - & $\frac{\sqrt{10}}{4}x(x^{2}-3y^{2})$ \\
  & $A_{2}$ [$B_{1u}$] & - & - & $\frac{\sqrt{10}}{4}y(3x^{2}-y^{2})$ \\
  & $E$ [$E_{1u}$]     & 1 [-] & $u$ & $\frac{\sqrt{6}}{4}x(5z^{2}-r^{2})$ \\
  &         &   & $v$ & $\frac{\sqrt{6}}{4}y(5z^{2}-r^{2})$ \\
  & $E$ [$E_{2u}$]        & 2 [-] & $u$ & $\frac{\sqrt{15}}{2}z(x^{2}-y^{2})$ \\
  &         &   & $v$ & $-\sqrt{15}xyz$ \\
\end{tabular}
\end{ruledtabular}
\end{spacing}
\end{table}

With this preliminary, the SAMB for a site cluster is obtained by evaluating $O_{l\xi}(\bm{r})$ at $\bm{r}=\bm{R}_{i}$ in the site cluster, i.e., we obtain the $N_{s}$-dimensional vector basis as
\begin{align}
\mathbb{Q}_{l\xi}^{\rm (s)}=(q_{1}^{(l\xi)},q_{2}^{(l\xi)},\cdots,q_{N_{s}}^{(l\xi)}),
\quad
q_{i}^{(l\xi)}=O_{l\xi}(\bm{R}_{i}),
\end{align}
where the black-board font is used to represent the orthonormal basis, and ``Q'' denotes the electric multipole indicating that it has an electric polar tensor property.
The superscript ``(s)'' indicates the SAMB for the site cluster.
We construct the SAMBs from the lowest rank as $l=0,1,2,\cdots$ until $N_{s}$ independent bases are obtained.
When the obtained SAMBs are not orthonormalized, we use the Gram-Schmidt method to orthonormalize them.
The obtained SAMBs are equivalent to the ordinary molecular orbitals consisting of spinless atomic $s$ orbitals at each site.

\subsubsection{Virtual cluster and mapping to original cluster}\label{sec:vcmapping}

Similarly to the case of the site cluster, the SAMB may be constructed for a bond cluster by evaluating $O_{l\xi}(\bm{r})$ at the bond center $\bm{r}=\bm{c}_{i}$.
However, there is a difficulty that $O_{l\xi}(\bm{r})$ sometimes gives useless results when a position of a bond center becomes the origin.
Moreover, in periodic crystals as discussed later, there is an ambiguity of the choice of the origin in a cluster.
There is an additional difficulty in the nonsymmorphic space group, i.e., position vectors of the symmetry-equivalent sites in a cluster have different distances from the origin whatever we choose.
These difficulties hamper us to construct the SAMB based on the spherical harmonics.

To avoid these difficulties, we introduce the virtual cluster in which the virtual sites ($\bm{r}_{1},\bm{r}_{2},\cdots,\bm{r}_{N_{g}}$; $\bm{r}_{1}$ can be arbitrarily chosen) are given by the $N_{g}$ (i.e., $N_{s}=N_{g}$) general points of the relevant point group~\cite{Suzuki_PhysRevB.99.174407}.
Then, we construct the SAMB with respect to the sites in the virtual cluster (indicated by the overline) as
\begin{align}
\overline{\mathbb{Q}}_{l\xi}=(v_{1}^{(l\xi)},v_{2}^{(l\xi)},\cdots,v_{N_{g}}^{(l\xi)}),
\quad
v_{g}^{(l\xi)}=O_{l\xi}(\bm{r}_{g}).
\end{align}
Note that the SAMB in the virtual cluster can be prepared in advance for 32 crystallographic point groups as they are independent from an original cluster.
For example, the orthonormalized SAMB in the virtual cluster of the C$_{\rm 3v}$ (3m1) point group is summarized in Table~\ref{tbl:vc_c3v}, in which the sites $\set{\bm{r}_{g}}$ are generated by the representative point, $\bm{r}_{1}=(1,-1,0)$.

\begin{table}
\caption{\label{tbl:vc_c3v}
Orthonormalized SAMB in the virtual cluster of the C$_{\rm 3v}$ (31m).
The cluster sites in the trigonal unit are given by $\set{\bm{r}_{g}}=\{(1,-1,0),(1,2,0),(-2,-1,0),(2,1,0),(-1,-2,0),(-1,1,0)\}$.
The indices $(l, \Gamma, n, \gamma)$ correspond to those in Table~\ref{tbl:harmonics_c3v}.
}
\begin{spacing}{1.5}
\begin{ruledtabular}
\begin{tabular}{ccccccc}
No. & $l$ & $\Gamma$ & $n$ & $\gamma$ & $\overline{\mathbb{Q}}_{l\xi}$ \\ \hline
1 & 0 & $A_{1}$ & - & - & $\frac{1}{\sqrt{6}}(1,1,1,1,1,1)$ \\ \hline
2 & 1 & $E$ & - & $u$ & $\frac{1}{2}(1,0,-1,1,0,-1)$ \\
3 &   &     & & $v$ & $\frac{1}{2\sqrt{3}}(-1,2,-1,1,-2,1)$ \\ \hline
4 & 2 & $E$ & 2 & $u$ & $\frac{1}{2\sqrt{3}}(1,-2,1,1,-2,1)$ \\
5 &   &     &   & $v$ & $\frac{1}{2}(1,0,-1,-1,0,1)$ \\ \hline
6 & 3 & $A_{2}$ & - & - & $\frac{1}{\sqrt{6}}(-1,-1,-1,1,1,1)$ \\
\end{tabular}
\end{ruledtabular}
\end{spacing}
\end{table}

Once we obtain a set of the SAMBs, $\set{\overline{\mathbb{Q}}_{l\xi}}$, the SAMB for the original site cluster can be obtained by mapping the virtual-cluster elements onto the original site-cluster ones as
\begin{align}
\mathbb{Q}_{l\xi}^{\rm (s)}=(q_{1}^{(l\xi)},q_{2}^{(l\xi)},\cdots,q_{N_{c}}^{(l\xi)}),
\quad
q_{i}^{(l\xi)}=\sum_{g}^{g(i)}v_{g}^{(l\xi)},
\end{align}
where the one-to-many correspondence (mapping) $g(i)$ is determined for the original site cluster.

Although the SAMB for an original bond cluster can be obtained in a similar way, we need special care for the bond direction.
When we express a symmetric-bond dependence such as a real hopping, e.g., $(c_{i}^{\dagger}c_{j}^{}+c_{j}^{\dagger}c_{i}^{})$, we can omit the directional property of bonds.
In this case, we construct the SAMB for a bond cluster in the same way as a site cluster as
\begin{align}
\mathbb{Q}_{l\xi}^{\rm (b)}=(c_{1}^{(l\xi)},c_{2}^{(l\xi)},\cdots,c_{N_{b}}^{(l\xi)}),
\quad
c_{i}^{(l\xi)}=\sum_{g}^{g(i)}v_{g}^{(l\xi)},
\end{align}
where the mapping $g(i)$ is determined for the original $N_{b}$ bondcluster, and the superscript ``(b)'' indicates the SAMB for the bond cluster.

On the other hand, when we consider an antisymmetric bond dependence such as an imaginary hopping, e.g., $i(c_{i}^{\dagger}c_{j}^{}-c_{j}^{\dagger}c_{i}^{})$, the directional property must be taken into account.
In this case, the SAMB for a bond cluster is given as
\begin{align}
\mathbb{T}_{l\xi}^{\rm (b)}=i(b_{1}^{(l\xi)},b_{2}^{(l\xi)},\cdots,b_{N_{b}}^{(l\xi)}),
\quad
b_{i}^{(l\xi)}=\sum_{g}^{g(i)}{\rm sgn}[g]v_{|g|}^{(l\xi)},
\end{align}
in order to satisfy the antisymmetric property of the bonds.
Here, we have attached the phase factor $i$ for later convenience.
The real $\mathbb{Q}_{l\xi}^{\rm (b)}$ and imaginary $\mathbb{T}_{l\xi}^{\rm (b)}$ SAMBs are always orthogonal to each other.
``T'' denotes the magnetic-toroidal multipole indicating that it has a magnetic polar tensor property.
The SAMB for the site/bond clusters in the case of the example shown in Fig.~\ref{fig:sb_cluster} is summarized in Table~\ref{tbl:sb_samb}.

\begin{table}
\caption{\label{tbl:sb_samb}
Orthonormalized SAMB for the example site/bond clusters shown in Fig.~\ref{fig:sb_cluster}.
The $Q$-type SAMB for A$\to$B bond cluster is the same as that for the virtual cluster in Table~\ref{tbl:vc_c3v}, while $T$-type is given by that multiplied by the phase factor $i$.
The indices $(l, \Gamma, n, \gamma)$ correspond to those in Table~\ref{tbl:harmonics_c3v}.
}
\begin{spacing}{1.5}
\begin{ruledtabular}
\begin{tabular}{cccccccccc}
$l$ & $\Gamma$ & $n$ & $\gamma$ & $\mathbb{Q}_{l\xi}^{\rm (s)}$ for A & $l$ & $\Gamma$ & $n$ & $\gamma$ & $\mathbb{Q}_{l\xi}^{\rm (s)}$ for B \\ \hline
0 & $A_{1}$ & - & - & $\frac{1}{\sqrt{3}}(1,1,1)$  & 0 & $A_{1}$ & - & - & $\frac{1}{\sqrt{3}}(1,1,1)$ \\
2 & $E$ & 2 & $u$ & $\frac{1}{\sqrt{6}}(1,-2,1)$ & 1 & $E$ & - & $u$ & $\frac{1}{\sqrt{6}}(2,-1,-1)$ \\
  &     & & $v$ & $\frac{1}{\sqrt{2}}(1,0,-1)$ &   &     &   & $v$ & $\frac{1}{\sqrt{2}}(0,1,-1)$ \\
\end{tabular}
\vspace{1mm}
\begin{tabular}{cccccccccc}
$l$ & $\Gamma$ & $n$ & $\gamma$ & $\mathbb{Q}_{l\xi}^{\rm (b)}$ for A$\to$A  & $l$ & $\Gamma$ & $n$ & $\gamma$ & $\mathbb{T}_{l\xi}^{\rm (b)}$ for A$\to$A \\ \hline
0 & $A_{1}$ & - & - & $\frac{1}{\sqrt{3}}(1,1,1)$   & 3 & $A_{2}$ & - & - & $\frac{i}{\sqrt{3}}(-1,-1,1)$ \\
1 & $E$ & - & $u$ & $\frac{1}{\sqrt{6}}(1,1,-2)$ & 1 & $E$ & - & $u$ & $\frac{i}{\sqrt{2}}(1,-1,0)$ \\
  &     & & $v$ & $\frac{1}{\sqrt{2}}(-1,1,0)$  &   &     & & $v$ & $\frac{i}{\sqrt{6}}(1,1,2)$ \\
\end{tabular}
\end{ruledtabular}
\end{spacing}
\end{table}

As explained the above, we can construct a set of the SAMBs for the original site- and bond clusters in terms of the polar tensors, $\mathbb{Q}_{l\xi}^{\rm (b)}$ and $\mathbb{T}_{l\xi}^{\rm (b)}$.
When the obtained SAMBs are not orthonormalized, we again use the Gram-Schmidt method.
The symbol $\mathbb{Y}_{l\xi}$ is used to refer to $\mathbb{Q}_{l\xi}^{\rm (s)}$, $\mathbb{Q}_{l\xi}^{\rm (b)}$, or $\mathbb{T}_{l\xi}^{\rm (b)}$ in the site/bond clusters, and we call them ``cluster SAMB''.

\subsubsection{SAMB for atomic degrees of freedom}

Next, we consider the SAMB for atomic degrees of freedom, which we call ``atomic SAMB''.
The complete set of the atomic SAMB has already been discussed in the literature~\cite{kusunose2020complete}, and they can be expressed in terms of electric (time-reversal even polar), magnetic (time-reversal odd axial), electric-toroidal (time-reversal even axial), and magnetic-toroidal (time-reversal odd polar) multipole bases, $\mathbb{Q}_{lm}^{\rm (a)}$, $\mathbb{M}_{lm}^{\rm (a)}$, $\mathbb{G}_{lm}^{\rm (a)}$, and $\mathbb{T}_{lm}^{\rm (a)}$ in the rotation group.
The superscript ``(a)'' indicates the atomic SAMB.
The symbol $\mathbb{X}_{lm}$ is used to refer to all of the four-type 
atomic SAMBs.

The spinful atomic SAMB can be obtained by the direct product of the spinless atomic SAMB $\mathbb{X}^{\rm (orb)}_{lm}$ and identity and Pauli matrices $\sigma_{sn}$ ($\sigma_{00}=\sigma_{0}$, $\sigma_{10}=\sigma_{z}$, $\sigma_{1\pm1}=\mp(\sigma_{x}\pm i\sigma_{y})/\sqrt{2}$) by using the addition rule of the angular momentum as
\begin{align}
&
\mathbb{X}_{lm}(s,k)=i^{s+k}
\sum_{n}
\braket{l+k,m-n;sn|lm}
\mathbb{X}_{l+k,m-n}^{\rm (orb)}\sigma_{sn},
\end{align}
where $\braket{l_{1},m_{1};l_{2},m_{2}|lm}$ is the Clebsch-Gordan (CG) coefficient. 
See Ref.~\cite{kusunose2020complete} for the expression of $\mathbb{X}^{\rm (orb)}_{lm}$.
Note that $\mathbb{X}_{lm}(0,0)=\mathbb{X}_{lm}^{\rm (orb)}\sigma_{0}$.

\begin{table}
\caption{\label{tbl:atom_samb}
Orthonormalized atomic SAMB.
We consider the spinless $s$ orbital in A sites, while $(p_{x}, p_{y}; p_{z})$ orbitals in B sites in the example of Fig.~\ref{fig:sb_cluster}.
Note that $p_{x},p_{y}$ ($p_{z}$) belong to $E$ ($A_{1}$) irrep.
As only $s=k=0$ (charge sector) SAMBs are active in the spinless Hilbert space, we omit $(s,k)$.
In $\braket{s|s}$ Hilbert space, there is only $\mathbb{Q}_{0,A_{1}}=(1)$.
In $\braket{s|p}$ Hilbert space, $\mathbb{T}_{l\xi}^{\rm (a)}$ is given by multiplying $\mathbb{Q}_{l\xi}^{\rm (a)}$ by the phase factor $i$.
The indices $(l, \Gamma, n, \gamma)$ correspond to those in Table~\ref{tbl:harmonics_c3v}.
}
\begin{spacing}{1.5}
\begin{ruledtabular}
\begin{tabular}{cccccccccc}
$l$ & $\Gamma$ & $n$ & $\gamma$ & $\mathbb{Q}_{l\xi}^{\rm (a)}$ for $\braket{p|p}$ & $l$ & $\Gamma$ & $n$ & $\gamma$ & $\mathbb{Q}_{l\xi}^{\rm (a)}$ for $\braket{p|p}$ \\ \hline
0 & $A_{1}$ & - & - & $\frac{1}{\sqrt{3}}\begin{pmatrix} 1 & 0 & 0 \\ 0 & 1 & 0 \\ 0 & 0 & 1 \end{pmatrix}$ &
2 & $A_{1}$ & - & - & $\frac{1}{\sqrt{6}}\begin{pmatrix} -1 & 0 & 0 \\ 0 & -1 & 0 \\ 0 & 0 & 2 \end{pmatrix}$ \\
2 & $E$ & 1 & $u$ & $\frac{1}{\sqrt{2}}\begin{pmatrix} 0 & 0 & 1 \\ 0 & 0 & 0 \\ 1 & 0 & 0 \end{pmatrix}$ &
2 & $E$ & 2 & $u$ & $\frac{1}{\sqrt{2}}\begin{pmatrix} 1 & 0 & 0 \\ 0 & -1 & 0 \\ 0 & 0 & 0 \end{pmatrix}$ \\
  &   &  & $v$ & $\frac{1}{\sqrt{2}}\begin{pmatrix} 0 & 0 & 0 \\ 0 & 0 & 1 \\ 0 & 1 & 0 \end{pmatrix}$ &
& & & $v$ & $-\frac{1}{\sqrt{2}}\begin{pmatrix} 0 & 1 & 0 \\ 1 & 0 & 0 \\ 0 & 0 & 0 \end{pmatrix}$ \\
\end{tabular}
\vspace{1mm}
\begin{tabular}{cccccccccc}
$l$ & $\Gamma$ & $n$ & $\gamma$ & $\mathbb{M}_{l\xi}^{\rm (a)}$ for $\braket{p|p}$ & $l$ & $\Gamma$ & $n$ & $\gamma$ & $\mathbb{Q}_{l\xi}^{\rm (a)}$ for $\braket{s|p}$ \\ \hline
1 & $A_{2}$ & - & - & $\frac{1}{\sqrt{2}}\begin{pmatrix} 0 & -i & 0 \\ i & 0 & 0 \\ 0 & 0 & 0 \end{pmatrix}$ & 1 & $A_{1}$ & - & - & $\begin{pmatrix} 0 & 0 & 1 \end{pmatrix}$ \\
1 & $E$ & - & $u$ & $\frac{1}{\sqrt{2}}\begin{pmatrix} 0 & 0 & -i \\ 0 & 0 & 0 \\ i & 0 & 0 \end{pmatrix}$ & 1 & $E$ & - & $u$ & $\begin{pmatrix} 1 & 0 & 0 \end{pmatrix}$ \\
 & & & $v$ & $\frac{1}{\sqrt{2}}\begin{pmatrix} 0 & 0 & 0 \\ 0 & 0 & -i \\ 0 & i & 0 \end{pmatrix}$ & & & & $v$ & $\begin{pmatrix} 0 & 1 & 0 \end{pmatrix}$ \\
\end{tabular}
\end{ruledtabular}
\end{spacing}
\end{table}

In evaluating the matrix elements of $\mathbb{X}_{lm}$, the orbital angular momentum of the bra and ket states can be different.
For instance, when we consider an electron hopping from $s$ orbital in A site to $p_{x}$ orbital in B site and vice versa, $s$-$p_{x}$ off-diagonal Hilbert space must be taken into account.
It should be emphasized that $\mathbb{X}_{lm}(s,k)$ must be treated independently between different (e.g., $\braket{s|s}$, $\braket{p|p}$, and $\braket{s|p}$) Hilbert spaces, even if all the indices $(X,l,m,s,k)$ are the same.

The atomic SAMB for point group can be obtained by means of the unitary matrix in Eq.~(\ref{eq:rtop}) as
\begin{align}
\mathbb{X}_{l\xi,sk}=\sum_{m}U_{m,\xi}^{(l)}\mathbb{X}_{lm}(s,k).
\end{align}
The formula to compute the matrix elements of $\mathbb{X}_{lm}(s,k)$ is summarized in Ref.~\cite{kusunose2020complete}, and those of $\mathbb{X}_{l\xi,sk}$ for point group can be obtained by the appropriate unitary transformation, $U_{m,\xi}^{(l)}$ as well. 
Then, $\mathbb{X}_{l\xi,sk}$ can be normalized straightforwardly.
The example of expressions of $\mathbb{X}_{l\xi} = \mathbb{X}_{l\xi,00}$ is given in Table~\ref{tbl:atom_samb}, where the spinless $s$ orbital ($p$-orbitals) are assumed at A (B) sites.

\subsubsection{Combined SAMB for atomic and site/bond cluster}

In the previous subsections, we have constructed the complete orthonormal SAMBs for the atomic degrees of freedom $\mathbb{X}_{l\xi,sk}$ and for the site/bond clusters, $\mathbb{Y}_{l\xi}$.
Here, we construct the SAMB by performing the irreducible decomposition of the direct product of these two SAMBs.

To this end, we begin with the addition rule of the spherical harmonic-like functions,
\begin{align}
Z_{lm}=(-i)^{l_{1}+l_{2}-l}\sum_{m_{1}m_{2}}\braket{l_{1}m_{1},l_{2}m_{2}|lm}X_{l_{1}m_{1}}Y_{l_{2}m_{2}},
\label{eq:addrule}
\end{align}
where $X_{lm}$ transforms like the spherical harmonics $Y_{lm}(\hat{\bm{r}})$ against a spatial rotation, and satisfies $X_{lm}^{\dagger}=(-1)^{m}X_{l-m}$.
$Y_{lm}$ and $Z_{lm}$ have the same properties as $X_{lm}$.
The phase factor has been introduced to satisfy $Z_{lm}^{\dagger}=Z_{l-m}$.

By considering consistency for the spatial and time-reversal parities, they must coincide with each other in both sides of Eq.~(\ref{eq:addrule}).
For the time-reversal parity, we introduce the time-reversal parity as a function of multipole type as
\begin{align}
t(Q)=t(G)=+1,
\quad
t(T)=t(M)=-1,
\end{align}
where $+1$ and $-1$ denote the time-reversal even and odd, respectively.
Then, the time-reversal selection rule is given by $\delta[t(X)t(Y),t(Z)]$, where $\delta[a,b]$ is the Kronecker's delta.

Similarly, we introduce the spatial parity function as
\begin{align}
p(Q)=p(T)=0,
\quad
p(G)=p(M)=1,
\end{align}
where $0$ and $1$ denote polar and axial, respectively.
Equation~(\ref{eq:addrule}) has finite value only for $|l_{1}-l_{2}|\leq l \leq l_{1}+l_{2}$, and when the difference between $l_{1}+l_{2}$ and $l$ is odd, the spatial parity of $Z$ becomes opposite to that of the product of $X$ and $Y$.
Hence, the spatial parity selection rule reads
\begin{align}
\delta[(l_{1}+l_{2}-l+p(X)+p(Y)-p(Z))\text{ mod 2},0].
\end{align}

By these considerations, we obtain the extended addition rule as
\begin{align}
\hat{\mathbb{Z}}_{lm}(s,k)=\sum_{m_{1}m_{2}}C_{lm}^{l_{1}m_{1},l_{2}m_{2}}(X,Y|Z)\,\mathbb{X}_{l_{1}m_{1}}(s,k)\otimes\mathbb{Y}_{l_{2}m_{2}},
\end{align}
with the ``CG'' coefficient,
\begin{multline}
C_{lm}^{l_{1}m_{1},l_{2}m_{2}}(X,Y|Z)=
(-i)^{l_{1}+l_{2}-l}\braket{l_{1}m_{1},l_{2}m_{2}|lm}
\\ \times
\Delta_{l}^{l_{1},l_{2}}(X,Y|Z),
\end{multline}
and
\begin{multline}
\Delta_{l}^{l_{1},l_{2}}(X,Y|Z)
=\delta[t(X)t(Y),t(Z)]
\\ \times
\delta[(l_{1}+l_{2}-l+p(X)+p(Y)-p(Z))\text{ mod 2},0].
\end{multline}
Then, by the unitary transformation from the rotation group to point group, we finally obtain the combined SAMB as
\begin{align}
\hat{\mathbb{Z}}_{l\xi,sk}=
\sum_{\xi_{1}\xi_{2}}C_{l\xi}^{l_{1}\xi_{1},l_{2}\xi_{2}}(X,Y|Z)\,\mathbb{X}_{l_{1}\xi_{1},sk}\!\otimes\mathbb{Y}_{l_{2}\xi_{2}},
\end{align}
where
\begin{multline}
C_{l\xi}^{l_{1}\xi_{1},l_{2}\xi_{2}}(X,Y|Z)=
\sum_{mm_{1}m_{2}}U_{m_{1},\xi_{1}}^{(l_{1})*}U_{m_{2},\xi_{2}}^{(l_{2})*}
\\ \times
C_{lm}^{l_{1}m_{1},l_{2}m_{2}}(X,Y|Z)U_{m,\xi}^{(l)}.
\end{multline}

If the obtained SAMBs are not orthonormalized, then the Gram-Schmidt method is applied.
Since $\mathbb{X}_{l_{1}\xi_{1},sk}$ and $\mathbb{Y}_{l_{2}\xi_{2}}$ are already orthonormalized, this is done by performing the Gram-Schmidt method only to the CG coefficients.
The example of the combined SAMB for the case of Fig.~\ref{fig:sb_cluster} with the spinless $s$ orbital ($p$ orbitals) at A (B) sites is given in Table~\ref{tbl:comb_samb}.

\begin{table}
\caption{\label{tbl:comb_samb}
Combined orthonormalized SAMBs belonging to the identity $A_{1}$ irrep.
The abbreviation, $[\mathbb{X}\otimes \mathbb{Y}]=(\mathbb{X}_{u}\otimes \mathbb{Y}_{u}+\mathbb{X}_{v}\otimes \mathbb{Y}_{v})/\sqrt{2}$ is used for E irrep.
As $\mathbb{Q}_{l\xi,sk}^{\rm (a)}$ is spinless, $s=0$ and $k=0$ are omitted.
The indices $(l, \Gamma, n, \gamma)$ correspond to those in Table~\ref{tbl:harmonics_c3v}.
}
\begin{ruledtabular}
\begin{tabular}{cccccccccc}
$l$ & $\Gamma$ & $n$ & $\gamma$ & $\hat{\mathbb{Q}}_{l\xi}$ for A($s$) & $l$ & $\Gamma$ & $n$ & $\gamma$ & $\hat{\mathbb{Q}}_{l\xi}$ for A($s$)$\to$A($s$) \\ \hline
0 & $A_{1}$ & - & - & $\mathbb{Q}_{0,A_{1}}^{\rm (a)}\otimes \mathbb{Q}_{0,A_{1}}^{\rm (s)}$ & 0 & $A_{1}$ & - & - & $\mathbb{Q}_{0,A_{1}}^{\rm (a)}\otimes\mathbb{Q}_{0,A_{1}}^{\rm (b)}$ \\
\end{tabular}
\vspace{1mm}
\begin{tabular}{cccccccccc}
$l$ & $\Gamma$ & $n$ & $\gamma$ & $\hat{\mathbb{Q}}_{l\xi}$ for B($p$) & $l$ & $\Gamma$ & $n$ & $\gamma$ & $\hat{\mathbb{Q}}_{l\xi}$ for A($s$)$\to$B($p$) \\ \hline
0 & $A_{1}$ & - & - & $\mathbb{Q}_{0,A_{1}}^{\rm (a)}\otimes \mathbb{Q}_{0,A_{1}}^{\rm (s)}$ & 0 & $A_{1}$ & - & - & $[\mathbb{Q}_{1,E}^{\rm (a)}\otimes \mathbb{Q}_{1,E}^{\rm (b)}]$ \\
1 & $A_{1}$ & - & - & $[\mathbb{Q}_{2,E,1}^{\rm (a)}\otimes \mathbb{Q}_{1,E}^{\rm (s)}]$ & 1 & $A_{1}$ & - & - & $\mathbb{Q}_{1,A_{1}}^{\rm (a)}\otimes \mathbb{Q}_{0,A_{1}}^{\rm (b)}$ \\
2 & $A_{1}$ & - & - & $\mathbb{Q}_{2,A_{1}}^{\rm (a)}\otimes \mathbb{Q}_{0,A_{1}}^{\rm (s)}$ & 3 & $A_{1}$ & 2 & - & $[\mathbb{Q}_{1,E}^{\rm (a)}\otimes \mathbb{Q}_{2,E,2}^{\rm (b)}]$ \\
3 & $A_{1}$ & 2 & - & $[\mathbb{Q}_{2,E,2}^{\rm (a)}\otimes \mathbb{Q}_{1,E}^{\rm (s)}]$ & \\
\end{tabular}
\end{ruledtabular}
\end{table}

\subsection{Periodic crystals}\label{sec:crystal}

In this subsection, we discuss the SAMB, $\hat{\mathbb{Z}}_{l\xi,sk}$, for periodic crystals.
The procedure is almost the same as that for isolated cluster systems described in the previous subsections.
Since the atomic SAMB $\mathbb{X}_{l\xi,sk}$ is irrelevant either to isolated cluster systems or periodic crystals, we only consider the SAMB of the site/bond clusters, $\mathbb{Y}_{l\xi}$.

In periodic crystals, the site/bond cluster is defined in a similar manner as described in Sec.~\ref{sec:sbcluster}.
Since there are translation operations, sites $\bm{R}_{g}'$ and bond centers $\bm{c}_{g}'$ must be shifted to the home unit cell in defining the site/bond cluster.

Then, let us introduce the associated point group for space group in question.
The associated point group is given by omitting the superscript of the space group in Sch\"onflies notation, e.g., C$_{\rm 3v}$ for C$_{\rm 3v}^{4}$ (\#159, P31c), and its symmetry operations are given by those of the relevant space group without the (partial) translations.
It should be noted that there is a one-to-one correspondence between the symmetry operations of the space group and those of its associated point group.
Through this associated point group, we can determine the one-to-many correspondence $g(i)$ between the site/bond cluster in the periodic crystals and virtual cluster by means of the symmetry operations, $g$.
Once we establish the one-to-many correspondence, we can construct the SAMB of the site/bond clusters $\mathbb{Y}_{l\xi}$, and hence the combined SAMB $\hat{\mathbb{Z}}_{l\xi,sk}$, in the same manner for isolated cluster systems.
It should be emphasized that there is essentially no ambiguity about how to choose site/bond clusters and their origin by this prescription.

\begin{table*}
\caption{\label{tbl:notation}
Notations used in this paper.
SAMB and CG are the abbreviations of Symmetry-Adapted Multipole Basis and Clebsch-Gordan, respectively.
}
\begin{spacing}{1.4}
\begin{ruledtabular}
\begin{tabular}{ll}
Symbol & Meaning \\ \hline
$Q,G,M,T$ & electric (E), electric-toroidal (ET), magnetic (M), and magnetic-toroidal (MT) multipoles \\
$l$, $m$ & rank ($0,1,2,\cdots$) and component ($-l,-l+1,\cdots,l$) \\
$\xi=(\Gamma,n,\gamma)$ & irrep., multiplicity, and component in point group \\
$s$, $k$ & charge ($s=0$, $k=0$) sector or spin ($s=1$, $k=-1,0,1$) sector for atomic SAMB \\
$O_{lm}(\bm{r})$ & normalized spherical harmonics, $\bm{r}=(x,y,z)$ \\
$O_{l\xi}(\bm{r})$ & symmetry-adapted harmonics in point group \\
$U_{m,\xi}^{(l)}$ & unitary matrix from $(l,m)$ to $(l,\xi)$ basis \\
\hline
$\overline{\mathbb{Q}}_{l\xi}$ & SAMB for virtual cluster \\
$\mathbb{Q}_{l\xi,sk}^{\rm (a)},\mathbb{G}_{l\xi,sk}^{\rm (a)},\mathbb{M}_{l\xi,sk}^{\rm (a)},\mathbb{T}_{l\xi,sk}^{\rm (a)}$ & SAMB for atomic degrees of freedom \\
$\mathbb{Q}_{l\xi}^{\rm (s)}$ & SAMB for site cluster \\
$\mathbb{Q}_{l\xi}^{\rm (b)},\mathbb{T}_{l\xi}^{\rm (b)}$ & SAMB for bond cluster (symmetric part, antisymmetric part) \\
$\mathbb{Q}_{l\xi}^{\rm (u)},\mathbb{T}_{l\xi}^{\rm (u)}$ & uniform component of $\mathbb{Q}_{l\xi}^{\rm (b)}(\bm{k}),\mathbb{T}_{l\xi}^{\rm (b)}(\bm{k})$ having off-diagonal matrix elements only \\
$\mathbb{Q}_{l\xi}^{\rm (k)}(\bm{k}),\mathbb{T}_{l\xi}^{\rm (k)}(\bm{k})$ & structure factor obtained from $\mathbb{Q}_{l\xi}^{\rm (b)},\mathbb{T}_{l\xi}^{\rm (b)}$ \\
$\hat{\mathbb{Q}}_{l\xi,sk},\hat{\mathbb{G}}_{l\xi,sk},\hat{\mathbb{M}}_{l\xi,sk},\hat{\mathbb{T}}_{l\xi,sk}$ & combined SAMB (isolated cluster systems) \\
$\hat{\mathbb{Q}}_{l\xi,sk}(\bm{k}),\hat{\mathbb{G}}_{l\xi,sk}(\bm{k}),$ & combined SAMB (periodic crystals) \\
$\hat{\mathbb{M}}_{l\xi,sk}(\bm{k}),\hat{\mathbb{T}}_{l\xi,sk}(\bm{k})$ & \\
\hline
$\mathbb{X}_{l\xi,sk}$ & SAMB for $\set{\mathbb{Q}^{\rm (a)}_{l\xi,sk},\mathbb{G}^{\rm (a)}_{l\xi,sk},\mathbb{M}^{\rm (a)}_{l\xi,sk},\mathbb{T}^{\rm (a)}_{l\xi,sk}}$ \\
$\mathbb{Y}_{l\xi}$ & SAMB for $\set{\mathbb{Q}_{l\xi}^{\rm (s)},\mathbb{Q}_{l\xi}^{\rm (b)},\mathbb{T}_{l\xi}^{\rm (b)}}$ \\
$\hat{\mathbb{Z}}_{l\xi,sk}$ & combined SAMB for $\mathbb{X}_{l_{1}\xi_{1},sk}$ and $\mathbb{Y}_{l_{2}\xi_{2}}$ \\
$\mathbb{U}_{l\xi}$ & SAMB for $\set{\mathbb{Q}_{l\xi}^{\rm (s)},\mathbb{Q}_{l\xi}^{\rm (u)},\mathbb{T}_{l\xi}^{\rm (u)}}$ (isolated cluster systems/periodic crystals) \\
$\mathbb{F}_{l\xi}(\bm{k})$ & SAMB for $\set{\mathbb{Q}_{l\xi}^{\rm (k)}(\bm{k}),\mathbb{T}_{l\xi}^{\rm (k)}(\bm{k}),1}$ (periodic crystals) \\
\hline
$C_{l\xi}^{l_{1}\xi_{1},l_{2}\xi_{2}}(X,Y|Z)$ & CG coefficient from $\mathbb{X}_{l_{1}\xi_{1},sk}$ and $\mathbb{Y}_{l_{2}\xi_{2}}$ to $\mathbb{Z}_{l\xi,sk}$ (isolated cluster systems/periodic crystals) \\
$C_{l\xi}^{l_{1}\xi_{1},l_{2}\xi_{2},l_{3}\xi_{3}}(X,U,F|Z)$ & CG coefficient from $\mathbb{X}_{l_{1}\xi_{1},sk}$, $\mathbb{U}_{l_{2}\xi_{2}}$, and $\mathbb{F}_{l_{3}\xi_{3}}(\bm{k})$ to $\mathbb{Z}_{l\xi,sk}(\bm{k})$ (periodic crystals) \\
\end{tabular}
\end{ruledtabular}
\end{spacing}
\end{table*}

\subsection{Full matrix form}\label{sec:fullmom}

\begin{figure}
\centering
\includegraphics[width=8.5cm]{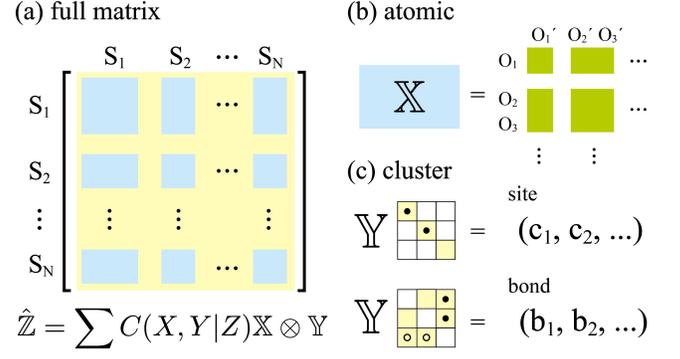}%
\caption{\label{fig:full_matrix}
(a) Full matrix form of SAMB $\hat{\mathbb{Z}}_{l\xi,sk}$, (b) atomic SAMB $\mathbb{X}_{l_{1}\xi_{1},sk}$ with respect to atomic orbitals in each site or bond, and (c) cluster SAMB $\mathbb{Y}_{l_{2}\xi_{2}}$ for each site/bond clusters.
}
\end{figure}

In the previous subsections, we have explained how to construct the SAMB both for isolated cluster systems and periodic crystals.
In general, the shapes of the obtained atomic SAMBs $\mathbb{X}_{l\xi,sk}$ are different from each other depending on the combination of the bra and ket states.
Moreover, the dimensions of the SAMBs $\mathbb{Y}_{l\xi}$ are different among site/bond clusters.
To obtain the full matrix form with respect to the total Hilbert space of the targeting system (see, Fig.~\ref{fig:full_matrix}), we carry out the rearrangement of the basis elements, and Fourier transformation for periodic crystals, in the following procedure.

\subsubsection{Conversion to full matrix form}

For the atomic orbitals, we consider the direct sum of the Hilbert spaces of all the relevant atomic orbitals, and assign each $\mathbb{X}_{l\xi,sk}$ for a given bra-ket states to an appropriate block of the matrix.
Similarly, we consider the Hilbert space of all of the relevant sites in the targeting system, and sum up each element $[\mathbb{Y}_{l\xi}](i)$ to the appropriate matrix element in the case of isolated cluster systems.

On the other hand, for periodic crystals, we must use momentum-space representation.
In order to transform the SAMB to the momentum space, we perform Fourier transformation for each bond in bond clusters.
Namely, the $i$th component of the SAMB is transformed as
\begin{align}
&
[\mathbb{Y}_{l\xi}](i)\to e^{-i\bm{k}\cdot\bm{b}_{i}}[\mathbb{Y}_{l\xi}](i).
\end{align}
Note that the complex conjugate of $e^{-i\bm{k}\cdot\bm{b}_{i}}[\mathbb{Y}_{l\xi}](i)$ is obtained by reverting the bond direction $\bm{b}_{i}\to -\bm{b}_{i}$, especially due to the phase factor $i$ in $\mathbb{T}_{l\xi}^{\rm (b)}$.
The expression of the SAMB in site clusters does not change in the momentum space.

By the above procedure, we obtain the full matrix form of $\mathbb{Y}_{l\xi}$ or its momentum-space representation, which we denote as $\mathbb{Y}_{l\xi}(\bm{k})$.
Then, the SAMB $\mathbb{Z}_{l\xi,sk}$ in the full matrix form is obtained by
\begin{align}
\hat{\mathbb{Z}}_{l\xi,sk}=\sum_{\xi_{1}\xi_{2}}C_{l\xi}^{l_{1}\xi_{1},l_{2}\xi_{2}}(X,Y|Z)\,\mathbb{X}_{l_{1}\xi_{1},sk}\!\otimes\mathbb{Y}_{l_{2}\xi_{2}},
\end{align}
or
\begin{align}
\hat{\mathbb{Z}}_{l\xi,sk}(\bm{k})=\sum_{\xi_{1}\xi_{2}}C_{l\xi}^{l_{1}\xi_{1},l_{2}\xi_{2}}(X,Y|Z)\,\mathbb{X}_{l_{1}\xi_{1},sk}\!\otimes\mathbb{Y}_{l_{2}\xi_{2}}(\bm{k}).
\end{align}
Here, the binary operator, $\otimes$, simply means the direct product of two Hermitian matrices, $\mathbb{X}_{l_{1}\xi_{1},sk}$ and $\mathbb{Y}_{l_{2}\xi_{2}}$.

\subsubsection{Conversion to structure-factor form}

The Hermitian full matrix form of $\mathbb{Y}_{l_{2}\xi_{2}}(\bm{k})$ can be further decomposed into $\bm{k}$-independent ``uniform matrix'' basis $\mathbb{U}_{l\xi}$ and ``structure factor'' $\mathbb{F}_{l\xi}(\bm{k})$ as follows.
Let us first introduce the momentum representation of $\mathbb{Q}_{l\xi}^{\rm (b)}$ and $\mathbb{T}_{l\xi}^{\rm (b)}$ as
\begin{align}
&
\mathbb{Q}_{l\xi}^{\rm (k)}(\bm{k})\equiv
\sqrt{2}\sum_{i}\cos(\bm{k}\cdot\bm{b}_{i})[\mathbb{Q}_{l\xi}^{\rm (b)}](i),
\cr&
\mathbb{T}_{l\xi}^{\rm (k)}(\bm{k})\equiv
-\sqrt{2}i\sum_{i}\sin(\bm{k}\cdot\bm{b}_{i})[\mathbb{T}_{l\xi}^{\rm (b)}](i).
\label{eq:stfactor}
\end{align}
Here, $\mathbb{Q}_{l\xi}^{\rm (k)}(\bm{k})$ and $\mathbb{T}_{l\xi}^{\rm (k)}(\bm{k})$ are real functions and transformed as $O_{l\xi}(\bm{k})$ for the symmetry operations, and the time-reversal operation is $\mathcal{T}\bigl[\mathbb{Q}_{l\xi}^{\rm (k)}(\bm{k})\bigr]=\mathbb{Q}_{l\xi}^{\rm (k)}(-\bm{k})$ and $\mathcal{T}\bigl[\mathbb{T}_{l\xi}^{\rm (k)}(\bm{k})\bigr]=-\mathbb{T}_{l\xi}^{\rm (k)}(-\bm{k})$.
We denote all of the symmetry-adapted structure factor (``structure SAMB'') as $\mathbb{F}_{l\xi}(\bm{k})$ either of $\mathbb{Q}_{l\xi}^{\rm (k)}(\bm{k})$, $\mathbb{T}_{l\xi}^{\rm (k)}(\bm{k})$ or unity.
Then, $\set{\mathbb{F}_{l\xi}(\bm{k})}$ constitutes the orthonormalized complete set, i.e.,
\begin{align}
&
\frac{1}{N_{0}}\sum_{\bm{k}}\mathbb{F}_{l_{1}\xi_{1}}^{\,*}(\bm{k})\mathbb{F}_{l_{2}\xi_{2}}^{'}(\bm{k})=\delta_{l_{1},l_{2}}\delta_{\xi_{1},\xi_{2}}\delta_{F,F'},
\\&
\sum_{Fl\xi}\mathbb{F}_{l\xi}^{}(\bm{k})\mathbb{F}_{l\xi}^{\,*}(\bm{k}')=\delta_{\bm{k},\bm{k}'},
\end{align}
where $N_{0}$ is the number of $\bm{k}$ points, and the summation is taken over the minimal periodic unit in which
\begin{align}
&
\frac{1}{N_{0}}\sum_{\bm{k}}\cos(\bm{k}\cdot\bm{b}_{i})\cos(\bm{k}\cdot\bm{b}_{j})
=\frac{1}{2}\delta_{ij},
\cr&
\frac{1}{N_{0}}\sum_{\bm{k}}\sin(\bm{k}\cdot\bm{b}_{i})\sin(\bm{k}\cdot\bm{b}_{j})
=\frac{1}{2}\delta_{ij},
\cr&
\frac{1}{N_{0}}\sum_{\bm{k}}\cos(\bm{k}\cdot\bm{b}_{i})\sin(\bm{k}\cdot\bm{b}_{j})=0.
\end{align}
It should be noted that the periodicity of $\mathbb{F}_{l\xi}(\bm{k})$ differs from the Brillouin zone of the system, unless the bond vectors are identical to the lattice vector in the primitive unit cell.

By using $\mathbb{F}_{l\xi}(\bm{k})$, we can reexpress $\mathbb{Y}_{l\xi}(\bm{k})$ as
\begin{align}
\mathbb{Y}_{l\xi}(\bm{k})=\sum_{l_{1}\xi_{1}l_{2}\xi_{2}}f^{l_{1}\xi_{1},l_{2}\xi_{2}}_{l\xi}(U,F)\,\mathbb{U}_{l_{1}\xi_{1}}\otimes\mathbb{F}_{l_{2}\xi_{2}}(\bm{k}),
\end{align}
where $\mathbb{U}_{l\xi}=\mathbb{Q}_{l\xi}^{\rm (s)}$ for site clusters or the uniform component $\mathbb{Y}_{l\xi}(\bm{k}=0)$ having only off-diagonal matrix elements for bond clusters, with appropriate normalization.
Here, $f^{l_{1}\xi_{1},l_{2}\xi_{2}}_{l\xi}(U,F)$ is the linear coefficient.
Note that $\mathbb{U}_{l\xi}=\mathbb{Y}_{l\xi}$ for isolated cluster systems.

By these prescriptions, we finally obtain the combined SAMB in terms of the uniform and structure SAMBs as
\begin{multline}
\hat{\mathbb{Z}}_{l\xi,sk}(\bm{k})=
\\
\sum_{\xi_{1},l_{3}\xi_{3}l_{4}\xi_{4}}C_{l\xi}^{l_{1}\xi_{1},l_{3}\xi_{3},l_{4}\xi_{4}}(X,U,F|Z)
\mathbb{X}_{l_{1}\xi_{1},sk}\!\otimes\mathbb{U}_{l_{3}\xi_{3}}\otimes\mathbb{F}_{l_{4}\xi_{4}}(\bm{k}),
\end{multline}
where
\begin{multline}
C_{l\xi}^{l_{1}\xi_{1},l_{3}\xi_{3},l_{4}\xi_{4}}(X,U,F|Z)\equiv
\\
\sum_{\xi_{2}} C_{l\xi}^{l_{1}\xi_{1},l_{2}\xi_{2}}(X,Y|Z)f_{l_{2}\xi_{2}}^{l_{3}\xi_{3},l_{4}\xi_{4}}(U,F).
\end{multline}
The notations and their meanings used in this paper are summarized in Table~\ref{tbl:notation}.

\section{Application to Graphene}\label{sec:graphene}

In this section, we demonstrate the symmetry-adapted TB modeling for graphene based on our method.
Then, using the obtained TB model, we determine the model parameters by optimizing them to reproduce the energy dispersion obtained by the DF computation.

\subsection{TB model based on SAMB}

\begin{table}
\caption{\label{tbl:graphene_so}
Symmetry operations (SOs) in D$_{\rm 6h}$.
}
\begin{ruledtabular}
\begin{tabular}{cccccccc}
No. & SO & No. & SO & No. & SO & No. & SO \\ \hline
1 & $1$ & 2 & $2_{001}$ & 3 & $2_{100}$ & 4 & $2_{010}$ \\
5 & $2_{110}$ & 6 & $2_{120}$ & 7 & $2_{210}$ & 8 & $2_{1-10}$ \\
9 & $3_{001}^{+}$ & 10 & $3_{001}^{-}$ & 11 & $6_{001}^{+}$ & 12 & $6_{001}^{-}$ \\
13 & $-1$ & 14 & m$_{100}$ & 15 & m$_{010}$ & 16 & m$_{110}$ \\
17 & m$_{001}$ & 18 & m$_{120}$ & 19 & m$_{210}$ & 20 & m$_{1-10}$ \\
21 & $-3_{001}^{+}$ & 22 & $-3_{001}^{-}$ & 23 & $-6_{001}^{+}$ & 24 & $-6_{001}^{-}$ \\
\end{tabular}
\end{ruledtabular}
\end{table}

Graphene has a honeycomb structure in space group $P6/mmm$ (\#191, D$_{\rm 6h}^{1}$).
The lattice constant is $a = 2.435$~{\AA} and the length of the vacuum layer along the $c$ axis is set as $c=4a$, and the unit vectors are given by $\bm{a}_{1}=(1,0,0)a$, $\bm{a}_{2}=(-1/2,\sqrt{3}/2,0)a$, and $\bm{a}_{3}=(0,0,1)c$.
The symmetry operations of D$_{\rm 6h}$ point group are given in Table~\ref{tbl:graphene_so}.

\begin{figure}
\centering
\includegraphics[width=8cm]{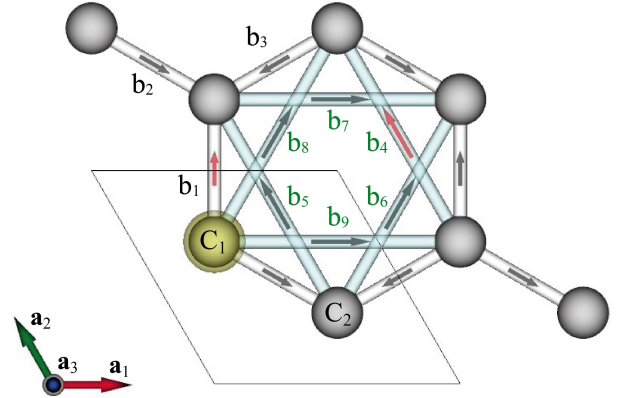}%
\caption{\label{fig:graphene}
Site cluster and nearest-neighbor and second-neighbor bond clusters in graphene.
The representative site and bonds are indicated by the yellow circle and red arrows, respectively.
}
\end{figure}

\begin{figure}[t!]
\centering
\includegraphics[width=8.5cm]{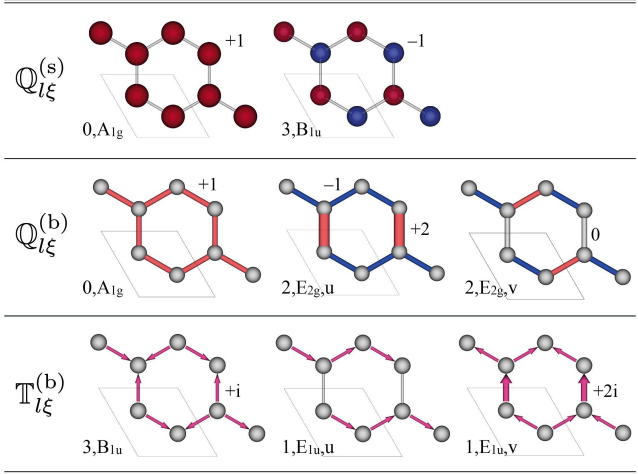}
\caption{\label{fig:graphene_mp}
Site cluster and the nearest-neighbor bond cluster SAMBs for graphene.
The color and size (width) represent the weight of the components in each SAMB.
}
\end{figure}

There are two C atoms in the unit cell, and we consider the spinless $p_{z}$ orbital at each C atom, and up to sixth-neighbor bonds.
The site cluster C, the bond clusters B$_{1}$ and B$_{2}$ for three nearest-neighbor and six second-neighbor bonds are summarized in Table~\ref{tbl:graphene_sb} and Fig.~\ref{fig:graphene}.
In constructing the virtual-cluster sites in D$_{\rm 6h}$, we have used the general point, $\bm{r}_{1}=(\sqrt{3}+1,\sqrt{3}-1,1)$ and its symmetry-operated points.
By the mapping procedure from the virtual cluster as explained in Sec.~\ref{sec:vcmapping}, we obtain the SAMB for the site/bond clusters in Table~\ref{tbl:graphene_sb_samb}.
The SAMBs for the clusters C and B$_{1}$ are schematically shown in Fig.~\ref{fig:graphene_mp}.

\begin{table}
\caption{\label{tbl:graphene_sb}
C-site cluster (C) and the nearest-neighbor (B$_{1}$) and second-neighbor (B$_{2}$) bond clusters in graphene and the one-to-many correspondences $g(i)$.
}
\begin{ruledtabular}
\begin{tabular}{ccc}
C & $\bm{R}_{i}$ & $g(i)$ \\ \hline
C$_{1}$ & $[\frac{1}{3},\frac{2}{3},0]$ & [1,6,7,8,9,10,14,15,16,17,23,24] \\
C$_{2}$ & $[\frac{2}{3},\frac{1}{3},0]$ & [2,3,4,5,11,12,13,18,19,20,21,22] \\
\end{tabular}
\vspace{1mm}
\begin{spacing}{1.3}
\begin{tabular}{ccc}
B$_{1}$ & $\bm{b}_{i}@\bm{c}_{i}$ & $g(i)$ \\ \hline
b$_{1}$ & $[\frac{1}{3},\frac{2}{3},0]@[\frac{1}{2},0,0]$ & [1,-2,-3,6,-13,14,17,-18] \\
b$_{2}$ & $[\frac{1}{3}, -\frac{1}{3}, 0]@[\frac{1}{2}, \frac{1}{2}, 0]$ & [-4,7,10,-11,15,-19,-22,23] \\
b$_{3}$ & $[-\frac{2}{3}, -\frac{1}{3}, 0]@[0, \frac{1}{2}, 0]$ & [-5,8,9,-12,16,-20,-21,24] \\
\end{tabular}
\vspace{1mm}
\begin{tabular}{ccc}
B$_{2}$ & $\bm{b}_{i}@\bm{c}_{i}$ & $g(i)$ \\ \hline
b$_{4}$ & $[0, 1, 0]@[\frac{1}{3}, \frac{1}{6}, 0]$ & [1,-7,-15,17] \\
b$_{5}$ & $[0, 1, 0]@[\frac{2}{3}, \frac{5}{6}, 0]$ & [-2,4,-13,19] \\
b$_{6}$ & $[1, 1, 0]@[\frac{1}{6}, \frac{5}{6}, 0]$ & [-3,12,-18,21] \\
b$_{7}$ & $[1, 0, 0]@[\frac{1}{6}, \frac{1}{3}, 0]$ & [5,-11,20,-22] \\
b$_{8}$ & $[1, 1, 0]@[\frac{5}{6}, \frac{1}{6}, 0]$ & [6,-9,14,-24] \\
b$_{9}$ & $[1, 0, 0]@[\frac{5}{6}, \frac{2}{3}, 0]$ & [-8,10,-16,23] \\
\end{tabular}
\end{spacing}
\end{ruledtabular}
\end{table}

\begin{table}
\caption{\label{tbl:graphene_sb_samb}
SAMB in C site cluster and B$_{1}$ and B$_{2}$ bond clusters.
}
\begin{ruledtabular}
\begin{tabular}{cccccccccc}
$l$ & $\Gamma$ & $n$ & $\gamma$ & $\mathbb{Q}_{l\xi}^{\rm (s)}$ for C & $l$ & $\Gamma$ & $n$ & $\gamma$ & $\mathbb{Q}_{l\xi}^{\rm (s)}$ for C \\ \hline
0 & $A_{1g}$ & - & - & $\frac{1}{\sqrt{2}}(1,1)$ & 3 & $B_{1u}$ & - & - & $\frac{1}{\sqrt{2}}(1,-1)$ \\
\end{tabular}
\vspace{1mm}
\begin{tabular}{cccccccccc}
$l$ & $\Gamma$ & $n$ & $\gamma$ & $\mathbb{Q}_{l\xi}^{\rm (b)}$ for B$_{1}$ & $l$ & $\Gamma$ & $n$ & $\gamma$ & $\mathbb{T}_{l\xi}^{\rm (b)}$ for B$_{1}$ \\ \hline
0 & $A_{1g}$ & - & - & $\frac{1}{\sqrt{3}}(1,1,1)$ & 1 & $E_{1u}$ & - & $u$ & $\frac{i}{\sqrt{2}}(0,1,-1)$ \\
2 & $E_{2g}$ & - & $u$ & $\frac{1}{\sqrt{6}}(2,-1,-1)$ & & & - & $v$ & $\frac{i}{\sqrt{6}}(2,-1,-1)$ \\
  & &  & $v$ & $\frac{1}{\sqrt{2}}(0,-1,1)$ & 3 & $B_{1u}$ & - & - & $\frac{i}{\sqrt{3}}(1,1,1)$ \\
\end{tabular}
\vspace{1mm}
\begin{tabular}{ccccc}
$l$ & $\Gamma$ & $n$ & $\gamma$ & $\mathbb{Q}_{l\xi}^{\rm (b)}$ for B$_{2}$ \\ \hline
0 & $A_{1g}$ & - & - & $\frac{1}{\sqrt{6}}(1,1,1,1,1,1)$ \\
1 & $E_{1u}$ & - & $u$ & $\frac{1}{2}(1,-1,1,0,-1,0)$ \\
& & & $v$ & $\frac{1}{2\sqrt{3}}(1,-1,-1,2,1,-2)$ \\
2 & $E_{2g}$ & - & $u$ & $\frac{1}{2\sqrt{3}}(1,1,1,-2,1,-2)$ \\
& & & $v$ & $\frac{1}{2}(-1,-1,1,0,1,0)$ \\
3 & $B_{1u}$ & - & - & $\frac{1}{\sqrt{6}}(1,-1,-1,-1,1,1)$ \\ \hline
$l$ & $\Gamma$ & $n$ & $\gamma$ & $\mathbb{T}_{l\xi}^{\rm (b)}$ for B$_{2}$ \\ \hline
1 & $E_{1u}$ & - & $u$ & $\frac{i}{2\sqrt{3}}(1,1,-1,-2,-1,-2)$ \\
 & & & $v$ & $-\frac{i}{2}(1,1,1,0,1,0)$ \\
2 & $E_{2g}$ & - & $u$ & $\frac{i}{2}(1,-1,-1,0,1,0)$ \\
 & & & $v$ & $\frac{i}{2\sqrt{6}}(1,-1,1,2,-1,-2)$ \\
3 & $B_{2u}$ & - & - & $\frac{i}{\sqrt{6}}(1,1,-1,1,-1,1)$ \\
6 & $A_{2g}$ & - & - & $\frac{i}{\sqrt{6}}(1,-1,1,-1,-1,1)$ \\
\end{tabular}
\end{ruledtabular}
\end{table}

In this example, since there is only one atomic SAMB for spinless $p_{z}$ bra-ket space,
\begin{align}
\mathbb{Q}_{0,A_{1g},(0,0)}^{\rm (a)}=(1),
\end{align}
the combined SAMB is always equivalent to $\mathbb{Y}_{l\xi}(\bm{k})$, i.e., $\hat{\mathbb{Z}}_{l\xi}(\bm{k})=\mathbb{Y}_{l\xi}(\bm{k})$.
Thus, we omit $\mathbb{Q}_{0,A_{1g},(0,0)}^{\rm (a)}$ hereafter.
Converting $\mathbb{Y}_{l\xi}$ to the momentum-space representation in $2\times 2$ full matrix form ($\ket{p_{z}@{\rm C}_{1}}, \ket{p_{z}@{\rm C}_{2}}$), we obtain $\mathbb{Y}_{l\xi}(\bm{k})=\hat{\mathbb{Z}}_{l\xi}(\bm{k})$ for the site cluster C and the nearest-neighbor bond cluster B$_{1}$ as
\begin{align}
&
\hat{\mathbb{Q}}_{0,A_{1g}}^{[{\rm C}]}=\frac{1}{\sqrt{2}}\begin{pmatrix}
1 & 0 \\
0 & 1 \\
\end{pmatrix},
\quad
\hat{\mathbb{Q}}_{3,B_{1u}}^{[{\rm C}]}=\frac{1}{\sqrt{2}}\begin{pmatrix}
1 & 0 \\
0 & -1 \\
\end{pmatrix},
\label{eq:gtb0}
\end{align}
\begin{align}
&
\hat{\mathbb{Q}}_{0,A_{1g}}^{[{\rm B}_{1}]}(\bm{k})=\frac{1}{\sqrt{6}}\begin{pmatrix}
0 & e^{-i\bm{k}\cdot\bm{b}_{1}}+e^{-i\bm{k}\cdot\bm{b}_{2}}+e^{-i\bm{k}\cdot\bm{b}_{3}} \\
{\rm c.c.} & 0 \\
\end{pmatrix},
\cr&
\hat{\mathbb{Q}}_{2,E_{2g},u}^{[{\rm B}_{1}]}(\bm{k})=\frac{1}{2\sqrt{3}}\begin{pmatrix}
0 & 2e^{-i\bm{k}\cdot\bm{b}_{1}}-e^{-i\bm{k}\cdot\bm{b}_{2}}-e^{-i\bm{k}\cdot\bm{b}_{3}} \\
{\rm c.c.} & 0 \\
\end{pmatrix},
\cr&
\hat{\mathbb{Q}}_{2,E_{2g},v}^{[{\rm B}_{1}]}(\bm{k})=\frac{1}{2}\begin{pmatrix}
0 & -e^{-i\bm{k}\cdot\bm{b}_{2}}+e^{-i\bm{k}\cdot\bm{b}_{3}} \\
{\rm c.c.} & 0 \\
\end{pmatrix},
\label{eq:gtb1}
\end{align}
\begin{align}
&
\hat{\mathbb{T}}_{1,E_{1u},u}^{[{\rm B}_{1}]}(\bm{k})=\frac{1}{2}\begin{pmatrix}
0 & ie^{-i\bm{k}\cdot\bm{b}_{2}}-ie^{i\bm{k}\cdot\bm{b}_{3}} \\
{\rm c.c.} & 0 \\
\end{pmatrix},
\cr&
\hat{\mathbb{T}}_{1,E_{1u},v}^{[{\rm B}_{1}]}(\bm{k})=\frac{1}{2\sqrt{3}}\begin{pmatrix}
0 & 2ie^{-i\bm{k}\cdot\bm{b}_{1}}-ie^{-i\bm{k}\cdot\bm{b}_{2}}-ie^{-i\bm{k}\cdot\bm{b}_{3}} \\
{\rm c.c.} & 0 \\
\end{pmatrix},
\cr&
\hat{\mathbb{T}}_{3,B_{1u}}^{[{\rm B}_{1}]}(\bm{k})=\frac{1}{\sqrt{6}}\begin{pmatrix}
0 & ie^{-i\bm{k}\cdot\bm{b}_{1}}+ie^{-i\bm{k}\cdot\bm{b}_{2}}+ie^{-i\bm{k}\cdot\bm{b}_{3}} \\
{\rm c.c.} & 0 \\
\end{pmatrix},
\label{eq:gtb2}
\cr&
\end{align}
where ``c.c.'' denotes the complex conjugate of the corresponding element in the upper-triangle of the matrix, and square brackets in the superscript indicate the relevant site/bond clusters.
We have omitted the superscripts, (s) and (b), as we have attached the cluster indices.
In the same procedure, the full matrix forms in the momentum representation for more than second-neighbor bond clusters, B$_{2}$, B$_{3}$, $\cdots$, may be obtained.

By selecting the identity representation $A_{1g}$, e.g., $\hat{\mathbb{Q}}_{0,A_{1g}}^{[{\rm C}]}$ and $\hat{\mathbb{Q}}_{0,A_{1g}}^{[{\rm B}_{1}]}(\bm{k})$, we obtain the TB Hamiltonian in terms of the combined SAMBs as
\begin{align}
H(\bm{k})&
=\sum_{j}z_{j}\hat{\mathbb{Z}}_{j}(\bm{k})
\cr&
=z_{1}\hat{\mathbb{Q}}_{0}^{[{\rm C}]}+z_{2}\hat{\mathbb{Q}}_{0}^{[{\rm B}_{1}]}(\bm{k})+z_{3}\hat{\mathbb{Q}}_{0}^{[{\rm B}_{2}]}(\bm{k})+\cdots,
\cr&
\label{eq:graphene_ham}
\end{align}
where the irrep. $A_{1g}$ has been omitted.
Here, $j=1,2,\cdots$ is the sequential number for the combined SAMBs, and the coefficient $z_{j}$ is the weight of each SAMB.
They will be determined by comparing the energy dispersion obtained from the TB model with that of the DF computation.

$\mathbb{Y}_{l\xi}(\bm{k})=\hat{\mathbb{Z}}_{l\xi}(\bm{k})$ can be further decomposed into the direct product of $\mathbb{U}_{l\xi}$ and $\mathbb{F}_{l\xi}(\bm{k})$.
In addition to the diagonal $\mathbb{Q}_{l\xi}^{\rm (s)}$ from the site cluster C, the uniform matrices $\mathbb{Q}_{l\xi}^{\rm (u)}$ and $\mathbb{T}_{l\xi}^{\rm (u)}$ are obtained by the off-diagonal $\mathbb{Q}_{l\xi}^{\rm (b)}(\bm{k})$ and $\mathbb{T}_{l\xi}^{\rm (b)}(\bm{k})$ with $\bm{k}=0$ and appropriate normalization as
\begin{align}
\mathbb{Q}_{0,A_{1g}}^{[{\rm B}_{1}]}=\frac{1}{\sqrt{2}}\begin{pmatrix}
0 & 1 \\
1 & 0 \\
\end{pmatrix},
\quad
\mathbb{T}_{3,B_{1u}}^{[{\rm B}_{1}]}=\frac{1}{\sqrt{2}}\begin{pmatrix}
0 & i \\
-i & 0 \\
\end{pmatrix}.
\label{eq:uniformmat}
\end{align}
Note that since Eqs.~(\ref{eq:gtb0}) and (\ref{eq:uniformmat}) form the complete set for $2\times 2$ space, the uniform matrices $\mathbb{Q}_{l\xi}^{\rm [B_{1}]}$ and $\mathbb{T}_{l\xi}^{\rm [B_{1}]}$ are common for all bond clusters.

\begin{figure}[t!]
\centering
\includegraphics[width=8.5cm]{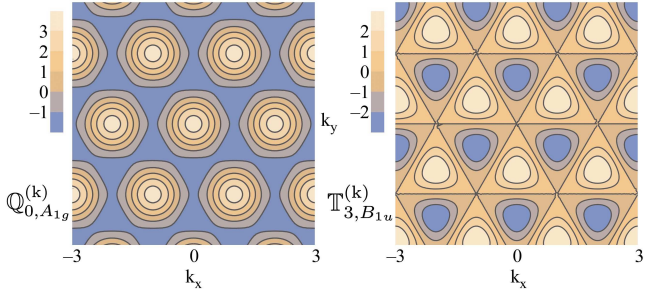}%
\caption{\label{fig:graphene_sf}
Symmetry-adapted structure factors for the nearest-neighbor bonds in graphene.
$k_{x}$ and $k_{y}$ are in unit of $2\pi/a$.
}
\end{figure}

By Eq.~(\ref{eq:stfactor}), the structure SAMBs are given by
\begin{align}
&
\mathbb{Q}_{0,A_{1g}}^{[{\rm B}_{1}]\rm (k)}(\bm{k})=\frac{2}{\sqrt{6}}(c_{1}+c_{2}+c_{3}),
\cr&
\mathbb{Q}_{2,E_{2g},u}^{[{\rm B}_{1}]\rm (k)}(\bm{k})=\frac{1}{\sqrt{3}}(2c_{1}-c_{2}-c_{3}),
\cr&
\mathbb{Q}_{2,E_{2g},v}^{[{\rm B}_{1}]\rm (k)}(\bm{k})=-c_{2}+c_{3},
\cr&
\mathbb{T}_{1,E_{1u},u}^{[{\rm B}_{1}]\rm (k)}(\bm{k})=s_{2}-s_{3},
\cr&
\mathbb{T}_{1,E_{1u},v}^{[{\rm B}_{1}]\rm (k)}(\bm{k})=\frac{1}{\sqrt{3}}(2s_{1}-s_{2}-s_{3}),
\cr&
\mathbb{T}_{3,B_{1u}}^{[{\rm B}_{1}]\rm (k)}(\bm{k})=\frac{2}{\sqrt{6}}(s_{1}+s_{2}+s_{3}),
\end{align}
where $c_{i}=\cos(\bm{k}\cdot\bm{b}_{i})$ and $s_{i}=\sin(\bm{k}\cdot\bm{b}_{i})$ with the reduced momentum $\bm{k}=2\pi(k_{1},k_{2},k_{3})$.
The expressions can be converted in terms of the Cartesian coordinate $k_{x}$ and $k_{y}$ by substituting $k_{1}=k_{x}a$ and $k_{2}=(-k_{x}a+\sqrt{3}k_{y}a)/2$.
The $\bm{k}$ dependence of $\mathbb{Q}_{0,A_{1g}}^{[{\rm B}_{1}]\rm (k)}(\bm{k})$ and $\mathbb{T}_{3,B_{1u}}^{[{\rm B}_{1}]\rm (k)}(\bm{k})$ are shown in Fig.~\ref{fig:graphene_sf}.

Using $\mathbb{U}_{l\xi}$ and $\mathbb{F}_{l\xi}(\bm{k})$, $\hat{\mathbb{Q}}_{0}^{[{\rm B}_{n}]}(\bm{k})$ for $n=1,3,4$ can be decomposed as
\begin{multline}
\hat{\mathbb{Q}}_{0}^{[{\rm B}_{n}]}(\bm{k})=\frac{1}{\sqrt{2}}\biggl[
\mathbb{Q}_{0,A_{1g}}^{[{\rm B}_{1}]}\otimes \mathbb{Q}_{0,A_{1g}}^{[{\rm B}_{n}]\rm (k)}(\bm{k})
\\
-\mathbb{T}_{3,B_{1u}}^{[{\rm B}_{1}}\otimes \mathbb{T}_{3,B_{1u}}^{[{\rm B}_{n}]\rm (k)}(\bm{k})
\biggr],
\end{multline}
and for $n=2,5,6$ as
\begin{align}
\hat{\mathbb{Q}}_{0}^{[{\rm B}_{n}]}(\bm{k})=
\mathbb{Q}_{0,A_{1g}}^{[{\rm C}]}\otimes \mathbb{Q}_{0,A_{1g}}^{[{\rm B}_{n}]\rm (k)}(\bm{k}).
\end{align}
Here, the structure factors for $n=3,4$ are given by
\begin{align}
&
\mathbb{Q}_{0,A_{1g}}^{[{\rm B}_{3}]{\rm (k)}}(\bm{k})=\frac{2}{\sqrt{6}}(c_{10}+c_{11}+c_{12}),
\cr&
\mathbb{Q}_{0,A_{1g}}^{[{\rm B}_{4}]{\rm (k)}}(\bm{k})=\frac{1}{\sqrt{3}}(c_{13}+c_{14}+c_{15}+c_{16}+c_{17}+c_{18}),
\cr&
\mathbb{T}_{3,B_{1u}}^{[{\rm B}_{3}]{\rm (k)}}(\bm{k})=\frac{2}{\sqrt{6}}(s_{10}+s_{11}+s_{12}),
\cr&
\mathbb{T}_{3,B_{1u}}^{[{\rm B}_{4}]{\rm (k)}}(\bm{k})=\frac{1}{\sqrt{3}}(s_{13}+s_{14}+s_{15}+s_{16}+s_{17}+s_{18}),
\cr&
\end{align}
and for $n=2,5,6$,
\begin{align}
&
\mathbb{Q}_{0,A_{1g}}^{[{\rm B}_{2}]{\rm (k)}}(\bm{k})=\frac{2}{\sqrt{6}}(c_{4}+c_{6}+c_{7}),
\cr&
\mathbb{Q}_{0,A_{1g}}^{[{\rm B}_{5}]{\rm (k)}}(\bm{k})=\frac{2}{\sqrt{6}}(c_{19}+c_{21}+c_{22}),
\cr&
\mathbb{Q}_{0,A_{1g}}^{[{\rm B}_{6}]{\rm (k)}}(\bm{k})=\frac{2}{\sqrt{6}}(c_{25}+c_{27}+c_{28}),
\end{align}
where the bond vectors up to sixth neighbors are shown in Table~\ref{tbl:graphene_bond}.

\begin{table}
\caption{\label{tbl:graphene_bond}
Bond vectors for graphene up to sixth neighbors.
}
\begin{spacing}{1.3}
\begin{ruledtabular}
\begin{tabular}{ccccccc}
$n$ & $\bm{b}_{i}$ & Vector & $\bm{b}_{i}$ & Vector & $\bm{b}_{i}$ & Vector \\ \hline
1 & $\bm{b}_{1}$ & $[\frac{1}{3}, \frac{2}{3}, 0]$ & $\bm{b}_{2}$ & $[\frac{1}{3}, -\frac{1}{3}, 0]$ & $\bm{b}_{3}$ & $[-\frac{2}{3}, -\frac{1}{3}, 0]$ \\ \hline
2 & $\bm{b}_{4}$ & $[0, 1, 0]$ & $\bm{b}_{5}$ & $=\bm{b}_{4}$ & $\bm{b}_{6}$ & $[1, 1, 0]$ \\
  & $\bm{b}_{7}$ & $[1, 0, 0]$ & $\bm{b}_{8}$ & $=\bm{b}_{6}$ & $\bm{b}_{9}$ & $=\bm{b}_{7}$ \\ \hline
3 & $\bm{b}_{10}$ & $[\frac{4}{3}, \frac{2}{3}, 0]$ & $\bm{b}_{11}$ & $[-\frac{2}{3}, \frac{2}{3}, 0]$ & $\bm{b}_{12}$ & $[-\frac{2}{3}, -\frac{4}{3}, 0]$ \\ \hline
4 & $\bm{b}_{13}$ & $[\frac{4}{3}, \frac{5}{3}, 0]$ & $\bm{b}_{14}$ & $[\frac{1}{3}, \frac{5}{3}, 0]$ & $\bm{b}_{15}$ & $[\frac{4}{3}, -\frac{1}{3}, 0]$ \\
  & $\bm{b}_{16}$ & $[-\frac{5}{3}, -\frac{4}{3}, 0]$ & $\bm{b}_{17}$ & $[-\frac{5}{3}, -\frac{1}{3}, 0]$ & $\bm{b}_{18}$ & $[\frac{1}{3}, -\frac{4}{3}, 0]$ \\ \hline
5 & $\bm{b}_{19}$ & $[1, 2, 0]$ & $\bm{b}_{20}$ & $=-\bm{b}_{19}$ & $\bm{b}_{21}$ & $[-1, 1, 0]$ \\
  & $\bm{b}_{22}$ & $[2, 1, 0]$ & $\bm{b}_{23}$ & $=-\bm{b}_{21}$ & $\bm{b}_{24}$ & $=-\bm{b}_{22}$ \\ \hline
6 & $\bm{b}_{25}$ & $[2, 2, 0]$ & $\bm{b}_{26}$ & $=\bm{b}_{25}$ & $\bm{b}_{27}$ & $[0, 2, 0]$ \\
  & $\bm{b}_{28}$ & $[2, 0, 0]$ & $\bm{b}_{29}$ & $=\bm{b}_{27}$ & $\bm{b}_{30}$ & $=\bm{b}_{28}$ \\
\end{tabular}
\end{ruledtabular}
\end{spacing}
\end{table}

By using the SAMB, the symmetry-breaking terms are classified according to point-group symmetry, which is useful to narrow down the possible order parameters in the phase transition.
For example, the mass term which lowers the symmetry from D$_{\rm 6h}$ to D$_{\rm 3h}$ is given by $\hat{\mathbb{Q}}_{3,B_{1u}}^{\rm [C]}$.
In the ordered phase, this term becomes the identity irrep.

Similarly, the Haldane's magnetic flux due to kinetic spin-orbit coupling is expressed as $\hat{\mathbb{M}}_{1,A_{2g}}^{\rm [B_{2}]}=\mathbb{Q}_{3,B_{1u}}^{\rm [C]}\otimes \mathbb{T}_{3,B_{2u}}^{\rm [B_{2}](k)}$ where the structure factor is $\mathbb{T}_{3,B_{2u}}^{\rm [B_{2}](k)}=\frac{2}{\sqrt{6}}(s_{5}-s_{6}+s_{7})$ which corresponds to the vortex like imaginary hopping in second-neighbor A-A or B-B bonds (See, Fig.~\ref{fig:graphene})~\cite{Haldane_PhysRevLett.61.2015}.

When the inversion symmetry is broken, e.g., by applying an electric field perpendicular to the plane, the Rashba term appears~\cite{liu2009spin}.
As the polar vector belongs to $A_{2u}$ irrep., we look for the SAMB belonging to $A_{2u}$. 
Although there is no SAMB belonging to $A_{2u}$ in the spinless Hilbert space, it can appear when taking into account the spin degree of freedom $(\sigma_x, \sigma_y, \sigma_z)$.
Considering the product decomposition, $A_{2u}=A_{2g}\otimes A_{1u}$ or $A_{2u}=E_{1g}\otimes E_{1u}$ and the irrep. of the spins as $\sigma_{z}$ ($A_{2g}$) and $(-\sigma_{y},\sigma_{x})$ ($E_{1g}$), we obtain the Rashba term for the nearest-neighbor bond cluster as
\begin{multline}
H_{\text{Rashba}}(\bm{k})=\frac{1}{2}\biggl[-\sigma_{y}\otimes\mathbb{Q}_{0,A_{1g}}^{\rm [B_{1}]}
\otimes\mathbb{T}_{1,E_{1u},u}^{\rm [B_{1}](k)}(\bm{k}) \\
+\sigma_{x}\otimes\mathbb{Q}_{0,A_{1g}}^{\rm [B_{1}]}
\otimes\mathbb{T}_{1,E_{1u},v}^{\rm [B_{1}](k)}(\bm{k})\biggr].
\end{multline}
In this way, the symmetry-breaking terms are easily classified in terms of the SAMB, and their $\bm{k}$ dependence is encoded in the structure SAMB, $\mathbb{F}_{l\xi}(\bm{k})$.

For obtaining deeper insights into the physical responses and the exploration of more efficient materials, it is highly desirable to achieve a microscopic understanding of the relevant mechanism and the essential parameters.
In this sense, the symmetry-adapted modeling as Eq.~(\ref{eq:graphene_ham}) is also useful to analyze various linear and nonlinear response functions, which bridges the gap between the phenomenological approaches and DF computations~\cite{Oiwa_PhysRevLett.129.116401, Hayami_PhysRevB.102.144441}.
The systematic analysis method for response functions proposed in Ref.~\cite{Oiwa_doi:10.7566/JPSJ.91.014701, Hayami_PhysRevB.102.144441} not only enables one to predict the possible responses but also extract essential parameters in a systematic manner by analyzing the indicators such as
\begin{align}
&
\Gamma_{\mu;\alpha}^{ij}={\rm Tr}\biggl[
A_{\mu}H^{i}B_{\alpha}H^{j}
\biggr],
\cr&
\Gamma_{\mu;\alpha,\beta}^{ijk}={\rm Tr}\biggl[
A_{\mu}H^{i}B_{\alpha}H^{j}B_{\beta}H^{k}
\biggr],
\end{align}
for linear and second-order nonlinear responses.
The corresponding response becomes active when the indicator is nonzero.
Here, $A$ and $B$ are the output and input operators in the responses, respectively, and $H^{i}$ is the $i$th power of the Hamiltonian matrices.
By using Eq.~(\ref{eq:graphene_ham}), the trace in the indicator is regarded as selecting the identity irrep. in the irreducible decomposition of the product of $A$, $B$, and $H^{i}$.
Thus, the combination of those operators giving the identity irrep. is nothing but the essential parameters of the response.
It can provide us with guidelines for future material design, beyond those obtained by the existing phenomenological approaches and DF calculations.

\subsection{Optimization of TB Model}

\begin{figure*}
\centering
\includegraphics[width=18cm]{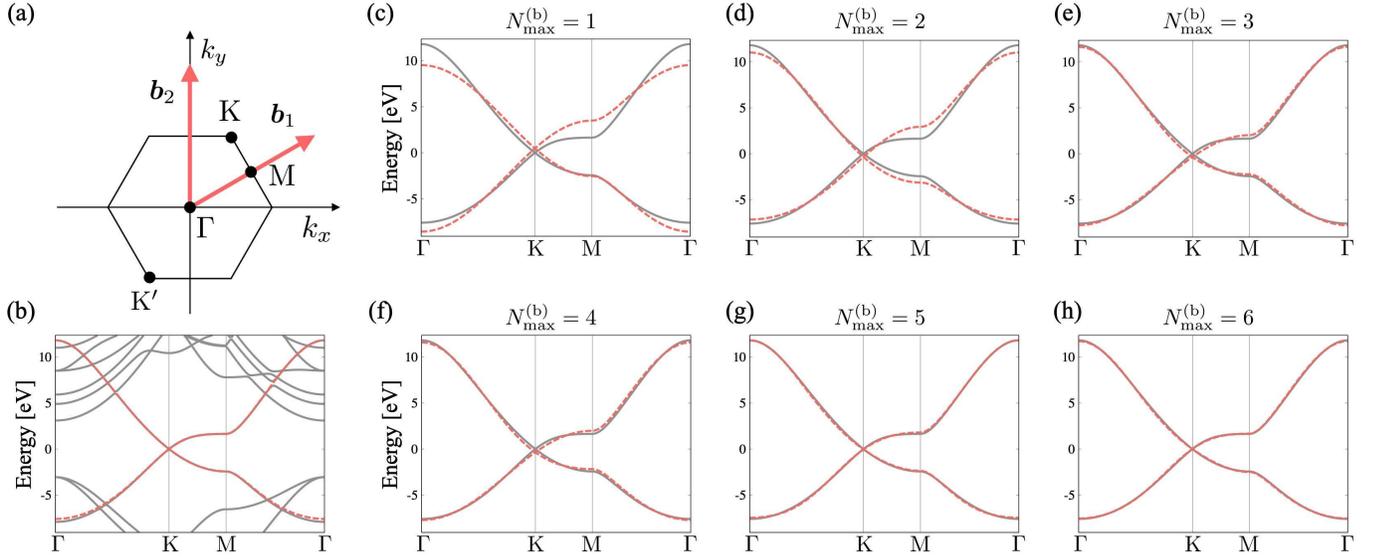}%
\caption{\label{fig:graphene_fit}
The comparisons of the band dispersion between the DF Wannier and SAMB TB models.
(a) The Brillouin zone and high-symmetry points, (b) the comparison of energy dispersions between DF (gray solid lines) and Wanner TB model (red dashed lines), (c)-(h) the comparison of energy dispersion between the DF Wannier TB model (gray solid lines) and our SAMB TB model up to $N_{\rm max}^{\rm (b)}$-neighbor bonds (red dashed lines).
The Fermi energy is set as the origin.
}
\end{figure*}

We have constructed the TB model for graphene in Eq.~(\ref{eq:graphene_ham}), and there are seven parameters $\bm{z}=(z_{1},z_{2},\cdots,z_{7})$ up to sixth-neighbor hoppings.
In order to optimize the model parameters $\bm{z}$, we compute the energy dispersion by DF computation.

For the DF computation, we have used the Quantum ESPRESSO open-software package~\cite{giannozzi2009quantum} with the Perdew-Zunger correlation functional~\cite{PhysRevB.23.5048} and the ultrasoft pseudopotential.
We have used the $\bm{k}$ grid, $(N_{1}, N_{2}, N_{3}) = (12, 12, 1)$, and the kinetic energy cutoff of the Kohn-Sham wave functions and convergence threshold are set as 30 Ry and 1$\times$10$^{-10}$ Ry, respectively.

\begin{figure}
\centering
\includegraphics[width=8.5cm]{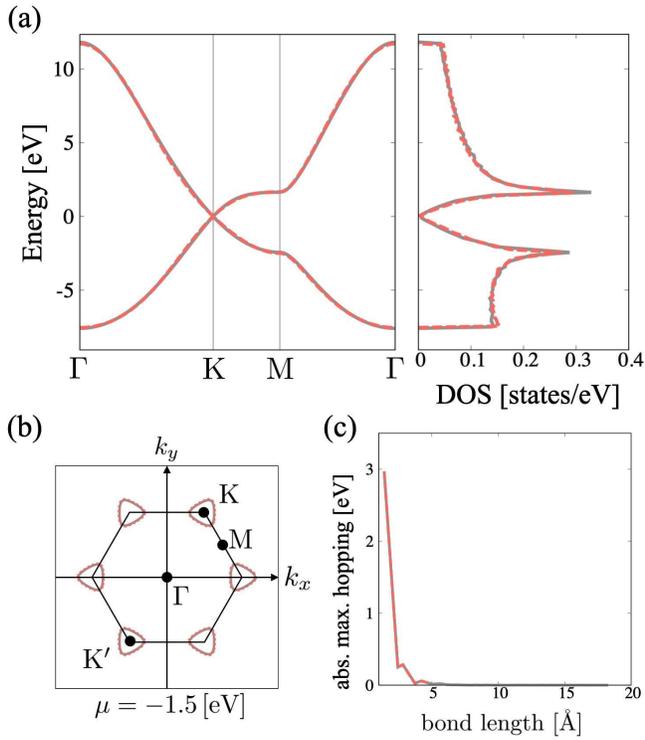}%
\caption{\label{fig:graphene_result}
(a) Comparison of the energy dispersion and DOS for the optimized SAMB TB with $N_{\rm max}^{\rm (b)}=6$.
(b) The isoenergy surface at $\mu=-1.5$ eV.
(c) The bond length dependence of the maximum strength of hoppings.
The gray solid (red dashed) lines represent the results of the DF Wannier (SAMB) TB model.
}
\end{figure}

The obtained electronic bands near the Fermi energy in the high-symmetry lines $\Gamma$-K-M-$\Gamma$ in Fig.~\ref{fig:graphene_fit}(a) are entangled as shown in Fig.~\ref{fig:graphene_fit}(b) (gray solid lines).
Thus, before optimizing the model parameters, two relevant bands near the Fermi energy must be disentangled.
For this purpose, we have used the Wannier90 open-source software package~\cite{mostofi2008wannier90,mostofi2014updated,pizzi2020wannier90}: The $p_{z}$ orbital is chosen for each C atom as the initial guess function, and the outer and inner energy windows are set as [$-30$, 12] eV and [$-3.0$, 2.6] eV, respectively.
Then, we obtain the $p_{z}$-like two Wannier orbitals and corresponding energy dispersions are indicated by red dashed lines as shown in Fig.~\ref{fig:graphene_fit}(b).

In the optimization process, we introduce the loss function as the dimensionless mean squared error of the normalized energy eigenvalues between the DF and our TB models~\cite{Oiwa_PhysRevLett.129.116401},
\begin{align}
L(\bm{z})=\frac{1}{N_{k}N_{n}}\sum_{n}\sum_{\bm{k}}\left(\frac{\epsilon_{n\bm{k}}(\bm{z})-\epsilon_{n\bm{k}}^{\rm DF}}{W}\right)^{2},
\end{align}
where $N_{k}=151$, $N_{n}=2$, and $W=21.40$ [eV] are the number of $\bm{k}$ points to evaluate the loss function, the size of the Hamiltonian, and band width, respectively.

In order to eliminate strong initial-guess dependence, we use the hidden layers in the neural network.
Namely, the relation between the input DF energy bands $\epsilon_{n\bm{k}}^{\rm DF}$ and the resultant energy bands of our TB model $\epsilon_{n\bm{k}}(\bm{z})$ is regarded as the nontrivial nonlinear system.
Then, we insert the hidden layers between the input and output energy bands to express flexibly this nonlinear relation, and apply the back-propagation error algorithm to optimize the model parameters $\bm{z}$ and hyper parameters in the hidden layers~\cite{rumelhart1986learning,lecun2015deep}.
It is turned out that extremely low initial-guess dependence was achieved.
We have chosen 50 $\bm{k}$ points in each line in the high-symmetry lines $\Gamma$-K-M-$\Gamma$, and used $N_{\rm h}=3$ hidden layers.
We have used the PyTorch package~\cite{NEURIPS2019_9015} and the Adam optimizer~\cite{kingma_adam_2015} with the learning rate $\alpha=0.1$.
The fixed maximum number of iterations $N_{\rm iter}=250$ is sufficient to reach convergence.
In the case of graphene, the construction of the SAMB and the optimization of the model parameters take within a minute by a standard laptop computer.
For more complicated systems, the computing cost increases mostly in the part of the construction of the SAMB, but it takes within 10 minutes for SrVO$_{3}$ and MoS$_{2}$ as shown in the supplementary materials.

The results of the best optimization with $N_{\rm max}^{\rm (b)}$-neighbor bonds are shown in Fig.~\ref{fig:graphene_fit}(c)-\ref{fig:graphene_fit}(h), where the convergence value of the loss function for Fig.~\ref{fig:graphene_fit}(h) is about $9.4\times 10^{-6}$.
With increase of $N_{\rm max}^{\rm (b)}$, the result gives better reproduction of the DF energy dispersion.
The optimization parameters for $N_{\rm max}^{\rm (b)}=6$ are obtained as
\begin{align}
&
z_{1}=-0.163,\,\,\, z_{2}=-7.274,\,\,\, z_{3}=0.880,\,\,\, z_{4}=-0.693, \cr&
z_{5}=0.0761,\,\,\, z_{6}=0.202,\,\,\, z_{7}=-0.080\,\,\,\text{[eV]}.
\end{align}
The band structure, density of states (DOS), isoenergy surface, and bond-length dependence of the maximum strength of hoppings are obtained using these optimized parameters as shown in Fig.~\ref{fig:graphene_result}.
The results of both models are in good agreement.
On the other hand, as shown in Fig.~\ref{fig:graphene_result}(c), the bond length dependence of the maximum strength of hoppings differs significantly for the two models.
In our SAMB TB model, the magnitude of the weight tends to decrease as the bond length increases, while that of the DF Wannier TB model shows long tail.
The advantage of the SAMB TB is that the hopping range can be freely chosen without losing the symmetry of the system, and the systematic comparison with different hopping range is possible as shown in Figs.~\ref{fig:graphene_fit}(c)-(h).

The similar analysis has been performed for a chiral nonsymmorphic system of Te~\cite{Oiwa_PhysRevLett.129.116401}, a typical orbitally degenerate system of SrVO$_{3}$, and the spin-orbital coupled metal of monolayer MoS$_{2}$.
We show the results for the latter two materials in the supplementary materials.

\section{Summary}

In this paper, we have developed a symmetry-adapted modeling procedure for molecules and crystals.
By constructing the SAMB set for atomic ($\mathbb{X}_{l\xi,sk}$) and molecular/crystal structural ($\mathbb{Y}_{l\xi}$) parts separately in terms of point-group harmonics, we express the final SAMBs denoted by $\hat{\mathbb{Z}}_{l\xi,sk}$ or $\hat{\mathbb{Z}}_{l\xi,sk}(\bm{k})$ as the irreducible decomposition of these products.
Since these SAMBs constitute a complete orthonormal basis set, they can describe any of electronic degrees of freedom in isolated cluster systems and periodic crystals.
Once we obtain the complete set of SAMBs, a physical system can be expressed in linear combination of these bases belonging to the identity irreducible representation of the system.
Moreover, the SAMBs other than the identity irreducible representations are all the candidates of possible order parameters, for which emergent macroscopic physical properties are easily predicted as they are already classified by the irreducible representation of the point-group symmetry.

We have demonstrated our method to electronic modeling of graphene as the simplest example, where the modeling parameters (linear coefficients of each SAMB) are optimized so as to reproduce the electronic structures given by the density-functional computation.
As compared with de-facto standard method of Wannier tight-binding modeling, the model obtained by our method satisfies rigorously the symmetry of the system, in which we can freely choose a range of hoppings.
This aspect is a strong advantage to compactify the relevant model to discuss various response functions and many-body effects with low computational cost.
Furthermore, meaning of physical operators is apparent since we describe any of atomic degrees of freedom in terms of atomic-orbital Hilbert space.

Although we have demonstrated our method only to electronic tight-binding modeling, our modeling procedure can also be utilized to decompose two-body multipolar interactions including the density-density one, magnetic exchange couplings including Dzyaloshinskii-Moriya type~\cite{Matsumoto_PhysRevB.101.224419, Matsumoto_PhysRevB.104.134420, Hayami_PhysRevB.105.014404}, and mechanical lattice model expressing the dynamical matrix of phonon~\cite{doi:10.7566/JPSJ.92.012001, tsunetsugu2023theory}.

Since the present SAMBs all belong to the root spherical harmonics in rotation group, various systems can be compared quantitatively with each other via the weight of spherical harmonics.
It is a crucial property of representation required for good descriptors in machine-learning based materials design~\cite{PhysRevB.87.184115,suzuki2023highperformance}.
Therefore, the present method provides us with fundamental basis to develop symmetry-based analysis for materials science.

\begin{acknowledgments}
The authors thank Yuki Yanagi, Megumi Yatsushiro, Hiroaki Ikeda, Ryotaro Arita, Yusuke Nomura, and Michito Suzuki for fruitful discussions.
This research was supported by JSPS KAKENHI Grants No.~JP21H01031, No.~JP21H01037, No.~JP22H04468, No.~JP22H00101, No.~JP22H01183, and by JST PRESTO (JPMJPR20L8).
\end{acknowledgments}

\bibliography{ref.bib}

\newpage

\begin{widetext}
\begin{center}
\LARGE\bf
Supplementary Materials
\end{center}
\vspace{3mm}
\begin{quote}
In this supplementary materials, we show other applications of our symmetry-adapted modeling method for more elaborate examples, SrVO$_{3}$ and MoS$_{2}$.
For each example, we explain the condition of DF computation, and give detailed information of SAMBs, and the results of the optimized energy dispersion.
\end{quote}
\vspace{1cm}

\section{Symmetry-Adapted Modeling for SrVO$_{3}$}\label{sec_usage_ex_srvo3}

In this section, we demonstrate our method to the bulk crystalline SrVO$_{3}$, a typical material and is often chosen as the benchmark in developing a new method associated with DF computation.
First, we show the condition of DF computation, and then we summarize the SAMBs of SrVO$_{3}$ up to sixth-neighbor hoppings that are automatically generated by the developed Python library, called ``MultiPie'', which will be open on the GitHub.

\subsection{DF computation for SrVO$_{3}$}
\label{sec_usage_ex_srvo3_dft}

\begin{figure}[h]
   \begin{center}
      \includegraphics[width=16.5cm]{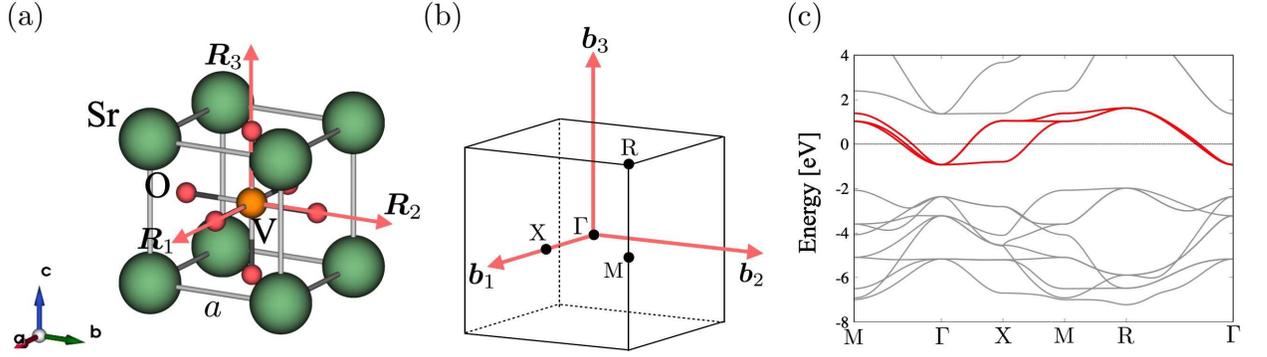}
      \caption{
         (a) Crystal structure and (b) Brillouin zone of bulk crystalline SrVO$_{3}$.
         (c) Band dispersion obtained from the DF computation, where the red solid lines represent the $t_{2g}$ orbitals of V atom.
         The Fermi energy is taken as the origin.
      }
      \label{fig_srvo3_dft_band}
   \end{center}
\end{figure}

The bulk crystalline SrVO$_{3}$ has the cubic structure including Sr, V, and O sublattices in a unit cell as shown in Fig.~\ref{fig_srvo3_dft_band}(a).
The space group of SrVO$_{3}$ is $Pm\bar{3}m$ (\#221, O$_{\rm h}^{1}$).
We set the lattice constant as $a = 3.8409$ \AA.
For the DF computation, we use the PBEsol exchange-correlation functional~\cite{Perdew_PBEsol_2008} and the PAW pseudopotential.
For the self-consistent functional (SCF) calculation, we use $(N_{1}, N_{2}, N_{3}) = (6, 6, 6)$ $\bm{k}$ grid, and the kinetic energy cutoff of the Kohn-Sham wave functions is 100 Ry, and the convergence threshold for the SCF calculation is 1$\times$10$^{-10}$ Ry.

As shown in Fig.~\ref{fig_srvo3_dft_band}(c), the bands near the Fermi level are isolated.
Therefore, we directly optimize our SAMB tight-binding (TB) model to those bands obtained from the DF computation without using the disentangling procedure embedded in Wannier90 code~\cite{mostofi2008wannier90,mostofi2014updated,pizzi2020wannier90}.
The isolated three electronic states near the Fermi level are mainly composed of the $t_{2g}$ orbitals of V atom as depicted by the solid red lines in Fig.~\ref{fig_srvo3_dft_band}(c).
Thus, we choose the $(d_{yz}, d_{zx}, d_{xy})$ orbitals of the V atom as basis functions for our TB model, and neglect the contributions of the Sr and O atoms.

\subsection{Symmetry-adapted multipole basis for SrVO$_{3}$}
\label{sec_usage_ex_srvo3_tb}

Here, we summarize the SAMB information for SrVO$_{3}$.
\begin{itemize}
\item The full Hilbert space of the model Hamiltonian is given by Table I, and its dimension is 3.
\begin{center}
\renewcommand{\arraystretch}{1.3}
\begin{longtable}{c|cc|cc|cc}
\caption{Hilbert space for full matrix.}
 \\
 \hline \hline
 & No. & ket & No. & ket & No. & ket \\ \hline \endfirsthead

\multicolumn{6}{l}{\tablename\ \thetable{}} \\
 \hline \hline
 & No. & ket & No. & ket & No. & ket \\ \hline \endhead

 \hline \hline
\multicolumn{6}{r}{\footnotesize\it continued ...} \\ \endfoot

 \hline \hline
\multicolumn{6}{r}{} \\ \endlastfoot

 & 1 & $d_{yz}$@V$_{1}$ & 2 & $d_{zx}$@V$_{1}$ & 3 & $d_{xy}$@V$_{1}$ \\
\end{longtable}
\end{center}

\item There are only one site cluster in this system as shown in Table II.
\begin{center}
\renewcommand{\arraystretch}{1.3}
\begin{longtable}{cc|c|l}
\caption{Site clusters.}
 \\
 \hline \hline
 & site & position & mapping \\ \hline \endfirsthead

\multicolumn{3}{l}{\tablename\ \thetable{}} \\
 \hline \hline
 & site & position & mapping \\ \hline \endhead

 \hline \hline
\multicolumn{3}{r}{\footnotesize\it continued ...} \\ \endfoot

 \hline \hline
\multicolumn{3}{r}{} \\ \endlastfoot

S$_{1}$ & V$_1$ & $\begin{pmatrix} 0 & 0 & 0 \end{pmatrix}$ & [1,2,3,4,5,6,7,8,9,10,11,12,13,14,15,16,17,18,19,20,21,22,23,24,\\ & & & 25,26,27,28,29,30,31,32,33,34,35,36,37,38,39,40,41,42,43,44,45,46,47,48] \\
\end{longtable}
\end{center}

\item There are 6 bond clusters up to sixth-neighbor V-V bonds as shown in Table III.
\begin{center}
\renewcommand{\arraystretch}{1.3}
\begin{longtable}{cc|cc|c|c|c|l}
\caption{Bond clusters.}
 \\
 \hline \hline
 & bond & tail & head & $n$ & \# & $\bm{b}@\bm{c}$ & mapping \\ \hline \endfirsthead

\multicolumn{7}{l}{\tablename\ \thetable{}} \\
 \hline \hline
 & bond & tail & head & $n$ & \# & $\bm{b}@\bm{c}$ & mapping \\ \hline \endhead

 \hline \hline
\multicolumn{7}{r}{\footnotesize\it continued ...} \\ \endfoot

 \hline \hline
\multicolumn{7}{r}{} \\ \endlastfoot

B$_{1}$ & b$_{1}$ & V$_{1}$ & V$_{1}$ & 1 & 1 & $\begin{pmatrix} 0 & 0 & 1 \end{pmatrix}@\begin{pmatrix} 0 & 0 & \frac{1}{2} \end{pmatrix}$ & [1,2,-3,-4,-5,-8,19,22,-25,-26,27,28,29,32,-43,-46] \\
& b$_{2}$ & V$_{1}$ & V$_{1}$ & 1 & 1 & $\begin{pmatrix} 1 & 0 & 0 \end{pmatrix}@\begin{pmatrix} \frac{1}{2} & 0 & 0 \end{pmatrix}$ & [6,-9,11,-12,13,-14,21,-24,-30,33,-35,36,-37,38,-45,48] \\
& b$_{3}$ & V$_{1}$ & V$_{1}$ & 1 & 1 & $\begin{pmatrix} 0 & 1 & 0 \end{pmatrix}@\begin{pmatrix} 0 & \frac{1}{2} & 0 \end{pmatrix}$ & [7,-10,15,16,-17,-18,-20,23,-31,34,-39,-40,41,42,44,-47] \\ \hline
B$_{2}$ & b$_{4}$ & V$_{1}$ & V$_{1}$ & 2 & 1 & $\begin{pmatrix} 0 & 1 & 1 \end{pmatrix}@\begin{pmatrix} 0 & \frac{1}{2} & \frac{1}{2} \end{pmatrix}$ & [1,-3,7,-10,-25,27,-31,34] \\
& b$_{5}$ & V$_{1}$ & V$_{1}$ & 2 & 1 & $\begin{pmatrix} 0 & 1 & -1 \end{pmatrix}@\begin{pmatrix} 0 & \frac{1}{2} & \frac{1}{2} \end{pmatrix}$ & [-2,4,-20,23,26,-28,44,-47] \\
& b$_{6}$ & V$_{1}$ & V$_{1}$ & 2 & 1 & $\begin{pmatrix} 1 & 0 & -1 \end{pmatrix}@\begin{pmatrix} \frac{1}{2} & 0 & \frac{1}{2} \end{pmatrix}$ & [5,-12,13,-19,-29,36,-37,43] \\
& b$_{7}$ & V$_{1}$ & V$_{1}$ & 2 & 1 & $\begin{pmatrix} 1 & -1 & 0 \end{pmatrix}@\begin{pmatrix} \frac{1}{2} & \frac{1}{2} & 0 \end{pmatrix}$ & [6,-16,18,-24,-30,40,-42,48] \\
& b$_{8}$ & V$_{1}$ & V$_{1}$ & 2 & 1 & $\begin{pmatrix} 1 & 0 & 1 \end{pmatrix}@\begin{pmatrix} \frac{1}{2} & 0 & \frac{1}{2} \end{pmatrix}$ & [-8,11,-14,22,32,-35,38,-46] \\
& b$_{9}$ & V$_{1}$ & V$_{1}$ & 2 & 1 & $\begin{pmatrix} 1 & 1 & 0 \end{pmatrix}@\begin{pmatrix} \frac{1}{2} & \frac{1}{2} & 0 \end{pmatrix}$ & [-9,15,-17,21,33,-39,41,-45] \\ \hline
B$_{3}$ & b$_{10}$ & V$_{1}$ & V$_{1}$ & 3 & 1 & $\begin{pmatrix} 1 & 1 & 1 \end{pmatrix}@\begin{pmatrix} \frac{1}{2} & \frac{1}{2} & \frac{1}{2} \end{pmatrix}$ & [1,-8,-9,-10,11,15,-25,32,33,34,-35,-39] \\
& b$_{11}$ & V$_{1}$ & V$_{1}$ & 3 & 1 & $\begin{pmatrix} 1 & 1 & -1 \end{pmatrix}@\begin{pmatrix} \frac{1}{2} & \frac{1}{2} & \frac{1}{2} \end{pmatrix}$ & [-2,5,-12,-17,21,23,26,-29,36,41,-45,-47] \\
& b$_{12}$ & V$_{1}$ & V$_{1}$ & 3 & 1 & $\begin{pmatrix} 1 & -1 & -1 \end{pmatrix}@\begin{pmatrix} \frac{1}{2} & \frac{1}{2} & \frac{1}{2} \end{pmatrix}$ & [3,-7,13,18,-19,-24,-27,31,-37,-42,43,48] \\
& b$_{13}$ & V$_{1}$ & V$_{1}$ & 3 & 1 & $\begin{pmatrix} 1 & -1 & 1 \end{pmatrix}@\begin{pmatrix} \frac{1}{2} & \frac{1}{2} & \frac{1}{2} \end{pmatrix}$ & [-4,6,-14,-16,20,22,28,-30,38,40,-44,-46] \\ \hline
B$_{4}$ & b$_{14}$ & V$_{1}$ & V$_{1}$ & 4 & 1 & $\begin{pmatrix} 2 & 0 & 0 \end{pmatrix}@\begin{pmatrix} 0 & 0 & 0 \end{pmatrix}$ & [1,-2,3,-4,-7,-10,20,23,-25,26,-27,28,31,34,-44,-47] \\
& b$_{15}$ & V$_{1}$ & V$_{1}$ & 4 & 1 & $\begin{pmatrix} 0 & 2 & 0 \end{pmatrix}@\begin{pmatrix} 0 & 0 & 0 \end{pmatrix}$ & [5,-8,11,-12,-13,14,19,-22,-29,32,-35,36,37,-38,-43,46] \\
& b$_{16}$ & V$_{1}$ & V$_{1}$ & 4 & 1 & $\begin{pmatrix} 0 & 0 & 2 \end{pmatrix}@\begin{pmatrix} 0 & 0 & 0 \end{pmatrix}$ & [6,-9,15,-16,17,-18,-21,24,-30,33,-39,40,-41,42,45,-48] \\ \hline
B$_{5}$ & b$_{17}$ & V$_{1}$ & V$_{1}$ & 5 & 1 & $\begin{pmatrix} 2 & 0 & 1 \end{pmatrix}@\begin{pmatrix} 0 & 0 & \frac{1}{2} \end{pmatrix}$ & [1,-4,-25,28] \\
& b$_{18}$ & V$_{1}$ & V$_{1}$ & 5 & 1 & $\begin{pmatrix} 2 & 0 & -1 \end{pmatrix}@\begin{pmatrix} 0 & 0 & \frac{1}{2} \end{pmatrix}$ & [-2,3,26,-27] \\
& b$_{19}$ & V$_{1}$ & V$_{1}$ & 5 & 1 & $\begin{pmatrix} 0 & 2 & -1 \end{pmatrix}@\begin{pmatrix} 0 & 0 & \frac{1}{2} \end{pmatrix}$ & [5,-22,-29,46] \\
& b$_{20}$ & V$_{1}$ & V$_{1}$ & 5 & 1 & $\begin{pmatrix} 1 & 0 & 2 \end{pmatrix}@\begin{pmatrix} \frac{1}{2} & 0 & 0 \end{pmatrix}$ & [6,-9,-30,33] \\
& b$_{21}$ & V$_{1}$ & V$_{1}$ & 5 & 1 & $\begin{pmatrix} 2 & -1 & 0 \end{pmatrix}@\begin{pmatrix} 0 & \frac{1}{2} & 0 \end{pmatrix}$ & [-7,20,31,-44] \\
& b$_{22}$ & V$_{1}$ & V$_{1}$ & 5 & 1 & $\begin{pmatrix} 0 & 2 & 1 \end{pmatrix}@\begin{pmatrix} 0 & 0 & \frac{1}{2} \end{pmatrix}$ & [-8,19,32,-43] \\
& b$_{23}$ & V$_{1}$ & V$_{1}$ & 5 & 1 & $\begin{pmatrix} 2 & 1 & 0 \end{pmatrix}@\begin{pmatrix} 0 & \frac{1}{2} & 0 \end{pmatrix}$ & [-10,23,34,-47] \\
& b$_{24}$ & V$_{1}$ & V$_{1}$ & 5 & 1 & $\begin{pmatrix} 1 & 2 & 0 \end{pmatrix}@\begin{pmatrix} \frac{1}{2} & 0 & 0 \end{pmatrix}$ & [11,-12,-35,36] \\
& b$_{25}$ & V$_{1}$ & V$_{1}$ & 5 & 1 & $\begin{pmatrix} 1 & -2 & 0 \end{pmatrix}@\begin{pmatrix} \frac{1}{2} & 0 & 0 \end{pmatrix}$ & [13,-14,-37,38] \\
& b$_{26}$ & V$_{1}$ & V$_{1}$ & 5 & 1 & $\begin{pmatrix} 0 & 1 & 2 \end{pmatrix}@\begin{pmatrix} 0 & \frac{1}{2} & 0 \end{pmatrix}$ & [15,-18,-39,42] \\
& b$_{27}$ & V$_{1}$ & V$_{1}$ & 5 & 1 & $\begin{pmatrix} 0 & 1 & -2 \end{pmatrix}@\begin{pmatrix} 0 & \frac{1}{2} & 0 \end{pmatrix}$ & [16,-17,-40,41] \\
& b$_{28}$ & V$_{1}$ & V$_{1}$ & 5 & 1 & $\begin{pmatrix} 1 & 0 & -2 \end{pmatrix}@\begin{pmatrix} \frac{1}{2} & 0 & 0 \end{pmatrix}$ & [21,-24,-45,48] \\ \hline
B$_{6}$ & b$_{29}$ & V$_{1}$ & V$_{1}$ & 6 & 1 & $\begin{pmatrix} 2 & 1 & 1 \end{pmatrix}@\begin{pmatrix} 0 & \frac{1}{2} & \frac{1}{2} \end{pmatrix}$ & [1,-10,-25,34] \\
& b$_{30}$ & V$_{1}$ & V$_{1}$ & 6 & 1 & $\begin{pmatrix} 2 & 1 & -1 \end{pmatrix}@\begin{pmatrix} 0 & \frac{1}{2} & \frac{1}{2} \end{pmatrix}$ & [-2,23,26,-47] \\
& b$_{31}$ & V$_{1}$ & V$_{1}$ & 6 & 1 & $\begin{pmatrix} 2 & -1 & -1 \end{pmatrix}@\begin{pmatrix} 0 & \frac{1}{2} & \frac{1}{2} \end{pmatrix}$ & [3,-7,-27,31] \\
& b$_{32}$ & V$_{1}$ & V$_{1}$ & 6 & 1 & $\begin{pmatrix} 2 & -1 & 1 \end{pmatrix}@\begin{pmatrix} 0 & \frac{1}{2} & \frac{1}{2} \end{pmatrix}$ & [-4,20,28,-44] \\
& b$_{33}$ & V$_{1}$ & V$_{1}$ & 6 & 1 & $\begin{pmatrix} 1 & 2 & -1 \end{pmatrix}@\begin{pmatrix} \frac{1}{2} & 0 & \frac{1}{2} \end{pmatrix}$ & [5,-12,-29,36] \\
& b$_{34}$ & V$_{1}$ & V$_{1}$ & 6 & 1 & $\begin{pmatrix} 1 & -1 & 2 \end{pmatrix}@\begin{pmatrix} \frac{1}{2} & \frac{1}{2} & 0 \end{pmatrix}$ & [6,-16,-30,40] \\
& b$_{35}$ & V$_{1}$ & V$_{1}$ & 6 & 1 & $\begin{pmatrix} 1 & 2 & 1 \end{pmatrix}@\begin{pmatrix} \frac{1}{2} & 0 & \frac{1}{2} \end{pmatrix}$ & [-8,11,32,-35] \\
& b$_{36}$ & V$_{1}$ & V$_{1}$ & 6 & 1 & $\begin{pmatrix} 1 & 1 & 2 \end{pmatrix}@\begin{pmatrix} \frac{1}{2} & \frac{1}{2} & 0 \end{pmatrix}$ & [-9,15,33,-39] \\
& b$_{37}$ & V$_{1}$ & V$_{1}$ & 6 & 1 & $\begin{pmatrix} 1 & -2 & -1 \end{pmatrix}@\begin{pmatrix} \frac{1}{2} & 0 & \frac{1}{2} \end{pmatrix}$ & [13,-19,-37,43] \\
& b$_{38}$ & V$_{1}$ & V$_{1}$ & 6 & 1 & $\begin{pmatrix} 1 & -2 & 1 \end{pmatrix}@\begin{pmatrix} \frac{1}{2} & 0 & \frac{1}{2} \end{pmatrix}$ & [-14,22,38,-46] \\
& b$_{39}$ & V$_{1}$ & V$_{1}$ & 6 & 1 & $\begin{pmatrix} 1 & 1 & -2 \end{pmatrix}@\begin{pmatrix} \frac{1}{2} & \frac{1}{2} & 0 \end{pmatrix}$ & [-17,21,41,-45] \\
& b$_{40}$ & V$_{1}$ & V$_{1}$ & 6 & 1 & $\begin{pmatrix} 1 & -1 & -2 \end{pmatrix}@\begin{pmatrix} \frac{1}{2} & \frac{1}{2} & 0 \end{pmatrix}$ & [18,-24,-42,48] \\
\end{longtable}
\end{center}

\item The SAMBs belonging to $A_{1g}$ irrep. are given as follows for which the bra-ket combination of atomic orbitals $M_{1}=\braket{d_{yz}, d_{zx}, d_{xy}|d_{yz}, d_{zx}, d_{xy}}$, and the site/bond clusters are indicated by square brackets.
There are 18 independent SAMBs in total.

\vspace{4mm}
\noindent \fbox{No. {1}} $\,\,\,\hat{\mathbb{Q}}_{0}^{(A_{1g})}$ [M$_{1}$,\,S$_{1}$]
\begin{align*}
\hat{\mathbb{Z}}_{1}=\mathbb{X}_{1}[\mathbb{Q}_{0}^{(a,A_{1g})}] \otimes\mathbb{Y}_{1}[\mathbb{Q}_{0}^{(s,A_{1g})}]
\end{align*}
\begin{align*}
\hat{\mathbb{Z}}_{1}(\bm{k})=\mathbb{X}_{1}[\mathbb{Q}_{0}^{(a,A_{1g})}] \otimes\mathbb{U}_{1}[\mathbb{Q}_{0}^{(s,A_{1g})}]
\end{align*}
\vspace{4mm}
\noindent \fbox{No. {2}} $\,\,\,\hat{\mathbb{Q}}_{0}^{(A_{1g})}$ [M$_{1}$,\,B$_{1}$]
\begin{align*}
\hat{\mathbb{Z}}_{2}=\mathbb{X}_{1}[\mathbb{Q}_{0}^{(a,A_{1g})}] \otimes\mathbb{Y}_{2}[\mathbb{Q}_{0}^{(b,A_{1g})}]
\end{align*}
\begin{align*}
\hat{\mathbb{Z}}_{2}(\bm{k})=\mathbb{X}_{1}[\mathbb{Q}_{0}^{(a,A_{1g})}] \otimes\mathbb{U}_{1}[\mathbb{Q}_{0}^{(s,A_{1g})}] \otimes\mathbb{F}_{1}[\mathbb{Q}_{0}^{(k,A_{1g})}]
\end{align*}
\vspace{4mm}
\noindent \fbox{No. {3}} $\,\,\,\hat{\mathbb{Q}}_{0}^{(A_{1g})}$ [M$_{1}$,\,B$_{1}$]
\begin{align*}
\hat{\mathbb{Z}}_{3}=\frac{\sqrt{2} \mathbb{X}_{2}[\mathbb{Q}_{2,0}^{(a,E_{g})}] \otimes\mathbb{Y}_{3}[\mathbb{Q}_{2,0}^{(b,E_{g})}]}{2} + \frac{\sqrt{2} \mathbb{X}_{3}[\mathbb{Q}_{2,1}^{(a,E_{g})}] \otimes\mathbb{Y}_{4}[\mathbb{Q}_{2,1}^{(b,E_{g})}]}{2}
\end{align*}
\begin{align*}
\hat{\mathbb{Z}}_{3}(\bm{k})=\frac{\sqrt{2} \mathbb{X}_{2}[\mathbb{Q}_{2,0}^{(a,E_{g})}] \otimes\mathbb{U}_{1}[\mathbb{Q}_{0}^{(s,A_{1g})}] \otimes\mathbb{F}_{2}[\mathbb{Q}_{2,0}^{(k,E_{g})}]}{2} + \frac{\sqrt{2} \mathbb{X}_{3}[\mathbb{Q}_{2,1}^{(a,E_{g})}] \otimes\mathbb{U}_{1}[\mathbb{Q}_{0}^{(s,A_{1g})}] \otimes\mathbb{F}_{3}[\mathbb{Q}_{2,1}^{(k,E_{g})}]}{2}
\end{align*}
\vspace{4mm}
\noindent \fbox{No. {4}} $\,\,\,\hat{\mathbb{Q}}_{0}^{(A_{1g})}$ [M$_{1}$,\,B$_{2}$]
\begin{align*}
\hat{\mathbb{Z}}_{4}=\mathbb{X}_{1}[\mathbb{Q}_{0}^{(a,A_{1g})}] \otimes\mathbb{Y}_{5}[\mathbb{Q}_{0}^{(b,A_{1g})}]
\end{align*}
\begin{align*}
\hat{\mathbb{Z}}_{4}(\bm{k})=\mathbb{X}_{1}[\mathbb{Q}_{0}^{(a,A_{1g})}] \otimes\mathbb{U}_{1}[\mathbb{Q}_{0}^{(s,A_{1g})}] \otimes\mathbb{F}_{4}[\mathbb{Q}_{0}^{(k,A_{1g})}]
\end{align*}
\vspace{4mm}
\noindent \fbox{No. {5}} $\,\,\,\hat{\mathbb{Q}}_{0}^{(A_{1g})}$ [M$_{1}$,\,B$_{2}$]
\begin{align*}
&
\hat{\mathbb{Z}}_{5}=\frac{\sqrt{5} \mathbb{X}_{2}[\mathbb{Q}_{2,0}^{(a,E_{g})}] \otimes\mathbb{Y}_{6}[\mathbb{Q}_{2,0}^{(b,E_{g})}]}{5} + \frac{\sqrt{5} \mathbb{X}_{3}[\mathbb{Q}_{2,1}^{(a,E_{g})}] \otimes\mathbb{Y}_{7}[\mathbb{Q}_{2,1}^{(b,E_{g})}]}{5} + \frac{\sqrt{5} \mathbb{X}_{4}[\mathbb{Q}_{2,0}^{(a,T_{2g})}] \otimes\mathbb{Y}_{8}[\mathbb{Q}_{2,0}^{(b,T_{2g})}]}{5}
\cr&\hspace{1cm}
 + \frac{\sqrt{5} \mathbb{X}_{5}[\mathbb{Q}_{2,1}^{(a,T_{2g})}] \otimes\mathbb{Y}_{9}[\mathbb{Q}_{2,1}^{(b,T_{2g})}]}{5} + \frac{\sqrt{5} \mathbb{X}_{6}[\mathbb{Q}_{2,2}^{(a,T_{2g})}] \otimes\mathbb{Y}_{10}[\mathbb{Q}_{2,2}^{(b,T_{2g})}]}{5}
\end{align*}
\begin{align*}
&
\hat{\mathbb{Z}}_{5}(\bm{k})=\frac{\sqrt{5} \mathbb{X}_{2}[\mathbb{Q}_{2,0}^{(a,E_{g})}] \otimes\mathbb{U}_{1}[\mathbb{Q}_{0}^{(s,A_{1g})}] \otimes\mathbb{F}_{5}[\mathbb{Q}_{2,0}^{(k,E_{g})}]}{5} + \frac{\sqrt{5} \mathbb{X}_{3}[\mathbb{Q}_{2,1}^{(a,E_{g})}] \otimes\mathbb{U}_{1}[\mathbb{Q}_{0}^{(s,A_{1g})}] \otimes\mathbb{F}_{6}[\mathbb{Q}_{2,1}^{(k,E_{g})}]}{5}
\cr&\hspace{1cm}
 + \frac{\sqrt{5} \mathbb{X}_{4}[\mathbb{Q}_{2,0}^{(a,T_{2g})}] \otimes\mathbb{U}_{1}[\mathbb{Q}_{0}^{(s,A_{1g})}] \otimes\mathbb{F}_{7}[\mathbb{Q}_{2,0}^{(k,T_{2g})}]}{5}
 + \frac{\sqrt{5} \mathbb{X}_{5}[\mathbb{Q}_{2,1}^{(a,T_{2g})}] \otimes\mathbb{U}_{1}[\mathbb{Q}_{0}^{(s,A_{1g})}] \otimes\mathbb{F}_{8}[\mathbb{Q}_{2,1}^{(k,T_{2g})}]}{5}
 \cr&\hspace{1cm}
 + \frac{\sqrt{5} \mathbb{X}_{6}[\mathbb{Q}_{2,2}^{(a,T_{2g})}] \otimes\mathbb{U}_{1}[\mathbb{Q}_{0}^{(s,A_{1g})}] \otimes\mathbb{F}_{9}[\mathbb{Q}_{2,2}^{(k,T_{2g})}]}{5}
\end{align*}
\vspace{4mm}
\noindent \fbox{No. {6}} $\,\,\,\hat{\mathbb{Q}}_{4}^{(A_{1g})}$ [M$_{1}$,\,B$_{2}$]
\begin{align*}
&
\hat{\mathbb{Z}}_{6}=\frac{\sqrt{30} \mathbb{X}_{2}[\mathbb{Q}_{2,0}^{(a,E_{g})}] \otimes\mathbb{Y}_{6}[\mathbb{Q}_{2,0}^{(b,E_{g})}]}{10} + \frac{\sqrt{30} \mathbb{X}_{3}[\mathbb{Q}_{2,1}^{(a,E_{g})}] \otimes\mathbb{Y}_{7}[\mathbb{Q}_{2,1}^{(b,E_{g})}]}{10} - \frac{\sqrt{30} \mathbb{X}_{4}[\mathbb{Q}_{2,0}^{(a,T_{2g})}] \otimes\mathbb{Y}_{8}[\mathbb{Q}_{2,0}^{(b,T_{2g})}]}{15}
\cr&\hspace{1cm}
 - \frac{\sqrt{30} \mathbb{X}_{5}[\mathbb{Q}_{2,1}^{(a,T_{2g})}] \otimes\mathbb{Y}_{9}[\mathbb{Q}_{2,1}^{(b,T_{2g})}]}{15} - \frac{\sqrt{30} \mathbb{X}_{6}[\mathbb{Q}_{2,2}^{(a,T_{2g})}] \otimes\mathbb{Y}_{10}[\mathbb{Q}_{2,2}^{(b,T_{2g})}]}{15}
\end{align*}
\begin{align*}
&
\hat{\mathbb{Z}}_{6}(\bm{k})=\frac{\sqrt{30} \mathbb{X}_{2}[\mathbb{Q}_{2,0}^{(a,E_{g})}] \otimes\mathbb{U}_{1}[\mathbb{Q}_{0}^{(s,A_{1g})}] \otimes\mathbb{F}_{5}[\mathbb{Q}_{2,0}^{(k,E_{g})}]}{10} + \frac{\sqrt{30} \mathbb{X}_{3}[\mathbb{Q}_{2,1}^{(a,E_{g})}] \otimes\mathbb{U}_{1}[\mathbb{Q}_{0}^{(s,A_{1g})}] \otimes\mathbb{F}_{6}[\mathbb{Q}_{2,1}^{(k,E_{g})}]}{10}
\cr&\hspace{1cm}
 - \frac{\sqrt{30} \mathbb{X}_{4}[\mathbb{Q}_{2,0}^{(a,T_{2g})}] \otimes\mathbb{U}_{1}[\mathbb{Q}_{0}^{(s,A_{1g})}] \otimes\mathbb{F}_{7}[\mathbb{Q}_{2,0}^{(k,T_{2g})}]}{15} - \frac{\sqrt{30} \mathbb{X}_{5}[\mathbb{Q}_{2,1}^{(a,T_{2g})}] \otimes\mathbb{U}_{1}[\mathbb{Q}_{0}^{(s,A_{1g})}] \otimes\mathbb{F}_{8}[\mathbb{Q}_{2,1}^{(k,T_{2g})}]}{15}
 \cr&\hspace{1cm}
 - \frac{\sqrt{30} \mathbb{X}_{6}[\mathbb{Q}_{2,2}^{(a,T_{2g})}] \otimes\mathbb{U}_{1}[\mathbb{Q}_{0}^{(s,A_{1g})}] \otimes\mathbb{F}_{9}[\mathbb{Q}_{2,2}^{(k,T_{2g})}]}{15}
\end{align*}
\vspace{4mm}
\noindent \fbox{No. {7}} $\,\,\,\hat{\mathbb{Q}}_{0}^{(A_{1g})}$ [M$_{1}$,\,B$_{3}$]
\begin{align*}
\hat{\mathbb{Z}}_{7}=\mathbb{X}_{1}[\mathbb{Q}_{0}^{(a,A_{1g})}] \otimes\mathbb{Y}_{11}[\mathbb{Q}_{0}^{(b,A_{1g})}]
\end{align*}
\begin{align*}
\hat{\mathbb{Z}}_{7}(\bm{k})=\mathbb{X}_{1}[\mathbb{Q}_{0}^{(a,A_{1g})}] \otimes\mathbb{U}_{1}[\mathbb{Q}_{0}^{(s,A_{1g})}] \otimes\mathbb{F}_{10}[\mathbb{Q}_{0}^{(k,A_{1g})}]
\end{align*}
\vspace{4mm}
\noindent \fbox{No. {8}} $\,\,\,\hat{\mathbb{Q}}_{0}^{(A_{1g})}$ [M$_{1}$,\,B$_{3}$]
\begin{align*}
\hat{\mathbb{Z}}_{8}=\frac{\sqrt{3} \mathbb{X}_{4}[\mathbb{Q}_{2,0}^{(a,T_{2g})}] \otimes\mathbb{Y}_{12}[\mathbb{Q}_{2,0}^{(b,T_{2g})}]}{3} + \frac{\sqrt{3} \mathbb{X}_{5}[\mathbb{Q}_{2,1}^{(a,T_{2g})}] \otimes\mathbb{Y}_{13}[\mathbb{Q}_{2,1}^{(b,T_{2g})}]}{3} + \frac{\sqrt{3} \mathbb{X}_{6}[\mathbb{Q}_{2,2}^{(a,T_{2g})}] \otimes\mathbb{Y}_{14}[\mathbb{Q}_{2,2}^{(b,T_{2g})}]}{3}
\end{align*}
\begin{align*}
&
\hat{\mathbb{Z}}_{8}(\bm{k})=\frac{\sqrt{3} \mathbb{X}_{4}[\mathbb{Q}_{2,0}^{(a,T_{2g})}] \otimes\mathbb{U}_{1}[\mathbb{Q}_{0}^{(s,A_{1g})}] \otimes\mathbb{F}_{11}[\mathbb{Q}_{2,0}^{(k,T_{2g})}]}{3} + \frac{\sqrt{3} \mathbb{X}_{5}[\mathbb{Q}_{2,1}^{(a,T_{2g})}] \otimes\mathbb{U}_{1}[\mathbb{Q}_{0}^{(s,A_{1g})}] \otimes\mathbb{F}_{12}[\mathbb{Q}_{2,1}^{(k,T_{2g})}]}{3}
\cr&\hspace{1cm}
 + \frac{\sqrt{3} \mathbb{X}_{6}[\mathbb{Q}_{2,2}^{(a,T_{2g})}] \otimes\mathbb{U}_{1}[\mathbb{Q}_{0}^{(s,A_{1g})}] \otimes\mathbb{F}_{13}[\mathbb{Q}_{2,2}^{(k,T_{2g})}]}{3}
\end{align*}
\vspace{4mm}
\noindent \fbox{No. {9}} $\,\,\,\hat{\mathbb{Q}}_{0}^{(A_{1g})}$ [M$_{1}$,\,B$_{4}$]
\begin{align*}
\hat{\mathbb{Z}}_{9}=\mathbb{X}_{1}[\mathbb{Q}_{0}^{(a,A_{1g})}] \otimes\mathbb{Y}_{15}[\mathbb{Q}_{0}^{(b,A_{1g})}]
\end{align*}
\begin{align*}
\hat{\mathbb{Z}}_{9}(\bm{k})=\mathbb{X}_{1}[\mathbb{Q}_{0}^{(a,A_{1g})}] \otimes\mathbb{U}_{1}[\mathbb{Q}_{0}^{(s,A_{1g})}] \otimes\mathbb{F}_{14}[\mathbb{Q}_{0}^{(k,A_{1g})}]
\end{align*}
\vspace{4mm}
\noindent \fbox{No. {10}} $\,\,\,\hat{\mathbb{Q}}_{0}^{(A_{1g})}$ [M$_{1}$,\,B$_{4}$]
\begin{align*}
\hat{\mathbb{Z}}_{10}=\frac{\sqrt{2} \mathbb{X}_{2}[\mathbb{Q}_{2,0}^{(a,E_{g})}] \otimes\mathbb{Y}_{16}[\mathbb{Q}_{2,0}^{(b,E_{g})}]}{2} + \frac{\sqrt{2} \mathbb{X}_{3}[\mathbb{Q}_{2,1}^{(a,E_{g})}] \otimes\mathbb{Y}_{17}[\mathbb{Q}_{2,1}^{(b,E_{g})}]}{2}
\end{align*}
\begin{align*}
\hat{\mathbb{Z}}_{10}(\bm{k})=\frac{\sqrt{2} \mathbb{X}_{2}[\mathbb{Q}_{2,0}^{(a,E_{g})}] \otimes\mathbb{U}_{1}[\mathbb{Q}_{0}^{(s,A_{1g})}] \otimes\mathbb{F}_{15}[\mathbb{Q}_{2,0}^{(k,E_{g})}]}{2} + \frac{\sqrt{2} \mathbb{X}_{3}[\mathbb{Q}_{2,1}^{(a,E_{g})}] \otimes\mathbb{U}_{1}[\mathbb{Q}_{0}^{(s,A_{1g})}] \otimes\mathbb{F}_{16}[\mathbb{Q}_{2,1}^{(k,E_{g})}]}{2}
\end{align*}
\vspace{4mm}
\noindent \fbox{No. {11}} $\,\,\,\hat{\mathbb{Q}}_{0}^{(A_{1g})}$ [M$_{1}$,\,B$_{5}$]
\begin{align*}
\hat{\mathbb{Z}}_{11}=\mathbb{X}_{1}[\mathbb{Q}_{0}^{(a,A_{1g})}] \otimes\mathbb{Y}_{18}[\mathbb{Q}_{0}^{(b,A_{1g})}]
\end{align*}
\begin{align*}
\hat{\mathbb{Z}}_{11}(\bm{k})=\mathbb{X}_{1}[\mathbb{Q}_{0}^{(a,A_{1g})}] \otimes\mathbb{U}_{1}[\mathbb{Q}_{0}^{(s,A_{1g})}] \otimes\mathbb{F}_{17}[\mathbb{Q}_{0}^{(k,A_{1g})}]
\end{align*}
\vspace{4mm}
\noindent \fbox{No. {12}} $\,\,\,\hat{\mathbb{Q}}_{0}^{(A_{1g})}$ [M$_{1}$,\,B$_{5}$]
\begin{align*}
&
\hat{\mathbb{Z}}_{12}=\frac{\sqrt{5} \mathbb{X}_{2}[\mathbb{Q}_{2,0}^{(a,E_{g})}] \otimes\mathbb{Y}_{19}[\mathbb{Q}_{2,0}^{(b,E_{g})}]}{5} + \frac{\sqrt{5} \mathbb{X}_{3}[\mathbb{Q}_{2,1}^{(a,E_{g})}] \otimes\mathbb{Y}_{20}[\mathbb{Q}_{2,1}^{(b,E_{g})}]}{5} + \frac{\sqrt{5} \mathbb{X}_{4}[\mathbb{Q}_{2,0}^{(a,T_{2g})}] \otimes\mathbb{Y}_{21}[\mathbb{Q}_{2,0}^{(b,T_{2g})}]}{5}
\cr&\hspace{1cm}
 + \frac{\sqrt{5} \mathbb{X}_{5}[\mathbb{Q}_{2,1}^{(a,T_{2g})}] \otimes\mathbb{Y}_{22}[\mathbb{Q}_{2,1}^{(b,T_{2g})}]}{5} + \frac{\sqrt{5} \mathbb{X}_{6}[\mathbb{Q}_{2,2}^{(a,T_{2g})}] \otimes\mathbb{Y}_{23}[\mathbb{Q}_{2,2}^{(b,T_{2g})}]}{5}
\end{align*}
\begin{align*}
&
\hat{\mathbb{Z}}_{12}(\bm{k})=\frac{\sqrt{5} \mathbb{X}_{2}[\mathbb{Q}_{2,0}^{(a,E_{g})}] \otimes\mathbb{U}_{1}[\mathbb{Q}_{0}^{(s,A_{1g})}] \otimes\mathbb{F}_{18}[\mathbb{Q}_{2,0}^{(k,E_{g})}]}{5} + \frac{\sqrt{5} \mathbb{X}_{3}[\mathbb{Q}_{2,1}^{(a,E_{g})}] \otimes\mathbb{U}_{1}[\mathbb{Q}_{0}^{(s,A_{1g})}] \otimes\mathbb{F}_{19}[\mathbb{Q}_{2,1}^{(k,E_{g})}]}{5}
\cr&\hspace{1cm}
 + \frac{\sqrt{5} \mathbb{X}_{4}[\mathbb{Q}_{2,0}^{(a,T_{2g})}] \otimes\mathbb{U}_{1}[\mathbb{Q}_{0}^{(s,A_{1g})}] \otimes\mathbb{F}_{20}[\mathbb{Q}_{2,0}^{(k,T_{2g})}]}{5} + \frac{\sqrt{5} \mathbb{X}_{5}[\mathbb{Q}_{2,1}^{(a,T_{2g})}] \otimes\mathbb{U}_{1}[\mathbb{Q}_{0}^{(s,A_{1g})}] \otimes\mathbb{F}_{21}[\mathbb{Q}_{2,1}^{(k,T_{2g})}]}{5}
 \cr&\hspace{1cm}
 + \frac{\sqrt{5} \mathbb{X}_{6}[\mathbb{Q}_{2,2}^{(a,T_{2g})}] \otimes\mathbb{U}_{1}[\mathbb{Q}_{0}^{(s,A_{1g})}] \otimes\mathbb{F}_{22}[\mathbb{Q}_{2,2}^{(k,T_{2g})}]}{5}
\end{align*}
\vspace{4mm}
\noindent \fbox{No. {13}} $\,\,\,\hat{\mathbb{Q}}_{4}^{(A_{1g})}$ [M$_{1}$,\,B$_{5}$]
\begin{align*}
&
\hat{\mathbb{Z}}_{13}=\frac{\sqrt{30} \mathbb{X}_{2}[\mathbb{Q}_{2,0}^{(a,E_{g})}] \otimes\mathbb{Y}_{19}[\mathbb{Q}_{2,0}^{(b,E_{g})}]}{10} + \frac{\sqrt{30} \mathbb{X}_{3}[\mathbb{Q}_{2,1}^{(a,E_{g})}] \otimes\mathbb{Y}_{20}[\mathbb{Q}_{2,1}^{(b,E_{g})}]}{10} - \frac{\sqrt{30} \mathbb{X}_{4}[\mathbb{Q}_{2,0}^{(a,T_{2g})}] \otimes\mathbb{Y}_{21}[\mathbb{Q}_{2,0}^{(b,T_{2g})}]}{15}
\cr&\hspace{1cm}
 - \frac{\sqrt{30} \mathbb{X}_{5}[\mathbb{Q}_{2,1}^{(a,T_{2g})}] \otimes\mathbb{Y}_{22}[\mathbb{Q}_{2,1}^{(b,T_{2g})}]}{15} - \frac{\sqrt{30} \mathbb{X}_{6}[\mathbb{Q}_{2,2}^{(a,T_{2g})}] \otimes\mathbb{Y}_{23}[\mathbb{Q}_{2,2}^{(b,T_{2g})}]}{15}
\end{align*}
\begin{align*}
&
\hat{\mathbb{Z}}_{13}(\bm{k})=\frac{\sqrt{30} \mathbb{X}_{2}[\mathbb{Q}_{2,0}^{(a,E_{g})}] \otimes\mathbb{U}_{1}[\mathbb{Q}_{0}^{(s,A_{1g})}] \otimes\mathbb{F}_{18}[\mathbb{Q}_{2,0}^{(k,E_{g})}]}{10} + \frac{\sqrt{30} \mathbb{X}_{3}[\mathbb{Q}_{2,1}^{(a,E_{g})}] \otimes\mathbb{U}_{1}[\mathbb{Q}_{0}^{(s,A_{1g})}] \otimes\mathbb{F}_{19}[\mathbb{Q}_{2,1}^{(k,E_{g})}]}{10}
\cr&\hspace{1cm}
 - \frac{\sqrt{30} \mathbb{X}_{4}[\mathbb{Q}_{2,0}^{(a,T_{2g})}] \otimes\mathbb{U}_{1}[\mathbb{Q}_{0}^{(s,A_{1g})}] \otimes\mathbb{F}_{20}[\mathbb{Q}_{2,0}^{(k,T_{2g})}]}{15} - \frac{\sqrt{30} \mathbb{X}_{5}[\mathbb{Q}_{2,1}^{(a,T_{2g})}] \otimes\mathbb{U}_{1}[\mathbb{Q}_{0}^{(s,A_{1g})}] \otimes\mathbb{F}_{21}[\mathbb{Q}_{2,1}^{(k,T_{2g})}]}{15}
 \cr&\hspace{1cm}
 - \frac{\sqrt{30} \mathbb{X}_{6}[\mathbb{Q}_{2,2}^{(a,T_{2g})}] \otimes\mathbb{U}_{1}[\mathbb{Q}_{0}^{(s,A_{1g})}] \otimes\mathbb{F}_{22}[\mathbb{Q}_{2,2}^{(k,T_{2g})}]}{15}
\end{align*}
\vspace{4mm}
\noindent \fbox{No. {14}} $\,\,\,\hat{\mathbb{Q}}_{4}^{(A_{1g})}$ [M$_{1}$,\,B$_{5}$]
\begin{align*}
\hat{\mathbb{Z}}_{14}=\frac{\sqrt{2} \mathbb{X}_{2}[\mathbb{Q}_{2,0}^{(a,E_{g})}] \otimes\mathbb{Y}_{24}[\mathbb{Q}_{4,0}^{(b,E_{g})}]}{2} + \frac{\sqrt{2} \mathbb{X}_{3}[\mathbb{Q}_{2,1}^{(a,E_{g})}] \otimes\mathbb{Y}_{25}[\mathbb{Q}_{4,1}^{(b,E_{g})}]}{2}
\end{align*}
\begin{align*}
\hat{\mathbb{Z}}_{14}(\bm{k})=\frac{\sqrt{2} \mathbb{X}_{2}[\mathbb{Q}_{2,0}^{(a,E_{g})}] \otimes\mathbb{U}_{1}[\mathbb{Q}_{0}^{(s,A_{1g})}] \otimes\mathbb{F}_{23}[\mathbb{Q}_{4,0}^{(k,E_{g})}]}{2} + \frac{\sqrt{2} \mathbb{X}_{3}[\mathbb{Q}_{2,1}^{(a,E_{g})}] \otimes\mathbb{U}_{1}[\mathbb{Q}_{0}^{(s,A_{1g})}] \otimes\mathbb{F}_{24}[\mathbb{Q}_{4,1}^{(k,E_{g})}]}{2}
\end{align*}
\vspace{4mm}
\noindent \fbox{No. {15}} $\,\,\,\hat{\mathbb{Q}}_{0}^{(A_{1g})}$ [M$_{1}$,\,B$_{6}$]
\begin{align*}
\hat{\mathbb{Z}}_{15}=\mathbb{X}_{1}[\mathbb{Q}_{0}^{(a,A_{1g})}] \otimes\mathbb{Y}_{26}[\mathbb{Q}_{0}^{(b,A_{1g})}]
\end{align*}
\begin{align*}
\hat{\mathbb{Z}}_{15}(\bm{k})=\mathbb{X}_{1}[\mathbb{Q}_{0}^{(a,A_{1g})}] \otimes\mathbb{U}_{1}[\mathbb{Q}_{0}^{(s,A_{1g})}] \otimes\mathbb{F}_{25}[\mathbb{Q}_{0}^{(k,A_{1g})}]
\end{align*}
\vspace{4mm}
\noindent \fbox{No. {16}} $\,\,\,\hat{\mathbb{Q}}_{0}^{(A_{1g})}$ [M$_{1}$,\,B$_{6}$]
\begin{align*}
&
\hat{\mathbb{Z}}_{16}=\frac{\sqrt{5} \mathbb{X}_{2}[\mathbb{Q}_{2,0}^{(a,E_{g})}] \otimes\mathbb{Y}_{27}[\mathbb{Q}_{2,0}^{(b,E_{g})}]}{5} + \frac{\sqrt{5} \mathbb{X}_{3}[\mathbb{Q}_{2,1}^{(a,E_{g})}] \otimes\mathbb{Y}_{28}[\mathbb{Q}_{2,1}^{(b,E_{g})}]}{5} + \frac{\sqrt{5} \mathbb{X}_{4}[\mathbb{Q}_{2,0}^{(a,T_{2g})}] \otimes\mathbb{Y}_{29}[\mathbb{Q}_{2,0}^{(b,T_{2g})}]}{5}
\cr&\hspace{1cm}
 + \frac{\sqrt{5} \mathbb{X}_{5}[\mathbb{Q}_{2,1}^{(a,T_{2g})}] \otimes\mathbb{Y}_{30}[\mathbb{Q}_{2,1}^{(b,T_{2g})}]}{5} + \frac{\sqrt{5} \mathbb{X}_{6}[\mathbb{Q}_{2,2}^{(a,T_{2g})}] \otimes\mathbb{Y}_{31}[\mathbb{Q}_{2,2}^{(b,T_{2g})}]}{5}
\end{align*}
\begin{align*}
&
\hat{\mathbb{Z}}_{16}(\bm{k})=\frac{\sqrt{5} \mathbb{X}_{2}[\mathbb{Q}_{2,0}^{(a,E_{g})}] \otimes\mathbb{U}_{1}[\mathbb{Q}_{0}^{(s,A_{1g})}] \otimes\mathbb{F}_{26}[\mathbb{Q}_{2,0}^{(k,E_{g})}]}{5} + \frac{\sqrt{5} \mathbb{X}_{3}[\mathbb{Q}_{2,1}^{(a,E_{g})}] \otimes\mathbb{U}_{1}[\mathbb{Q}_{0}^{(s,A_{1g})}] \otimes\mathbb{F}_{27}[\mathbb{Q}_{2,1}^{(k,E_{g})}]}{5}
\cr&\hspace{1cm}
 + \frac{\sqrt{5} \mathbb{X}_{4}[\mathbb{Q}_{2,0}^{(a,T_{2g})}] \otimes\mathbb{U}_{1}[\mathbb{Q}_{0}^{(s,A_{1g})}] \otimes\mathbb{F}_{28}[\mathbb{Q}_{2,0}^{(k,T_{2g})}]}{5} + \frac{\sqrt{5} \mathbb{X}_{5}[\mathbb{Q}_{2,1}^{(a,T_{2g})}] \otimes\mathbb{U}_{1}[\mathbb{Q}_{0}^{(s,A_{1g})}] \otimes\mathbb{F}_{29}[\mathbb{Q}_{2,1}^{(k,T_{2g})}]}{5}
 \cr&\hspace{1cm}
 + \frac{\sqrt{5} \mathbb{X}_{6}[\mathbb{Q}_{2,2}^{(a,T_{2g})}] \otimes\mathbb{U}_{1}[\mathbb{Q}_{0}^{(s,A_{1g})}] \otimes\mathbb{F}_{30}[\mathbb{Q}_{2,2}^{(k,T_{2g})}]}{5}
\end{align*}
\vspace{4mm}
\noindent \fbox{No. {17}} $\,\,\,\hat{\mathbb{Q}}_{4}^{(A_{1g})}$ [M$_{1}$,\,B$_{6}$]
\begin{align*}
&
\hat{\mathbb{Z}}_{17}=\frac{\sqrt{30} \mathbb{X}_{2}[\mathbb{Q}_{2,0}^{(a,E_{g})}] \otimes\mathbb{Y}_{27}[\mathbb{Q}_{2,0}^{(b,E_{g})}]}{10} + \frac{\sqrt{30} \mathbb{X}_{3}[\mathbb{Q}_{2,1}^{(a,E_{g})}] \otimes\mathbb{Y}_{28}[\mathbb{Q}_{2,1}^{(b,E_{g})}]}{10} - \frac{\sqrt{30} \mathbb{X}_{4}[\mathbb{Q}_{2,0}^{(a,T_{2g})}] \otimes\mathbb{Y}_{29}[\mathbb{Q}_{2,0}^{(b,T_{2g})}]}{15}
\cr&\hspace{1cm}
 - \frac{\sqrt{30} \mathbb{X}_{5}[\mathbb{Q}_{2,1}^{(a,T_{2g})}] \otimes\mathbb{Y}_{30}[\mathbb{Q}_{2,1}^{(b,T_{2g})}]}{15} - \frac{\sqrt{30} \mathbb{X}_{6}[\mathbb{Q}_{2,2}^{(a,T_{2g})}] \otimes\mathbb{Y}_{31}[\mathbb{Q}_{2,2}^{(b,T_{2g})}]}{15}
\end{align*}
\begin{align*}
&
\hat{\mathbb{Z}}_{17}(\bm{k})=\frac{\sqrt{30} \mathbb{X}_{2}[\mathbb{Q}_{2,0}^{(a,E_{g})}] \otimes\mathbb{U}_{1}[\mathbb{Q}_{0}^{(s,A_{1g})}] \otimes\mathbb{F}_{26}[\mathbb{Q}_{2,0}^{(k,E_{g})}]}{10} + \frac{\sqrt{30} \mathbb{X}_{3}[\mathbb{Q}_{2,1}^{(a,E_{g})}] \otimes\mathbb{U}_{1}[\mathbb{Q}_{0}^{(s,A_{1g})}] \otimes\mathbb{F}_{27}[\mathbb{Q}_{2,1}^{(k,E_{g})}]}{10}
\cr&\hspace{1cm}
 - \frac{\sqrt{30} \mathbb{X}_{4}[\mathbb{Q}_{2,0}^{(a,T_{2g})}] \otimes\mathbb{U}_{1}[\mathbb{Q}_{0}^{(s,A_{1g})}] \otimes\mathbb{F}_{28}[\mathbb{Q}_{2,0}^{(k,T_{2g})}]}{15} - \frac{\sqrt{30} \mathbb{X}_{5}[\mathbb{Q}_{2,1}^{(a,T_{2g})}] \otimes\mathbb{U}_{1}[\mathbb{Q}_{0}^{(s,A_{1g})}] \otimes\mathbb{F}_{29}[\mathbb{Q}_{2,1}^{(k,T_{2g})}]}{15}
 \cr&\hspace{1cm}
 - \frac{\sqrt{30} \mathbb{X}_{6}[\mathbb{Q}_{2,2}^{(a,T_{2g})}] \otimes\mathbb{U}_{1}[\mathbb{Q}_{0}^{(s,A_{1g})}] \otimes\mathbb{F}_{30}[\mathbb{Q}_{2,2}^{(k,T_{2g})}]}{15}
\end{align*}
\vspace{4mm}
\noindent \fbox{No. {18}} $\,\,\,\hat{\mathbb{Q}}_{4}^{(A_{1g})}$ [M$_{1}$,\,B$_{6}$]
\begin{align*}
\hat{\mathbb{Z}}_{18}=- \frac{\sqrt{3} \mathbb{X}_{4}[\mathbb{Q}_{2,0}^{(a,T_{2g})}] \otimes\mathbb{Y}_{32}[\mathbb{Q}_{4,0}^{(b,T_{2g})}]}{3} - \frac{\sqrt{3} \mathbb{X}_{5}[\mathbb{Q}_{2,1}^{(a,T_{2g})}] \otimes\mathbb{Y}_{33}[\mathbb{Q}_{4,1}^{(b,T_{2g})}]}{3} - \frac{\sqrt{3} \mathbb{X}_{6}[\mathbb{Q}_{2,2}^{(a,T_{2g})}] \otimes\mathbb{Y}_{34}[\mathbb{Q}_{4,2}^{(b,T_{2g})}]}{3}
\end{align*}
\begin{align*}
&
\hat{\mathbb{Z}}_{18}(\bm{k})=- \frac{\sqrt{3} \mathbb{X}_{4}[\mathbb{Q}_{2,0}^{(a,T_{2g})}] \otimes\mathbb{U}_{1}[\mathbb{Q}_{0}^{(s,A_{1g})}] \otimes\mathbb{F}_{31}[\mathbb{Q}_{4,0}^{(k,T_{2g})}]}{3} - \frac{\sqrt{3} \mathbb{X}_{5}[\mathbb{Q}_{2,1}^{(a,T_{2g})}] \otimes\mathbb{U}_{1}[\mathbb{Q}_{0}^{(s,A_{1g})}] \otimes\mathbb{F}_{32}[\mathbb{Q}_{4,1}^{(k,T_{2g})}]}{3}
\cr&\hspace{1cm}
 - \frac{\sqrt{3} \mathbb{X}_{6}[\mathbb{Q}_{2,2}^{(a,T_{2g})}] \otimes\mathbb{U}_{1}[\mathbb{Q}_{0}^{(s,A_{1g})}] \otimes\mathbb{F}_{33}[\mathbb{Q}_{4,2}^{(k,T_{2g})}]}{3}
\end{align*}

\item The atomic SAMBs are given in Table IV.

\begin{center}
\renewcommand{\arraystretch}{1.3}
\begin{longtable}{c|c|c|c}
\caption{Atomic SAMB.}
 \\
 \hline \hline
symbol & type & group & form \\ \hline \endfirsthead

\multicolumn{3}{l}{\tablename\ \thetable{}} \\
 \hline \hline
symbol & type & group & form \\ \hline \endhead

 \hline \hline
\multicolumn{3}{r}{\footnotesize\it continued ...} \\ \endfoot

 \hline \hline
\multicolumn{3}{r}{} \\ \endlastfoot

$ \mathbb{X}_{1} $ & $\mathbb{Q}_{0}^{(a,A_{1g})}$ & M$_{1}$ & $\begin{pmatrix} \frac{\sqrt{3}}{3} & 0 & 0 \\ 0 & \frac{\sqrt{3}}{3} & 0 \\ 0 & 0 & \frac{\sqrt{3}}{3} \end{pmatrix}$ \\
$ \mathbb{X}_{2} $ & $\mathbb{Q}_{2,0}^{(a,E_{g})}$ & M$_{1}$ & $\begin{pmatrix} \frac{\sqrt{6}}{6} & 0 & 0 \\ 0 & \frac{\sqrt{6}}{6} & 0 \\ 0 & 0 & - \frac{\sqrt{6}}{3} \end{pmatrix}$ \\
$ \mathbb{X}_{3} $ & $\mathbb{Q}_{2,1}^{(a,E_{g})}$ & M$_{1}$ & $\begin{pmatrix} - \frac{\sqrt{2}}{2} & 0 & 0 \\ 0 & \frac{\sqrt{2}}{2} & 0 \\ 0 & 0 & 0 \end{pmatrix}$ \\
$ \mathbb{X}_{4} $ & $\mathbb{Q}_{2,0}^{(a,T_{2g})}$ & M$_{1}$ & $\begin{pmatrix} 0 & 0 & 0 \\ 0 & 0 & \frac{\sqrt{2}}{2} \\ 0 & \frac{\sqrt{2}}{2} & 0 \end{pmatrix}$ \\
$ \mathbb{X}_{5} $ & $\mathbb{Q}_{2,1}^{(a,T_{2g})}$ & M$_{1}$ & $\begin{pmatrix} 0 & 0 & \frac{\sqrt{2}}{2} \\ 0 & 0 & 0 \\ \frac{\sqrt{2}}{2} & 0 & 0 \end{pmatrix}$ \\
$ \mathbb{X}_{6} $ & $\mathbb{Q}_{2,2}^{(a,T_{2g})}$ & M$_{1}$ & $\begin{pmatrix} 0 & \frac{\sqrt{2}}{2} & 0 \\ \frac{\sqrt{2}}{2} & 0 & 0 \\ 0 & 0 & 0 \end{pmatrix}$ \\
\end{longtable}
\end{center}

\item The site/bond cluster SAMBs are given in Table V.

\begin{center}
\renewcommand{\arraystretch}{1.3}
\begin{longtable}{c|c|c|c}
\caption{Cluster SAMB.}
 \\
 \hline \hline
symbol & type & cluster & form \\ \hline \endfirsthead

\multicolumn{3}{l}{\tablename\ \thetable{}} \\
 \hline \hline
symbol & type & cluster & form \\ \hline \endhead

 \hline \hline
\multicolumn{3}{r}{\footnotesize\it continued ...} \\ \endfoot

 \hline \hline
\multicolumn{3}{r}{} \\ \endlastfoot

$ \mathbb{Y}_{1} $ & $\mathbb{Q}_{0}^{(s,A_{1g})}$ & S$_{1}$ & $\begin{pmatrix} 1 \end{pmatrix}$ \\ \hline
$ \mathbb{Y}_{2} $ & $\mathbb{Q}_{0}^{(b,A_{1g})}$ & B$_{1}$ & $\begin{pmatrix} \frac{\sqrt{3}}{3} & \frac{\sqrt{3}}{3} & \frac{\sqrt{3}}{3} \end{pmatrix}$ \\
$ \mathbb{Y}_{3} $ & $\mathbb{Q}_{2,0}^{(b,E_{g})}$ & B$_{1}$ & $\begin{pmatrix} - \frac{\sqrt{6}}{3} & \frac{\sqrt{6}}{6} & \frac{\sqrt{6}}{6} \end{pmatrix}$ \\
$ \mathbb{Y}_{4} $ & $\mathbb{Q}_{2,1}^{(b,E_{g})}$ & B$_{1}$ & $\begin{pmatrix} 0 & - \frac{\sqrt{2}}{2} & \frac{\sqrt{2}}{2} \end{pmatrix}$ \\ \hline
$ \mathbb{Y}_{5} $ & $\mathbb{Q}_{0}^{(b,A_{1g})}$ & B$_{2}$ & $\begin{pmatrix} \frac{\sqrt{6}}{6} & \frac{\sqrt{6}}{6} & \frac{\sqrt{6}}{6} & \frac{\sqrt{6}}{6} & \frac{\sqrt{6}}{6} & \frac{\sqrt{6}}{6} \end{pmatrix}$ \\
$ \mathbb{Y}_{6} $ & $\mathbb{Q}_{2,0}^{(b,E_{g})}$ & B$_{2}$ & $\begin{pmatrix} - \frac{\sqrt{3}}{6} & - \frac{\sqrt{3}}{6} & - \frac{\sqrt{3}}{6} & \frac{\sqrt{3}}{3} & - \frac{\sqrt{3}}{6} & \frac{\sqrt{3}}{3} \end{pmatrix}$ \\
$ \mathbb{Y}_{7} $ & $\mathbb{Q}_{2,1}^{(b,E_{g})}$ & B$_{2}$ & $\begin{pmatrix} \frac{1}{2} & \frac{1}{2} & - \frac{1}{2} & 0 & - \frac{1}{2} & 0 \end{pmatrix}$ \\
$ \mathbb{Y}_{8} $ & $\mathbb{Q}_{2,0}^{(b,T_{2g})}$ & B$_{2}$ & $\begin{pmatrix} \frac{\sqrt{2}}{2} & - \frac{\sqrt{2}}{2} & 0 & 0 & 0 & 0 \end{pmatrix}$ \\
$ \mathbb{Y}_{9} $ & $\mathbb{Q}_{2,1}^{(b,T_{2g})}$ & B$_{2}$ & $\begin{pmatrix} 0 & 0 & - \frac{\sqrt{2}}{2} & 0 & \frac{\sqrt{2}}{2} & 0 \end{pmatrix}$ \\
$ \mathbb{Y}_{10} $ & $\mathbb{Q}_{2,2}^{(b,T_{2g})}$ & B$_{2}$ & $\begin{pmatrix} 0 & 0 & 0 & - \frac{\sqrt{2}}{2} & 0 & \frac{\sqrt{2}}{2} \end{pmatrix}$ \\ \hline
$ \mathbb{Y}_{11} $ & $\mathbb{Q}_{0}^{(b,A_{1g})}$ & B$_{3}$ & $\begin{pmatrix} \frac{1}{2} & \frac{1}{2} & \frac{1}{2} & \frac{1}{2} \end{pmatrix}$ \\
$ \mathbb{Y}_{12} $ & $\mathbb{Q}_{2,0}^{(b,T_{2g})}$ & B$_{3}$ & $\begin{pmatrix} \frac{1}{2} & - \frac{1}{2} & \frac{1}{2} & - \frac{1}{2} \end{pmatrix}$ \\
$ \mathbb{Y}_{13} $ & $\mathbb{Q}_{2,1}^{(b,T_{2g})}$ & B$_{3}$ & $\begin{pmatrix} \frac{1}{2} & - \frac{1}{2} & - \frac{1}{2} & \frac{1}{2} \end{pmatrix}$ \\
$ \mathbb{Y}_{14} $ & $\mathbb{Q}_{2,2}^{(b,T_{2g})}$ & B$_{3}$ & $\begin{pmatrix} \frac{1}{2} & \frac{1}{2} & - \frac{1}{2} & - \frac{1}{2} \end{pmatrix}$ \\ \hline
$ \mathbb{Y}_{15} $ & $\mathbb{Q}_{0}^{(b,A_{1g})}$ & B$_{4}$ & $\begin{pmatrix} \frac{\sqrt{3}}{3} & \frac{\sqrt{3}}{3} & \frac{\sqrt{3}}{3} \end{pmatrix}$ \\
$ \mathbb{Y}_{16} $ & $\mathbb{Q}_{2,0}^{(b,E_{g})}$ & B$_{4}$ & $\begin{pmatrix} - \frac{\sqrt{6}}{6} & - \frac{\sqrt{6}}{6} & \frac{\sqrt{6}}{3} \end{pmatrix}$ \\
$ \mathbb{Y}_{17} $ & $\mathbb{Q}_{2,1}^{(b,E_{g})}$ & B$_{4}$ & $\begin{pmatrix} \frac{\sqrt{2}}{2} & - \frac{\sqrt{2}}{2} & 0 \end{pmatrix}$ \\ \hline
$ \mathbb{Y}_{18} $ & $\mathbb{Q}_{0}^{(b,A_{1g})}$ & B$_{5}$ & $\begin{pmatrix} \frac{\sqrt{3}}{6} & \frac{\sqrt{3}}{6} & \frac{\sqrt{3}}{6} & \frac{\sqrt{3}}{6} & \frac{\sqrt{3}}{6} & \frac{\sqrt{3}}{6} & \frac{\sqrt{3}}{6} & \frac{\sqrt{3}}{6} & \frac{\sqrt{3}}{6} & \frac{\sqrt{3}}{6} & \frac{\sqrt{3}}{6} & \frac{\sqrt{3}}{6} \end{pmatrix}$ \\
$ \mathbb{Y}_{19} $ & $\mathbb{Q}_{2,0}^{(b,E_{g})}$ & B$_{5}$ & $\begin{pmatrix} - \frac{11 \sqrt{6}}{84} & - \frac{11 \sqrt{6}}{84} & - \frac{11 \sqrt{6}}{84} & \frac{13 \sqrt{6}}{84} & - \frac{\sqrt{6}}{42} & - \frac{11 \sqrt{6}}{84} & - \frac{\sqrt{6}}{42} & - \frac{\sqrt{6}}{42} & - \frac{\sqrt{6}}{42} & \frac{13 \sqrt{6}}{84} & \frac{13 \sqrt{6}}{84} & \frac{13 \sqrt{6}}{84} \end{pmatrix}$ \\
$ \mathbb{Y}_{20} $ & $\mathbb{Q}_{2,1}^{(b,E_{g})}$ & B$_{5}$ & $\begin{pmatrix} \frac{5 \sqrt{2}}{28} & \frac{5 \sqrt{2}}{28} & - \frac{5 \sqrt{2}}{28} & - \frac{3 \sqrt{2}}{28} & \frac{2 \sqrt{2}}{7} & - \frac{5 \sqrt{2}}{28} & \frac{2 \sqrt{2}}{7} & - \frac{2 \sqrt{2}}{7} & - \frac{2 \sqrt{2}}{7} & \frac{3 \sqrt{2}}{28} & \frac{3 \sqrt{2}}{28} & - \frac{3 \sqrt{2}}{28} \end{pmatrix}$ \\
$ \mathbb{Y}_{21} $ & $\mathbb{Q}_{2,0}^{(b,T_{2g})}$ & B$_{5}$ & $\begin{pmatrix} 0 & 0 & - \frac{1}{2} & 0 & 0 & \frac{1}{2} & 0 & 0 & 0 & \frac{1}{2} & - \frac{1}{2} & 0 \end{pmatrix}$ \\
$ \mathbb{Y}_{22} $ & $\mathbb{Q}_{2,1}^{(b,T_{2g})}$ & B$_{5}$ & $\begin{pmatrix} \frac{1}{2} & - \frac{1}{2} & 0 & \frac{1}{2} & 0 & 0 & 0 & 0 & 0 & 0 & 0 & - \frac{1}{2} \end{pmatrix}$ \\
$ \mathbb{Y}_{23} $ & $\mathbb{Q}_{2,2}^{(b,T_{2g})}$ & B$_{5}$ & $\begin{pmatrix} 0 & 0 & 0 & 0 & - \frac{1}{2} & 0 & \frac{1}{2} & \frac{1}{2} & - \frac{1}{2} & 0 & 0 & 0 \end{pmatrix}$ \\
$ \mathbb{Y}_{24} $ & $\mathbb{Q}_{4,0}^{(b,E_{g})}$ & B$_{5}$ & $\begin{pmatrix} \frac{5 \sqrt{2}}{28} & \frac{5 \sqrt{2}}{28} & \frac{5 \sqrt{2}}{28} & \frac{3 \sqrt{2}}{28} & - \frac{2 \sqrt{2}}{7} & \frac{5 \sqrt{2}}{28} & - \frac{2 \sqrt{2}}{7} & - \frac{2 \sqrt{2}}{7} & - \frac{2 \sqrt{2}}{7} & \frac{3 \sqrt{2}}{28} & \frac{3 \sqrt{2}}{28} & \frac{3 \sqrt{2}}{28} \end{pmatrix}$ \\
$ \mathbb{Y}_{25} $ & $\mathbb{Q}_{4,1}^{(b,E_{g})}$ & B$_{5}$ & $\begin{pmatrix} \frac{11 \sqrt{6}}{84} & \frac{11 \sqrt{6}}{84} & - \frac{11 \sqrt{6}}{84} & \frac{13 \sqrt{6}}{84} & - \frac{\sqrt{6}}{42} & - \frac{11 \sqrt{6}}{84} & - \frac{\sqrt{6}}{42} & \frac{\sqrt{6}}{42} & \frac{\sqrt{6}}{42} & - \frac{13 \sqrt{6}}{84} & - \frac{13 \sqrt{6}}{84} & \frac{13 \sqrt{6}}{84} \end{pmatrix}$ \\ \hline
$ \mathbb{Y}_{26} $ & $\mathbb{Q}_{0}^{(b,A_{1g})}$ & B$_{6}$ & $\begin{pmatrix} \frac{\sqrt{3}}{6} & \frac{\sqrt{3}}{6} & \frac{\sqrt{3}}{6} & \frac{\sqrt{3}}{6} & \frac{\sqrt{3}}{6} & \frac{\sqrt{3}}{6} & \frac{\sqrt{3}}{6} & \frac{\sqrt{3}}{6} & \frac{\sqrt{3}}{6} & \frac{\sqrt{3}}{6} & \frac{\sqrt{3}}{6} & \frac{\sqrt{3}}{6} \end{pmatrix}$ \\
$ \mathbb{Y}_{27} $ & $\mathbb{Q}_{2,0}^{(b,E_{g})}$ & B$_{6}$ & $\begin{pmatrix} - \frac{\sqrt{6}}{12} & - \frac{\sqrt{6}}{12} & - \frac{\sqrt{6}}{12} & - \frac{\sqrt{6}}{12} & - \frac{\sqrt{6}}{12} & \frac{\sqrt{6}}{6} & - \frac{\sqrt{6}}{12} & \frac{\sqrt{6}}{6} & - \frac{\sqrt{6}}{12} & - \frac{\sqrt{6}}{12} & \frac{\sqrt{6}}{6} & \frac{\sqrt{6}}{6} \end{pmatrix}$ \\
$ \mathbb{Y}_{28} $ & $\mathbb{Q}_{2,1}^{(b,E_{g})}$ & B$_{6}$ & $\begin{pmatrix} \frac{\sqrt{2}}{4} & \frac{\sqrt{2}}{4} & \frac{\sqrt{2}}{4} & \frac{\sqrt{2}}{4} & - \frac{\sqrt{2}}{4} & 0 & - \frac{\sqrt{2}}{4} & 0 & - \frac{\sqrt{2}}{4} & - \frac{\sqrt{2}}{4} & 0 & 0 \end{pmatrix}$ \\
$ \mathbb{Y}_{29} $ & $\mathbb{Q}_{2,0}^{(b,T_{2g})}$ & B$_{6}$ & $\begin{pmatrix} \frac{\sqrt{178}}{89} & - \frac{\sqrt{178}}{89} & \frac{\sqrt{178}}{89} & - \frac{\sqrt{178}}{89} & - \frac{9 \sqrt{178}}{356} & - \frac{9 \sqrt{178}}{356} & \frac{9 \sqrt{178}}{356} & \frac{9 \sqrt{178}}{356} & \frac{9 \sqrt{178}}{356} & - \frac{9 \sqrt{178}}{356} & - \frac{9 \sqrt{178}}{356} & \frac{9 \sqrt{178}}{356} \end{pmatrix}$ \\
$ \mathbb{Y}_{30} $ & $\mathbb{Q}_{2,1}^{(b,T_{2g})}$ & B$_{6}$ & $\begin{pmatrix} \frac{9 \sqrt{178}}{356} & - \frac{9 \sqrt{178}}{356} & - \frac{9 \sqrt{178}}{356} & \frac{9 \sqrt{178}}{356} & - \frac{\sqrt{178}}{89} & \frac{9 \sqrt{178}}{356} & \frac{\sqrt{178}}{89} & \frac{9 \sqrt{178}}{356} & - \frac{\sqrt{178}}{89} & \frac{\sqrt{178}}{89} & - \frac{9 \sqrt{178}}{356} & - \frac{9 \sqrt{178}}{356} \end{pmatrix}$ \\
$ \mathbb{Y}_{31} $ & $\mathbb{Q}_{2,2}^{(b,T_{2g})}$ & B$_{6}$ & $\begin{pmatrix} \frac{9 \sqrt{178}}{356} & \frac{9 \sqrt{178}}{356} & - \frac{9 \sqrt{178}}{356} & - \frac{9 \sqrt{178}}{356} & \frac{9 \sqrt{178}}{356} & - \frac{\sqrt{178}}{89} & \frac{9 \sqrt{178}}{356} & \frac{\sqrt{178}}{89} & - \frac{9 \sqrt{178}}{356} & - \frac{9 \sqrt{178}}{356} & \frac{\sqrt{178}}{89} & - \frac{\sqrt{178}}{89} \end{pmatrix}$ \\
$ \mathbb{Y}_{32} $ & $\mathbb{Q}_{4,0}^{(b,T_{2g})}$ & B$_{6}$ & $\begin{pmatrix} \frac{9 \sqrt{89}}{178} & - \frac{9 \sqrt{89}}{178} & \frac{9 \sqrt{89}}{178} & - \frac{9 \sqrt{89}}{178} & \frac{\sqrt{89}}{89} & \frac{\sqrt{89}}{89} & - \frac{\sqrt{89}}{89} & - \frac{\sqrt{89}}{89} & - \frac{\sqrt{89}}{89} & \frac{\sqrt{89}}{89} & \frac{\sqrt{89}}{89} & - \frac{\sqrt{89}}{89} \end{pmatrix}$ \\
$ \mathbb{Y}_{33} $ & $\mathbb{Q}_{4,1}^{(b,T_{2g})}$ & B$_{6}$ & $\begin{pmatrix} - \frac{\sqrt{89}}{89} & \frac{\sqrt{89}}{89} & \frac{\sqrt{89}}{89} & - \frac{\sqrt{89}}{89} & - \frac{9 \sqrt{89}}{178} & - \frac{\sqrt{89}}{89} & \frac{9 \sqrt{89}}{178} & - \frac{\sqrt{89}}{89} & - \frac{9 \sqrt{89}}{178} & \frac{9 \sqrt{89}}{178} & \frac{\sqrt{89}}{89} & \frac{\sqrt{89}}{89} \end{pmatrix}$ \\
$ \mathbb{Y}_{34} $ & $\mathbb{Q}_{4,2}^{(b,T_{2g})}$ & B$_{6}$ & $\begin{pmatrix} - \frac{\sqrt{89}}{89} & - \frac{\sqrt{89}}{89} & \frac{\sqrt{89}}{89} & \frac{\sqrt{89}}{89} & - \frac{\sqrt{89}}{89} & - \frac{9 \sqrt{89}}{178} & - \frac{\sqrt{89}}{89} & \frac{9 \sqrt{89}}{178} & \frac{\sqrt{89}}{89} & \frac{\sqrt{89}}{89} & \frac{9 \sqrt{89}}{178} & - \frac{9 \sqrt{89}}{178} \end{pmatrix}$ \\
\end{longtable}
\end{center}

\item The uniform SAMBs are given in Table VI.

\begin{center}
\renewcommand{\arraystretch}{1.3}
\begin{longtable}{c|c|c|c}
\caption{Uniform SAMB.}
 \\
 \hline \hline
symbol & type & cluster & form \\ \hline \endfirsthead

\multicolumn{3}{l}{\tablename\ \thetable{}} \\
 \hline \hline
symbol & type & cluster & form \\ \hline \endhead

 \hline \hline
\multicolumn{3}{r}{\footnotesize\it continued ...} \\ \endfoot

 \hline \hline
\multicolumn{3}{r}{} \\ \endlastfoot

$ \mathbb{U}_{1} $ & $\mathbb{Q}_{0}^{(s,A_{1g})}$ & S$_{1}$ & $\begin{pmatrix} 1 \end{pmatrix}$ \\
\end{longtable}
\end{center}

\item The structure SAMBs are given in Table VII.

\begin{center}
\renewcommand{\arraystretch}{1.3}
\begin{longtable}{c|c|c|c}
\caption{Structure SAMB.}
 \\
 \hline \hline
symbol & type & cluster & form \\ \hline \endfirsthead

\multicolumn{3}{l}{\tablename\ \thetable{}} \\
 \hline \hline
symbol & type & cluster & form \\ \hline \endhead

 \hline \hline
\multicolumn{3}{r}{\footnotesize\it continued ...} \\ \endfoot

 \hline \hline
\multicolumn{3}{r}{} \\ \endlastfoot

$ \mathbb{F}_{1} $ & $\mathbb{Q}_{0}^{(k,A_{1g})}$ & B$_{1}$ & $\frac{\sqrt{6} c_{001}}{3} + \frac{\sqrt{6} c_{002}}{3} + \frac{\sqrt{6} c_{003}}{3}$ \\
$ \mathbb{F}_{2} $ & $\mathbb{Q}_{2,0}^{(k,E_{g})}$ & B$_{1}$ & $- \frac{2 \sqrt{3} c_{001}}{3} + \frac{\sqrt{3} c_{002}}{3} + \frac{\sqrt{3} c_{003}}{3}$ \\
$ \mathbb{F}_{3} $ & $\mathbb{Q}_{2,1}^{(k,E_{g})}$ & B$_{1}$ & $- c_{002} + c_{003}$ \\ \hline
$ \mathbb{F}_{4} $ & $\mathbb{Q}_{0}^{(k,A_{1g})}$ & B$_{2}$ & $\frac{\sqrt{3} c_{004}}{3} + \frac{\sqrt{3} c_{005}}{3} + \frac{\sqrt{3} c_{006}}{3} + \frac{\sqrt{3} c_{007}}{3} + \frac{\sqrt{3} c_{008}}{3} + \frac{\sqrt{3} c_{009}}{3}$ \\
$ \mathbb{F}_{5} $ & $\mathbb{Q}_{2,0}^{(k,E_{g})}$ & B$_{2}$ & $- \frac{\sqrt{6} c_{004}}{6} - \frac{\sqrt{6} c_{005}}{6} - \frac{\sqrt{6} c_{006}}{6} + \frac{\sqrt{6} c_{007}}{3} - \frac{\sqrt{6} c_{008}}{6} + \frac{\sqrt{6} c_{009}}{3}$ \\
$ \mathbb{F}_{6} $ & $\mathbb{Q}_{2,1}^{(k,E_{g})}$ & B$_{2}$ & $\frac{\sqrt{2} c_{004}}{2} + \frac{\sqrt{2} c_{005}}{2} - \frac{\sqrt{2} c_{006}}{2} - \frac{\sqrt{2} c_{008}}{2}$ \\
$ \mathbb{F}_{7} $ & $\mathbb{Q}_{2,0}^{(k,T_{2g})}$ & B$_{2}$ & $c_{004} - c_{005}$ \\
$ \mathbb{F}_{8} $ & $\mathbb{Q}_{2,1}^{(k,T_{2g})}$ & B$_{2}$ & $- c_{006} + c_{008}$ \\
$ \mathbb{F}_{9} $ & $\mathbb{Q}_{2,2}^{(k,T_{2g})}$ & B$_{2}$ & $- c_{007} + c_{009}$ \\ \hline
$ \mathbb{F}_{10} $ & $\mathbb{Q}_{0}^{(k,A_{1g})}$ & B$_{3}$ & $\frac{\sqrt{2} c_{010}}{2} + \frac{\sqrt{2} c_{011}}{2} + \frac{\sqrt{2} c_{012}}{2} + \frac{\sqrt{2} c_{013}}{2}$ \\
$ \mathbb{F}_{11} $ & $\mathbb{Q}_{2,0}^{(k,T_{2g})}$ & B$_{3}$ & $\frac{\sqrt{2} c_{010}}{2} - \frac{\sqrt{2} c_{011}}{2} + \frac{\sqrt{2} c_{012}}{2} - \frac{\sqrt{2} c_{013}}{2}$ \\
$ \mathbb{F}_{12} $ & $\mathbb{Q}_{2,1}^{(k,T_{2g})}$ & B$_{3}$ & $\frac{\sqrt{2} c_{010}}{2} - \frac{\sqrt{2} c_{011}}{2} - \frac{\sqrt{2} c_{012}}{2} + \frac{\sqrt{2} c_{013}}{2}$ \\
$ \mathbb{F}_{13} $ & $\mathbb{Q}_{2,2}^{(k,T_{2g})}$ & B$_{3}$ & $\frac{\sqrt{2} c_{010}}{2} + \frac{\sqrt{2} c_{011}}{2} - \frac{\sqrt{2} c_{012}}{2} - \frac{\sqrt{2} c_{013}}{2}$ \\ \hline
$ \mathbb{F}_{14} $ & $\mathbb{Q}_{0}^{(k,A_{1g})}$ & B$_{4}$ & $\frac{\sqrt{6} c_{014}}{3} + \frac{\sqrt{6} c_{015}}{3} + \frac{\sqrt{6} c_{016}}{3}$ \\
$ \mathbb{F}_{15} $ & $\mathbb{Q}_{2,0}^{(k,E_{g})}$ & B$_{4}$ & $- \frac{\sqrt{3} c_{014}}{3} - \frac{\sqrt{3} c_{015}}{3} + \frac{2 \sqrt{3} c_{016}}{3}$ \\
$ \mathbb{F}_{16} $ & $\mathbb{Q}_{2,1}^{(k,E_{g})}$ & B$_{4}$ & $c_{014} - c_{015}$ \\ \hline
$ \mathbb{F}_{17} $ & $\mathbb{Q}_{0}^{(k,A_{1g})}$ & B$_{5}$ & $\frac{\sqrt{6} c_{017}}{6} + \frac{\sqrt{6} c_{018}}{6} + \frac{\sqrt{6} c_{019}}{6} + \frac{\sqrt{6} c_{020}}{6} + \frac{\sqrt{6} c_{021}}{6} + \frac{\sqrt{6} c_{022}}{6}$ \\&&&
$ + \frac{\sqrt{6} c_{023}}{6} + \frac{\sqrt{6} c_{024}}{6} + \frac{\sqrt{6} c_{025}}{6} + \frac{\sqrt{6} c_{026}}{6} + \frac{\sqrt{6} c_{027}}{6} + \frac{\sqrt{6} c_{028}}{6}$ \\
$ \mathbb{F}_{18} $ & $\mathbb{Q}_{2,0}^{(k,E_{g})}$ & B$_{5}$ & $- \frac{11 \sqrt{3} c_{017}}{42} - \frac{11 \sqrt{3} c_{018}}{42} - \frac{11 \sqrt{3} c_{019}}{42} + \frac{13 \sqrt{3} c_{020}}{42} - \frac{\sqrt{3} c_{021}}{21} - \frac{11 \sqrt{3} c_{022}}{42}$ \\&&&
$ - \frac{\sqrt{3} c_{023}}{21} - \frac{\sqrt{3} c_{024}}{21} - \frac{\sqrt{3} c_{025}}{21} + \frac{13 \sqrt{3} c_{026}}{42} + \frac{13 \sqrt{3} c_{027}}{42} + \frac{13 \sqrt{3} c_{028}}{42}$ \\
$ \mathbb{F}_{19} $ & $\mathbb{Q}_{2,1}^{(k,E_{g})}$ & B$_{5}$ & $\frac{5 c_{017}}{14} + \frac{5 c_{018}}{14} - \frac{5 c_{019}}{14} - \frac{3 c_{020}}{14} + \frac{4 c_{021}}{7} - \frac{5 c_{022}}{14}$ \\&&&
$ + \frac{4 c_{023}}{7} - \frac{4 c_{024}}{7} - \frac{4 c_{025}}{7} + \frac{3 c_{026}}{14} + \frac{3 c_{027}}{14} - \frac{3 c_{028}}{14}$ \\
$ \mathbb{F}_{20} $ & $\mathbb{Q}_{2,0}^{(k,T_{2g})}$ & B$_{5}$ & $- \frac{\sqrt{2} c_{019}}{2} + \frac{\sqrt{2} c_{022}}{2} + \frac{\sqrt{2} c_{026}}{2} - \frac{\sqrt{2} c_{027}}{2}$ \\
$ \mathbb{F}_{21} $ & $\mathbb{Q}_{2,1}^{(k,T_{2g})}$ & B$_{5}$ & $\frac{\sqrt{2} c_{017}}{2} - \frac{\sqrt{2} c_{018}}{2} + \frac{\sqrt{2} c_{020}}{2} - \frac{\sqrt{2} c_{028}}{2}$ \\
$ \mathbb{F}_{22} $ & $\mathbb{Q}_{2,2}^{(k,T_{2g})}$ & B$_{5}$ & $- \frac{\sqrt{2} c_{021}}{2} + \frac{\sqrt{2} c_{023}}{2} + \frac{\sqrt{2} c_{024}}{2} - \frac{\sqrt{2} c_{025}}{2}$ \\
$ \mathbb{F}_{23} $ & $\mathbb{Q}_{4,0}^{(k,E_{g})}$ & B$_{5}$ & $\frac{5 c_{017}}{14} + \frac{5 c_{018}}{14} + \frac{5 c_{019}}{14} + \frac{3 c_{020}}{14} - \frac{4 c_{021}}{7} + \frac{5 c_{022}}{14}$ \\&&&
$ - \frac{4 c_{023}}{7} - \frac{4 c_{024}}{7} - \frac{4 c_{025}}{7} + \frac{3 c_{026}}{14} + \frac{3 c_{027}}{14} + \frac{3 c_{028}}{14}$ \\
$ \mathbb{F}_{24} $ & $\mathbb{Q}_{4,1}^{(k,E_{g})}$ & B$_{5}$ & $\frac{11 \sqrt{3} c_{017}}{42} + \frac{11 \sqrt{3} c_{018}}{42} - \frac{11 \sqrt{3} c_{019}}{42} + \frac{13 \sqrt{3} c_{020}}{42} - \frac{\sqrt{3} c_{021}}{21} - \frac{11 \sqrt{3} c_{022}}{42}$ \\&&&
$ - \frac{\sqrt{3} c_{023}}{21} + \frac{\sqrt{3} c_{024}}{21} + \frac{\sqrt{3} c_{025}}{21} - \frac{13 \sqrt{3} c_{026}}{42} - \frac{13 \sqrt{3} c_{027}}{42} + \frac{13 \sqrt{3} c_{028}}{42}$ \\ \hline
$ \mathbb{F}_{25} $ & $\mathbb{Q}_{0}^{(k,A_{1g})}$ & B$_{6}$ & $\frac{\sqrt{6} c_{029}}{6} + \frac{\sqrt{6} c_{030}}{6} + \frac{\sqrt{6} c_{031}}{6} + \frac{\sqrt{6} c_{032}}{6} + \frac{\sqrt{6} c_{033}}{6} + \frac{\sqrt{6} c_{034}}{6}$ \\&&&
$ + \frac{\sqrt{6} c_{035}}{6} + \frac{\sqrt{6} c_{036}}{6} + \frac{\sqrt{6} c_{037}}{6} + \frac{\sqrt{6} c_{038}}{6} + \frac{\sqrt{6} c_{039}}{6} + \frac{\sqrt{6} c_{040}}{6}$ \\
$ \mathbb{F}_{26} $ & $\mathbb{Q}_{2,0}^{(k,E_{g})}$ & B$_{6}$ & $- \frac{\sqrt{3} c_{029}}{6} - \frac{\sqrt{3} c_{030}}{6} - \frac{\sqrt{3} c_{031}}{6} - \frac{\sqrt{3} c_{032}}{6} - \frac{\sqrt{3} c_{033}}{6} + \frac{\sqrt{3} c_{034}}{3}$ \\&&&
$ - \frac{\sqrt{3} c_{035}}{6} + \frac{\sqrt{3} c_{036}}{3} - \frac{\sqrt{3} c_{037}}{6} - \frac{\sqrt{3} c_{038}}{6} + \frac{\sqrt{3} c_{039}}{3} + \frac{\sqrt{3} c_{040}}{3}$ \\
$ \mathbb{F}_{27} $ & $\mathbb{Q}_{2,1}^{(k,E_{g})}$ & B$_{6}$ & $\frac{c_{029}}{2} + \frac{c_{030}}{2} + \frac{c_{031}}{2} + \frac{c_{032}}{2} - \frac{c_{033}}{2} - \frac{c_{035}}{2} - \frac{c_{037}}{2} - \frac{c_{038}}{2}$ \\
$ \mathbb{F}_{28} $ & $\mathbb{Q}_{2,0}^{(k,T_{2g})}$ & B$_{6}$ & $\frac{2 \sqrt{89} c_{029}}{89} - \frac{2 \sqrt{89} c_{030}}{89} + \frac{2 \sqrt{89} c_{031}}{89} - \frac{2 \sqrt{89} c_{032}}{89} - \frac{9 \sqrt{89} c_{033}}{178} - \frac{9 \sqrt{89} c_{034}}{178}$ \\&&&
$ + \frac{9 \sqrt{89} c_{035}}{178} + \frac{9 \sqrt{89} c_{036}}{178} + \frac{9 \sqrt{89} c_{037}}{178} - \frac{9 \sqrt{89} c_{038}}{178} - \frac{9 \sqrt{89} c_{039}}{178} + \frac{9 \sqrt{89} c_{040}}{178}$ \\
$ \mathbb{F}_{29} $ & $\mathbb{Q}_{2,1}^{(k,T_{2g})}$ & B$_{6}$ & $\frac{9 \sqrt{89} c_{029}}{178} - \frac{9 \sqrt{89} c_{030}}{178} - \frac{9 \sqrt{89} c_{031}}{178} + \frac{9 \sqrt{89} c_{032}}{178} - \frac{2 \sqrt{89} c_{033}}{89} + \frac{9 \sqrt{89} c_{034}}{178}$ \\&&&
$ + \frac{2 \sqrt{89} c_{035}}{89} + \frac{9 \sqrt{89} c_{036}}{178} - \frac{2 \sqrt{89} c_{037}}{89} + \frac{2 \sqrt{89} c_{038}}{89} - \frac{9 \sqrt{89} c_{039}}{178} - \frac{9 \sqrt{89} c_{040}}{178}$ \\
$ \mathbb{F}_{30} $ & $\mathbb{Q}_{2,2}^{(k,T_{2g})}$ & B$_{6}$ & $\frac{9 \sqrt{89} c_{029}}{178} + \frac{9 \sqrt{89} c_{030}}{178} - \frac{9 \sqrt{89} c_{031}}{178} - \frac{9 \sqrt{89} c_{032}}{178} + \frac{9 \sqrt{89} c_{033}}{178} - \frac{2 \sqrt{89} c_{034}}{89}$ \\&&&
$ + \frac{9 \sqrt{89} c_{035}}{178} + \frac{2 \sqrt{89} c_{036}}{89} - \frac{9 \sqrt{89} c_{037}}{178} - \frac{9 \sqrt{89} c_{038}}{178} + \frac{2 \sqrt{89} c_{039}}{89} - \frac{2 \sqrt{89} c_{040}}{89}$ \\
$ \mathbb{F}_{31} $ & $\mathbb{Q}_{4,0}^{(k,T_{2g})}$ & B$_{6}$ & $\frac{9 \sqrt{178} c_{029}}{178} - \frac{9 \sqrt{178} c_{030}}{178} + \frac{9 \sqrt{178} c_{031}}{178} - \frac{9 \sqrt{178} c_{032}}{178} + \frac{\sqrt{178} c_{033}}{89} + \frac{\sqrt{178} c_{034}}{89}$ \\&&&
$ - \frac{\sqrt{178} c_{035}}{89} - \frac{\sqrt{178} c_{036}}{89} - \frac{\sqrt{178} c_{037}}{89} + \frac{\sqrt{178} c_{038}}{89} + \frac{\sqrt{178} c_{039}}{89} - \frac{\sqrt{178} c_{040}}{89}$ \\
$ \mathbb{F}_{32} $ & $\mathbb{Q}_{4,1}^{(k,T_{2g})}$ & B$_{6}$ & $- \frac{\sqrt{178} c_{029}}{89} + \frac{\sqrt{178} c_{030}}{89} + \frac{\sqrt{178} c_{031}}{89} - \frac{\sqrt{178} c_{032}}{89} - \frac{9 \sqrt{178} c_{033}}{178} - \frac{\sqrt{178} c_{034}}{89}$ \\&&&
$ + \frac{9 \sqrt{178} c_{035}}{178} - \frac{\sqrt{178} c_{036}}{89} - \frac{9 \sqrt{178} c_{037}}{178} + \frac{9 \sqrt{178} c_{038}}{178} + \frac{\sqrt{178} c_{039}}{89} + \frac{\sqrt{178} c_{040}}{89}$ \\
$ \mathbb{F}_{33} $ & $\mathbb{Q}_{4,2}^{(k,T_{2g})}$ & B$_{6}$ & $- \frac{\sqrt{178} c_{029}}{89} - \frac{\sqrt{178} c_{030}}{89} + \frac{\sqrt{178} c_{031}}{89} + \frac{\sqrt{178} c_{032}}{89} - \frac{\sqrt{178} c_{033}}{89} - \frac{9 \sqrt{178} c_{034}}{178}$ \\&&&
$ - \frac{\sqrt{178} c_{035}}{89} + \frac{9 \sqrt{178} c_{036}}{178} + \frac{\sqrt{178} c_{037}}{89} + \frac{\sqrt{178} c_{038}}{89} + \frac{9 \sqrt{178} c_{039}}{178} - \frac{9 \sqrt{178} c_{040}}{178}$ \\
\end{longtable}
\end{center}

\item The relevant polar harmonics are summarized in Table VIII.

\begin{center}
\renewcommand{\arraystretch}{1.3}
\begin{longtable}{ccccccc}
\caption{Polar harmonics.}
 \\
 \hline \hline
No. & symbol & rank & irrep. & mul. & comp. & form \\ \hline \endfirsthead

\multicolumn{6}{l}{\tablename\ \thetable{}} \\
 \hline \hline
No. & symbol & rank & irrep. & mul. & comp. & form \\ \hline \endhead

 \hline \hline
\multicolumn{6}{r}{\footnotesize\it continued ...} \\ \endfoot

 \hline \hline
\multicolumn{6}{r}{} \\ \endlastfoot

$ 1 $ & $ \mathbb{Q}_{0}^{(A_{1g})} $ & $ 0 $ & $ A_{1g} $ & $ - $ & $ - $ & $ 1 $ \\ \hline
$ 2 $ & $ \mathbb{Q}_{2,0}^{(E_{g})} $ & $ 2 $ & $ E_{g} $ & $ - $ & $ 0 $ & $ - \frac{x^{2}}{2} - \frac{y^{2}}{2} + z^{2} $ \\
$ 3 $ & $ \mathbb{Q}_{2,1}^{(E_{g})} $ & $ 2 $ & $ E_{g} $ & $ - $ & $ 1 $ & $ \frac{\sqrt{3} \left(x - y\right) \left(x + y\right)}{2} $ \\
$ 4 $ & $ \mathbb{Q}_{2,0}^{(T_{2g})} $ & $ 2 $ & $ T_{2g} $ & $ - $ & $ 0 $ & $ \sqrt{3} y z $ \\
$ 5 $ & $ \mathbb{Q}_{2,1}^{(T_{2g})} $ & $ 2 $ & $ T_{2g} $ & $ - $ & $ 1 $ & $ \sqrt{3} x z $ \\
$ 6 $ & $ \mathbb{Q}_{2,2}^{(T_{2g})} $ & $ 2 $ & $ T_{2g} $ & $ - $ & $ 2 $ & $ \sqrt{3} x y $ \\ \hline
$ 7 $ & $ \mathbb{Q}_{4,0}^{(E_{g})} $ & $ 4 $ & $ E_{g} $ & $ - $ & $ 0 $ & $ - \frac{\sqrt{15} \left(x^{4} - 12 x^{2} y^{2} + 6 x^{2} z^{2} + y^{4} + 6 y^{2} z^{2} - 2 z^{4}\right)}{12} $ \\
$ 8 $ & $ \mathbb{Q}_{4,1}^{(E_{g})} $ & $ 4 $ & $ E_{g} $ & $ - $ & $ 1 $ & $ \frac{\sqrt{5} \left(x - y\right) \left(x + y\right) \left(x^{2} + y^{2} - 6 z^{2}\right)}{4} $ \\
$ 9 $ & $ \mathbb{Q}_{4,0}^{(T_{2g})} $ & $ 4 $ & $ T_{2g} $ & $ - $ & $ 0 $ & $ \frac{\sqrt{5} y z \left(6 x^{2} - y^{2} - z^{2}\right)}{2} $ \\
$ 10 $ & $ \mathbb{Q}_{4,1}^{(T_{2g})} $ & $ 4 $ & $ T_{2g} $ & $ - $ & $ 1 $ & $ - \frac{\sqrt{5} x z \left(x^{2} - 6 y^{2} + z^{2}\right)}{2} $ \\
$ 11 $ & $ \mathbb{Q}_{4,2}^{(T_{2g})} $ & $ 4 $ & $ T_{2g} $ & $ - $ & $ 2 $ & $ - \frac{\sqrt{5} x y \left(x^{2} + y^{2} - 6 z^{2}\right)}{2} $ \\
\end{longtable}
\end{center}

\end{itemize}

\subsection{Optimization and resultant energy dispersion}
\label{sec_usage_ex_srvo3_fit}

In this section, we show the results of the parameter optimization.
To this end, we choose the high symmetry lines M$-\Gamma-$X$-$M$-$R$-\Gamma$, and $50$ $\bm{k}$ points in each line are used to evaluate the loss function.
In the optimization process, we use the PyTorch package~\cite{NEURIPS2019_9015} with $N_{\rm h}=4$ hidden layers, and the Adam optimizer~\cite{kingma_adam_2015} with the learning rate $\alpha=0.01$.
The maximum number of iterations is fixed as $N_{\rm iter} = 250$ that is sufficient to reach convergence.
The number of the optimization parameters including the hyper-parameters in the hidden layers is about 270,000.
We perform 50 optimizations with different random initial parameters in order to investigate the initial-guess dependence.

\begin{figure}[htbp]
   \begin{center}
      \includegraphics[width=16.5cm]{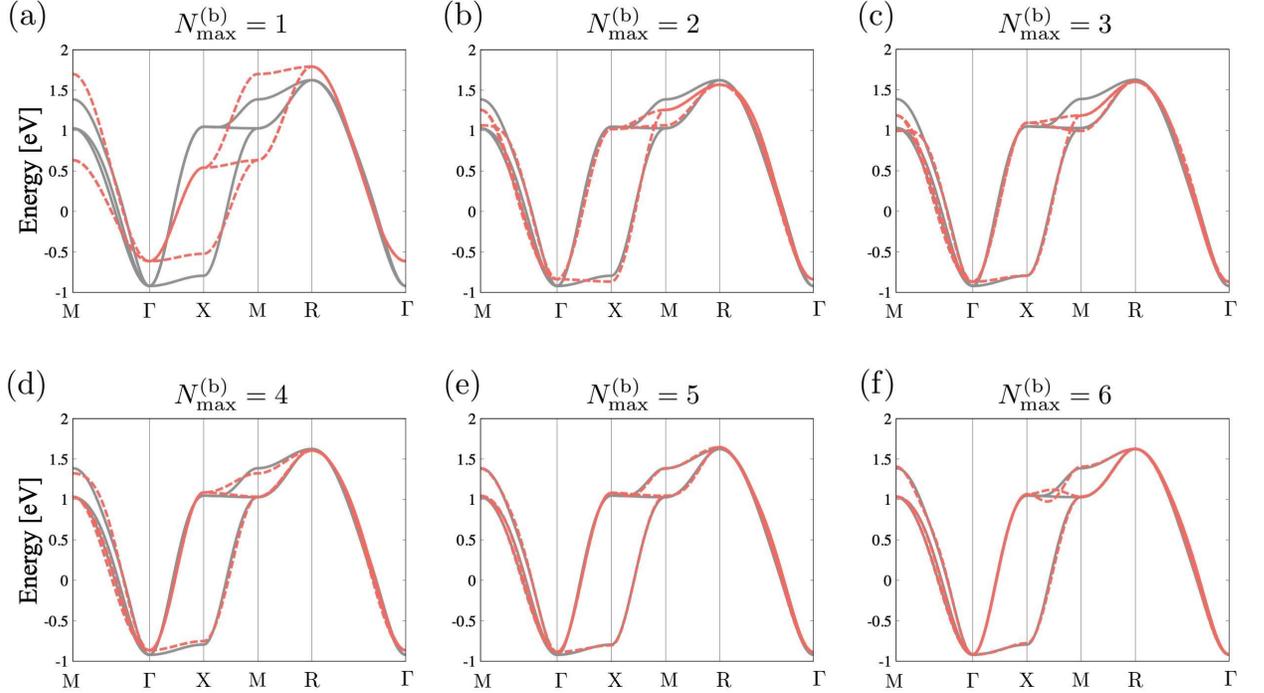}
        \caption{
         The comparisons of the band dispersion between the DF computation (solid gray lines) and our TB models (dashed red lines).
         (a)-(f) The optimized results obtained by using the maximum number of neighboring bonds $N_{\rm max} =$1-6.
         The Fermi energy is taken as the origin.
        }
        \label{fig_srvo3_opt_band_comp_ML}
   \end{center}
\end{figure}

Figure~\ref{fig_srvo3_opt_band_comp_ML} shows the results of parameter optimization.
As the maximum number of neighboring bond $N_{\rm max}^{\rm (b)}$ increases, the quality of the optimization are increased gradually.
In particular, when $N_{\rm max}^{\rm (b)} \geq 5$, the obtained TB model reproduces the DF band dispersions with high accuracy less than $10^{-4}$.
Note that the optimization based on the least-square method results in poor accuracy whatever we choose the initial guess.
The optimized model parameters $z_{j}$ [eV] are given by
\begin{align}
&
z_{1}=22.633, \,\,\, z_{2}=-0.799, \,\,\, z_{3}=0.475, \,\,\, z_{4}=-0.147, \,\,\,
z_{5}=0.223, \,\,\, z_{6}=0.202, \,\,\, z_{7}=-0.032,\,\,\, z_{8}=-0.0223,\cr&
z_{9}=0.0123,\,\,\, z_{10}=0.029,\,\,\, z_{11}=-0.031,\,\,\, z_{12}=0.0005,\,\,\,
z_{13}=0.0001,\,\,\, z_{14}=-0.062,\,\,\, z_{15}=-0.015,\,\,\, z_{16}=-0.015,\cr&
z_{17}=-0.022,\,\,\, z_{18}=-0.006.
\end{align}

\begin{figure}[htbp]
   \begin{center}
   \includegraphics[width=8cm]{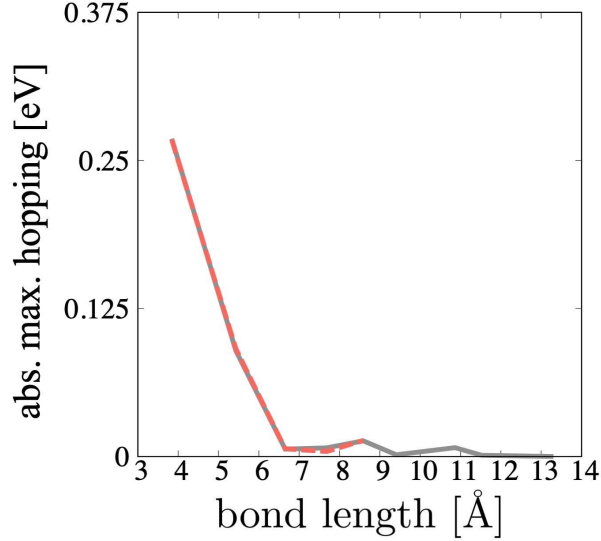}
      \caption{
         The comparison of the bond length dependence of the absolute maximum value of the hopping parameters ($2 \leq j \leq 18$) in eV units between the Wannier TB model (solid gray lines) and our TB model (dashed red lines) with $N_{\rm h} = 4$ and $N_{\rm max}^{\rm (b)} = 5$.
      }
      \label{fig_srvo3_bond_dep_Nh4_Nb5}
   \end{center}
\end{figure}

As shown in Fig.~\ref{fig_srvo3_bond_dep_Nh4_Nb5}, the magnitude of the hopping parameters of our TB model decreases for further neighbor hoppings.
It should be emphasized that the number of parameters is much less than that of the Wannier TB model.

\section{Symmetry-Adapted Modeling for MoS$_{2}$}\label{sec_usage_ex_mos2}

\subsection{DF computation for monolayer MoS$_{2}$}
\label{sec_usage_ex_mos2_dft}

\begin{figure}[h]
   \begin{center}
      \includegraphics[width=16.5cm]{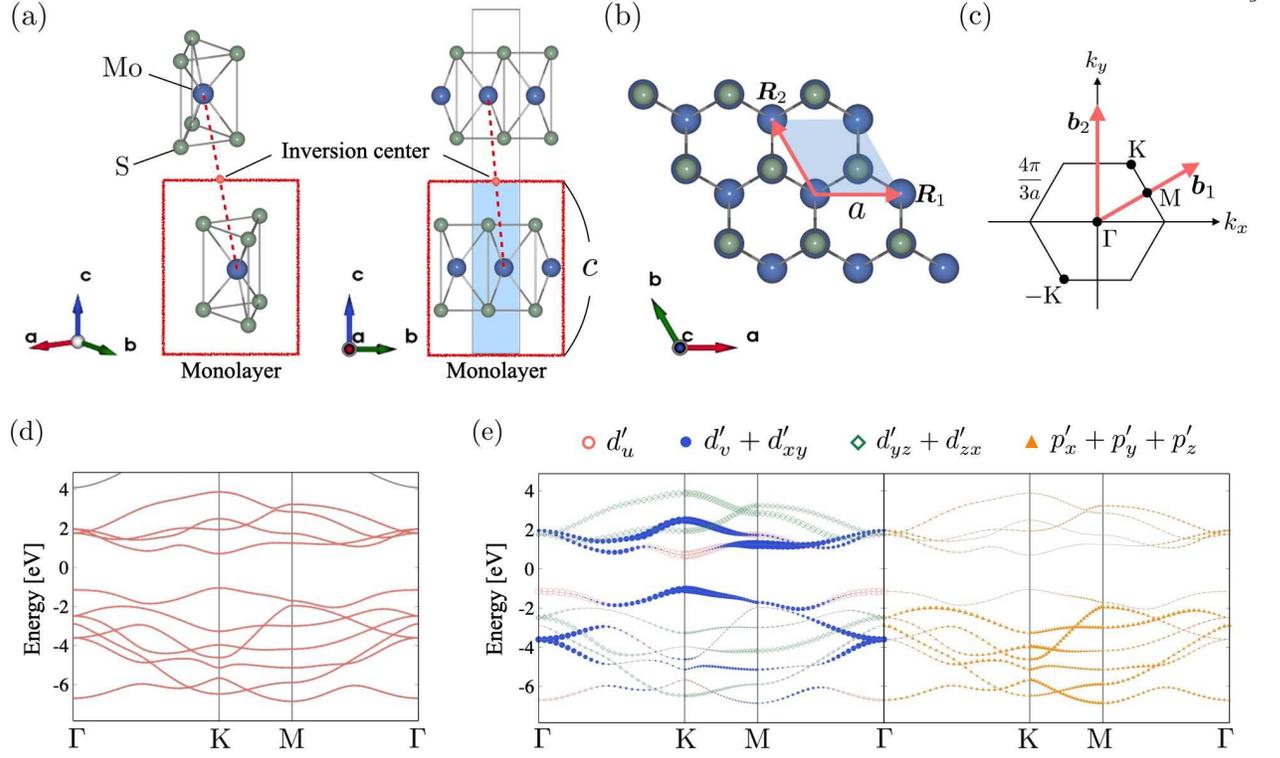}
      \caption{
         Crystal structure of (a) bulk and (b) monolayer MoS$_{2}$ where blue shaded area represents the unit cell.
         (c) The Brillouin zone of the monolayer MoS$_{2}$.
         (d) Band dispersion obtained from the DF computation (solid gray lines) and the Wannier TB model (dashed red lines).
         The Fermi energy is taken as the origin.
         (e) The Wannier orbitals dependence of the band dispersion.
         The red, blue, and green points represent the contributions of the $d_{u}'$, $d_{v}' + d_{xy}'$, and $d_{yz}' + d_{zx}'$ orbitals of Mo atom, while the orange points denote the contributions of the 3 $p$ orbitals, $p_{x}'+p_{y}'+p_{z}'$, of S atoms.
         The size of the points represents the magnitude of contribution of each orbital.
      }
      \label{fig_mos2_dft_band}
   \end{center}
\end{figure}

Next, we present the example for monolayer MoS$_{2}$.
The bulk MoS$_{2}$ consists of two units, and each unit is composed of one Mo atom located at the center of six S atoms at the corners of the triangular prism.
The triangular prism constitutes a building block of a MoS$_{2}$ monolayer~\cite{Cappelluti_MoS2_TB_2013}.
The bulk MoS$_{2}$ has the inversion center between two monolayers as shown in Fig.~\ref{fig_mos2_dft_band}(a), whereas the monolayer MoS$_{2}$ lacks the inversion symmetry as shown in Fig.~\ref{fig_mos2_dft_band}(b).
The space group of monolayer MoS$_{2}$ is $P\bar{6}m2$ (\#187, D$_{\rm 3h}^{1}$).

We set the lattice constant as $a = 3.1661$ \AA, and the length of the vacuum layer along the $c$ axis is chosen as $4 \times a$.
The Mo atom is located at the origin Mo $= (0,0,0)$, and the two S atoms are located at S$_{1} = (2/3, 1/3, z)$ and S$_{2} = (2/3, 1/3, -z)$ with $z = 0.12425$.
For the DF computation, we use the PBE exchange-correlation functional~\cite{Perdew_Burke_Ernzerhof_PBE_1996} and the PAW pseudopotential as in the previous section.
The kinetic energy cutoff of the Kohn-Sham wave functions is 50 Ry, and the convergence threshold for the SCF computation is set as 1$\times$10$^{-10}$ Ry.
We use $(N_{1}, N_{2}, N_{3}) = (12, 12, 1)$ grid.

As shown in Fig.~\ref{fig_mos2_dft_band}(d), the bands near the Fermi level are entangled.
Therefore, we use Wannier90 code to disentangle them, and the resultant band dispersions are represented by the solid red lines, which reproduce well the DF band dispersion.
Figure~\ref{fig_mos2_dft_band} (e) represents the Wannier orbitals dependence of the band dispersion.

The Mo five atomic $d$ orbitals split into A$_{1}'$ ($d_{u}$), E$'$ ($d_{v}, d_{xy}$), and E$''$ ($d_{yz}, d_{zx}$) orbitals owing to the trigonal prismatic structure of S atoms.
The $xy$ mirror symmetry hybridizes the A$_{1}'$ and E$'$ orbitals, giving rise to the direct band gap at the K point.
The top valance band and two bottom conduction bands are mainly composed of the A$_{1}'$ and E$'$ orbitals, whereas E$''$ orbitals have less contribution near the Fermi level.
On the other hand, the S three $p$ orbitals contribute to the six bottom valence bands.
We choose five atomic $d$ orbitals $(d_{u}, d_{v}, d_{yz}, d_{zx}, d_{xy})$ for Mo atom and three atomic $p$ orbitals $(p_{x}, p_{y}, p_{z})$ for two S atoms in order to construct SAMBs.
Note that the spin-orbit coupling is neglected in what follows.

\subsection{Symmetry-adapted multipole basis for MoS$_{2}$}
\label{sec_usage_ex_mos2_tb}

Here, we summarize the SAMB information for monolayer MoS$_{2}$ in the same manner as the previous section.

\begin{itemize}

\item The full Hilbert space of the model Hamiltonian is given by Table IX, and its dimension is 11.
\begin{center}
\renewcommand{\arraystretch}{1.3}
\begin{longtable}{c|cc|cc|cc|cc|cc}
\caption{Hilbert space for full matrix.}
 \\
 \hline \hline
 & No. & ket & No. & ket & No. & ket & No. & ket & No. & ket \\ \hline \endfirsthead

\multicolumn{10}{l}{\tablename\ \thetable{}} \\
 \hline \hline
 & No. & ket & No. & ket & No. & ket & No. & ket & No. & ket \\ \hline \endhead

 \hline \hline
\multicolumn{10}{r}{\footnotesize\it continued ...} \\ \endfoot

 \hline \hline
\multicolumn{10}{r}{} \\ \endlastfoot

 & 1 & $d_{u}$@Mo$_{1}$ & 2 & $d_{v}$@Mo$_{1}$ & 3 & $d_{yz}$@Mo$_{1}$ & 4 & $d_{zx}$@Mo$_{1}$ & 5 & $d_{xy}$@Mo$_{1}$ \\
& 6 & $p_{x}$@S$_{1}$ & 7 & $p_{y}$@S$_{1}$ & 8 & $p_{z}$@S$_{1}$ & 9 & $p_{x}$@S$_{2}$ & 10 & $p_{y}$@S$_{2}$ \\
& 11 & $p_{z}$@S$_{2}$ &  &  &  &  &  &  &  &  \\
\end{longtable}
\end{center}

\item The site cluster of Mo atom and S atoms are given in Table X.
\begin{center}
\renewcommand{\arraystretch}{1.3}
\begin{longtable}{cc|c|l}
\caption{Site clusters.}
 \\
 \hline \hline
 & site & position & mapping \\ \hline \endfirsthead

\multicolumn{3}{l}{\tablename\ \thetable{}} \\
 \hline \hline
 & site & position & mapping \\ \hline \endhead

 \hline \hline
\multicolumn{3}{r}{\footnotesize\it continued ...} \\ \endfoot

 \hline \hline
\multicolumn{3}{r}{} \\ \endlastfoot

S$_{1}$ & Mo$_1$ & $\begin{pmatrix} 0 & 0 & 0 \end{pmatrix}$ & [1,2,3,4,5,6,7,8,9,10,11,12] \\ \hline
S$_{2}$ & S$_1$ & $\begin{pmatrix} \frac{2}{3} & \frac{1}{3} & 0.12425 \end{pmatrix}$ & [1,5,6,7,8,9] \\
& S$_2$ & $\begin{pmatrix} \frac{2}{3} & \frac{1}{3} & 0.87575 \end{pmatrix}$ & [2,3,4,10,11,12] \\
\end{longtable}
\end{center}

\item There are 9 bond clusters up to 3rd neighbor Mo-Mo, Mo-S, S-S bonds as shown in Table XI.

\begin{center}
\renewcommand{\arraystretch}{1.3}
\begin{longtable}{cc|cc|c|c|c|l}
\caption{Bond clusters.}
 \\
 \hline \hline
 & bond & tail & head & $n$ & \# & $\bm{b}@\bm{c}$ & mapping \\ \hline \endfirsthead

\multicolumn{7}{l}{\tablename\ \thetable{}} \\
 \hline \hline
 & bond & tail & head & $n$ & \# & $\bm{b}@\bm{c}$ & mapping \\ \hline \endhead

 \hline \hline
\multicolumn{7}{r}{\footnotesize\it continued ...} \\ \endfoot

 \hline \hline
\multicolumn{7}{r}{} \\ \endlastfoot

B$_{1}$ & b$_{1}$ & Mo$_{1}$ & Mo$_{1}$ & 1 & 1 & $\begin{pmatrix} 0 & 1 & 0 \end{pmatrix}@\begin{pmatrix} 0 & \frac{1}{2} & 0 \end{pmatrix}$ & [1,-3,-8,10] \\
& b$_{2}$ & Mo$_{1}$ & Mo$_{1}$ & 1 & 1 & $\begin{pmatrix} 1 & 1 & 0 \end{pmatrix}@\begin{pmatrix} \frac{1}{2} & \frac{1}{2} & 0 \end{pmatrix}$ & [2,-5,7,-12] \\
& b$_{3}$ & Mo$_{1}$ & Mo$_{1}$ & 1 & 1 & $\begin{pmatrix} 1 & 0 & 0 \end{pmatrix}@\begin{pmatrix} \frac{1}{2} & 0 & 0 \end{pmatrix}$ & [-4,6,-9,11] \\ \hline
B$_{2}$ & b$_{4}$ & Mo$_{1}$ & Mo$_{1}$ & 2 & 1 & $\begin{pmatrix} 1 & 2 & 0 \end{pmatrix}@\begin{pmatrix} \frac{1}{2} & 0 & 0 \end{pmatrix}$ & [1,2,7,10] \\
& b$_{5}$ & Mo$_{1}$ & Mo$_{1}$ & 2 & 1 & $\begin{pmatrix} 1 & -1 & 0 \end{pmatrix}@\begin{pmatrix} \frac{1}{2} & \frac{1}{2} & 0 \end{pmatrix}$ & [3,6,8,11] \\
& b$_{6}$ & Mo$_{1}$ & Mo$_{1}$ & 2 & 1 & $\begin{pmatrix} -2 & -1 & 0 \end{pmatrix}@\begin{pmatrix} 0 & \frac{1}{2} & 0 \end{pmatrix}$ & [4,5,9,12] \\ \hline
B$_{3}$ & b$_{7}$ & Mo$_{1}$ & Mo$_{1}$ & 3 & 1 & $\begin{pmatrix} 2 & 2 & 0 \end{pmatrix}@\begin{pmatrix} 0 & 0 & 0 \end{pmatrix}$ & [1,-4,-9,10] \\
& b$_{8}$ & Mo$_{1}$ & Mo$_{1}$ & 3 & 1 & $\begin{pmatrix} 0 & 2 & 0 \end{pmatrix}@\begin{pmatrix} 0 & 0 & 0 \end{pmatrix}$ & [2,-6,7,-11] \\
& b$_{9}$ & Mo$_{1}$ & Mo$_{1}$ & 3 & 1 & $\begin{pmatrix} 2 & 0 & 0 \end{pmatrix}@\begin{pmatrix} 0 & 0 & 0 \end{pmatrix}$ & [3,-5,8,-12] \\ \hline
B$_{4}$ & b$_{10}$ & Mo$_{1}$ & S$_{1}$ & 1 & 1 & $\begin{pmatrix} \frac{2}{3} & \frac{1}{3} & 0.12425 \end{pmatrix}@\begin{pmatrix} \frac{1}{3} & \frac{1}{6} & 0.062125 \end{pmatrix}$ & [1,8] \\
& b$_{11}$ & Mo$_{1}$ & S$_{2}$ & 1 & 1 & $\begin{pmatrix} - \frac{1}{3} & \frac{1}{3} & -0.12425 \end{pmatrix}@\begin{pmatrix} \frac{5}{6} & \frac{1}{6} & 0.937875 \end{pmatrix}$ & [2,12] \\
& b$_{12}$ & Mo$_{1}$ & S$_{2}$ & 1 & 1 & $\begin{pmatrix} \frac{2}{3} & \frac{1}{3} & -0.12425 \end{pmatrix}@\begin{pmatrix} \frac{1}{3} & \frac{1}{6} & 0.937875 \end{pmatrix}$ & [3,10] \\
& b$_{13}$ & Mo$_{1}$ & S$_{2}$ & 1 & 1 & $\begin{pmatrix} - \frac{1}{3} & - \frac{2}{3} & -0.12425 \end{pmatrix}@\begin{pmatrix} \frac{5}{6} & \frac{2}{3} & 0.937875 \end{pmatrix}$ & [4,11] \\
& b$_{14}$ & Mo$_{1}$ & S$_{1}$ & 1 & 1 & $\begin{pmatrix} - \frac{1}{3} & \frac{1}{3} & 0.12425 \end{pmatrix}@\begin{pmatrix} \frac{5}{6} & \frac{1}{6} & 0.062125 \end{pmatrix}$ & [5,7] \\
& b$_{15}$ & Mo$_{1}$ & S$_{1}$ & 1 & 1 & $\begin{pmatrix} - \frac{1}{3} & - \frac{2}{3} & 0.12425 \end{pmatrix}@\begin{pmatrix} \frac{5}{6} & \frac{2}{3} & 0.062125 \end{pmatrix}$ & [6,9] \\ \hline
B$_{5}$ & b$_{16}$ & Mo$_{1}$ & S$_{1}$ & 2 & 1 & $\begin{pmatrix} \frac{2}{3} & \frac{4}{3} & 0.12425 \end{pmatrix}@\begin{pmatrix} \frac{1}{3} & \frac{2}{3} & 0.062125 \end{pmatrix}$ & [1,7] \\
& b$_{17}$ & Mo$_{1}$ & S$_{2}$ & 2 & 1 & $\begin{pmatrix} \frac{2}{3} & \frac{4}{3} & -0.12425 \end{pmatrix}@\begin{pmatrix} \frac{1}{3} & \frac{2}{3} & 0.937875 \end{pmatrix}$ & [2,10] \\
& b$_{18}$ & Mo$_{1}$ & S$_{2}$ & 2 & 1 & $\begin{pmatrix} \frac{2}{3} & - \frac{2}{3} & -0.12425 \end{pmatrix}@\begin{pmatrix} \frac{1}{3} & \frac{2}{3} & 0.937875 \end{pmatrix}$ & [3,11] \\
& b$_{19}$ & Mo$_{1}$ & S$_{2}$ & 2 & 1 & $\begin{pmatrix} - \frac{4}{3} & - \frac{2}{3} & -0.12425 \end{pmatrix}@\begin{pmatrix} \frac{1}{3} & \frac{2}{3} & 0.937875 \end{pmatrix}$ & [4,12] \\
& b$_{20}$ & Mo$_{1}$ & S$_{1}$ & 2 & 1 & $\begin{pmatrix} - \frac{4}{3} & - \frac{2}{3} & 0.12425 \end{pmatrix}@\begin{pmatrix} \frac{1}{3} & \frac{2}{3} & 0.062125 \end{pmatrix}$ & [5,9] \\
& b$_{21}$ & Mo$_{1}$ & S$_{1}$ & 2 & 1 & $\begin{pmatrix} \frac{2}{3} & - \frac{2}{3} & 0.12425 \end{pmatrix}@\begin{pmatrix} \frac{1}{3} & \frac{2}{3} & 0.062125 \end{pmatrix}$ & [6,8] \\ \hline
B$_{6}$ & b$_{22}$ & Mo$_{1}$ & S$_{1}$ & 3 & 1 & $\begin{pmatrix} \frac{5}{3} & \frac{1}{3} & 0.12425 \end{pmatrix}@\begin{pmatrix} \frac{5}{6} & \frac{1}{6} & 0.062125 \end{pmatrix}$ & [1] \\
& b$_{23}$ & Mo$_{1}$ & S$_{2}$ & 3 & 1 & $\begin{pmatrix} - \frac{4}{3} & \frac{1}{3} & -0.12425 \end{pmatrix}@\begin{pmatrix} \frac{1}{3} & \frac{1}{6} & 0.937875 \end{pmatrix}$ & [2] \\
& b$_{24}$ & Mo$_{1}$ & S$_{2}$ & 3 & 1 & $\begin{pmatrix} \frac{5}{3} & \frac{4}{3} & -0.12425 \end{pmatrix}@\begin{pmatrix} \frac{5}{6} & \frac{2}{3} & 0.937875 \end{pmatrix}$ & [3] \\
& b$_{25}$ & Mo$_{1}$ & S$_{2}$ & 3 & 1 & $\begin{pmatrix} - \frac{1}{3} & - \frac{5}{3} & -0.12425 \end{pmatrix}@\begin{pmatrix} \frac{5}{6} & \frac{1}{6} & 0.937875 \end{pmatrix}$ & [4] \\
& b$_{26}$ & Mo$_{1}$ & S$_{1}$ & 3 & 1 & $\begin{pmatrix} - \frac{1}{3} & \frac{4}{3} & 0.12425 \end{pmatrix}@\begin{pmatrix} \frac{5}{6} & \frac{2}{3} & 0.062125 \end{pmatrix}$ & [5] \\
& b$_{27}$ & Mo$_{1}$ & S$_{1}$ & 3 & 1 & $\begin{pmatrix} - \frac{4}{3} & - \frac{5}{3} & 0.12425 \end{pmatrix}@\begin{pmatrix} \frac{1}{3} & \frac{1}{6} & 0.062125 \end{pmatrix}$ & [6] \\
& b$_{28}$ & Mo$_{1}$ & S$_{1}$ & 3 & 1 & $\begin{pmatrix} - \frac{4}{3} & \frac{1}{3} & 0.12425 \end{pmatrix}@\begin{pmatrix} \frac{1}{3} & \frac{1}{6} & 0.062125 \end{pmatrix}$ & [7] \\
& b$_{29}$ & Mo$_{1}$ & S$_{1}$ & 3 & 1 & $\begin{pmatrix} \frac{5}{3} & \frac{4}{3} & 0.12425 \end{pmatrix}@\begin{pmatrix} \frac{5}{6} & \frac{2}{3} & 0.062125 \end{pmatrix}$ & [8] \\
& b$_{30}$ & Mo$_{1}$ & S$_{1}$ & 3 & 1 & $\begin{pmatrix} - \frac{1}{3} & - \frac{5}{3} & 0.12425 \end{pmatrix}@\begin{pmatrix} \frac{5}{6} & \frac{1}{6} & 0.062125 \end{pmatrix}$ & [9] \\
& b$_{31}$ & Mo$_{1}$ & S$_{2}$ & 3 & 1 & $\begin{pmatrix} \frac{5}{3} & \frac{1}{3} & -0.12425 \end{pmatrix}@\begin{pmatrix} \frac{5}{6} & \frac{1}{6} & 0.937875 \end{pmatrix}$ & [10] \\
& b$_{32}$ & Mo$_{1}$ & S$_{2}$ & 3 & 1 & $\begin{pmatrix} - \frac{4}{3} & - \frac{5}{3} & -0.12425 \end{pmatrix}@\begin{pmatrix} \frac{1}{3} & \frac{1}{6} & 0.937875 \end{pmatrix}$ & [11] \\
& b$_{33}$ & Mo$_{1}$ & S$_{2}$ & 3 & 1 & $\begin{pmatrix} - \frac{1}{3} & \frac{4}{3} & -0.12425 \end{pmatrix}@\begin{pmatrix} \frac{5}{6} & \frac{2}{3} & 0.937875 \end{pmatrix}$ & [12] \\ \hline
B$_{7}$ & b$_{34}$ & S$_{1}$ & S$_{1}$ & 1 & 1 & $\begin{pmatrix} 1 & 0 & 0 \end{pmatrix}@\begin{pmatrix} \frac{1}{6} & \frac{1}{3} & 0.12425 \end{pmatrix}$ & [1,-7] \\
& b$_{35}$ & S$_{2}$ & S$_{2}$ & 1 & 1 & $\begin{pmatrix} 1 & 0 & 0 \end{pmatrix}@\begin{pmatrix} \frac{1}{6} & \frac{1}{3} & 0.87575 \end{pmatrix}$ & [-2,10] \\
& b$_{36}$ & S$_{2}$ & S$_{2}$ & 1 & 1 & $\begin{pmatrix} 1 & 1 & 0 \end{pmatrix}@\begin{pmatrix} \frac{1}{6} & \frac{5}{6} & 0.87575 \end{pmatrix}$ & [3,-11] \\
& b$_{37}$ & S$_{2}$ & S$_{2}$ & 1 & 1 & $\begin{pmatrix} 0 & 1 & 0 \end{pmatrix}@\begin{pmatrix} \frac{2}{3} & \frac{5}{6} & 0.87575 \end{pmatrix}$ & [-4,12] \\
& b$_{38}$ & S$_{1}$ & S$_{1}$ & 1 & 1 & $\begin{pmatrix} 0 & 1 & 0 \end{pmatrix}@\begin{pmatrix} \frac{2}{3} & \frac{5}{6} & 0.12425 \end{pmatrix}$ & [5,-9] \\
& b$_{39}$ & S$_{1}$ & S$_{1}$ & 1 & 1 & $\begin{pmatrix} 1 & 1 & 0 \end{pmatrix}@\begin{pmatrix} \frac{1}{6} & \frac{5}{6} & 0.12425 \end{pmatrix}$ & [-6,8] \\ \hline
B$_{8}$ & b$_{40}$ & S$_{1}$ & S$_{2}$ & 2 & 1 & $\begin{pmatrix} 0 & 0 & -0.2485 \end{pmatrix}@\begin{pmatrix} \frac{2}{3} & \frac{1}{3} & 0 \end{pmatrix}$ & [1,-2,-3,-4,5,6,7,8,9,-10,-11,-12] \\ \hline
B$_{9}$ & b$_{41}$ & S$_{1}$ & S$_{1}$ & 3 & 1 & $\begin{pmatrix} 1 & 2 & 0 \end{pmatrix}@\begin{pmatrix} \frac{1}{6} & \frac{1}{3} & 0.12425 \end{pmatrix}$ & [1,7] \\
& b$_{42}$ & S$_{2}$ & S$_{2}$ & 3 & 1 & $\begin{pmatrix} 1 & 2 & 0 \end{pmatrix}@\begin{pmatrix} \frac{1}{6} & \frac{1}{3} & 0.87575 \end{pmatrix}$ & [2,10] \\
& b$_{43}$ & S$_{2}$ & S$_{2}$ & 3 & 1 & $\begin{pmatrix} 1 & -1 & 0 \end{pmatrix}@\begin{pmatrix} \frac{1}{6} & \frac{5}{6} & 0.87575 \end{pmatrix}$ & [3,11] \\
& b$_{44}$ & S$_{2}$ & S$_{2}$ & 3 & 1 & $\begin{pmatrix} -2 & -1 & 0 \end{pmatrix}@\begin{pmatrix} \frac{2}{3} & \frac{5}{6} & 0.87575 \end{pmatrix}$ & [4,12] \\
& b$_{45}$ & S$_{1}$ & S$_{1}$ & 3 & 1 & $\begin{pmatrix} -2 & -1 & 0 \end{pmatrix}@\begin{pmatrix} \frac{2}{3} & \frac{5}{6} & 0.12425 \end{pmatrix}$ & [5,9] \\
& b$_{46}$ & S$_{1}$ & S$_{1}$ & 3 & 1 & $\begin{pmatrix} 1 & -1 & 0 \end{pmatrix}@\begin{pmatrix} \frac{1}{6} & \frac{5}{6} & 0.12425 \end{pmatrix}$ & [6,8] \\
\end{longtable}
\end{center}

\item The SAMBs belonging to $A_{1}'$ irrep. are given as follows for which the bra-ket combination of atomic orbitals $M_{1} = \braket{d_{u},d_{v},d_{yz},d_{zx},d_{xy}|d_{u},d_{v},d_{yz},d_{zx},d_{xy}}$, $M_{2} = \braket{p_{x},p_{y},p_{z}|p_{x},p_{y},p_{z}}$, and $M_{3} = \braket{d_{u},d_{v},d_{yz},d_{zx},d_{xy}|p_{x},p_{y},p_{z}}$, and the site/bond clusters are indicated by square brackets. There are 74 independent SAMBs in total.
Here, we omit the decomposition of the uniform and structure SAMBs.

\vspace{4mm}
\noindent \fbox{No. {1}} $\,\,\,\hat{\mathbb{Q}}_{0}^{(A_{1}^{\prime})}$ [M$_{1}$,\,S$_{1}$]
\begin{align*}
\hat{\mathbb{Z}}_{1}=\mathbb{X}_{1}[\mathbb{Q}_{0}^{(a,A_{1}^{\prime})}] \otimes\mathbb{Y}_{1}[\mathbb{Q}_{0}^{(s,A_{1}^{\prime})}]
\end{align*}
\begin{align*}
\hat{\mathbb{Z}}_{1}(\bm{k})=\mathbb{X}_{1}[\mathbb{Q}_{0}^{(a,A_{1}^{\prime})}] \otimes\mathbb{U}_{1}[\mathbb{Q}_{0}^{(s,A_{1}^{\prime})}]
\end{align*}
\vspace{4mm}
\noindent \fbox{No. {2}} $\,\,\,\hat{\mathbb{Q}}_{2}^{(A_{1}^{\prime})}$ [M$_{1}$,\,S$_{1}$]
\begin{align*}
\hat{\mathbb{Z}}_{2}=\mathbb{X}_{2}[\mathbb{Q}_{2}^{(a,A_{1}^{\prime})}] \otimes\mathbb{Y}_{1}[\mathbb{Q}_{0}^{(s,A_{1}^{\prime})}]
\end{align*}
\begin{align*}
\hat{\mathbb{Z}}_{2}(\bm{k})=\mathbb{X}_{2}[\mathbb{Q}_{2}^{(a,A_{1}^{\prime})}] \otimes\mathbb{U}_{1}[\mathbb{Q}_{0}^{(s,A_{1}^{\prime})}]
\end{align*}
\vspace{4mm}
\noindent \fbox{No. {3}} $\,\,\,\hat{\mathbb{Q}}_{4}^{(A_{1}^{\prime})}$ [M$_{1}$,\,S$_{1}$]
\begin{align*}
\hat{\mathbb{Z}}_{3}=\mathbb{X}_{3}[\mathbb{Q}_{4}^{(a,A_{1}^{\prime})}] \otimes\mathbb{Y}_{1}[\mathbb{Q}_{0}^{(s,A_{1}^{\prime})}]
\end{align*}
\begin{align*}
\hat{\mathbb{Z}}_{3}(\bm{k})=\mathbb{X}_{3}[\mathbb{Q}_{4}^{(a,A_{1}^{\prime})}] \otimes\mathbb{U}_{1}[\mathbb{Q}_{0}^{(s,A_{1}^{\prime})}]
\end{align*}
\vspace{4mm}
\noindent \fbox{No. {4}} $\,\,\,\hat{\mathbb{Q}}_{0}^{(A_{1}^{\prime})}$ [M$_{2}$,\,S$_{2}$]
\begin{align*}
\hat{\mathbb{Z}}_{4}=\mathbb{X}_{14}[\mathbb{Q}_{0}^{(a,A_{1}^{\prime})}] \otimes\mathbb{Y}_{2}[\mathbb{Q}_{0}^{(s,A_{1}^{\prime})}]
\end{align*}
\begin{align*}
\hat{\mathbb{Z}}_{4}(\bm{k})=\mathbb{X}_{14}[\mathbb{Q}_{0}^{(a,A_{1}^{\prime})}] \otimes\mathbb{U}_{2}[\mathbb{Q}_{0}^{(s,A_{1}^{\prime})}]
\end{align*}
\vspace{4mm}
\noindent \fbox{No. {5}} $\,\,\,\hat{\mathbb{Q}}_{2}^{(A_{1}^{\prime})}$ [M$_{2}$,\,S$_{2}$]
\begin{align*}
\hat{\mathbb{Z}}_{5}=\mathbb{X}_{15}[\mathbb{Q}_{2}^{(a,A_{1}^{\prime})}] \otimes\mathbb{Y}_{2}[\mathbb{Q}_{0}^{(s,A_{1}^{\prime})}]
\end{align*}
\begin{align*}
\hat{\mathbb{Z}}_{5}(\bm{k})=\mathbb{X}_{15}[\mathbb{Q}_{2}^{(a,A_{1}^{\prime})}] \otimes\mathbb{U}_{2}[\mathbb{Q}_{0}^{(s,A_{1}^{\prime})}]
\end{align*}
\vspace{4mm}
\noindent \fbox{No. {6}} $\,\,\,\hat{\mathbb{Q}}_{0}^{(A_{1}^{\prime})}$ [M$_{1}$,\,B$_{1}$]
\begin{align*}
\hat{\mathbb{Z}}_{6}=\mathbb{X}_{1}[\mathbb{Q}_{0}^{(a,A_{1}^{\prime})}] \otimes\mathbb{Y}_{3}[\mathbb{Q}_{0}^{(b,A_{1}^{\prime})}]
\end{align*}
\begin{align*}
\hat{\mathbb{Z}}_{6}(\bm{k})=\mathbb{X}_{1}[\mathbb{Q}_{0}^{(a,A_{1}^{\prime})}] \otimes\mathbb{U}_{1}[\mathbb{Q}_{0}^{(s,A_{1}^{\prime})}] \otimes\mathbb{F}_{1}[\mathbb{Q}_{0}^{(k,A_{1}^{\prime})}]
\end{align*}
\vspace{4mm}
\noindent \fbox{No. {7}} $\,\,\,\hat{\mathbb{Q}}_{2}^{(A_{1}^{\prime})}$ [M$_{1}$,\,B$_{1}$]
\begin{align*}
\hat{\mathbb{Z}}_{7}=\mathbb{X}_{2}[\mathbb{Q}_{2}^{(a,A_{1}^{\prime})}] \otimes\mathbb{Y}_{3}[\mathbb{Q}_{0}^{(b,A_{1}^{\prime})}]
\end{align*}
\begin{align*}
\hat{\mathbb{Z}}_{7}(\bm{k})=\mathbb{X}_{2}[\mathbb{Q}_{2}^{(a,A_{1}^{\prime})}] \otimes\mathbb{U}_{1}[\mathbb{Q}_{0}^{(s,A_{1}^{\prime})}] \otimes\mathbb{F}_{1}[\mathbb{Q}_{0}^{(k,A_{1}^{\prime})}]
\end{align*}
\vspace{4mm}
\noindent \fbox{No. {8}} $\,\,\,\hat{\mathbb{Q}}_{3}^{(A_{1}^{\prime})}$ [M$_{1}$,\,B$_{1}$]
\begin{align*}
\hat{\mathbb{Z}}_{8}=- \frac{\sqrt{2} \mathbb{X}_{4}[\mathbb{Q}_{2,0}^{(a,E^{\prime})}] \otimes\mathbb{Y}_{4}[\mathbb{Q}_{1,0}^{(b,E^{\prime})}]}{2} - \frac{\sqrt{2} \mathbb{X}_{5}[\mathbb{Q}_{2,1}^{(a,E^{\prime})}] \otimes\mathbb{Y}_{5}[\mathbb{Q}_{1,1}^{(b,E^{\prime})}]}{2}
\end{align*}
\begin{align*}
\hat{\mathbb{Z}}_{8}(\bm{k})=- \frac{\sqrt{2} \mathbb{X}_{4}[\mathbb{Q}_{2,0}^{(a,E^{\prime})}] \otimes\mathbb{U}_{1}[\mathbb{Q}_{0}^{(s,A_{1}^{\prime})}] \otimes\mathbb{F}_{2}[\mathbb{Q}_{1,0}^{(k,E^{\prime})}]}{2} - \frac{\sqrt{2} \mathbb{X}_{5}[\mathbb{Q}_{2,1}^{(a,E^{\prime})}] \otimes\mathbb{U}_{1}[\mathbb{Q}_{0}^{(s,A_{1}^{\prime})}] \otimes\mathbb{F}_{3}[\mathbb{Q}_{1,1}^{(k,E^{\prime})}]}{2}
\end{align*}
\vspace{4mm}
\noindent \fbox{No. {9}} $\,\,\,\hat{\mathbb{Q}}_{4}^{(A_{1}^{\prime})}$ [M$_{1}$,\,B$_{1}$]
\begin{align*}
\hat{\mathbb{Z}}_{9}=\mathbb{X}_{3}[\mathbb{Q}_{4}^{(a,A_{1}^{\prime})}] \otimes\mathbb{Y}_{3}[\mathbb{Q}_{0}^{(b,A_{1}^{\prime})}]
\end{align*}
\begin{align*}
\hat{\mathbb{Z}}_{9}(\bm{k})=\mathbb{X}_{3}[\mathbb{Q}_{4}^{(a,A_{1}^{\prime})}] \otimes\mathbb{U}_{1}[\mathbb{Q}_{0}^{(s,A_{1}^{\prime})}] \otimes\mathbb{F}_{1}[\mathbb{Q}_{0}^{(k,A_{1}^{\prime})}]
\end{align*}
\vspace{4mm}
\noindent \fbox{No. {10}} $\,\,\,\hat{\mathbb{Q}}_{3}^{(A_{1}^{\prime})}$ [M$_{1}$,\,B$_{1}$]
\begin{align*}
&
\hat{\mathbb{Z}}_{10}=\frac{\sqrt{406} \mathbb{X}_{6}[\mathbb{Q}_{4,0}^{(a,E^{\prime},1)}] \otimes\mathbb{Y}_{4}[\mathbb{Q}_{1,0}^{(b,E^{\prime})}]}{29} + \frac{\sqrt{406} \mathbb{X}_{7}[\mathbb{Q}_{4,1}^{(a,E^{\prime},1)}] \otimes\mathbb{Y}_{5}[\mathbb{Q}_{1,1}^{(b,E^{\prime})}]}{29} + \frac{\sqrt{58} \mathbb{X}_{8}[\mathbb{Q}_{4,0}^{(a,E^{\prime},2)}] \otimes\mathbb{Y}_{4}[\mathbb{Q}_{1,0}^{(b,E^{\prime})}]}{58}
\cr&\hspace{1cm}
 + \frac{\sqrt{58} \mathbb{X}_{9}[\mathbb{Q}_{4,1}^{(a,E^{\prime},2)}] \otimes\mathbb{Y}_{5}[\mathbb{Q}_{1,1}^{(b,E^{\prime})}]}{58}
\end{align*}
\begin{align*}
&
\hat{\mathbb{Z}}_{10}(\bm{k})=\frac{\sqrt{406} \mathbb{X}_{6}[\mathbb{Q}_{4,0}^{(a,E^{\prime},1)}] \otimes\mathbb{U}_{1}[\mathbb{Q}_{0}^{(s,A_{1}^{\prime})}] \otimes\mathbb{F}_{2}[\mathbb{Q}_{1,0}^{(k,E^{\prime})}]}{29} + \frac{\sqrt{406} \mathbb{X}_{7}[\mathbb{Q}_{4,1}^{(a,E^{\prime},1)}] \otimes\mathbb{U}_{1}[\mathbb{Q}_{0}^{(s,A_{1}^{\prime})}] \otimes\mathbb{F}_{3}[\mathbb{Q}_{1,1}^{(k,E^{\prime})}]}{29}
\cr&\hspace{1cm}
 + \frac{\sqrt{58} \mathbb{X}_{8}[\mathbb{Q}_{4,0}^{(a,E^{\prime},2)}] \otimes\mathbb{U}_{1}[\mathbb{Q}_{0}^{(s,A_{1}^{\prime})}] \otimes\mathbb{F}_{2}[\mathbb{Q}_{1,0}^{(k,E^{\prime})}]}{58} + \frac{\sqrt{58} \mathbb{X}_{9}[\mathbb{Q}_{4,1}^{(a,E^{\prime},2)}] \otimes\mathbb{U}_{1}[\mathbb{Q}_{0}^{(s,A_{1}^{\prime})}] \otimes\mathbb{F}_{3}[\mathbb{Q}_{1,1}^{(k,E^{\prime})}]}{58}
\end{align*}
\vspace{4mm}
\noindent \fbox{No. {11}} $\,\,\,\hat{\mathbb{G}}_{4}^{(A_{1}^{\prime})}$ [M$_{1}$,\,B$_{1}$]
\begin{align*}
&
\hat{\mathbb{Z}}_{11}=- \frac{\sqrt{58} \mathbb{X}_{6}[\mathbb{Q}_{4,0}^{(a,E^{\prime},1)}] \otimes\mathbb{Y}_{4}[\mathbb{Q}_{1,0}^{(b,E^{\prime})}]}{58} - \frac{\sqrt{58} \mathbb{X}_{7}[\mathbb{Q}_{4,1}^{(a,E^{\prime},1)}] \otimes\mathbb{Y}_{5}[\mathbb{Q}_{1,1}^{(b,E^{\prime})}]}{58} + \frac{\sqrt{406} \mathbb{X}_{8}[\mathbb{Q}_{4,0}^{(a,E^{\prime},2)}] \otimes\mathbb{Y}_{4}[\mathbb{Q}_{1,0}^{(b,E^{\prime})}]}{29}
\cr&\hspace{1cm}
 + \frac{\sqrt{406} \mathbb{X}_{9}[\mathbb{Q}_{4,1}^{(a,E^{\prime},2)}] \otimes\mathbb{Y}_{5}[\mathbb{Q}_{1,1}^{(b,E^{\prime})}]}{29}
\end{align*}
\begin{align*}
&
\hat{\mathbb{Z}}_{11}(\bm{k})=- \frac{\sqrt{58} \mathbb{X}_{6}[\mathbb{Q}_{4,0}^{(a,E^{\prime},1)}] \otimes\mathbb{U}_{1}[\mathbb{Q}_{0}^{(s,A_{1}^{\prime})}] \otimes\mathbb{F}_{2}[\mathbb{Q}_{1,0}^{(k,E^{\prime})}]}{58} - \frac{\sqrt{58} \mathbb{X}_{7}[\mathbb{Q}_{4,1}^{(a,E^{\prime},1)}] \otimes\mathbb{U}_{1}[\mathbb{Q}_{0}^{(s,A_{1}^{\prime})}] \otimes\mathbb{F}_{3}[\mathbb{Q}_{1,1}^{(k,E^{\prime})}]}{58}
\cr&\hspace{1cm}
 + \frac{\sqrt{406} \mathbb{X}_{8}[\mathbb{Q}_{4,0}^{(a,E^{\prime},2)}] \otimes\mathbb{U}_{1}[\mathbb{Q}_{0}^{(s,A_{1}^{\prime})}] \otimes\mathbb{F}_{2}[\mathbb{Q}_{1,0}^{(k,E^{\prime})}]}{29} + \frac{\sqrt{406} \mathbb{X}_{9}[\mathbb{Q}_{4,1}^{(a,E^{\prime},2)}] \otimes\mathbb{U}_{1}[\mathbb{Q}_{0}^{(s,A_{1}^{\prime})}] \otimes\mathbb{F}_{3}[\mathbb{Q}_{1,1}^{(k,E^{\prime})}]}{29}
\end{align*}
\vspace{4mm}
\noindent \fbox{No. {12}} $\,\,\,\hat{\mathbb{Q}}_{3}^{(A_{1}^{\prime})}$ [M$_{1}$,\,B$_{1}$]
\begin{align*}
\hat{\mathbb{Z}}_{12}=\mathbb{X}_{10}[\mathbb{M}_{1}^{(a,A_{2}^{\prime})}] \otimes\mathbb{Y}_{8}[\mathbb{T}_{3}^{(b,A_{2}^{\prime})}]
\end{align*}
\begin{align*}
\hat{\mathbb{Z}}_{12}(\bm{k})=\mathbb{X}_{10}[\mathbb{M}_{1}^{(a,A_{2}^{\prime})}] \otimes\mathbb{U}_{1}[\mathbb{Q}_{0}^{(s,A_{1}^{\prime})}] \otimes\mathbb{F}_{6}[\mathbb{T}_{3}^{(k,A_{2}^{\prime})}]
\end{align*}
\vspace{4mm}
\noindent \fbox{No. {13}} $\,\,\,\hat{\mathbb{Q}}_{3}^{(A_{1}^{\prime})}$ [M$_{1}$,\,B$_{1}$]
\begin{align*}
\hat{\mathbb{Z}}_{13}=\frac{\sqrt{2} \mathbb{X}_{12}[\mathbb{M}_{3,0}^{(a,E^{\prime})}] \otimes\mathbb{Y}_{6}[\mathbb{T}_{1,0}^{(b,E^{\prime})}]}{2} + \frac{\sqrt{2} \mathbb{X}_{13}[\mathbb{M}_{3,1}^{(a,E^{\prime})}] \otimes\mathbb{Y}_{7}[\mathbb{T}_{1,1}^{(b,E^{\prime})}]}{2}
\end{align*}
\begin{align*}
\hat{\mathbb{Z}}_{13}(\bm{k})=\frac{\sqrt{2} \mathbb{X}_{12}[\mathbb{M}_{3,0}^{(a,E^{\prime})}] \otimes\mathbb{U}_{1}[\mathbb{Q}_{0}^{(s,A_{1}^{\prime})}] \otimes\mathbb{F}_{4}[\mathbb{T}_{1,0}^{(k,E^{\prime})}]}{2} + \frac{\sqrt{2} \mathbb{X}_{13}[\mathbb{M}_{3,1}^{(a,E^{\prime})}] \otimes\mathbb{U}_{1}[\mathbb{Q}_{0}^{(s,A_{1}^{\prime})}] \otimes\mathbb{F}_{5}[\mathbb{T}_{1,1}^{(k,E^{\prime})}]}{2}
\end{align*}
\vspace{4mm}
\noindent \fbox{No. {14}} $\,\,\,\hat{\mathbb{Q}}_{3}^{(A_{1}^{\prime})}$ [M$_{1}$,\,B$_{1}$]
\begin{align*}
\hat{\mathbb{Z}}_{14}=- \mathbb{X}_{11}[\mathbb{M}_{3}^{(a,A_{2}^{\prime})}] \otimes\mathbb{Y}_{8}[\mathbb{T}_{3}^{(b,A_{2}^{\prime})}]
\end{align*}
\begin{align*}
\hat{\mathbb{Z}}_{14}(\bm{k})=- \mathbb{X}_{11}[\mathbb{M}_{3}^{(a,A_{2}^{\prime})}] \otimes\mathbb{U}_{1}[\mathbb{Q}_{0}^{(s,A_{1}^{\prime})}] \otimes\mathbb{F}_{6}[\mathbb{T}_{3}^{(k,A_{2}^{\prime})}]
\end{align*}
\vspace{4mm}
\noindent \fbox{No. {15}} $\,\,\,\hat{\mathbb{Q}}_{0}^{(A_{1}^{\prime})}$ [M$_{1}$,\,B$_{2}$]
\begin{align*}
\hat{\mathbb{Z}}_{15}=\mathbb{X}_{1}[\mathbb{Q}_{0}^{(a,A_{1}^{\prime})}] \otimes\mathbb{Y}_{9}[\mathbb{Q}_{0}^{(b,A_{1}^{\prime})}]
\end{align*}
\begin{align*}
\hat{\mathbb{Z}}_{15}(\bm{k})=\mathbb{X}_{1}[\mathbb{Q}_{0}^{(a,A_{1}^{\prime})}] \otimes\mathbb{U}_{1}[\mathbb{Q}_{0}^{(s,A_{1}^{\prime})}] \otimes\mathbb{F}_{7}[\mathbb{Q}_{0}^{(k,A_{1}^{\prime})}]
\end{align*}
\vspace{4mm}
\noindent \fbox{No. {16}} $\,\,\,\hat{\mathbb{Q}}_{2}^{(A_{1}^{\prime})}$ [M$_{1}$,\,B$_{2}$]
\begin{align*}
\hat{\mathbb{Z}}_{16}=\mathbb{X}_{2}[\mathbb{Q}_{2}^{(a,A_{1}^{\prime})}] \otimes\mathbb{Y}_{9}[\mathbb{Q}_{0}^{(b,A_{1}^{\prime})}]
\end{align*}
\begin{align*}
\hat{\mathbb{Z}}_{16}(\bm{k})=\mathbb{X}_{2}[\mathbb{Q}_{2}^{(a,A_{1}^{\prime})}] \otimes\mathbb{U}_{1}[\mathbb{Q}_{0}^{(s,A_{1}^{\prime})}] \otimes\mathbb{F}_{7}[\mathbb{Q}_{0}^{(k,A_{1}^{\prime})}]
\end{align*}
\vspace{4mm}
\noindent \fbox{No. {17}} $\,\,\,\hat{\mathbb{Q}}_{3}^{(A_{1}^{\prime})}$ [M$_{1}$,\,B$_{2}$]
\begin{align*}
\hat{\mathbb{Z}}_{17}=- \frac{\sqrt{2} \mathbb{X}_{4}[\mathbb{Q}_{2,0}^{(a,E^{\prime})}] \otimes\mathbb{Y}_{10}[\mathbb{Q}_{1,0}^{(b,E^{\prime})}]}{2} - \frac{\sqrt{2} \mathbb{X}_{5}[\mathbb{Q}_{2,1}^{(a,E^{\prime})}] \otimes\mathbb{Y}_{11}[\mathbb{Q}_{1,1}^{(b,E^{\prime})}]}{2}
\end{align*}
\begin{align*}
\hat{\mathbb{Z}}_{17}(\bm{k})=- \frac{\sqrt{2} \mathbb{X}_{4}[\mathbb{Q}_{2,0}^{(a,E^{\prime})}] \otimes\mathbb{U}_{1}[\mathbb{Q}_{0}^{(s,A_{1}^{\prime})}] \otimes\mathbb{F}_{8}[\mathbb{Q}_{1,0}^{(k,E^{\prime})}]}{2} - \frac{\sqrt{2} \mathbb{X}_{5}[\mathbb{Q}_{2,1}^{(a,E^{\prime})}] \otimes\mathbb{U}_{1}[\mathbb{Q}_{0}^{(s,A_{1}^{\prime})}] \otimes\mathbb{F}_{9}[\mathbb{Q}_{1,1}^{(k,E^{\prime})}]}{2}
\end{align*}
\vspace{4mm}
\noindent \fbox{No. {18}} $\,\,\,\hat{\mathbb{Q}}_{4}^{(A_{1}^{\prime})}$ [M$_{1}$,\,B$_{2}$]
\begin{align*}
\hat{\mathbb{Z}}_{18}=\mathbb{X}_{3}[\mathbb{Q}_{4}^{(a,A_{1}^{\prime})}] \otimes\mathbb{Y}_{9}[\mathbb{Q}_{0}^{(b,A_{1}^{\prime})}]
\end{align*}
\begin{align*}
\hat{\mathbb{Z}}_{18}(\bm{k})=\mathbb{X}_{3}[\mathbb{Q}_{4}^{(a,A_{1}^{\prime})}] \otimes\mathbb{U}_{1}[\mathbb{Q}_{0}^{(s,A_{1}^{\prime})}] \otimes\mathbb{F}_{7}[\mathbb{Q}_{0}^{(k,A_{1}^{\prime})}]
\end{align*}
\vspace{4mm}
\noindent \fbox{No. {19}} $\,\,\,\hat{\mathbb{Q}}_{3}^{(A_{1}^{\prime})}$ [M$_{1}$,\,B$_{2}$]
\begin{align*}
&
\hat{\mathbb{Z}}_{19}=\frac{\sqrt{406} \mathbb{X}_{6}[\mathbb{Q}_{4,0}^{(a,E^{\prime},1)}] \otimes\mathbb{Y}_{10}[\mathbb{Q}_{1,0}^{(b,E^{\prime})}]}{29} + \frac{\sqrt{406} \mathbb{X}_{7}[\mathbb{Q}_{4,1}^{(a,E^{\prime},1)}] \otimes\mathbb{Y}_{11}[\mathbb{Q}_{1,1}^{(b,E^{\prime})}]}{29} + \frac{\sqrt{58} \mathbb{X}_{8}[\mathbb{Q}_{4,0}^{(a,E^{\prime},2)}] \otimes\mathbb{Y}_{10}[\mathbb{Q}_{1,0}^{(b,E^{\prime})}]}{58}
\cr&\hspace{1cm}
 + \frac{\sqrt{58} \mathbb{X}_{9}[\mathbb{Q}_{4,1}^{(a,E^{\prime},2)}] \otimes\mathbb{Y}_{11}[\mathbb{Q}_{1,1}^{(b,E^{\prime})}]}{58}
\end{align*}
\begin{align*}
&
\hat{\mathbb{Z}}_{19}(\bm{k})=\frac{\sqrt{406} \mathbb{X}_{6}[\mathbb{Q}_{4,0}^{(a,E^{\prime},1)}] \otimes\mathbb{U}_{1}[\mathbb{Q}_{0}^{(s,A_{1}^{\prime})}] \otimes\mathbb{F}_{8}[\mathbb{Q}_{1,0}^{(k,E^{\prime})}]}{29} + \frac{\sqrt{406} \mathbb{X}_{7}[\mathbb{Q}_{4,1}^{(a,E^{\prime},1)}] \otimes\mathbb{U}_{1}[\mathbb{Q}_{0}^{(s,A_{1}^{\prime})}] \otimes\mathbb{F}_{9}[\mathbb{Q}_{1,1}^{(k,E^{\prime})}]}{29}
\cr&\hspace{1cm}
 + \frac{\sqrt{58} \mathbb{X}_{8}[\mathbb{Q}_{4,0}^{(a,E^{\prime},2)}] \otimes\mathbb{U}_{1}[\mathbb{Q}_{0}^{(s,A_{1}^{\prime})}] \otimes\mathbb{F}_{8}[\mathbb{Q}_{1,0}^{(k,E^{\prime})}]}{58} + \frac{\sqrt{58} \mathbb{X}_{9}[\mathbb{Q}_{4,1}^{(a,E^{\prime},2)}] \otimes\mathbb{U}_{1}[\mathbb{Q}_{0}^{(s,A_{1}^{\prime})}] \otimes\mathbb{F}_{9}[\mathbb{Q}_{1,1}^{(k,E^{\prime})}]}{58}
\end{align*}
\vspace{4mm}
\noindent \fbox{No. {20}} $\,\,\,\hat{\mathbb{G}}_{4}^{(A_{1}^{\prime})}$ [M$_{1}$,\,B$_{2}$]
\begin{align*}
&
\hat{\mathbb{Z}}_{20}=- \frac{\sqrt{58} \mathbb{X}_{6}[\mathbb{Q}_{4,0}^{(a,E^{\prime},1)}] \otimes\mathbb{Y}_{10}[\mathbb{Q}_{1,0}^{(b,E^{\prime})}]}{58} - \frac{\sqrt{58} \mathbb{X}_{7}[\mathbb{Q}_{4,1}^{(a,E^{\prime},1)}] \otimes\mathbb{Y}_{11}[\mathbb{Q}_{1,1}^{(b,E^{\prime})}]}{58} + \frac{\sqrt{406} \mathbb{X}_{8}[\mathbb{Q}_{4,0}^{(a,E^{\prime},2)}] \otimes\mathbb{Y}_{10}[\mathbb{Q}_{1,0}^{(b,E^{\prime})}]}{29}
\cr&\hspace{1cm}
 + \frac{\sqrt{406} \mathbb{X}_{9}[\mathbb{Q}_{4,1}^{(a,E^{\prime},2)}] \otimes\mathbb{Y}_{11}[\mathbb{Q}_{1,1}^{(b,E^{\prime})}]}{29}
\end{align*}
\begin{align*}
&
\hat{\mathbb{Z}}_{20}(\bm{k})=- \frac{\sqrt{58} \mathbb{X}_{6}[\mathbb{Q}_{4,0}^{(a,E^{\prime},1)}] \otimes\mathbb{U}_{1}[\mathbb{Q}_{0}^{(s,A_{1}^{\prime})}] \otimes\mathbb{F}_{8}[\mathbb{Q}_{1,0}^{(k,E^{\prime})}]}{58} - \frac{\sqrt{58} \mathbb{X}_{7}[\mathbb{Q}_{4,1}^{(a,E^{\prime},1)}] \otimes\mathbb{U}_{1}[\mathbb{Q}_{0}^{(s,A_{1}^{\prime})}] \otimes\mathbb{F}_{9}[\mathbb{Q}_{1,1}^{(k,E^{\prime})}]}{58}
\cr&\hspace{1cm}
 + \frac{\sqrt{406} \mathbb{X}_{8}[\mathbb{Q}_{4,0}^{(a,E^{\prime},2)}] \otimes\mathbb{U}_{1}[\mathbb{Q}_{0}^{(s,A_{1}^{\prime})}] \otimes\mathbb{F}_{8}[\mathbb{Q}_{1,0}^{(k,E^{\prime})}]}{29} + \frac{\sqrt{406} \mathbb{X}_{9}[\mathbb{Q}_{4,1}^{(a,E^{\prime},2)}] \otimes\mathbb{U}_{1}[\mathbb{Q}_{0}^{(s,A_{1}^{\prime})}] \otimes\mathbb{F}_{9}[\mathbb{Q}_{1,1}^{(k,E^{\prime})}]}{29}
\end{align*}
\vspace{4mm}
\noindent \fbox{No. {21}} $\,\,\,\hat{\mathbb{Q}}_{3}^{(A_{1}^{\prime})}$ [M$_{1}$,\,B$_{2}$]
\begin{align*}
\hat{\mathbb{Z}}_{21}=\frac{\sqrt{2} \mathbb{X}_{12}[\mathbb{M}_{3,0}^{(a,E^{\prime})}] \otimes\mathbb{Y}_{12}[\mathbb{T}_{1,0}^{(b,E^{\prime})}]}{2} + \frac{\sqrt{2} \mathbb{X}_{13}[\mathbb{M}_{3,1}^{(a,E^{\prime})}] \otimes\mathbb{Y}_{13}[\mathbb{T}_{1,1}^{(b,E^{\prime})}]}{2}
\end{align*}
\begin{align*}
\hat{\mathbb{Z}}_{21}(\bm{k})=\frac{\sqrt{2} \mathbb{X}_{12}[\mathbb{M}_{3,0}^{(a,E^{\prime})}] \otimes\mathbb{U}_{1}[\mathbb{Q}_{0}^{(s,A_{1}^{\prime})}] \otimes\mathbb{F}_{10}[\mathbb{T}_{1,0}^{(k,E^{\prime})}]}{2} + \frac{\sqrt{2} \mathbb{X}_{13}[\mathbb{M}_{3,1}^{(a,E^{\prime})}] \otimes\mathbb{U}_{1}[\mathbb{Q}_{0}^{(s,A_{1}^{\prime})}] \otimes\mathbb{F}_{11}[\mathbb{T}_{1,1}^{(k,E^{\prime})}]}{2}
\end{align*}
\vspace{4mm}
\noindent \fbox{No. {22}} $\,\,\,\hat{\mathbb{Q}}_{0}^{(A_{1}^{\prime})}$ [M$_{1}$,\,B$_{3}$]
\begin{align*}
\hat{\mathbb{Z}}_{22}=\mathbb{X}_{1}[\mathbb{Q}_{0}^{(a,A_{1}^{\prime})}] \otimes\mathbb{Y}_{14}[\mathbb{Q}_{0}^{(b,A_{1}^{\prime})}]
\end{align*}
\begin{align*}
\hat{\mathbb{Z}}_{22}(\bm{k})=\mathbb{X}_{1}[\mathbb{Q}_{0}^{(a,A_{1}^{\prime})}] \otimes\mathbb{U}_{1}[\mathbb{Q}_{0}^{(s,A_{1}^{\prime})}] \otimes\mathbb{F}_{12}[\mathbb{Q}_{0}^{(k,A_{1}^{\prime})}]
\end{align*}
\vspace{4mm}
\noindent \fbox{No. {23}} $\,\,\,\hat{\mathbb{Q}}_{2}^{(A_{1}^{\prime})}$ [M$_{1}$,\,B$_{3}$]
\begin{align*}
\hat{\mathbb{Z}}_{23}=\mathbb{X}_{2}[\mathbb{Q}_{2}^{(a,A_{1}^{\prime})}] \otimes\mathbb{Y}_{14}[\mathbb{Q}_{0}^{(b,A_{1}^{\prime})}]
\end{align*}
\begin{align*}
\hat{\mathbb{Z}}_{23}(\bm{k})=\mathbb{X}_{2}[\mathbb{Q}_{2}^{(a,A_{1}^{\prime})}] \otimes\mathbb{U}_{1}[\mathbb{Q}_{0}^{(s,A_{1}^{\prime})}] \otimes\mathbb{F}_{12}[\mathbb{Q}_{0}^{(k,A_{1}^{\prime})}]
\end{align*}
\vspace{4mm}
\noindent \fbox{No. {24}} $\,\,\,\hat{\mathbb{Q}}_{0}^{(A_{1}^{\prime})}$ [M$_{1}$,\,B$_{3}$]
\begin{align*}
\hat{\mathbb{Z}}_{24}=\frac{\sqrt{2} \mathbb{X}_{4}[\mathbb{Q}_{2,0}^{(a,E^{\prime})}] \otimes\mathbb{Y}_{15}[\mathbb{Q}_{2,0}^{(b,E^{\prime})}]}{2} + \frac{\sqrt{2} \mathbb{X}_{5}[\mathbb{Q}_{2,1}^{(a,E^{\prime})}] \otimes\mathbb{Y}_{16}[\mathbb{Q}_{2,1}^{(b,E^{\prime})}]}{2}
\end{align*}
\begin{align*}
\hat{\mathbb{Z}}_{24}(\bm{k})=\frac{\sqrt{2} \mathbb{X}_{4}[\mathbb{Q}_{2,0}^{(a,E^{\prime})}] \otimes\mathbb{U}_{1}[\mathbb{Q}_{0}^{(s,A_{1}^{\prime})}] \otimes\mathbb{F}_{13}[\mathbb{Q}_{2,0}^{(k,E^{\prime})}]}{2} + \frac{\sqrt{2} \mathbb{X}_{5}[\mathbb{Q}_{2,1}^{(a,E^{\prime})}] \otimes\mathbb{U}_{1}[\mathbb{Q}_{0}^{(s,A_{1}^{\prime})}] \otimes\mathbb{F}_{14}[\mathbb{Q}_{2,1}^{(k,E^{\prime})}]}{2}
\end{align*}
\vspace{4mm}
\noindent \fbox{No. {25}} $\,\,\,\hat{\mathbb{Q}}_{4}^{(A_{1}^{\prime})}$ [M$_{1}$,\,B$_{3}$]
\begin{align*}
\hat{\mathbb{Z}}_{25}=\mathbb{X}_{3}[\mathbb{Q}_{4}^{(a,A_{1}^{\prime})}] \otimes\mathbb{Y}_{14}[\mathbb{Q}_{0}^{(b,A_{1}^{\prime})}]
\end{align*}
\begin{align*}
\hat{\mathbb{Z}}_{25}(\bm{k})=\mathbb{X}_{3}[\mathbb{Q}_{4}^{(a,A_{1}^{\prime})}] \otimes\mathbb{U}_{1}[\mathbb{Q}_{0}^{(s,A_{1}^{\prime})}] \otimes\mathbb{F}_{12}[\mathbb{Q}_{0}^{(k,A_{1}^{\prime})}]
\end{align*}
\vspace{4mm}
\noindent \fbox{No. {26}} $\,\,\,\hat{\mathbb{Q}}_{2}^{(A_{1}^{\prime})}$ [M$_{1}$,\,B$_{3}$]
\begin{align*}
\hat{\mathbb{Z}}_{26}=\frac{\sqrt{2} \mathbb{X}_{8}[\mathbb{Q}_{4,0}^{(a,E^{\prime},2)}] \otimes\mathbb{Y}_{15}[\mathbb{Q}_{2,0}^{(b,E^{\prime})}]}{2} + \frac{\sqrt{2} \mathbb{X}_{9}[\mathbb{Q}_{4,1}^{(a,E^{\prime},2)}] \otimes\mathbb{Y}_{16}[\mathbb{Q}_{2,1}^{(b,E^{\prime})}]}{2}
\end{align*}
\begin{align*}
\hat{\mathbb{Z}}_{26}(\bm{k})=\frac{\sqrt{2} \mathbb{X}_{8}[\mathbb{Q}_{4,0}^{(a,E^{\prime},2)}] \otimes\mathbb{U}_{1}[\mathbb{Q}_{0}^{(s,A_{1}^{\prime})}] \otimes\mathbb{F}_{13}[\mathbb{Q}_{2,0}^{(k,E^{\prime})}]}{2} + \frac{\sqrt{2} \mathbb{X}_{9}[\mathbb{Q}_{4,1}^{(a,E^{\prime},2)}] \otimes\mathbb{U}_{1}[\mathbb{Q}_{0}^{(s,A_{1}^{\prime})}] \otimes\mathbb{F}_{14}[\mathbb{Q}_{2,1}^{(k,E^{\prime})}]}{2}
\end{align*}
\vspace{4mm}
\noindent \fbox{No. {27}} $\,\,\,\hat{\mathbb{Q}}_{6}^{(A_{1}^{\prime},2)}$ [M$_{1}$,\,B$_{3}$]
\begin{align*}
\hat{\mathbb{Z}}_{27}=\frac{\sqrt{2} \mathbb{X}_{6}[\mathbb{Q}_{4,0}^{(a,E^{\prime},1)}] \otimes\mathbb{Y}_{15}[\mathbb{Q}_{2,0}^{(b,E^{\prime})}]}{2} + \frac{\sqrt{2} \mathbb{X}_{7}[\mathbb{Q}_{4,1}^{(a,E^{\prime},1)}] \otimes\mathbb{Y}_{16}[\mathbb{Q}_{2,1}^{(b,E^{\prime})}]}{2}
\end{align*}
\begin{align*}
\hat{\mathbb{Z}}_{27}(\bm{k})=\frac{\sqrt{2} \mathbb{X}_{6}[\mathbb{Q}_{4,0}^{(a,E^{\prime},1)}] \otimes\mathbb{U}_{1}[\mathbb{Q}_{0}^{(s,A_{1}^{\prime})}] \otimes\mathbb{F}_{13}[\mathbb{Q}_{2,0}^{(k,E^{\prime})}]}{2} + \frac{\sqrt{2} \mathbb{X}_{7}[\mathbb{Q}_{4,1}^{(a,E^{\prime},1)}] \otimes\mathbb{U}_{1}[\mathbb{Q}_{0}^{(s,A_{1}^{\prime})}] \otimes\mathbb{F}_{14}[\mathbb{Q}_{2,1}^{(k,E^{\prime})}]}{2}
\end{align*}
\vspace{4mm}
\noindent \fbox{No. {28}} $\,\,\,\hat{\mathbb{Q}}_{3}^{(A_{1}^{\prime})}$ [M$_{1}$,\,B$_{3}$]
\begin{align*}
\hat{\mathbb{Z}}_{28}=\mathbb{X}_{10}[\mathbb{M}_{1}^{(a,A_{2}^{\prime})}] \otimes\mathbb{Y}_{19}[\mathbb{T}_{3}^{(b,A_{2}^{\prime})}]
\end{align*}
\begin{align*}
\hat{\mathbb{Z}}_{28}(\bm{k})=\mathbb{X}_{10}[\mathbb{M}_{1}^{(a,A_{2}^{\prime})}] \otimes\mathbb{U}_{1}[\mathbb{Q}_{0}^{(s,A_{1}^{\prime})}] \otimes\mathbb{F}_{17}[\mathbb{T}_{3}^{(k,A_{2}^{\prime})}]
\end{align*}
\vspace{4mm}
\noindent \fbox{No. {29}} $\,\,\,\hat{\mathbb{Q}}_{3}^{(A_{1}^{\prime})}$ [M$_{1}$,\,B$_{3}$]
\begin{align*}
\hat{\mathbb{Z}}_{29}=\frac{\sqrt{2} \mathbb{X}_{12}[\mathbb{M}_{3,0}^{(a,E^{\prime})}] \otimes\mathbb{Y}_{17}[\mathbb{T}_{1,0}^{(b,E^{\prime})}]}{2} + \frac{\sqrt{2} \mathbb{X}_{13}[\mathbb{M}_{3,1}^{(a,E^{\prime})}] \otimes\mathbb{Y}_{18}[\mathbb{T}_{1,1}^{(b,E^{\prime})}]}{2}
\end{align*}
\begin{align*}
\hat{\mathbb{Z}}_{29}(\bm{k})=\frac{\sqrt{2} \mathbb{X}_{12}[\mathbb{M}_{3,0}^{(a,E^{\prime})}] \otimes\mathbb{U}_{1}[\mathbb{Q}_{0}^{(s,A_{1}^{\prime})}] \otimes\mathbb{F}_{15}[\mathbb{T}_{1,0}^{(k,E^{\prime})}]}{2} + \frac{\sqrt{2} \mathbb{X}_{13}[\mathbb{M}_{3,1}^{(a,E^{\prime})}] \otimes\mathbb{U}_{1}[\mathbb{Q}_{0}^{(s,A_{1}^{\prime})}] \otimes\mathbb{F}_{16}[\mathbb{T}_{1,1}^{(k,E^{\prime})}]}{2}
\end{align*}
\vspace{4mm}
\noindent \fbox{No. {30}} $\,\,\,\hat{\mathbb{Q}}_{3}^{(A_{1}^{\prime})}$ [M$_{1}$,\,B$_{3}$]
\begin{align*}
\hat{\mathbb{Z}}_{30}=- \mathbb{X}_{11}[\mathbb{M}_{3}^{(a,A_{2}^{\prime})}] \otimes\mathbb{Y}_{19}[\mathbb{T}_{3}^{(b,A_{2}^{\prime})}]
\end{align*}
\begin{align*}
\hat{\mathbb{Z}}_{30}(\bm{k})=- \mathbb{X}_{11}[\mathbb{M}_{3}^{(a,A_{2}^{\prime})}] \otimes\mathbb{U}_{1}[\mathbb{Q}_{0}^{(s,A_{1}^{\prime})}] \otimes\mathbb{F}_{17}[\mathbb{T}_{3}^{(k,A_{2}^{\prime})}]
\end{align*}
\vspace{4mm}
\noindent \fbox{No. {31}} $\,\,\,\hat{\mathbb{Q}}_{0}^{(A_{1}^{\prime})}$ [M$_{3}$,\,B$_{4}$]
\begin{align*}
\hat{\mathbb{Z}}_{31}=\frac{\sqrt{3} \mathbb{X}_{23}[\mathbb{Q}_{1}^{(a,A_{2}^{\prime\prime})}] \otimes\mathbb{Y}_{21}[\mathbb{Q}_{1}^{(b,A_{2}^{\prime\prime})}]}{3} + \frac{\sqrt{3} \mathbb{X}_{25}[\mathbb{Q}_{1,0}^{(a,E^{\prime})}] \otimes\mathbb{Y}_{22}[\mathbb{Q}_{1,0}^{(b,E^{\prime})}]}{3} + \frac{\sqrt{3} \mathbb{X}_{26}[\mathbb{Q}_{1,1}^{(a,E^{\prime})}] \otimes\mathbb{Y}_{23}[\mathbb{Q}_{1,1}^{(b,E^{\prime})}]}{3}
\end{align*}
\begin{align*}
&
\hat{\mathbb{Z}}_{31}(\bm{k})=\frac{\sqrt{3} \mathbb{X}_{23}[\mathbb{Q}_{1}^{(a,A_{2}^{\prime\prime})}] \otimes\mathbb{U}_{4}[\mathbb{Q}_{0}^{(u,A_{1}^{\prime})}] \otimes\mathbb{F}_{19}[\mathbb{Q}_{1}^{(k,A_{2}^{\prime\prime})}]}{6} + \frac{\sqrt{3} \mathbb{X}_{23}[\mathbb{Q}_{1}^{(a,A_{2}^{\prime\prime})}] \otimes\mathbb{U}_{5}[\mathbb{Q}_{1}^{(u,A_{2}^{\prime\prime})}] \otimes\mathbb{F}_{18}[\mathbb{Q}_{0}^{(k,A_{1}^{\prime})}]}{6}
\cr&\hspace{1cm}
 - \frac{\sqrt{3} \mathbb{X}_{23}[\mathbb{Q}_{1}^{(a,A_{2}^{\prime\prime})}] \otimes\mathbb{U}_{6}[\mathbb{T}_{0}^{(u,A_{1}^{\prime})}] \otimes\mathbb{F}_{25}[\mathbb{T}_{1}^{(k,A_{2}^{\prime\prime})}]}{6} - \frac{\sqrt{3} \mathbb{X}_{23}[\mathbb{Q}_{1}^{(a,A_{2}^{\prime\prime})}] \otimes\mathbb{U}_{7}[\mathbb{T}_{1}^{(u,A_{2}^{\prime\prime})}] \otimes\mathbb{F}_{24}[\mathbb{T}_{0}^{(k,A_{1}^{\prime})}]}{6}
 \cr&\hspace{1cm}
 + \frac{\sqrt{3} \mathbb{X}_{25}[\mathbb{Q}_{1,0}^{(a,E^{\prime})}] \otimes\mathbb{U}_{4}[\mathbb{Q}_{0}^{(u,A_{1}^{\prime})}] \otimes\mathbb{F}_{20}[\mathbb{Q}_{1,0}^{(k,E^{\prime})}]}{6} + \frac{\sqrt{3} \mathbb{X}_{25}[\mathbb{Q}_{1,0}^{(a,E^{\prime})}] \otimes\mathbb{U}_{5}[\mathbb{Q}_{1}^{(u,A_{2}^{\prime\prime})}] \otimes\mathbb{F}_{22}[\mathbb{Q}_{2,0}^{(k,E^{\prime\prime})}]}{6}
 \cr&\hspace{1cm}
 - \frac{\sqrt{3} \mathbb{X}_{25}[\mathbb{Q}_{1,0}^{(a,E^{\prime})}] \otimes\mathbb{U}_{6}[\mathbb{T}_{0}^{(u,A_{1}^{\prime})}] \otimes\mathbb{F}_{26}[\mathbb{T}_{1,0}^{(k,E^{\prime})}]}{6} - \frac{\sqrt{3} \mathbb{X}_{25}[\mathbb{Q}_{1,0}^{(a,E^{\prime})}] \otimes\mathbb{U}_{7}[\mathbb{T}_{1}^{(u,A_{2}^{\prime\prime})}] \otimes\mathbb{F}_{28}[\mathbb{T}_{2,0}^{(k,E^{\prime\prime})}]}{6}
 \cr&\hspace{1cm}
 + \frac{\sqrt{3} \mathbb{X}_{26}[\mathbb{Q}_{1,1}^{(a,E^{\prime})}] \otimes\mathbb{U}_{4}[\mathbb{Q}_{0}^{(u,A_{1}^{\prime})}] \otimes\mathbb{F}_{21}[\mathbb{Q}_{1,1}^{(k,E^{\prime})}]}{6} + \frac{\sqrt{3} \mathbb{X}_{26}[\mathbb{Q}_{1,1}^{(a,E^{\prime})}] \otimes\mathbb{U}_{5}[\mathbb{Q}_{1}^{(u,A_{2}^{\prime\prime})}] \otimes\mathbb{F}_{23}[\mathbb{Q}_{2,1}^{(k,E^{\prime\prime})}]}{6}
 \cr&\hspace{1cm}
 - \frac{\sqrt{3} \mathbb{X}_{26}[\mathbb{Q}_{1,1}^{(a,E^{\prime})}] \otimes\mathbb{U}_{6}[\mathbb{T}_{0}^{(u,A_{1}^{\prime})}] \otimes\mathbb{F}_{27}[\mathbb{T}_{1,1}^{(k,E^{\prime})}]}{6} - \frac{\sqrt{3} \mathbb{X}_{26}[\mathbb{Q}_{1,1}^{(a,E^{\prime})}] \otimes\mathbb{U}_{7}[\mathbb{T}_{1}^{(u,A_{2}^{\prime\prime})}] \otimes\mathbb{F}_{29}[\mathbb{T}_{2,1}^{(k,E^{\prime\prime})}]}{6}
\end{align*}
\vspace{4mm}
\noindent \fbox{No. {32}} $\,\,\,\hat{\mathbb{Q}}_{2}^{(A_{1}^{\prime})}$ [M$_{3}$,\,B$_{4}$]
\begin{align*}
\hat{\mathbb{Z}}_{32}=\frac{\sqrt{6} \mathbb{X}_{23}[\mathbb{Q}_{1}^{(a,A_{2}^{\prime\prime})}] \otimes\mathbb{Y}_{21}[\mathbb{Q}_{1}^{(b,A_{2}^{\prime\prime})}]}{3} - \frac{\sqrt{6} \mathbb{X}_{25}[\mathbb{Q}_{1,0}^{(a,E^{\prime})}] \otimes\mathbb{Y}_{22}[\mathbb{Q}_{1,0}^{(b,E^{\prime})}]}{6} - \frac{\sqrt{6} \mathbb{X}_{26}[\mathbb{Q}_{1,1}^{(a,E^{\prime})}] \otimes\mathbb{Y}_{23}[\mathbb{Q}_{1,1}^{(b,E^{\prime})}]}{6}
\end{align*}
\begin{align*}
&
\hat{\mathbb{Z}}_{32}(\bm{k})=\frac{\sqrt{6} \mathbb{X}_{23}[\mathbb{Q}_{1}^{(a,A_{2}^{\prime\prime})}] \otimes\mathbb{U}_{4}[\mathbb{Q}_{0}^{(u,A_{1}^{\prime})}] \otimes\mathbb{F}_{19}[\mathbb{Q}_{1}^{(k,A_{2}^{\prime\prime})}]}{6} + \frac{\sqrt{6} \mathbb{X}_{23}[\mathbb{Q}_{1}^{(a,A_{2}^{\prime\prime})}] \otimes\mathbb{U}_{5}[\mathbb{Q}_{1}^{(u,A_{2}^{\prime\prime})}] \otimes\mathbb{F}_{18}[\mathbb{Q}_{0}^{(k,A_{1}^{\prime})}]}{6}
\cr&\hspace{1cm}
 - \frac{\sqrt{6} \mathbb{X}_{23}[\mathbb{Q}_{1}^{(a,A_{2}^{\prime\prime})}] \otimes\mathbb{U}_{6}[\mathbb{T}_{0}^{(u,A_{1}^{\prime})}] \otimes\mathbb{F}_{25}[\mathbb{T}_{1}^{(k,A_{2}^{\prime\prime})}]}{6} - \frac{\sqrt{6} \mathbb{X}_{23}[\mathbb{Q}_{1}^{(a,A_{2}^{\prime\prime})}] \otimes\mathbb{U}_{7}[\mathbb{T}_{1}^{(u,A_{2}^{\prime\prime})}] \otimes\mathbb{F}_{24}[\mathbb{T}_{0}^{(k,A_{1}^{\prime})}]}{6}
 \cr&\hspace{1cm}
 - \frac{\sqrt{6} \mathbb{X}_{25}[\mathbb{Q}_{1,0}^{(a,E^{\prime})}] \otimes\mathbb{U}_{4}[\mathbb{Q}_{0}^{(u,A_{1}^{\prime})}] \otimes\mathbb{F}_{20}[\mathbb{Q}_{1,0}^{(k,E^{\prime})}]}{12} - \frac{\sqrt{6} \mathbb{X}_{25}[\mathbb{Q}_{1,0}^{(a,E^{\prime})}] \otimes\mathbb{U}_{5}[\mathbb{Q}_{1}^{(u,A_{2}^{\prime\prime})}] \otimes\mathbb{F}_{22}[\mathbb{Q}_{2,0}^{(k,E^{\prime\prime})}]}{12}
 \cr&\hspace{1cm}
 + \frac{\sqrt{6} \mathbb{X}_{25}[\mathbb{Q}_{1,0}^{(a,E^{\prime})}] \otimes\mathbb{U}_{6}[\mathbb{T}_{0}^{(u,A_{1}^{\prime})}] \otimes\mathbb{F}_{26}[\mathbb{T}_{1,0}^{(k,E^{\prime})}]}{12} + \frac{\sqrt{6} \mathbb{X}_{25}[\mathbb{Q}_{1,0}^{(a,E^{\prime})}] \otimes\mathbb{U}_{7}[\mathbb{T}_{1}^{(u,A_{2}^{\prime\prime})}] \otimes\mathbb{F}_{28}[\mathbb{T}_{2,0}^{(k,E^{\prime\prime})}]}{12}
 \cr&\hspace{1cm}
 - \frac{\sqrt{6} \mathbb{X}_{26}[\mathbb{Q}_{1,1}^{(a,E^{\prime})}] \otimes\mathbb{U}_{4}[\mathbb{Q}_{0}^{(u,A_{1}^{\prime})}] \otimes\mathbb{F}_{21}[\mathbb{Q}_{1,1}^{(k,E^{\prime})}]}{12} - \frac{\sqrt{6} \mathbb{X}_{26}[\mathbb{Q}_{1,1}^{(a,E^{\prime})}] \otimes\mathbb{U}_{5}[\mathbb{Q}_{1}^{(u,A_{2}^{\prime\prime})}] \otimes\mathbb{F}_{23}[\mathbb{Q}_{2,1}^{(k,E^{\prime\prime})}]}{12}
 \cr&\hspace{1cm}
 + \frac{\sqrt{6} \mathbb{X}_{26}[\mathbb{Q}_{1,1}^{(a,E^{\prime})}] \otimes\mathbb{U}_{6}[\mathbb{T}_{0}^{(u,A_{1}^{\prime})}] \otimes\mathbb{F}_{27}[\mathbb{T}_{1,1}^{(k,E^{\prime})}]}{12} + \frac{\sqrt{6} \mathbb{X}_{26}[\mathbb{Q}_{1,1}^{(a,E^{\prime})}] \otimes\mathbb{U}_{7}[\mathbb{T}_{1}^{(u,A_{2}^{\prime\prime})}] \otimes\mathbb{F}_{29}[\mathbb{T}_{2,1}^{(k,E^{\prime\prime})}]}{12}
\end{align*}
\vspace{4mm}
\noindent \fbox{No. {33}} $\,\,\,\hat{\mathbb{Q}}_{3}^{(A_{1}^{\prime})}$ [M$_{3}$,\,B$_{4}$]
\begin{align*}
\hat{\mathbb{Z}}_{33}=\mathbb{X}_{31}[\mathbb{Q}_{3}^{(a,A_{1}^{\prime})}] \otimes\mathbb{Y}_{20}[\mathbb{Q}_{0}^{(b,A_{1}^{\prime})}]
\end{align*}
\begin{align*}
&
\hat{\mathbb{Z}}_{33}(\bm{k})=\frac{\mathbb{X}_{31}[\mathbb{Q}_{3}^{(a,A_{1}^{\prime})}] \otimes\mathbb{U}_{4}[\mathbb{Q}_{0}^{(u,A_{1}^{\prime})}] \otimes\mathbb{F}_{18}[\mathbb{Q}_{0}^{(k,A_{1}^{\prime})}]}{2} + \frac{\mathbb{X}_{31}[\mathbb{Q}_{3}^{(a,A_{1}^{\prime})}] \otimes\mathbb{U}_{5}[\mathbb{Q}_{1}^{(u,A_{2}^{\prime\prime})}] \otimes\mathbb{F}_{19}[\mathbb{Q}_{1}^{(k,A_{2}^{\prime\prime})}]}{2}
\cr&\hspace{1cm}
 - \frac{\mathbb{X}_{31}[\mathbb{Q}_{3}^{(a,A_{1}^{\prime})}] \otimes\mathbb{U}_{6}[\mathbb{T}_{0}^{(u,A_{1}^{\prime})}] \otimes\mathbb{F}_{24}[\mathbb{T}_{0}^{(k,A_{1}^{\prime})}]}{2} - \frac{\mathbb{X}_{31}[\mathbb{Q}_{3}^{(a,A_{1}^{\prime})}] \otimes\mathbb{U}_{7}[\mathbb{T}_{1}^{(u,A_{2}^{\prime\prime})}] \otimes\mathbb{F}_{25}[\mathbb{T}_{1}^{(k,A_{2}^{\prime\prime})}]}{2}
\end{align*}
\vspace{4mm}
\noindent \fbox{No. {34}} $\,\,\,\hat{\mathbb{Q}}_{2}^{(A_{1}^{\prime})}$ [M$_{3}$,\,B$_{4}$]
\begin{align*}
\hat{\mathbb{Z}}_{34}=\frac{\sqrt{21} \mathbb{X}_{24}[\mathbb{Q}_{3}^{(a,A_{2}^{\prime\prime})}] \otimes\mathbb{Y}_{21}[\mathbb{Q}_{1}^{(b,A_{2}^{\prime\prime})}]}{7} + \frac{\sqrt{14} \mathbb{X}_{27}[\mathbb{Q}_{3,0}^{(a,E^{\prime})}] \otimes\mathbb{Y}_{22}[\mathbb{Q}_{1,0}^{(b,E^{\prime})}]}{7} + \frac{\sqrt{14} \mathbb{X}_{28}[\mathbb{Q}_{3,1}^{(a,E^{\prime})}] \otimes\mathbb{Y}_{23}[\mathbb{Q}_{1,1}^{(b,E^{\prime})}]}{7}
\end{align*}
\begin{align*}
&
\hat{\mathbb{Z}}_{34}(\bm{k})=\frac{\sqrt{21} \mathbb{X}_{24}[\mathbb{Q}_{3}^{(a,A_{2}^{\prime\prime})}] \otimes\mathbb{U}_{4}[\mathbb{Q}_{0}^{(u,A_{1}^{\prime})}] \otimes\mathbb{F}_{19}[\mathbb{Q}_{1}^{(k,A_{2}^{\prime\prime})}]}{14} + \frac{\sqrt{21} \mathbb{X}_{24}[\mathbb{Q}_{3}^{(a,A_{2}^{\prime\prime})}] \otimes\mathbb{U}_{5}[\mathbb{Q}_{1}^{(u,A_{2}^{\prime\prime})}] \otimes\mathbb{F}_{18}[\mathbb{Q}_{0}^{(k,A_{1}^{\prime})}]}{14}
\cr&\hspace{1cm}
 - \frac{\sqrt{21} \mathbb{X}_{24}[\mathbb{Q}_{3}^{(a,A_{2}^{\prime\prime})}] \otimes\mathbb{U}_{6}[\mathbb{T}_{0}^{(u,A_{1}^{\prime})}] \otimes\mathbb{F}_{25}[\mathbb{T}_{1}^{(k,A_{2}^{\prime\prime})}]}{14} - \frac{\sqrt{21} \mathbb{X}_{24}[\mathbb{Q}_{3}^{(a,A_{2}^{\prime\prime})}] \otimes\mathbb{U}_{7}[\mathbb{T}_{1}^{(u,A_{2}^{\prime\prime})}] \otimes\mathbb{F}_{24}[\mathbb{T}_{0}^{(k,A_{1}^{\prime})}]}{14}
 \cr&\hspace{1cm}
 + \frac{\sqrt{14} \mathbb{X}_{27}[\mathbb{Q}_{3,0}^{(a,E^{\prime})}] \otimes\mathbb{U}_{4}[\mathbb{Q}_{0}^{(u,A_{1}^{\prime})}] \otimes\mathbb{F}_{20}[\mathbb{Q}_{1,0}^{(k,E^{\prime})}]}{14} + \frac{\sqrt{14} \mathbb{X}_{27}[\mathbb{Q}_{3,0}^{(a,E^{\prime})}] \otimes\mathbb{U}_{5}[\mathbb{Q}_{1}^{(u,A_{2}^{\prime\prime})}] \otimes\mathbb{F}_{22}[\mathbb{Q}_{2,0}^{(k,E^{\prime\prime})}]}{14}
 \cr&\hspace{1cm}
 - \frac{\sqrt{14} \mathbb{X}_{27}[\mathbb{Q}_{3,0}^{(a,E^{\prime})}] \otimes\mathbb{U}_{6}[\mathbb{T}_{0}^{(u,A_{1}^{\prime})}] \otimes\mathbb{F}_{26}[\mathbb{T}_{1,0}^{(k,E^{\prime})}]}{14} - \frac{\sqrt{14} \mathbb{X}_{27}[\mathbb{Q}_{3,0}^{(a,E^{\prime})}] \otimes\mathbb{U}_{7}[\mathbb{T}_{1}^{(u,A_{2}^{\prime\prime})}] \otimes\mathbb{F}_{28}[\mathbb{T}_{2,0}^{(k,E^{\prime\prime})}]}{14}
 \cr&\hspace{1cm}
 + \frac{\sqrt{14} \mathbb{X}_{28}[\mathbb{Q}_{3,1}^{(a,E^{\prime})}] \otimes\mathbb{U}_{4}[\mathbb{Q}_{0}^{(u,A_{1}^{\prime})}] \otimes\mathbb{F}_{21}[\mathbb{Q}_{1,1}^{(k,E^{\prime})}]}{14} + \frac{\sqrt{14} \mathbb{X}_{28}[\mathbb{Q}_{3,1}^{(a,E^{\prime})}] \otimes\mathbb{U}_{5}[\mathbb{Q}_{1}^{(u,A_{2}^{\prime\prime})}] \otimes\mathbb{F}_{23}[\mathbb{Q}_{2,1}^{(k,E^{\prime\prime})}]}{14}
 \cr&\hspace{1cm}
 - \frac{\sqrt{14} \mathbb{X}_{28}[\mathbb{Q}_{3,1}^{(a,E^{\prime})}] \otimes\mathbb{U}_{6}[\mathbb{T}_{0}^{(u,A_{1}^{\prime})}] \otimes\mathbb{F}_{27}[\mathbb{T}_{1,1}^{(k,E^{\prime})}]}{14} - \frac{\sqrt{14} \mathbb{X}_{28}[\mathbb{Q}_{3,1}^{(a,E^{\prime})}] \otimes\mathbb{U}_{7}[\mathbb{T}_{1}^{(u,A_{2}^{\prime\prime})}] \otimes\mathbb{F}_{29}[\mathbb{T}_{2,1}^{(k,E^{\prime\prime})}]}{14}
\end{align*}
\vspace{4mm}
\noindent \fbox{No. {35}} $\,\,\,\hat{\mathbb{Q}}_{4}^{(A_{1}^{\prime})}$ [M$_{3}$,\,B$_{4}$]
\begin{align*}
\hat{\mathbb{Z}}_{35}=\frac{2 \sqrt{7} \mathbb{X}_{24}[\mathbb{Q}_{3}^{(a,A_{2}^{\prime\prime})}] \otimes\mathbb{Y}_{21}[\mathbb{Q}_{1}^{(b,A_{2}^{\prime\prime})}]}{7} - \frac{\sqrt{42} \mathbb{X}_{27}[\mathbb{Q}_{3,0}^{(a,E^{\prime})}] \otimes\mathbb{Y}_{22}[\mathbb{Q}_{1,0}^{(b,E^{\prime})}]}{14} - \frac{\sqrt{42} \mathbb{X}_{28}[\mathbb{Q}_{3,1}^{(a,E^{\prime})}] \otimes\mathbb{Y}_{23}[\mathbb{Q}_{1,1}^{(b,E^{\prime})}]}{14}
\end{align*}
\begin{align*}
&
\hat{\mathbb{Z}}_{35}(\bm{k})=\frac{\sqrt{7} \mathbb{X}_{24}[\mathbb{Q}_{3}^{(a,A_{2}^{\prime\prime})}] \otimes\mathbb{U}_{4}[\mathbb{Q}_{0}^{(u,A_{1}^{\prime})}] \otimes\mathbb{F}_{19}[\mathbb{Q}_{1}^{(k,A_{2}^{\prime\prime})}]}{7} + \frac{\sqrt{7} \mathbb{X}_{24}[\mathbb{Q}_{3}^{(a,A_{2}^{\prime\prime})}] \otimes\mathbb{U}_{5}[\mathbb{Q}_{1}^{(u,A_{2}^{\prime\prime})}] \otimes\mathbb{F}_{18}[\mathbb{Q}_{0}^{(k,A_{1}^{\prime})}]}{7}
\cr&\hspace{1cm}
 - \frac{\sqrt{7} \mathbb{X}_{24}[\mathbb{Q}_{3}^{(a,A_{2}^{\prime\prime})}] \otimes\mathbb{U}_{6}[\mathbb{T}_{0}^{(u,A_{1}^{\prime})}] \otimes\mathbb{F}_{25}[\mathbb{T}_{1}^{(k,A_{2}^{\prime\prime})}]}{7} - \frac{\sqrt{7} \mathbb{X}_{24}[\mathbb{Q}_{3}^{(a,A_{2}^{\prime\prime})}] \otimes\mathbb{U}_{7}[\mathbb{T}_{1}^{(u,A_{2}^{\prime\prime})}] \otimes\mathbb{F}_{24}[\mathbb{T}_{0}^{(k,A_{1}^{\prime})}]}{7}
 \cr&\hspace{1cm}
 - \frac{\sqrt{42} \mathbb{X}_{27}[\mathbb{Q}_{3,0}^{(a,E^{\prime})}] \otimes\mathbb{U}_{4}[\mathbb{Q}_{0}^{(u,A_{1}^{\prime})}] \otimes\mathbb{F}_{20}[\mathbb{Q}_{1,0}^{(k,E^{\prime})}]}{28} - \frac{\sqrt{42} \mathbb{X}_{27}[\mathbb{Q}_{3,0}^{(a,E^{\prime})}] \otimes\mathbb{U}_{5}[\mathbb{Q}_{1}^{(u,A_{2}^{\prime\prime})}] \otimes\mathbb{F}_{22}[\mathbb{Q}_{2,0}^{(k,E^{\prime\prime})}]}{28}
 \cr&\hspace{1cm}
 + \frac{\sqrt{42} \mathbb{X}_{27}[\mathbb{Q}_{3,0}^{(a,E^{\prime})}] \otimes\mathbb{U}_{6}[\mathbb{T}_{0}^{(u,A_{1}^{\prime})}] \otimes\mathbb{F}_{26}[\mathbb{T}_{1,0}^{(k,E^{\prime})}]}{28} + \frac{\sqrt{42} \mathbb{X}_{27}[\mathbb{Q}_{3,0}^{(a,E^{\prime})}] \otimes\mathbb{U}_{7}[\mathbb{T}_{1}^{(u,A_{2}^{\prime\prime})}] \otimes\mathbb{F}_{28}[\mathbb{T}_{2,0}^{(k,E^{\prime\prime})}]}{28}
 \cr&\hspace{1cm}
 - \frac{\sqrt{42} \mathbb{X}_{28}[\mathbb{Q}_{3,1}^{(a,E^{\prime})}] \otimes\mathbb{U}_{4}[\mathbb{Q}_{0}^{(u,A_{1}^{\prime})}] \otimes\mathbb{F}_{21}[\mathbb{Q}_{1,1}^{(k,E^{\prime})}]}{28} - \frac{\sqrt{42} \mathbb{X}_{28}[\mathbb{Q}_{3,1}^{(a,E^{\prime})}] \otimes\mathbb{U}_{5}[\mathbb{Q}_{1}^{(u,A_{2}^{\prime\prime})}] \otimes\mathbb{F}_{23}[\mathbb{Q}_{2,1}^{(k,E^{\prime\prime})}]}{28}
 \cr&\hspace{1cm}
 + \frac{\sqrt{42} \mathbb{X}_{28}[\mathbb{Q}_{3,1}^{(a,E^{\prime})}] \otimes\mathbb{U}_{6}[\mathbb{T}_{0}^{(u,A_{1}^{\prime})}] \otimes\mathbb{F}_{27}[\mathbb{T}_{1,1}^{(k,E^{\prime})}]}{28} + \frac{\sqrt{42} \mathbb{X}_{28}[\mathbb{Q}_{3,1}^{(a,E^{\prime})}] \otimes\mathbb{U}_{7}[\mathbb{T}_{1}^{(u,A_{2}^{\prime\prime})}] \otimes\mathbb{F}_{29}[\mathbb{T}_{2,1}^{(k,E^{\prime\prime})}]}{28}
\end{align*}
\vspace{4mm}
\noindent \fbox{No. {36}} $\,\,\,\hat{\mathbb{Q}}_{3}^{(A_{1}^{\prime})}$ [M$_{3}$,\,B$_{4}$]
\begin{align*}
\hat{\mathbb{Z}}_{36}=- \frac{\sqrt{2} \mathbb{X}_{33}[\mathbb{Q}_{3,0}^{(a,E^{\prime\prime})}] \otimes\mathbb{Y}_{24}[\mathbb{Q}_{2,0}^{(b,E^{\prime\prime})}]}{2} - \frac{\sqrt{2} \mathbb{X}_{34}[\mathbb{Q}_{3,1}^{(a,E^{\prime\prime})}] \otimes\mathbb{Y}_{25}[\mathbb{Q}_{2,1}^{(b,E^{\prime\prime})}]}{2}
\end{align*}
\begin{align*}
&
\hat{\mathbb{Z}}_{36}(\bm{k})=- \frac{\sqrt{2} \mathbb{X}_{33}[\mathbb{Q}_{3,0}^{(a,E^{\prime\prime})}] \otimes\mathbb{U}_{4}[\mathbb{Q}_{0}^{(u,A_{1}^{\prime})}] \otimes\mathbb{F}_{22}[\mathbb{Q}_{2,0}^{(k,E^{\prime\prime})}]}{4} - \frac{\sqrt{2} \mathbb{X}_{33}[\mathbb{Q}_{3,0}^{(a,E^{\prime\prime})}] \otimes\mathbb{U}_{5}[\mathbb{Q}_{1}^{(u,A_{2}^{\prime\prime})}] \otimes\mathbb{F}_{20}[\mathbb{Q}_{1,0}^{(k,E^{\prime})}]}{4}
\cr&\hspace{1cm}
 + \frac{\sqrt{2} \mathbb{X}_{33}[\mathbb{Q}_{3,0}^{(a,E^{\prime\prime})}] \otimes\mathbb{U}_{6}[\mathbb{T}_{0}^{(u,A_{1}^{\prime})}] \otimes\mathbb{F}_{28}[\mathbb{T}_{2,0}^{(k,E^{\prime\prime})}]}{4} + \frac{\sqrt{2} \mathbb{X}_{33}[\mathbb{Q}_{3,0}^{(a,E^{\prime\prime})}] \otimes\mathbb{U}_{7}[\mathbb{T}_{1}^{(u,A_{2}^{\prime\prime})}] \otimes\mathbb{F}_{26}[\mathbb{T}_{1,0}^{(k,E^{\prime})}]}{4}
 \cr&\hspace{1cm}
 - \frac{\sqrt{2} \mathbb{X}_{34}[\mathbb{Q}_{3,1}^{(a,E^{\prime\prime})}] \otimes\mathbb{U}_{4}[\mathbb{Q}_{0}^{(u,A_{1}^{\prime})}] \otimes\mathbb{F}_{23}[\mathbb{Q}_{2,1}^{(k,E^{\prime\prime})}]}{4} - \frac{\sqrt{2} \mathbb{X}_{34}[\mathbb{Q}_{3,1}^{(a,E^{\prime\prime})}] \otimes\mathbb{U}_{5}[\mathbb{Q}_{1}^{(u,A_{2}^{\prime\prime})}] \otimes\mathbb{F}_{21}[\mathbb{Q}_{1,1}^{(k,E^{\prime})}]}{4}
 \cr&\hspace{1cm}
 + \frac{\sqrt{2} \mathbb{X}_{34}[\mathbb{Q}_{3,1}^{(a,E^{\prime\prime})}] \otimes\mathbb{U}_{6}[\mathbb{T}_{0}^{(u,A_{1}^{\prime})}] \otimes\mathbb{F}_{29}[\mathbb{T}_{2,1}^{(k,E^{\prime\prime})}]}{4} + \frac{\sqrt{2} \mathbb{X}_{34}[\mathbb{Q}_{3,1}^{(a,E^{\prime\prime})}] \otimes\mathbb{U}_{7}[\mathbb{T}_{1}^{(u,A_{2}^{\prime\prime})}] \otimes\mathbb{F}_{27}[\mathbb{T}_{1,1}^{(k,E^{\prime})}]}{4}
\end{align*}
\vspace{4mm}
\noindent \fbox{No. {37}} $\,\,\,\hat{\mathbb{Q}}_{2}^{(A_{1}^{\prime})}$ [M$_{3}$,\,B$_{4}$]
\begin{align*}
\hat{\mathbb{Z}}_{37}=\frac{\sqrt{2} \mathbb{X}_{29}[\mathbb{M}_{2,0}^{(a,E^{\prime})}] \otimes\mathbb{Y}_{26}[\mathbb{T}_{1,0}^{(b,E^{\prime})}]}{2} + \frac{\sqrt{2} \mathbb{X}_{30}[\mathbb{M}_{2,1}^{(a,E^{\prime})}] \otimes\mathbb{Y}_{27}[\mathbb{T}_{1,1}^{(b,E^{\prime})}]}{2}
\end{align*}
\begin{align*}
&
\hat{\mathbb{Z}}_{37}(\bm{k})=\frac{\sqrt{2} \mathbb{X}_{29}[\mathbb{M}_{2,0}^{(a,E^{\prime})}] \otimes\mathbb{U}_{4}[\mathbb{Q}_{0}^{(u,A_{1}^{\prime})}] \otimes\mathbb{F}_{26}[\mathbb{T}_{1,0}^{(k,E^{\prime})}]}{4} + \frac{\sqrt{2} \mathbb{X}_{29}[\mathbb{M}_{2,0}^{(a,E^{\prime})}] \otimes\mathbb{U}_{5}[\mathbb{Q}_{1}^{(u,A_{2}^{\prime\prime})}] \otimes\mathbb{F}_{28}[\mathbb{T}_{2,0}^{(k,E^{\prime\prime})}]}{4}
\cr&\hspace{1cm}
 + \frac{\sqrt{2} \mathbb{X}_{29}[\mathbb{M}_{2,0}^{(a,E^{\prime})}] \otimes\mathbb{U}_{6}[\mathbb{T}_{0}^{(u,A_{1}^{\prime})}] \otimes\mathbb{F}_{20}[\mathbb{Q}_{1,0}^{(k,E^{\prime})}]}{4} + \frac{\sqrt{2} \mathbb{X}_{29}[\mathbb{M}_{2,0}^{(a,E^{\prime})}] \otimes\mathbb{U}_{7}[\mathbb{T}_{1}^{(u,A_{2}^{\prime\prime})}] \otimes\mathbb{F}_{22}[\mathbb{Q}_{2,0}^{(k,E^{\prime\prime})}]}{4}
 \cr&\hspace{1cm}
 + \frac{\sqrt{2} \mathbb{X}_{30}[\mathbb{M}_{2,1}^{(a,E^{\prime})}] \otimes\mathbb{U}_{4}[\mathbb{Q}_{0}^{(u,A_{1}^{\prime})}] \otimes\mathbb{F}_{27}[\mathbb{T}_{1,1}^{(k,E^{\prime})}]}{4} + \frac{\sqrt{2} \mathbb{X}_{30}[\mathbb{M}_{2,1}^{(a,E^{\prime})}] \otimes\mathbb{U}_{5}[\mathbb{Q}_{1}^{(u,A_{2}^{\prime\prime})}] \otimes\mathbb{F}_{29}[\mathbb{T}_{2,1}^{(k,E^{\prime\prime})}]}{4}
 \cr&\hspace{1cm}
 + \frac{\sqrt{2} \mathbb{X}_{30}[\mathbb{M}_{2,1}^{(a,E^{\prime})}] \otimes\mathbb{U}_{6}[\mathbb{T}_{0}^{(u,A_{1}^{\prime})}] \otimes\mathbb{F}_{21}[\mathbb{Q}_{1,1}^{(k,E^{\prime})}]}{4} + \frac{\sqrt{2} \mathbb{X}_{30}[\mathbb{M}_{2,1}^{(a,E^{\prime})}] \otimes\mathbb{U}_{7}[\mathbb{T}_{1}^{(u,A_{2}^{\prime\prime})}] \otimes\mathbb{F}_{23}[\mathbb{Q}_{2,1}^{(k,E^{\prime\prime})}]}{4}
\end{align*}
\vspace{4mm}
\noindent \fbox{No. {38}} $\,\,\,\hat{\mathbb{Q}}_{3}^{(A_{1}^{\prime})}$ [M$_{3}$,\,B$_{4}$]
\begin{align*}
\hat{\mathbb{Z}}_{38}=- \frac{\sqrt{2} \mathbb{X}_{35}[\mathbb{M}_{2,0}^{(a,E^{\prime\prime})}] \otimes\mathbb{Y}_{28}[\mathbb{T}_{2,0}^{(b,E^{\prime\prime})}]}{2} - \frac{\sqrt{2} \mathbb{X}_{36}[\mathbb{M}_{2,1}^{(a,E^{\prime\prime})}] \otimes\mathbb{Y}_{29}[\mathbb{T}_{2,1}^{(b,E^{\prime\prime})}]}{2}
\end{align*}
\begin{align*}
&
\hat{\mathbb{Z}}_{38}(\bm{k})=- \frac{\sqrt{2} \mathbb{X}_{35}[\mathbb{M}_{2,0}^{(a,E^{\prime\prime})}] \otimes\mathbb{U}_{4}[\mathbb{Q}_{0}^{(u,A_{1}^{\prime})}] \otimes\mathbb{F}_{28}[\mathbb{T}_{2,0}^{(k,E^{\prime\prime})}]}{4} - \frac{\sqrt{2} \mathbb{X}_{35}[\mathbb{M}_{2,0}^{(a,E^{\prime\prime})}] \otimes\mathbb{U}_{5}[\mathbb{Q}_{1}^{(u,A_{2}^{\prime\prime})}] \otimes\mathbb{F}_{26}[\mathbb{T}_{1,0}^{(k,E^{\prime})}]}{4}
\cr&\hspace{1cm}
 - \frac{\sqrt{2} \mathbb{X}_{35}[\mathbb{M}_{2,0}^{(a,E^{\prime\prime})}] \otimes\mathbb{U}_{6}[\mathbb{T}_{0}^{(u,A_{1}^{\prime})}] \otimes\mathbb{F}_{22}[\mathbb{Q}_{2,0}^{(k,E^{\prime\prime})}]}{4} - \frac{\sqrt{2} \mathbb{X}_{35}[\mathbb{M}_{2,0}^{(a,E^{\prime\prime})}] \otimes\mathbb{U}_{7}[\mathbb{T}_{1}^{(u,A_{2}^{\prime\prime})}] \otimes\mathbb{F}_{20}[\mathbb{Q}_{1,0}^{(k,E^{\prime})}]}{4}
 \cr&\hspace{1cm}
 - \frac{\sqrt{2} \mathbb{X}_{36}[\mathbb{M}_{2,1}^{(a,E^{\prime\prime})}] \otimes\mathbb{U}_{4}[\mathbb{Q}_{0}^{(u,A_{1}^{\prime})}] \otimes\mathbb{F}_{29}[\mathbb{T}_{2,1}^{(k,E^{\prime\prime})}]}{4} - \frac{\sqrt{2} \mathbb{X}_{36}[\mathbb{M}_{2,1}^{(a,E^{\prime\prime})}] \otimes\mathbb{U}_{5}[\mathbb{Q}_{1}^{(u,A_{2}^{\prime\prime})}] \otimes\mathbb{F}_{27}[\mathbb{T}_{1,1}^{(k,E^{\prime})}]}{4}
 \cr&\hspace{1cm}
 - \frac{\sqrt{2} \mathbb{X}_{36}[\mathbb{M}_{2,1}^{(a,E^{\prime\prime})}] \otimes\mathbb{U}_{6}[\mathbb{T}_{0}^{(u,A_{1}^{\prime})}] \otimes\mathbb{F}_{23}[\mathbb{Q}_{2,1}^{(k,E^{\prime\prime})}]}{4} - \frac{\sqrt{2} \mathbb{X}_{36}[\mathbb{M}_{2,1}^{(a,E^{\prime\prime})}] \otimes\mathbb{U}_{7}[\mathbb{T}_{1}^{(u,A_{2}^{\prime\prime})}] \otimes\mathbb{F}_{21}[\mathbb{Q}_{1,1}^{(k,E^{\prime})}]}{4}
\end{align*}
\vspace{4mm}
\noindent \fbox{No. {39}} $\,\,\,\hat{\mathbb{Q}}_{0}^{(A_{1}^{\prime})}$ [M$_{3}$,\,B$_{5}$]
\begin{align*}
\hat{\mathbb{Z}}_{39}=\frac{\sqrt{3} \mathbb{X}_{23}[\mathbb{Q}_{1}^{(a,A_{2}^{\prime\prime})}] \otimes\mathbb{Y}_{31}[\mathbb{Q}_{1}^{(b,A_{2}^{\prime\prime})}]}{3} + \frac{\sqrt{3} \mathbb{X}_{25}[\mathbb{Q}_{1,0}^{(a,E^{\prime})}] \otimes\mathbb{Y}_{32}[\mathbb{Q}_{1,0}^{(b,E^{\prime})}]}{3} + \frac{\sqrt{3} \mathbb{X}_{26}[\mathbb{Q}_{1,1}^{(a,E^{\prime})}] \otimes\mathbb{Y}_{33}[\mathbb{Q}_{1,1}^{(b,E^{\prime})}]}{3}
\end{align*}
\begin{align*}
&
\hat{\mathbb{Z}}_{39}(\bm{k})=\frac{\sqrt{3} \mathbb{X}_{23}[\mathbb{Q}_{1}^{(a,A_{2}^{\prime\prime})}] \otimes\mathbb{U}_{4}[\mathbb{Q}_{0}^{(u,A_{1}^{\prime})}] \otimes\mathbb{F}_{31}[\mathbb{Q}_{1}^{(k,A_{2}^{\prime\prime})}]}{6} + \frac{\sqrt{3} \mathbb{X}_{23}[\mathbb{Q}_{1}^{(a,A_{2}^{\prime\prime})}] \otimes\mathbb{U}_{5}[\mathbb{Q}_{1}^{(u,A_{2}^{\prime\prime})}] \otimes\mathbb{F}_{30}[\mathbb{Q}_{0}^{(k,A_{1}^{\prime})}]}{6}
\cr&\hspace{1cm}
 - \frac{\sqrt{3} \mathbb{X}_{23}[\mathbb{Q}_{1}^{(a,A_{2}^{\prime\prime})}] \otimes\mathbb{U}_{6}[\mathbb{T}_{0}^{(u,A_{1}^{\prime})}] \otimes\mathbb{F}_{37}[\mathbb{T}_{1}^{(k,A_{2}^{\prime\prime})}]}{6} - \frac{\sqrt{3} \mathbb{X}_{23}[\mathbb{Q}_{1}^{(a,A_{2}^{\prime\prime})}] \otimes\mathbb{U}_{7}[\mathbb{T}_{1}^{(u,A_{2}^{\prime\prime})}] \otimes\mathbb{F}_{36}[\mathbb{T}_{0}^{(k,A_{1}^{\prime})}]}{6}
 \cr&\hspace{1cm}
 + \frac{\sqrt{3} \mathbb{X}_{25}[\mathbb{Q}_{1,0}^{(a,E^{\prime})}] \otimes\mathbb{U}_{4}[\mathbb{Q}_{0}^{(u,A_{1}^{\prime})}] \otimes\mathbb{F}_{32}[\mathbb{Q}_{1,0}^{(k,E^{\prime})}]}{6} + \frac{\sqrt{3} \mathbb{X}_{25}[\mathbb{Q}_{1,0}^{(a,E^{\prime})}] \otimes\mathbb{U}_{5}[\mathbb{Q}_{1}^{(u,A_{2}^{\prime\prime})}] \otimes\mathbb{F}_{34}[\mathbb{Q}_{2,0}^{(k,E^{\prime\prime})}]}{6}
 \cr&\hspace{1cm}
 - \frac{\sqrt{3} \mathbb{X}_{25}[\mathbb{Q}_{1,0}^{(a,E^{\prime})}] \otimes\mathbb{U}_{6}[\mathbb{T}_{0}^{(u,A_{1}^{\prime})}] \otimes\mathbb{F}_{38}[\mathbb{T}_{1,0}^{(k,E^{\prime})}]}{6} - \frac{\sqrt{3} \mathbb{X}_{25}[\mathbb{Q}_{1,0}^{(a,E^{\prime})}] \otimes\mathbb{U}_{7}[\mathbb{T}_{1}^{(u,A_{2}^{\prime\prime})}] \otimes\mathbb{F}_{40}[\mathbb{T}_{2,0}^{(k,E^{\prime\prime})}]}{6}
 \cr&\hspace{1cm}
 + \frac{\sqrt{3} \mathbb{X}_{26}[\mathbb{Q}_{1,1}^{(a,E^{\prime})}] \otimes\mathbb{U}_{4}[\mathbb{Q}_{0}^{(u,A_{1}^{\prime})}] \otimes\mathbb{F}_{33}[\mathbb{Q}_{1,1}^{(k,E^{\prime})}]}{6} + \frac{\sqrt{3} \mathbb{X}_{26}[\mathbb{Q}_{1,1}^{(a,E^{\prime})}] \otimes\mathbb{U}_{5}[\mathbb{Q}_{1}^{(u,A_{2}^{\prime\prime})}] \otimes\mathbb{F}_{35}[\mathbb{Q}_{2,1}^{(k,E^{\prime\prime})}]}{6}
 \cr&\hspace{1cm}
 - \frac{\sqrt{3} \mathbb{X}_{26}[\mathbb{Q}_{1,1}^{(a,E^{\prime})}] \otimes\mathbb{U}_{6}[\mathbb{T}_{0}^{(u,A_{1}^{\prime})}] \otimes\mathbb{F}_{39}[\mathbb{T}_{1,1}^{(k,E^{\prime})}]}{6} - \frac{\sqrt{3} \mathbb{X}_{26}[\mathbb{Q}_{1,1}^{(a,E^{\prime})}] \otimes\mathbb{U}_{7}[\mathbb{T}_{1}^{(u,A_{2}^{\prime\prime})}] \otimes\mathbb{F}_{41}[\mathbb{T}_{2,1}^{(k,E^{\prime\prime})}]}{6}
\end{align*}
\vspace{4mm}
\noindent \fbox{No. {40}} $\,\,\,\hat{\mathbb{Q}}_{2}^{(A_{1}^{\prime})}$ [M$_{3}$,\,B$_{5}$]
\begin{align*}
\hat{\mathbb{Z}}_{40}=\frac{\sqrt{6} \mathbb{X}_{23}[\mathbb{Q}_{1}^{(a,A_{2}^{\prime\prime})}] \otimes\mathbb{Y}_{31}[\mathbb{Q}_{1}^{(b,A_{2}^{\prime\prime})}]}{3} - \frac{\sqrt{6} \mathbb{X}_{25}[\mathbb{Q}_{1,0}^{(a,E^{\prime})}] \otimes\mathbb{Y}_{32}[\mathbb{Q}_{1,0}^{(b,E^{\prime})}]}{6} - \frac{\sqrt{6} \mathbb{X}_{26}[\mathbb{Q}_{1,1}^{(a,E^{\prime})}] \otimes\mathbb{Y}_{33}[\mathbb{Q}_{1,1}^{(b,E^{\prime})}]}{6}
\end{align*}
\begin{align*}
&
\hat{\mathbb{Z}}_{40}(\bm{k})=\frac{\sqrt{6} \mathbb{X}_{23}[\mathbb{Q}_{1}^{(a,A_{2}^{\prime\prime})}] \otimes\mathbb{U}_{4}[\mathbb{Q}_{0}^{(u,A_{1}^{\prime})}] \otimes\mathbb{F}_{31}[\mathbb{Q}_{1}^{(k,A_{2}^{\prime\prime})}]}{6} + \frac{\sqrt{6} \mathbb{X}_{23}[\mathbb{Q}_{1}^{(a,A_{2}^{\prime\prime})}] \otimes\mathbb{U}_{5}[\mathbb{Q}_{1}^{(u,A_{2}^{\prime\prime})}] \otimes\mathbb{F}_{30}[\mathbb{Q}_{0}^{(k,A_{1}^{\prime})}]}{6}
\cr&\hspace{1cm}
 - \frac{\sqrt{6} \mathbb{X}_{23}[\mathbb{Q}_{1}^{(a,A_{2}^{\prime\prime})}] \otimes\mathbb{U}_{6}[\mathbb{T}_{0}^{(u,A_{1}^{\prime})}] \otimes\mathbb{F}_{37}[\mathbb{T}_{1}^{(k,A_{2}^{\prime\prime})}]}{6} - \frac{\sqrt{6} \mathbb{X}_{23}[\mathbb{Q}_{1}^{(a,A_{2}^{\prime\prime})}] \otimes\mathbb{U}_{7}[\mathbb{T}_{1}^{(u,A_{2}^{\prime\prime})}] \otimes\mathbb{F}_{36}[\mathbb{T}_{0}^{(k,A_{1}^{\prime})}]}{6}
 \cr&\hspace{1cm}
 - \frac{\sqrt{6} \mathbb{X}_{25}[\mathbb{Q}_{1,0}^{(a,E^{\prime})}] \otimes\mathbb{U}_{4}[\mathbb{Q}_{0}^{(u,A_{1}^{\prime})}] \otimes\mathbb{F}_{32}[\mathbb{Q}_{1,0}^{(k,E^{\prime})}]}{12} - \frac{\sqrt{6} \mathbb{X}_{25}[\mathbb{Q}_{1,0}^{(a,E^{\prime})}] \otimes\mathbb{U}_{5}[\mathbb{Q}_{1}^{(u,A_{2}^{\prime\prime})}] \otimes\mathbb{F}_{34}[\mathbb{Q}_{2,0}^{(k,E^{\prime\prime})}]}{12}
 \cr&\hspace{1cm}
 + \frac{\sqrt{6} \mathbb{X}_{25}[\mathbb{Q}_{1,0}^{(a,E^{\prime})}] \otimes\mathbb{U}_{6}[\mathbb{T}_{0}^{(u,A_{1}^{\prime})}] \otimes\mathbb{F}_{38}[\mathbb{T}_{1,0}^{(k,E^{\prime})}]}{12} + \frac{\sqrt{6} \mathbb{X}_{25}[\mathbb{Q}_{1,0}^{(a,E^{\prime})}] \otimes\mathbb{U}_{7}[\mathbb{T}_{1}^{(u,A_{2}^{\prime\prime})}] \otimes\mathbb{F}_{40}[\mathbb{T}_{2,0}^{(k,E^{\prime\prime})}]}{12}
 \cr&\hspace{1cm}
 - \frac{\sqrt{6} \mathbb{X}_{26}[\mathbb{Q}_{1,1}^{(a,E^{\prime})}] \otimes\mathbb{U}_{4}[\mathbb{Q}_{0}^{(u,A_{1}^{\prime})}] \otimes\mathbb{F}_{33}[\mathbb{Q}_{1,1}^{(k,E^{\prime})}]}{12} - \frac{\sqrt{6} \mathbb{X}_{26}[\mathbb{Q}_{1,1}^{(a,E^{\prime})}] \otimes\mathbb{U}_{5}[\mathbb{Q}_{1}^{(u,A_{2}^{\prime\prime})}] \otimes\mathbb{F}_{35}[\mathbb{Q}_{2,1}^{(k,E^{\prime\prime})}]}{12}
 \cr&\hspace{1cm}
 + \frac{\sqrt{6} \mathbb{X}_{26}[\mathbb{Q}_{1,1}^{(a,E^{\prime})}] \otimes\mathbb{U}_{6}[\mathbb{T}_{0}^{(u,A_{1}^{\prime})}] \otimes\mathbb{F}_{39}[\mathbb{T}_{1,1}^{(k,E^{\prime})}]}{12} + \frac{\sqrt{6} \mathbb{X}_{26}[\mathbb{Q}_{1,1}^{(a,E^{\prime})}] \otimes\mathbb{U}_{7}[\mathbb{T}_{1}^{(u,A_{2}^{\prime\prime})}] \otimes\mathbb{F}_{41}[\mathbb{T}_{2,1}^{(k,E^{\prime\prime})}]}{12}
\end{align*}
\vspace{4mm}
\noindent \fbox{No. {41}} $\,\,\,\hat{\mathbb{Q}}_{3}^{(A_{1}^{\prime})}$ [M$_{3}$,\,B$_{5}$]
\begin{align*}
\hat{\mathbb{Z}}_{41}=\mathbb{X}_{31}[\mathbb{Q}_{3}^{(a,A_{1}^{\prime})}] \otimes\mathbb{Y}_{30}[\mathbb{Q}_{0}^{(b,A_{1}^{\prime})}]
\end{align*}
\begin{align*}
&
\hat{\mathbb{Z}}_{41}(\bm{k})=\frac{\mathbb{X}_{31}[\mathbb{Q}_{3}^{(a,A_{1}^{\prime})}] \otimes\mathbb{U}_{4}[\mathbb{Q}_{0}^{(u,A_{1}^{\prime})}] \otimes\mathbb{F}_{30}[\mathbb{Q}_{0}^{(k,A_{1}^{\prime})}]}{2} + \frac{\mathbb{X}_{31}[\mathbb{Q}_{3}^{(a,A_{1}^{\prime})}] \otimes\mathbb{U}_{5}[\mathbb{Q}_{1}^{(u,A_{2}^{\prime\prime})}] \otimes\mathbb{F}_{31}[\mathbb{Q}_{1}^{(k,A_{2}^{\prime\prime})}]}{2}
\cr&\hspace{1cm}
 - \frac{\mathbb{X}_{31}[\mathbb{Q}_{3}^{(a,A_{1}^{\prime})}] \otimes\mathbb{U}_{6}[\mathbb{T}_{0}^{(u,A_{1}^{\prime})}] \otimes\mathbb{F}_{36}[\mathbb{T}_{0}^{(k,A_{1}^{\prime})}]}{2} - \frac{\mathbb{X}_{31}[\mathbb{Q}_{3}^{(a,A_{1}^{\prime})}] \otimes\mathbb{U}_{7}[\mathbb{T}_{1}^{(u,A_{2}^{\prime\prime})}] \otimes\mathbb{F}_{37}[\mathbb{T}_{1}^{(k,A_{2}^{\prime\prime})}]}{2}
\end{align*}
\vspace{4mm}
\noindent \fbox{No. {42}} $\,\,\,\hat{\mathbb{Q}}_{2}^{(A_{1}^{\prime})}$ [M$_{3}$,\,B$_{5}$]
\begin{align*}
\hat{\mathbb{Z}}_{42}=\frac{\sqrt{21} \mathbb{X}_{24}[\mathbb{Q}_{3}^{(a,A_{2}^{\prime\prime})}] \otimes\mathbb{Y}_{31}[\mathbb{Q}_{1}^{(b,A_{2}^{\prime\prime})}]}{7} + \frac{\sqrt{14} \mathbb{X}_{27}[\mathbb{Q}_{3,0}^{(a,E^{\prime})}] \otimes\mathbb{Y}_{32}[\mathbb{Q}_{1,0}^{(b,E^{\prime})}]}{7} + \frac{\sqrt{14} \mathbb{X}_{28}[\mathbb{Q}_{3,1}^{(a,E^{\prime})}] \otimes\mathbb{Y}_{33}[\mathbb{Q}_{1,1}^{(b,E^{\prime})}]}{7}
\end{align*}
\begin{align*}
&
\hat{\mathbb{Z}}_{42}(\bm{k})=\frac{\sqrt{21} \mathbb{X}_{24}[\mathbb{Q}_{3}^{(a,A_{2}^{\prime\prime})}] \otimes\mathbb{U}_{4}[\mathbb{Q}_{0}^{(u,A_{1}^{\prime})}] \otimes\mathbb{F}_{31}[\mathbb{Q}_{1}^{(k,A_{2}^{\prime\prime})}]}{14} + \frac{\sqrt{21} \mathbb{X}_{24}[\mathbb{Q}_{3}^{(a,A_{2}^{\prime\prime})}] \otimes\mathbb{U}_{5}[\mathbb{Q}_{1}^{(u,A_{2}^{\prime\prime})}] \otimes\mathbb{F}_{30}[\mathbb{Q}_{0}^{(k,A_{1}^{\prime})}]}{14}
\cr&\hspace{1cm}
 - \frac{\sqrt{21} \mathbb{X}_{24}[\mathbb{Q}_{3}^{(a,A_{2}^{\prime\prime})}] \otimes\mathbb{U}_{6}[\mathbb{T}_{0}^{(u,A_{1}^{\prime})}] \otimes\mathbb{F}_{37}[\mathbb{T}_{1}^{(k,A_{2}^{\prime\prime})}]}{14} - \frac{\sqrt{21} \mathbb{X}_{24}[\mathbb{Q}_{3}^{(a,A_{2}^{\prime\prime})}] \otimes\mathbb{U}_{7}[\mathbb{T}_{1}^{(u,A_{2}^{\prime\prime})}] \otimes\mathbb{F}_{36}[\mathbb{T}_{0}^{(k,A_{1}^{\prime})}]}{14}
 \cr&\hspace{1cm}
 + \frac{\sqrt{14} \mathbb{X}_{27}[\mathbb{Q}_{3,0}^{(a,E^{\prime})}] \otimes\mathbb{U}_{4}[\mathbb{Q}_{0}^{(u,A_{1}^{\prime})}] \otimes\mathbb{F}_{32}[\mathbb{Q}_{1,0}^{(k,E^{\prime})}]}{14} + \frac{\sqrt{14} \mathbb{X}_{27}[\mathbb{Q}_{3,0}^{(a,E^{\prime})}] \otimes\mathbb{U}_{5}[\mathbb{Q}_{1}^{(u,A_{2}^{\prime\prime})}] \otimes\mathbb{F}_{34}[\mathbb{Q}_{2,0}^{(k,E^{\prime\prime})}]}{14}
 \cr&\hspace{1cm}
 - \frac{\sqrt{14} \mathbb{X}_{27}[\mathbb{Q}_{3,0}^{(a,E^{\prime})}] \otimes\mathbb{U}_{6}[\mathbb{T}_{0}^{(u,A_{1}^{\prime})}] \otimes\mathbb{F}_{38}[\mathbb{T}_{1,0}^{(k,E^{\prime})}]}{14} - \frac{\sqrt{14} \mathbb{X}_{27}[\mathbb{Q}_{3,0}^{(a,E^{\prime})}] \otimes\mathbb{U}_{7}[\mathbb{T}_{1}^{(u,A_{2}^{\prime\prime})}] \otimes\mathbb{F}_{40}[\mathbb{T}_{2,0}^{(k,E^{\prime\prime})}]}{14}
 \cr&\hspace{1cm}
 + \frac{\sqrt{14} \mathbb{X}_{28}[\mathbb{Q}_{3,1}^{(a,E^{\prime})}] \otimes\mathbb{U}_{4}[\mathbb{Q}_{0}^{(u,A_{1}^{\prime})}] \otimes\mathbb{F}_{33}[\mathbb{Q}_{1,1}^{(k,E^{\prime})}]}{14} + \frac{\sqrt{14} \mathbb{X}_{28}[\mathbb{Q}_{3,1}^{(a,E^{\prime})}] \otimes\mathbb{U}_{5}[\mathbb{Q}_{1}^{(u,A_{2}^{\prime\prime})}] \otimes\mathbb{F}_{35}[\mathbb{Q}_{2,1}^{(k,E^{\prime\prime})}]}{14}
 \cr&\hspace{1cm}
 - \frac{\sqrt{14} \mathbb{X}_{28}[\mathbb{Q}_{3,1}^{(a,E^{\prime})}] \otimes\mathbb{U}_{6}[\mathbb{T}_{0}^{(u,A_{1}^{\prime})}] \otimes\mathbb{F}_{39}[\mathbb{T}_{1,1}^{(k,E^{\prime})}]}{14} - \frac{\sqrt{14} \mathbb{X}_{28}[\mathbb{Q}_{3,1}^{(a,E^{\prime})}] \otimes\mathbb{U}_{7}[\mathbb{T}_{1}^{(u,A_{2}^{\prime\prime})}] \otimes\mathbb{F}_{41}[\mathbb{T}_{2,1}^{(k,E^{\prime\prime})}]}{14}
\end{align*}
\vspace{4mm}
\noindent \fbox{No. {43}} $\,\,\,\hat{\mathbb{Q}}_{4}^{(A_{1}^{\prime})}$ [M$_{3}$,\,B$_{5}$]
\begin{align*}
\hat{\mathbb{Z}}_{43}=\frac{2 \sqrt{7} \mathbb{X}_{24}[\mathbb{Q}_{3}^{(a,A_{2}^{\prime\prime})}] \otimes\mathbb{Y}_{31}[\mathbb{Q}_{1}^{(b,A_{2}^{\prime\prime})}]}{7} - \frac{\sqrt{42} \mathbb{X}_{27}[\mathbb{Q}_{3,0}^{(a,E^{\prime})}] \otimes\mathbb{Y}_{32}[\mathbb{Q}_{1,0}^{(b,E^{\prime})}]}{14} - \frac{\sqrt{42} \mathbb{X}_{28}[\mathbb{Q}_{3,1}^{(a,E^{\prime})}] \otimes\mathbb{Y}_{33}[\mathbb{Q}_{1,1}^{(b,E^{\prime})}]}{14}
\end{align*}
\begin{align*}
&
\hat{\mathbb{Z}}_{43}(\bm{k})=\frac{\sqrt{7} \mathbb{X}_{24}[\mathbb{Q}_{3}^{(a,A_{2}^{\prime\prime})}] \otimes\mathbb{U}_{4}[\mathbb{Q}_{0}^{(u,A_{1}^{\prime})}] \otimes\mathbb{F}_{31}[\mathbb{Q}_{1}^{(k,A_{2}^{\prime\prime})}]}{7} + \frac{\sqrt{7} \mathbb{X}_{24}[\mathbb{Q}_{3}^{(a,A_{2}^{\prime\prime})}] \otimes\mathbb{U}_{5}[\mathbb{Q}_{1}^{(u,A_{2}^{\prime\prime})}] \otimes\mathbb{F}_{30}[\mathbb{Q}_{0}^{(k,A_{1}^{\prime})}]}{7}
\cr&\hspace{1cm}
 - \frac{\sqrt{7} \mathbb{X}_{24}[\mathbb{Q}_{3}^{(a,A_{2}^{\prime\prime})}] \otimes\mathbb{U}_{6}[\mathbb{T}_{0}^{(u,A_{1}^{\prime})}] \otimes\mathbb{F}_{37}[\mathbb{T}_{1}^{(k,A_{2}^{\prime\prime})}]}{7} - \frac{\sqrt{7} \mathbb{X}_{24}[\mathbb{Q}_{3}^{(a,A_{2}^{\prime\prime})}] \otimes\mathbb{U}_{7}[\mathbb{T}_{1}^{(u,A_{2}^{\prime\prime})}] \otimes\mathbb{F}_{36}[\mathbb{T}_{0}^{(k,A_{1}^{\prime})}]}{7}
 \cr&\hspace{1cm}
 - \frac{\sqrt{42} \mathbb{X}_{27}[\mathbb{Q}_{3,0}^{(a,E^{\prime})}] \otimes\mathbb{U}_{4}[\mathbb{Q}_{0}^{(u,A_{1}^{\prime})}] \otimes\mathbb{F}_{32}[\mathbb{Q}_{1,0}^{(k,E^{\prime})}]}{28} - \frac{\sqrt{42} \mathbb{X}_{27}[\mathbb{Q}_{3,0}^{(a,E^{\prime})}] \otimes\mathbb{U}_{5}[\mathbb{Q}_{1}^{(u,A_{2}^{\prime\prime})}] \otimes\mathbb{F}_{34}[\mathbb{Q}_{2,0}^{(k,E^{\prime\prime})}]}{28}
 \cr&\hspace{1cm}
 + \frac{\sqrt{42} \mathbb{X}_{27}[\mathbb{Q}_{3,0}^{(a,E^{\prime})}] \otimes\mathbb{U}_{6}[\mathbb{T}_{0}^{(u,A_{1}^{\prime})}] \otimes\mathbb{F}_{38}[\mathbb{T}_{1,0}^{(k,E^{\prime})}]}{28} + \frac{\sqrt{42} \mathbb{X}_{27}[\mathbb{Q}_{3,0}^{(a,E^{\prime})}] \otimes\mathbb{U}_{7}[\mathbb{T}_{1}^{(u,A_{2}^{\prime\prime})}] \otimes\mathbb{F}_{40}[\mathbb{T}_{2,0}^{(k,E^{\prime\prime})}]}{28}
 \cr&\hspace{1cm}
 - \frac{\sqrt{42} \mathbb{X}_{28}[\mathbb{Q}_{3,1}^{(a,E^{\prime})}] \otimes\mathbb{U}_{4}[\mathbb{Q}_{0}^{(u,A_{1}^{\prime})}] \otimes\mathbb{F}_{33}[\mathbb{Q}_{1,1}^{(k,E^{\prime})}]}{28} - \frac{\sqrt{42} \mathbb{X}_{28}[\mathbb{Q}_{3,1}^{(a,E^{\prime})}] \otimes\mathbb{U}_{5}[\mathbb{Q}_{1}^{(u,A_{2}^{\prime\prime})}] \otimes\mathbb{F}_{35}[\mathbb{Q}_{2,1}^{(k,E^{\prime\prime})}]}{28}
 \cr&\hspace{1cm}
 + \frac{\sqrt{42} \mathbb{X}_{28}[\mathbb{Q}_{3,1}^{(a,E^{\prime})}] \otimes\mathbb{U}_{6}[\mathbb{T}_{0}^{(u,A_{1}^{\prime})}] \otimes\mathbb{F}_{39}[\mathbb{T}_{1,1}^{(k,E^{\prime})}]}{28} + \frac{\sqrt{42} \mathbb{X}_{28}[\mathbb{Q}_{3,1}^{(a,E^{\prime})}] \otimes\mathbb{U}_{7}[\mathbb{T}_{1}^{(u,A_{2}^{\prime\prime})}] \otimes\mathbb{F}_{41}[\mathbb{T}_{2,1}^{(k,E^{\prime\prime})}]}{28}
\end{align*}
\vspace{4mm}
\noindent \fbox{No. {44}} $\,\,\,\hat{\mathbb{Q}}_{3}^{(A_{1}^{\prime})}$ [M$_{3}$,\,B$_{5}$]
\begin{align*}
\hat{\mathbb{Z}}_{44}=- \frac{\sqrt{2} \mathbb{X}_{33}[\mathbb{Q}_{3,0}^{(a,E^{\prime\prime})}] \otimes\mathbb{Y}_{34}[\mathbb{Q}_{2,0}^{(b,E^{\prime\prime})}]}{2} - \frac{\sqrt{2} \mathbb{X}_{34}[\mathbb{Q}_{3,1}^{(a,E^{\prime\prime})}] \otimes\mathbb{Y}_{35}[\mathbb{Q}_{2,1}^{(b,E^{\prime\prime})}]}{2}
\end{align*}
\begin{align*}
&
\hat{\mathbb{Z}}_{44}(\bm{k})=- \frac{\sqrt{2} \mathbb{X}_{33}[\mathbb{Q}_{3,0}^{(a,E^{\prime\prime})}] \otimes\mathbb{U}_{4}[\mathbb{Q}_{0}^{(u,A_{1}^{\prime})}] \otimes\mathbb{F}_{34}[\mathbb{Q}_{2,0}^{(k,E^{\prime\prime})}]}{4} - \frac{\sqrt{2} \mathbb{X}_{33}[\mathbb{Q}_{3,0}^{(a,E^{\prime\prime})}] \otimes\mathbb{U}_{5}[\mathbb{Q}_{1}^{(u,A_{2}^{\prime\prime})}] \otimes\mathbb{F}_{32}[\mathbb{Q}_{1,0}^{(k,E^{\prime})}]}{4}
\cr&\hspace{1cm}
 + \frac{\sqrt{2} \mathbb{X}_{33}[\mathbb{Q}_{3,0}^{(a,E^{\prime\prime})}] \otimes\mathbb{U}_{6}[\mathbb{T}_{0}^{(u,A_{1}^{\prime})}] \otimes\mathbb{F}_{40}[\mathbb{T}_{2,0}^{(k,E^{\prime\prime})}]}{4} + \frac{\sqrt{2} \mathbb{X}_{33}[\mathbb{Q}_{3,0}^{(a,E^{\prime\prime})}] \otimes\mathbb{U}_{7}[\mathbb{T}_{1}^{(u,A_{2}^{\prime\prime})}] \otimes\mathbb{F}_{38}[\mathbb{T}_{1,0}^{(k,E^{\prime})}]}{4}
 \cr&\hspace{1cm}
 - \frac{\sqrt{2} \mathbb{X}_{34}[\mathbb{Q}_{3,1}^{(a,E^{\prime\prime})}] \otimes\mathbb{U}_{4}[\mathbb{Q}_{0}^{(u,A_{1}^{\prime})}] \otimes\mathbb{F}_{35}[\mathbb{Q}_{2,1}^{(k,E^{\prime\prime})}]}{4} - \frac{\sqrt{2} \mathbb{X}_{34}[\mathbb{Q}_{3,1}^{(a,E^{\prime\prime})}] \otimes\mathbb{U}_{5}[\mathbb{Q}_{1}^{(u,A_{2}^{\prime\prime})}] \otimes\mathbb{F}_{33}[\mathbb{Q}_{1,1}^{(k,E^{\prime})}]}{4}
 \cr&\hspace{1cm}
 + \frac{\sqrt{2} \mathbb{X}_{34}[\mathbb{Q}_{3,1}^{(a,E^{\prime\prime})}] \otimes\mathbb{U}_{6}[\mathbb{T}_{0}^{(u,A_{1}^{\prime})}] \otimes\mathbb{F}_{41}[\mathbb{T}_{2,1}^{(k,E^{\prime\prime})}]}{4} + \frac{\sqrt{2} \mathbb{X}_{34}[\mathbb{Q}_{3,1}^{(a,E^{\prime\prime})}] \otimes\mathbb{U}_{7}[\mathbb{T}_{1}^{(u,A_{2}^{\prime\prime})}] \otimes\mathbb{F}_{39}[\mathbb{T}_{1,1}^{(k,E^{\prime})}]}{4}
\end{align*}
\vspace{4mm}
\noindent \fbox{No. {45}} $\,\,\,\hat{\mathbb{Q}}_{2}^{(A_{1}^{\prime})}$ [M$_{3}$,\,B$_{5}$]
\begin{align*}
\hat{\mathbb{Z}}_{45}=\frac{\sqrt{2} \mathbb{X}_{29}[\mathbb{M}_{2,0}^{(a,E^{\prime})}] \otimes\mathbb{Y}_{36}[\mathbb{T}_{1,0}^{(b,E^{\prime})}]}{2} + \frac{\sqrt{2} \mathbb{X}_{30}[\mathbb{M}_{2,1}^{(a,E^{\prime})}] \otimes\mathbb{Y}_{37}[\mathbb{T}_{1,1}^{(b,E^{\prime})}]}{2}
\end{align*}
\begin{align*}
&
\hat{\mathbb{Z}}_{45}(\bm{k})=\frac{\sqrt{2} \mathbb{X}_{29}[\mathbb{M}_{2,0}^{(a,E^{\prime})}] \otimes\mathbb{U}_{4}[\mathbb{Q}_{0}^{(u,A_{1}^{\prime})}] \otimes\mathbb{F}_{38}[\mathbb{T}_{1,0}^{(k,E^{\prime})}]}{4} + \frac{\sqrt{2} \mathbb{X}_{29}[\mathbb{M}_{2,0}^{(a,E^{\prime})}] \otimes\mathbb{U}_{5}[\mathbb{Q}_{1}^{(u,A_{2}^{\prime\prime})}] \otimes\mathbb{F}_{40}[\mathbb{T}_{2,0}^{(k,E^{\prime\prime})}]}{4}
\cr&\hspace{1cm}
 + \frac{\sqrt{2} \mathbb{X}_{29}[\mathbb{M}_{2,0}^{(a,E^{\prime})}] \otimes\mathbb{U}_{6}[\mathbb{T}_{0}^{(u,A_{1}^{\prime})}] \otimes\mathbb{F}_{32}[\mathbb{Q}_{1,0}^{(k,E^{\prime})}]}{4} + \frac{\sqrt{2} \mathbb{X}_{29}[\mathbb{M}_{2,0}^{(a,E^{\prime})}] \otimes\mathbb{U}_{7}[\mathbb{T}_{1}^{(u,A_{2}^{\prime\prime})}] \otimes\mathbb{F}_{34}[\mathbb{Q}_{2,0}^{(k,E^{\prime\prime})}]}{4}
 \cr&\hspace{1cm}
 + \frac{\sqrt{2} \mathbb{X}_{30}[\mathbb{M}_{2,1}^{(a,E^{\prime})}] \otimes\mathbb{U}_{4}[\mathbb{Q}_{0}^{(u,A_{1}^{\prime})}] \otimes\mathbb{F}_{39}[\mathbb{T}_{1,1}^{(k,E^{\prime})}]}{4} + \frac{\sqrt{2} \mathbb{X}_{30}[\mathbb{M}_{2,1}^{(a,E^{\prime})}] \otimes\mathbb{U}_{5}[\mathbb{Q}_{1}^{(u,A_{2}^{\prime\prime})}] \otimes\mathbb{F}_{41}[\mathbb{T}_{2,1}^{(k,E^{\prime\prime})}]}{4}
 \cr&\hspace{1cm}
 + \frac{\sqrt{2} \mathbb{X}_{30}[\mathbb{M}_{2,1}^{(a,E^{\prime})}] \otimes\mathbb{U}_{6}[\mathbb{T}_{0}^{(u,A_{1}^{\prime})}] \otimes\mathbb{F}_{33}[\mathbb{Q}_{1,1}^{(k,E^{\prime})}]}{4} + \frac{\sqrt{2} \mathbb{X}_{30}[\mathbb{M}_{2,1}^{(a,E^{\prime})}] \otimes\mathbb{U}_{7}[\mathbb{T}_{1}^{(u,A_{2}^{\prime\prime})}] \otimes\mathbb{F}_{35}[\mathbb{Q}_{2,1}^{(k,E^{\prime\prime})}]}{4}
\end{align*}
\vspace{4mm}
\noindent \fbox{No. {46}} $\,\,\,\hat{\mathbb{Q}}_{3}^{(A_{1}^{\prime})}$ [M$_{3}$,\,B$_{5}$]
\begin{align*}
\hat{\mathbb{Z}}_{46}=- \frac{\sqrt{2} \mathbb{X}_{35}[\mathbb{M}_{2,0}^{(a,E^{\prime\prime})}] \otimes\mathbb{Y}_{38}[\mathbb{T}_{2,0}^{(b,E^{\prime\prime})}]}{2} - \frac{\sqrt{2} \mathbb{X}_{36}[\mathbb{M}_{2,1}^{(a,E^{\prime\prime})}] \otimes\mathbb{Y}_{39}[\mathbb{T}_{2,1}^{(b,E^{\prime\prime})}]}{2}
\end{align*}
\begin{align*}
&
\hat{\mathbb{Z}}_{46}(\bm{k})=- \frac{\sqrt{2} \mathbb{X}_{35}[\mathbb{M}_{2,0}^{(a,E^{\prime\prime})}] \otimes\mathbb{U}_{4}[\mathbb{Q}_{0}^{(u,A_{1}^{\prime})}] \otimes\mathbb{F}_{40}[\mathbb{T}_{2,0}^{(k,E^{\prime\prime})}]}{4} - \frac{\sqrt{2} \mathbb{X}_{35}[\mathbb{M}_{2,0}^{(a,E^{\prime\prime})}] \otimes\mathbb{U}_{5}[\mathbb{Q}_{1}^{(u,A_{2}^{\prime\prime})}] \otimes\mathbb{F}_{38}[\mathbb{T}_{1,0}^{(k,E^{\prime})}]}{4}
\cr&\hspace{1cm}
 - \frac{\sqrt{2} \mathbb{X}_{35}[\mathbb{M}_{2,0}^{(a,E^{\prime\prime})}] \otimes\mathbb{U}_{6}[\mathbb{T}_{0}^{(u,A_{1}^{\prime})}] \otimes\mathbb{F}_{34}[\mathbb{Q}_{2,0}^{(k,E^{\prime\prime})}]}{4} - \frac{\sqrt{2} \mathbb{X}_{35}[\mathbb{M}_{2,0}^{(a,E^{\prime\prime})}] \otimes\mathbb{U}_{7}[\mathbb{T}_{1}^{(u,A_{2}^{\prime\prime})}] \otimes\mathbb{F}_{32}[\mathbb{Q}_{1,0}^{(k,E^{\prime})}]}{4}
 \cr&\hspace{1cm}
 - \frac{\sqrt{2} \mathbb{X}_{36}[\mathbb{M}_{2,1}^{(a,E^{\prime\prime})}] \otimes\mathbb{U}_{4}[\mathbb{Q}_{0}^{(u,A_{1}^{\prime})}] \otimes\mathbb{F}_{41}[\mathbb{T}_{2,1}^{(k,E^{\prime\prime})}]}{4} - \frac{\sqrt{2} \mathbb{X}_{36}[\mathbb{M}_{2,1}^{(a,E^{\prime\prime})}] \otimes\mathbb{U}_{5}[\mathbb{Q}_{1}^{(u,A_{2}^{\prime\prime})}] \otimes\mathbb{F}_{39}[\mathbb{T}_{1,1}^{(k,E^{\prime})}]}{4}
 \cr&\hspace{1cm}
 - \frac{\sqrt{2} \mathbb{X}_{36}[\mathbb{M}_{2,1}^{(a,E^{\prime\prime})}] \otimes\mathbb{U}_{6}[\mathbb{T}_{0}^{(u,A_{1}^{\prime})}] \otimes\mathbb{F}_{35}[\mathbb{Q}_{2,1}^{(k,E^{\prime\prime})}]}{4} - \frac{\sqrt{2} \mathbb{X}_{36}[\mathbb{M}_{2,1}^{(a,E^{\prime\prime})}] \otimes\mathbb{U}_{7}[\mathbb{T}_{1}^{(u,A_{2}^{\prime\prime})}] \otimes\mathbb{F}_{33}[\mathbb{Q}_{1,1}^{(k,E^{\prime})}]}{4}
\end{align*}
\vspace{4mm}
\noindent \fbox{No. {47}} $\,\,\,\hat{\mathbb{Q}}_{0}^{(A_{1}^{\prime})}$ [M$_{3}$,\,B$_{6}$]
\begin{align*}
\hat{\mathbb{Z}}_{47}=\frac{\sqrt{3} \mathbb{X}_{23}[\mathbb{Q}_{1}^{(a,A_{2}^{\prime\prime})}] \otimes\mathbb{Y}_{41}[\mathbb{Q}_{1}^{(b,A_{2}^{\prime\prime})}]}{3} + \frac{\sqrt{3} \mathbb{X}_{25}[\mathbb{Q}_{1,0}^{(a,E^{\prime})}] \otimes\mathbb{Y}_{42}[\mathbb{Q}_{1,0}^{(b,E^{\prime})}]}{3} + \frac{\sqrt{3} \mathbb{X}_{26}[\mathbb{Q}_{1,1}^{(a,E^{\prime})}] \otimes\mathbb{Y}_{43}[\mathbb{Q}_{1,1}^{(b,E^{\prime})}]}{3}
\end{align*}
\begin{align*}
&
\hat{\mathbb{Z}}_{47}(\bm{k})=\frac{\sqrt{3} \mathbb{X}_{23}[\mathbb{Q}_{1}^{(a,A_{2}^{\prime\prime})}] \otimes\mathbb{U}_{4}[\mathbb{Q}_{0}^{(u,A_{1}^{\prime})}] \otimes\mathbb{F}_{43}[\mathbb{Q}_{1}^{(k,A_{2}^{\prime\prime})}]}{6} + \frac{\sqrt{3} \mathbb{X}_{23}[\mathbb{Q}_{1}^{(a,A_{2}^{\prime\prime})}] \otimes\mathbb{U}_{5}[\mathbb{Q}_{1}^{(u,A_{2}^{\prime\prime})}] \otimes\mathbb{F}_{42}[\mathbb{Q}_{0}^{(k,A_{1}^{\prime})}]}{6}
\cr&\hspace{1cm}
 - \frac{\sqrt{3} \mathbb{X}_{23}[\mathbb{Q}_{1}^{(a,A_{2}^{\prime\prime})}] \otimes\mathbb{U}_{6}[\mathbb{T}_{0}^{(u,A_{1}^{\prime})}] \otimes\mathbb{F}_{55}[\mathbb{T}_{1}^{(k,A_{2}^{\prime\prime})}]}{6} - \frac{\sqrt{3} \mathbb{X}_{23}[\mathbb{Q}_{1}^{(a,A_{2}^{\prime\prime})}] \otimes\mathbb{U}_{7}[\mathbb{T}_{1}^{(u,A_{2}^{\prime\prime})}] \otimes\mathbb{F}_{54}[\mathbb{T}_{0}^{(k,A_{1}^{\prime})}]}{6}
 \cr&\hspace{1cm}
 + \frac{\sqrt{3} \mathbb{X}_{25}[\mathbb{Q}_{1,0}^{(a,E^{\prime})}] \otimes\mathbb{U}_{4}[\mathbb{Q}_{0}^{(u,A_{1}^{\prime})}] \otimes\mathbb{F}_{44}[\mathbb{Q}_{1,0}^{(k,E^{\prime})}]}{6} + \frac{\sqrt{3} \mathbb{X}_{25}[\mathbb{Q}_{1,0}^{(a,E^{\prime})}] \otimes\mathbb{U}_{5}[\mathbb{Q}_{1}^{(u,A_{2}^{\prime\prime})}] \otimes\mathbb{F}_{48}[\mathbb{Q}_{2,0}^{(k,E^{\prime\prime})}]}{6}
 \cr&\hspace{1cm}
 - \frac{\sqrt{3} \mathbb{X}_{25}[\mathbb{Q}_{1,0}^{(a,E^{\prime})}] \otimes\mathbb{U}_{6}[\mathbb{T}_{0}^{(u,A_{1}^{\prime})}] \otimes\mathbb{F}_{56}[\mathbb{T}_{1,0}^{(k,E^{\prime})}]}{6} - \frac{\sqrt{3} \mathbb{X}_{25}[\mathbb{Q}_{1,0}^{(a,E^{\prime})}] \otimes\mathbb{U}_{7}[\mathbb{T}_{1}^{(u,A_{2}^{\prime\prime})}] \otimes\mathbb{F}_{60}[\mathbb{T}_{2,0}^{(k,E^{\prime\prime})}]}{6}
 \cr&\hspace{1cm}
 + \frac{\sqrt{3} \mathbb{X}_{26}[\mathbb{Q}_{1,1}^{(a,E^{\prime})}] \otimes\mathbb{U}_{4}[\mathbb{Q}_{0}^{(u,A_{1}^{\prime})}] \otimes\mathbb{F}_{45}[\mathbb{Q}_{1,1}^{(k,E^{\prime})}]}{6} + \frac{\sqrt{3} \mathbb{X}_{26}[\mathbb{Q}_{1,1}^{(a,E^{\prime})}] \otimes\mathbb{U}_{5}[\mathbb{Q}_{1}^{(u,A_{2}^{\prime\prime})}] \otimes\mathbb{F}_{49}[\mathbb{Q}_{2,1}^{(k,E^{\prime\prime})}]}{6}
 \cr&\hspace{1cm}
 - \frac{\sqrt{3} \mathbb{X}_{26}[\mathbb{Q}_{1,1}^{(a,E^{\prime})}] \otimes\mathbb{U}_{6}[\mathbb{T}_{0}^{(u,A_{1}^{\prime})}] \otimes\mathbb{F}_{57}[\mathbb{T}_{1,1}^{(k,E^{\prime})}]}{6} - \frac{\sqrt{3} \mathbb{X}_{26}[\mathbb{Q}_{1,1}^{(a,E^{\prime})}] \otimes\mathbb{U}_{7}[\mathbb{T}_{1}^{(u,A_{2}^{\prime\prime})}] \otimes\mathbb{F}_{61}[\mathbb{T}_{2,1}^{(k,E^{\prime\prime})}]}{6}
\end{align*}
\vspace{4mm}
\noindent \fbox{No. {48}} $\,\,\,\hat{\mathbb{Q}}_{2}^{(A_{1}^{\prime})}$ [M$_{3}$,\,B$_{6}$]
\begin{align*}
\hat{\mathbb{Z}}_{48}=\frac{\sqrt{6} \mathbb{X}_{23}[\mathbb{Q}_{1}^{(a,A_{2}^{\prime\prime})}] \otimes\mathbb{Y}_{41}[\mathbb{Q}_{1}^{(b,A_{2}^{\prime\prime})}]}{3} - \frac{\sqrt{6} \mathbb{X}_{25}[\mathbb{Q}_{1,0}^{(a,E^{\prime})}] \otimes\mathbb{Y}_{42}[\mathbb{Q}_{1,0}^{(b,E^{\prime})}]}{6} - \frac{\sqrt{6} \mathbb{X}_{26}[\mathbb{Q}_{1,1}^{(a,E^{\prime})}] \otimes\mathbb{Y}_{43}[\mathbb{Q}_{1,1}^{(b,E^{\prime})}]}{6}
\end{align*}
\begin{align*}
&
\hat{\mathbb{Z}}_{48}(\bm{k})=\frac{\sqrt{6} \mathbb{X}_{23}[\mathbb{Q}_{1}^{(a,A_{2}^{\prime\prime})}] \otimes\mathbb{U}_{4}[\mathbb{Q}_{0}^{(u,A_{1}^{\prime})}] \otimes\mathbb{F}_{43}[\mathbb{Q}_{1}^{(k,A_{2}^{\prime\prime})}]}{6} + \frac{\sqrt{6} \mathbb{X}_{23}[\mathbb{Q}_{1}^{(a,A_{2}^{\prime\prime})}] \otimes\mathbb{U}_{5}[\mathbb{Q}_{1}^{(u,A_{2}^{\prime\prime})}] \otimes\mathbb{F}_{42}[\mathbb{Q}_{0}^{(k,A_{1}^{\prime})}]}{6}
\cr&\hspace{1cm}
 - \frac{\sqrt{6} \mathbb{X}_{23}[\mathbb{Q}_{1}^{(a,A_{2}^{\prime\prime})}] \otimes\mathbb{U}_{6}[\mathbb{T}_{0}^{(u,A_{1}^{\prime})}] \otimes\mathbb{F}_{55}[\mathbb{T}_{1}^{(k,A_{2}^{\prime\prime})}]}{6} - \frac{\sqrt{6} \mathbb{X}_{23}[\mathbb{Q}_{1}^{(a,A_{2}^{\prime\prime})}] \otimes\mathbb{U}_{7}[\mathbb{T}_{1}^{(u,A_{2}^{\prime\prime})}] \otimes\mathbb{F}_{54}[\mathbb{T}_{0}^{(k,A_{1}^{\prime})}]}{6}
 \cr&\hspace{1cm}
 - \frac{\sqrt{6} \mathbb{X}_{25}[\mathbb{Q}_{1,0}^{(a,E^{\prime})}] \otimes\mathbb{U}_{4}[\mathbb{Q}_{0}^{(u,A_{1}^{\prime})}] \otimes\mathbb{F}_{44}[\mathbb{Q}_{1,0}^{(k,E^{\prime})}]}{12} - \frac{\sqrt{6} \mathbb{X}_{25}[\mathbb{Q}_{1,0}^{(a,E^{\prime})}] \otimes\mathbb{U}_{5}[\mathbb{Q}_{1}^{(u,A_{2}^{\prime\prime})}] \otimes\mathbb{F}_{48}[\mathbb{Q}_{2,0}^{(k,E^{\prime\prime})}]}{12}
 \cr&\hspace{1cm}
 + \frac{\sqrt{6} \mathbb{X}_{25}[\mathbb{Q}_{1,0}^{(a,E^{\prime})}] \otimes\mathbb{U}_{6}[\mathbb{T}_{0}^{(u,A_{1}^{\prime})}] \otimes\mathbb{F}_{56}[\mathbb{T}_{1,0}^{(k,E^{\prime})}]}{12} + \frac{\sqrt{6} \mathbb{X}_{25}[\mathbb{Q}_{1,0}^{(a,E^{\prime})}] \otimes\mathbb{U}_{7}[\mathbb{T}_{1}^{(u,A_{2}^{\prime\prime})}] \otimes\mathbb{F}_{60}[\mathbb{T}_{2,0}^{(k,E^{\prime\prime})}]}{12}
 \cr&\hspace{1cm}
 - \frac{\sqrt{6} \mathbb{X}_{26}[\mathbb{Q}_{1,1}^{(a,E^{\prime})}] \otimes\mathbb{U}_{4}[\mathbb{Q}_{0}^{(u,A_{1}^{\prime})}] \otimes\mathbb{F}_{45}[\mathbb{Q}_{1,1}^{(k,E^{\prime})}]}{12} - \frac{\sqrt{6} \mathbb{X}_{26}[\mathbb{Q}_{1,1}^{(a,E^{\prime})}] \otimes\mathbb{U}_{5}[\mathbb{Q}_{1}^{(u,A_{2}^{\prime\prime})}] \otimes\mathbb{F}_{49}[\mathbb{Q}_{2,1}^{(k,E^{\prime\prime})}]}{12}
 \cr&\hspace{1cm}
 + \frac{\sqrt{6} \mathbb{X}_{26}[\mathbb{Q}_{1,1}^{(a,E^{\prime})}] \otimes\mathbb{U}_{6}[\mathbb{T}_{0}^{(u,A_{1}^{\prime})}] \otimes\mathbb{F}_{57}[\mathbb{T}_{1,1}^{(k,E^{\prime})}]}{12} + \frac{\sqrt{6} \mathbb{X}_{26}[\mathbb{Q}_{1,1}^{(a,E^{\prime})}] \otimes\mathbb{U}_{7}[\mathbb{T}_{1}^{(u,A_{2}^{\prime\prime})}] \otimes\mathbb{F}_{61}[\mathbb{T}_{2,1}^{(k,E^{\prime\prime})}]}{12}
\end{align*}
\vspace{4mm}
\noindent \fbox{No. {49}} $\,\,\,\hat{\mathbb{Q}}_{3}^{(A_{1}^{\prime})}$ [M$_{3}$,\,B$_{6}$]
\begin{align*}
\hat{\mathbb{Z}}_{49}=- \frac{\sqrt{2} \mathbb{X}_{25}[\mathbb{Q}_{1,0}^{(a,E^{\prime})}] \otimes\mathbb{Y}_{44}[\mathbb{Q}_{2,0}^{(b,E^{\prime})}]}{2} - \frac{\sqrt{2} \mathbb{X}_{26}[\mathbb{Q}_{1,1}^{(a,E^{\prime})}] \otimes\mathbb{Y}_{45}[\mathbb{Q}_{2,1}^{(b,E^{\prime})}]}{2}
\end{align*}
\begin{align*}
&
\hat{\mathbb{Z}}_{49}(\bm{k})=- \frac{\sqrt{2} \mathbb{X}_{25}[\mathbb{Q}_{1,0}^{(a,E^{\prime})}] \otimes\mathbb{U}_{4}[\mathbb{Q}_{0}^{(u,A_{1}^{\prime})}] \otimes\mathbb{F}_{46}[\mathbb{Q}_{2,0}^{(k,E^{\prime})}]}{4} - \frac{\sqrt{2} \mathbb{X}_{25}[\mathbb{Q}_{1,0}^{(a,E^{\prime})}] \otimes\mathbb{U}_{5}[\mathbb{Q}_{1}^{(u,A_{2}^{\prime\prime})}] \otimes\mathbb{F}_{51}[\mathbb{Q}_{3,0}^{(k,E^{\prime\prime})}]}{4}
\cr&\hspace{1cm}
 + \frac{\sqrt{2} \mathbb{X}_{25}[\mathbb{Q}_{1,0}^{(a,E^{\prime})}] \otimes\mathbb{U}_{6}[\mathbb{T}_{0}^{(u,A_{1}^{\prime})}] \otimes\mathbb{F}_{58}[\mathbb{T}_{2,0}^{(k,E^{\prime})}]}{4} + \frac{\sqrt{2} \mathbb{X}_{25}[\mathbb{Q}_{1,0}^{(a,E^{\prime})}] \otimes\mathbb{U}_{7}[\mathbb{T}_{1}^{(u,A_{2}^{\prime\prime})}] \otimes\mathbb{F}_{63}[\mathbb{T}_{3,0}^{(k,E^{\prime\prime})}]}{4}
 \cr&\hspace{1cm}
 - \frac{\sqrt{2} \mathbb{X}_{26}[\mathbb{Q}_{1,1}^{(a,E^{\prime})}] \otimes\mathbb{U}_{4}[\mathbb{Q}_{0}^{(u,A_{1}^{\prime})}] \otimes\mathbb{F}_{47}[\mathbb{Q}_{2,1}^{(k,E^{\prime})}]}{4} - \frac{\sqrt{2} \mathbb{X}_{26}[\mathbb{Q}_{1,1}^{(a,E^{\prime})}] \otimes\mathbb{U}_{5}[\mathbb{Q}_{1}^{(u,A_{2}^{\prime\prime})}] \otimes\mathbb{F}_{52}[\mathbb{Q}_{3,1}^{(k,E^{\prime\prime})}]}{4}
 \cr&\hspace{1cm}
 + \frac{\sqrt{2} \mathbb{X}_{26}[\mathbb{Q}_{1,1}^{(a,E^{\prime})}] \otimes\mathbb{U}_{6}[\mathbb{T}_{0}^{(u,A_{1}^{\prime})}] \otimes\mathbb{F}_{59}[\mathbb{T}_{2,1}^{(k,E^{\prime})}]}{4} + \frac{\sqrt{2} \mathbb{X}_{26}[\mathbb{Q}_{1,1}^{(a,E^{\prime})}] \otimes\mathbb{U}_{7}[\mathbb{T}_{1}^{(u,A_{2}^{\prime\prime})}] \otimes\mathbb{F}_{64}[\mathbb{T}_{3,1}^{(k,E^{\prime\prime})}]}{4}
\end{align*}
\vspace{4mm}
\noindent \fbox{No. {50}} $\,\,\,\hat{\mathbb{Q}}_{3}^{(A_{1}^{\prime})}$ [M$_{3}$,\,B$_{6}$]
\begin{align*}
\hat{\mathbb{Z}}_{50}=\mathbb{X}_{31}[\mathbb{Q}_{3}^{(a,A_{1}^{\prime})}] \otimes\mathbb{Y}_{40}[\mathbb{Q}_{0}^{(b,A_{1}^{\prime})}]
\end{align*}
\begin{align*}
&
\hat{\mathbb{Z}}_{50}(\bm{k})=\frac{\mathbb{X}_{31}[\mathbb{Q}_{3}^{(a,A_{1}^{\prime})}] \otimes\mathbb{U}_{4}[\mathbb{Q}_{0}^{(u,A_{1}^{\prime})}] \otimes\mathbb{F}_{42}[\mathbb{Q}_{0}^{(k,A_{1}^{\prime})}]}{2} + \frac{\mathbb{X}_{31}[\mathbb{Q}_{3}^{(a,A_{1}^{\prime})}] \otimes\mathbb{U}_{5}[\mathbb{Q}_{1}^{(u,A_{2}^{\prime\prime})}] \otimes\mathbb{F}_{43}[\mathbb{Q}_{1}^{(k,A_{2}^{\prime\prime})}]}{2}
\cr&\hspace{1cm}
 - \frac{\mathbb{X}_{31}[\mathbb{Q}_{3}^{(a,A_{1}^{\prime})}] \otimes\mathbb{U}_{6}[\mathbb{T}_{0}^{(u,A_{1}^{\prime})}] \otimes\mathbb{F}_{54}[\mathbb{T}_{0}^{(k,A_{1}^{\prime})}]}{2} - \frac{\mathbb{X}_{31}[\mathbb{Q}_{3}^{(a,A_{1}^{\prime})}] \otimes\mathbb{U}_{7}[\mathbb{T}_{1}^{(u,A_{2}^{\prime\prime})}] \otimes\mathbb{F}_{55}[\mathbb{T}_{1}^{(k,A_{2}^{\prime\prime})}]}{2}
\end{align*}
\vspace{4mm}
\noindent \fbox{No. {51}} $\,\,\,\hat{\mathbb{Q}}_{2}^{(A_{1}^{\prime})}$ [M$_{3}$,\,B$_{6}$]
\begin{align*}
\hat{\mathbb{Z}}_{51}=\frac{\sqrt{21} \mathbb{X}_{24}[\mathbb{Q}_{3}^{(a,A_{2}^{\prime\prime})}] \otimes\mathbb{Y}_{41}[\mathbb{Q}_{1}^{(b,A_{2}^{\prime\prime})}]}{7} + \frac{\sqrt{14} \mathbb{X}_{27}[\mathbb{Q}_{3,0}^{(a,E^{\prime})}] \otimes\mathbb{Y}_{42}[\mathbb{Q}_{1,0}^{(b,E^{\prime})}]}{7} + \frac{\sqrt{14} \mathbb{X}_{28}[\mathbb{Q}_{3,1}^{(a,E^{\prime})}] \otimes\mathbb{Y}_{43}[\mathbb{Q}_{1,1}^{(b,E^{\prime})}]}{7}
\end{align*}
\begin{align*}
&
\hat{\mathbb{Z}}_{51}(\bm{k})=\frac{\sqrt{21} \mathbb{X}_{24}[\mathbb{Q}_{3}^{(a,A_{2}^{\prime\prime})}] \otimes\mathbb{U}_{4}[\mathbb{Q}_{0}^{(u,A_{1}^{\prime})}] \otimes\mathbb{F}_{43}[\mathbb{Q}_{1}^{(k,A_{2}^{\prime\prime})}]}{14} + \frac{\sqrt{21} \mathbb{X}_{24}[\mathbb{Q}_{3}^{(a,A_{2}^{\prime\prime})}] \otimes\mathbb{U}_{5}[\mathbb{Q}_{1}^{(u,A_{2}^{\prime\prime})}] \otimes\mathbb{F}_{42}[\mathbb{Q}_{0}^{(k,A_{1}^{\prime})}]}{14}
\cr&\hspace{1cm}
 - \frac{\sqrt{21} \mathbb{X}_{24}[\mathbb{Q}_{3}^{(a,A_{2}^{\prime\prime})}] \otimes\mathbb{U}_{6}[\mathbb{T}_{0}^{(u,A_{1}^{\prime})}] \otimes\mathbb{F}_{55}[\mathbb{T}_{1}^{(k,A_{2}^{\prime\prime})}]}{14} - \frac{\sqrt{21} \mathbb{X}_{24}[\mathbb{Q}_{3}^{(a,A_{2}^{\prime\prime})}] \otimes\mathbb{U}_{7}[\mathbb{T}_{1}^{(u,A_{2}^{\prime\prime})}] \otimes\mathbb{F}_{54}[\mathbb{T}_{0}^{(k,A_{1}^{\prime})}]}{14}
 \cr&\hspace{1cm}
 + \frac{\sqrt{14} \mathbb{X}_{27}[\mathbb{Q}_{3,0}^{(a,E^{\prime})}] \otimes\mathbb{U}_{4}[\mathbb{Q}_{0}^{(u,A_{1}^{\prime})}] \otimes\mathbb{F}_{44}[\mathbb{Q}_{1,0}^{(k,E^{\prime})}]}{14} + \frac{\sqrt{14} \mathbb{X}_{27}[\mathbb{Q}_{3,0}^{(a,E^{\prime})}] \otimes\mathbb{U}_{5}[\mathbb{Q}_{1}^{(u,A_{2}^{\prime\prime})}] \otimes\mathbb{F}_{48}[\mathbb{Q}_{2,0}^{(k,E^{\prime\prime})}]}{14}
 \cr&\hspace{1cm}
 - \frac{\sqrt{14} \mathbb{X}_{27}[\mathbb{Q}_{3,0}^{(a,E^{\prime})}] \otimes\mathbb{U}_{6}[\mathbb{T}_{0}^{(u,A_{1}^{\prime})}] \otimes\mathbb{F}_{56}[\mathbb{T}_{1,0}^{(k,E^{\prime})}]}{14} - \frac{\sqrt{14} \mathbb{X}_{27}[\mathbb{Q}_{3,0}^{(a,E^{\prime})}] \otimes\mathbb{U}_{7}[\mathbb{T}_{1}^{(u,A_{2}^{\prime\prime})}] \otimes\mathbb{F}_{60}[\mathbb{T}_{2,0}^{(k,E^{\prime\prime})}]}{14}
 \cr&\hspace{1cm}
 + \frac{\sqrt{14} \mathbb{X}_{28}[\mathbb{Q}_{3,1}^{(a,E^{\prime})}] \otimes\mathbb{U}_{4}[\mathbb{Q}_{0}^{(u,A_{1}^{\prime})}] \otimes\mathbb{F}_{45}[\mathbb{Q}_{1,1}^{(k,E^{\prime})}]}{14} + \frac{\sqrt{14} \mathbb{X}_{28}[\mathbb{Q}_{3,1}^{(a,E^{\prime})}] \otimes\mathbb{U}_{5}[\mathbb{Q}_{1}^{(u,A_{2}^{\prime\prime})}] \otimes\mathbb{F}_{49}[\mathbb{Q}_{2,1}^{(k,E^{\prime\prime})}]}{14}
 \cr&\hspace{1cm}
 - \frac{\sqrt{14} \mathbb{X}_{28}[\mathbb{Q}_{3,1}^{(a,E^{\prime})}] \otimes\mathbb{U}_{6}[\mathbb{T}_{0}^{(u,A_{1}^{\prime})}] \otimes\mathbb{F}_{57}[\mathbb{T}_{1,1}^{(k,E^{\prime})}]}{14} - \frac{\sqrt{14} \mathbb{X}_{28}[\mathbb{Q}_{3,1}^{(a,E^{\prime})}] \otimes\mathbb{U}_{7}[\mathbb{T}_{1}^{(u,A_{2}^{\prime\prime})}] \otimes\mathbb{F}_{61}[\mathbb{T}_{2,1}^{(k,E^{\prime\prime})}]}{14}
\end{align*}
\vspace{4mm}
\noindent \fbox{No. {52}} $\,\,\,\hat{\mathbb{Q}}_{4}^{(A_{1}^{\prime})}$ [M$_{3}$,\,B$_{6}$]
\begin{align*}
\hat{\mathbb{Z}}_{52}=\frac{2 \sqrt{7} \mathbb{X}_{24}[\mathbb{Q}_{3}^{(a,A_{2}^{\prime\prime})}] \otimes\mathbb{Y}_{41}[\mathbb{Q}_{1}^{(b,A_{2}^{\prime\prime})}]}{7} - \frac{\sqrt{42} \mathbb{X}_{27}[\mathbb{Q}_{3,0}^{(a,E^{\prime})}] \otimes\mathbb{Y}_{42}[\mathbb{Q}_{1,0}^{(b,E^{\prime})}]}{14} - \frac{\sqrt{42} \mathbb{X}_{28}[\mathbb{Q}_{3,1}^{(a,E^{\prime})}] \otimes\mathbb{Y}_{43}[\mathbb{Q}_{1,1}^{(b,E^{\prime})}]}{14}
\end{align*}
\begin{align*}
&
\hat{\mathbb{Z}}_{52}(\bm{k})=\frac{\sqrt{7} \mathbb{X}_{24}[\mathbb{Q}_{3}^{(a,A_{2}^{\prime\prime})}] \otimes\mathbb{U}_{4}[\mathbb{Q}_{0}^{(u,A_{1}^{\prime})}] \otimes\mathbb{F}_{43}[\mathbb{Q}_{1}^{(k,A_{2}^{\prime\prime})}]}{7} + \frac{\sqrt{7} \mathbb{X}_{24}[\mathbb{Q}_{3}^{(a,A_{2}^{\prime\prime})}] \otimes\mathbb{U}_{5}[\mathbb{Q}_{1}^{(u,A_{2}^{\prime\prime})}] \otimes\mathbb{F}_{42}[\mathbb{Q}_{0}^{(k,A_{1}^{\prime})}]}{7}
\cr&\hspace{1cm}
 - \frac{\sqrt{7} \mathbb{X}_{24}[\mathbb{Q}_{3}^{(a,A_{2}^{\prime\prime})}] \otimes\mathbb{U}_{6}[\mathbb{T}_{0}^{(u,A_{1}^{\prime})}] \otimes\mathbb{F}_{55}[\mathbb{T}_{1}^{(k,A_{2}^{\prime\prime})}]}{7} - \frac{\sqrt{7} \mathbb{X}_{24}[\mathbb{Q}_{3}^{(a,A_{2}^{\prime\prime})}] \otimes\mathbb{U}_{7}[\mathbb{T}_{1}^{(u,A_{2}^{\prime\prime})}] \otimes\mathbb{F}_{54}[\mathbb{T}_{0}^{(k,A_{1}^{\prime})}]}{7}
 \cr&\hspace{1cm}
 - \frac{\sqrt{42} \mathbb{X}_{27}[\mathbb{Q}_{3,0}^{(a,E^{\prime})}] \otimes\mathbb{U}_{4}[\mathbb{Q}_{0}^{(u,A_{1}^{\prime})}] \otimes\mathbb{F}_{44}[\mathbb{Q}_{1,0}^{(k,E^{\prime})}]}{28} - \frac{\sqrt{42} \mathbb{X}_{27}[\mathbb{Q}_{3,0}^{(a,E^{\prime})}] \otimes\mathbb{U}_{5}[\mathbb{Q}_{1}^{(u,A_{2}^{\prime\prime})}] \otimes\mathbb{F}_{48}[\mathbb{Q}_{2,0}^{(k,E^{\prime\prime})}]}{28}
 \cr&\hspace{1cm}
 + \frac{\sqrt{42} \mathbb{X}_{27}[\mathbb{Q}_{3,0}^{(a,E^{\prime})}] \otimes\mathbb{U}_{6}[\mathbb{T}_{0}^{(u,A_{1}^{\prime})}] \otimes\mathbb{F}_{56}[\mathbb{T}_{1,0}^{(k,E^{\prime})}]}{28} + \frac{\sqrt{42} \mathbb{X}_{27}[\mathbb{Q}_{3,0}^{(a,E^{\prime})}] \otimes\mathbb{U}_{7}[\mathbb{T}_{1}^{(u,A_{2}^{\prime\prime})}] \otimes\mathbb{F}_{60}[\mathbb{T}_{2,0}^{(k,E^{\prime\prime})}]}{28}
 \cr&\hspace{1cm}
 - \frac{\sqrt{42} \mathbb{X}_{28}[\mathbb{Q}_{3,1}^{(a,E^{\prime})}] \otimes\mathbb{U}_{4}[\mathbb{Q}_{0}^{(u,A_{1}^{\prime})}] \otimes\mathbb{F}_{45}[\mathbb{Q}_{1,1}^{(k,E^{\prime})}]}{28} - \frac{\sqrt{42} \mathbb{X}_{28}[\mathbb{Q}_{3,1}^{(a,E^{\prime})}] \otimes\mathbb{U}_{5}[\mathbb{Q}_{1}^{(u,A_{2}^{\prime\prime})}] \otimes\mathbb{F}_{49}[\mathbb{Q}_{2,1}^{(k,E^{\prime\prime})}]}{28}
 \cr&\hspace{1cm}
 + \frac{\sqrt{42} \mathbb{X}_{28}[\mathbb{Q}_{3,1}^{(a,E^{\prime})}] \otimes\mathbb{U}_{6}[\mathbb{T}_{0}^{(u,A_{1}^{\prime})}] \otimes\mathbb{F}_{57}[\mathbb{T}_{1,1}^{(k,E^{\prime})}]}{28} + \frac{\sqrt{42} \mathbb{X}_{28}[\mathbb{Q}_{3,1}^{(a,E^{\prime})}] \otimes\mathbb{U}_{7}[\mathbb{T}_{1}^{(u,A_{2}^{\prime\prime})}] \otimes\mathbb{F}_{61}[\mathbb{T}_{2,1}^{(k,E^{\prime\prime})}]}{28}
\end{align*}
\vspace{4mm}
\noindent \fbox{No. {53}} $\,\,\,\hat{\mathbb{Q}}_{3}^{(A_{1}^{\prime})}$ [M$_{3}$,\,B$_{6}$]
\begin{align*}
&
\hat{\mathbb{Z}}_{53}=\frac{\sqrt{7} \mathbb{X}_{27}[\mathbb{Q}_{3,0}^{(a,E^{\prime})}] \otimes\mathbb{Y}_{44}[\mathbb{Q}_{2,0}^{(b,E^{\prime})}]}{7} + \frac{\sqrt{7} \mathbb{X}_{28}[\mathbb{Q}_{3,1}^{(a,E^{\prime})}] \otimes\mathbb{Y}_{45}[\mathbb{Q}_{2,1}^{(b,E^{\prime})}]}{7} - \frac{\sqrt{70} \mathbb{X}_{33}[\mathbb{Q}_{3,0}^{(a,E^{\prime\prime})}] \otimes\mathbb{Y}_{46}[\mathbb{Q}_{2,0}^{(b,E^{\prime\prime})}]}{14}
\cr&\hspace{1cm}
 - \frac{\sqrt{70} \mathbb{X}_{34}[\mathbb{Q}_{3,1}^{(a,E^{\prime\prime})}] \otimes\mathbb{Y}_{47}[\mathbb{Q}_{2,1}^{(b,E^{\prime\prime})}]}{14}
\end{align*}
\begin{align*}
&
\hat{\mathbb{Z}}_{53}(\bm{k})=\frac{\sqrt{7} \mathbb{X}_{27}[\mathbb{Q}_{3,0}^{(a,E^{\prime})}] \otimes\mathbb{U}_{4}[\mathbb{Q}_{0}^{(u,A_{1}^{\prime})}] \otimes\mathbb{F}_{46}[\mathbb{Q}_{2,0}^{(k,E^{\prime})}]}{14} + \frac{\sqrt{7} \mathbb{X}_{27}[\mathbb{Q}_{3,0}^{(a,E^{\prime})}] \otimes\mathbb{U}_{5}[\mathbb{Q}_{1}^{(u,A_{2}^{\prime\prime})}] \otimes\mathbb{F}_{51}[\mathbb{Q}_{3,0}^{(k,E^{\prime\prime})}]}{14}
\cr&\hspace{1cm}
 - \frac{\sqrt{7} \mathbb{X}_{27}[\mathbb{Q}_{3,0}^{(a,E^{\prime})}] \otimes\mathbb{U}_{6}[\mathbb{T}_{0}^{(u,A_{1}^{\prime})}] \otimes\mathbb{F}_{58}[\mathbb{T}_{2,0}^{(k,E^{\prime})}]}{14} - \frac{\sqrt{7} \mathbb{X}_{27}[\mathbb{Q}_{3,0}^{(a,E^{\prime})}] \otimes\mathbb{U}_{7}[\mathbb{T}_{1}^{(u,A_{2}^{\prime\prime})}] \otimes\mathbb{F}_{63}[\mathbb{T}_{3,0}^{(k,E^{\prime\prime})}]}{14}
 \cr&\hspace{1cm}
 + \frac{\sqrt{7} \mathbb{X}_{28}[\mathbb{Q}_{3,1}^{(a,E^{\prime})}] \otimes\mathbb{U}_{4}[\mathbb{Q}_{0}^{(u,A_{1}^{\prime})}] \otimes\mathbb{F}_{47}[\mathbb{Q}_{2,1}^{(k,E^{\prime})}]}{14} + \frac{\sqrt{7} \mathbb{X}_{28}[\mathbb{Q}_{3,1}^{(a,E^{\prime})}] \otimes\mathbb{U}_{5}[\mathbb{Q}_{1}^{(u,A_{2}^{\prime\prime})}] \otimes\mathbb{F}_{52}[\mathbb{Q}_{3,1}^{(k,E^{\prime\prime})}]}{14}
 \cr&\hspace{1cm}
 - \frac{\sqrt{7} \mathbb{X}_{28}[\mathbb{Q}_{3,1}^{(a,E^{\prime})}] \otimes\mathbb{U}_{6}[\mathbb{T}_{0}^{(u,A_{1}^{\prime})}] \otimes\mathbb{F}_{59}[\mathbb{T}_{2,1}^{(k,E^{\prime})}]}{14} - \frac{\sqrt{7} \mathbb{X}_{28}[\mathbb{Q}_{3,1}^{(a,E^{\prime})}] \otimes\mathbb{U}_{7}[\mathbb{T}_{1}^{(u,A_{2}^{\prime\prime})}] \otimes\mathbb{F}_{64}[\mathbb{T}_{3,1}^{(k,E^{\prime\prime})}]}{14}
 \cr&\hspace{1cm}
 - \frac{\sqrt{70} \mathbb{X}_{33}[\mathbb{Q}_{3,0}^{(a,E^{\prime\prime})}] \otimes\mathbb{U}_{4}[\mathbb{Q}_{0}^{(u,A_{1}^{\prime})}] \otimes\mathbb{F}_{48}[\mathbb{Q}_{2,0}^{(k,E^{\prime\prime})}]}{28} - \frac{\sqrt{70} \mathbb{X}_{33}[\mathbb{Q}_{3,0}^{(a,E^{\prime\prime})}] \otimes\mathbb{U}_{5}[\mathbb{Q}_{1}^{(u,A_{2}^{\prime\prime})}] \otimes\mathbb{F}_{44}[\mathbb{Q}_{1,0}^{(k,E^{\prime})}]}{28}
 \cr&\hspace{1cm}
 + \frac{\sqrt{70} \mathbb{X}_{33}[\mathbb{Q}_{3,0}^{(a,E^{\prime\prime})}] \otimes\mathbb{U}_{6}[\mathbb{T}_{0}^{(u,A_{1}^{\prime})}] \otimes\mathbb{F}_{60}[\mathbb{T}_{2,0}^{(k,E^{\prime\prime})}]}{28} + \frac{\sqrt{70} \mathbb{X}_{33}[\mathbb{Q}_{3,0}^{(a,E^{\prime\prime})}] \otimes\mathbb{U}_{7}[\mathbb{T}_{1}^{(u,A_{2}^{\prime\prime})}] \otimes\mathbb{F}_{56}[\mathbb{T}_{1,0}^{(k,E^{\prime})}]}{28}
 \cr&\hspace{1cm}
 - \frac{\sqrt{70} \mathbb{X}_{34}[\mathbb{Q}_{3,1}^{(a,E^{\prime\prime})}] \otimes\mathbb{U}_{4}[\mathbb{Q}_{0}^{(u,A_{1}^{\prime})}] \otimes\mathbb{F}_{49}[\mathbb{Q}_{2,1}^{(k,E^{\prime\prime})}]}{28} - \frac{\sqrt{70} \mathbb{X}_{34}[\mathbb{Q}_{3,1}^{(a,E^{\prime\prime})}] \otimes\mathbb{U}_{5}[\mathbb{Q}_{1}^{(u,A_{2}^{\prime\prime})}] \otimes\mathbb{F}_{45}[\mathbb{Q}_{1,1}^{(k,E^{\prime})}]}{28}
 \cr&\hspace{1cm}
 + \frac{\sqrt{70} \mathbb{X}_{34}[\mathbb{Q}_{3,1}^{(a,E^{\prime\prime})}] \otimes\mathbb{U}_{6}[\mathbb{T}_{0}^{(u,A_{1}^{\prime})}] \otimes\mathbb{F}_{61}[\mathbb{T}_{2,1}^{(k,E^{\prime\prime})}]}{28} + \frac{\sqrt{70} \mathbb{X}_{34}[\mathbb{Q}_{3,1}^{(a,E^{\prime\prime})}] \otimes\mathbb{U}_{7}[\mathbb{T}_{1}^{(u,A_{2}^{\prime\prime})}] \otimes\mathbb{F}_{57}[\mathbb{T}_{1,1}^{(k,E^{\prime})}]}{28}
\end{align*}
\vspace{4mm}
\noindent \fbox{No. {54}} $\,\,\,\hat{\mathbb{G}}_{4}^{(A_{1}^{\prime})}$ [M$_{3}$,\,B$_{6}$]
\begin{align*}
&
\hat{\mathbb{Z}}_{54}=\frac{\sqrt{70} \mathbb{X}_{27}[\mathbb{Q}_{3,0}^{(a,E^{\prime})}] \otimes\mathbb{Y}_{44}[\mathbb{Q}_{2,0}^{(b,E^{\prime})}]}{14} + \frac{\sqrt{70} \mathbb{X}_{28}[\mathbb{Q}_{3,1}^{(a,E^{\prime})}] \otimes\mathbb{Y}_{45}[\mathbb{Q}_{2,1}^{(b,E^{\prime})}]}{14} + \frac{\sqrt{7} \mathbb{X}_{33}[\mathbb{Q}_{3,0}^{(a,E^{\prime\prime})}] \otimes\mathbb{Y}_{46}[\mathbb{Q}_{2,0}^{(b,E^{\prime\prime})}]}{7}
\cr&\hspace{1cm}
 + \frac{\sqrt{7} \mathbb{X}_{34}[\mathbb{Q}_{3,1}^{(a,E^{\prime\prime})}] \otimes\mathbb{Y}_{47}[\mathbb{Q}_{2,1}^{(b,E^{\prime\prime})}]}{7}
\end{align*}
\begin{align*}
&
\hat{\mathbb{Z}}_{54}(\bm{k})=\frac{\sqrt{70} \mathbb{X}_{27}[\mathbb{Q}_{3,0}^{(a,E^{\prime})}] \otimes\mathbb{U}_{4}[\mathbb{Q}_{0}^{(u,A_{1}^{\prime})}] \otimes\mathbb{F}_{46}[\mathbb{Q}_{2,0}^{(k,E^{\prime})}]}{28} + \frac{\sqrt{70} \mathbb{X}_{27}[\mathbb{Q}_{3,0}^{(a,E^{\prime})}] \otimes\mathbb{U}_{5}[\mathbb{Q}_{1}^{(u,A_{2}^{\prime\prime})}] \otimes\mathbb{F}_{51}[\mathbb{Q}_{3,0}^{(k,E^{\prime\prime})}]}{28}
\cr&\hspace{1cm}
 - \frac{\sqrt{70} \mathbb{X}_{27}[\mathbb{Q}_{3,0}^{(a,E^{\prime})}] \otimes\mathbb{U}_{6}[\mathbb{T}_{0}^{(u,A_{1}^{\prime})}] \otimes\mathbb{F}_{58}[\mathbb{T}_{2,0}^{(k,E^{\prime})}]}{28} - \frac{\sqrt{70} \mathbb{X}_{27}[\mathbb{Q}_{3,0}^{(a,E^{\prime})}] \otimes\mathbb{U}_{7}[\mathbb{T}_{1}^{(u,A_{2}^{\prime\prime})}] \otimes\mathbb{F}_{63}[\mathbb{T}_{3,0}^{(k,E^{\prime\prime})}]}{28}
 \cr&\hspace{1cm}
 + \frac{\sqrt{70} \mathbb{X}_{28}[\mathbb{Q}_{3,1}^{(a,E^{\prime})}] \otimes\mathbb{U}_{4}[\mathbb{Q}_{0}^{(u,A_{1}^{\prime})}] \otimes\mathbb{F}_{47}[\mathbb{Q}_{2,1}^{(k,E^{\prime})}]}{28} + \frac{\sqrt{70} \mathbb{X}_{28}[\mathbb{Q}_{3,1}^{(a,E^{\prime})}] \otimes\mathbb{U}_{5}[\mathbb{Q}_{1}^{(u,A_{2}^{\prime\prime})}] \otimes\mathbb{F}_{52}[\mathbb{Q}_{3,1}^{(k,E^{\prime\prime})}]}{28}
 \cr&\hspace{1cm}
 - \frac{\sqrt{70} \mathbb{X}_{28}[\mathbb{Q}_{3,1}^{(a,E^{\prime})}] \otimes\mathbb{U}_{6}[\mathbb{T}_{0}^{(u,A_{1}^{\prime})}] \otimes\mathbb{F}_{59}[\mathbb{T}_{2,1}^{(k,E^{\prime})}]}{28} - \frac{\sqrt{70} \mathbb{X}_{28}[\mathbb{Q}_{3,1}^{(a,E^{\prime})}] \otimes\mathbb{U}_{7}[\mathbb{T}_{1}^{(u,A_{2}^{\prime\prime})}] \otimes\mathbb{F}_{64}[\mathbb{T}_{3,1}^{(k,E^{\prime\prime})}]}{28}
 \cr&\hspace{1cm}
 + \frac{\sqrt{7} \mathbb{X}_{33}[\mathbb{Q}_{3,0}^{(a,E^{\prime\prime})}] \otimes\mathbb{U}_{4}[\mathbb{Q}_{0}^{(u,A_{1}^{\prime})}] \otimes\mathbb{F}_{48}[\mathbb{Q}_{2,0}^{(k,E^{\prime\prime})}]}{14} + \frac{\sqrt{7} \mathbb{X}_{33}[\mathbb{Q}_{3,0}^{(a,E^{\prime\prime})}] \otimes\mathbb{U}_{5}[\mathbb{Q}_{1}^{(u,A_{2}^{\prime\prime})}] \otimes\mathbb{F}_{44}[\mathbb{Q}_{1,0}^{(k,E^{\prime})}]}{14}
 \cr&\hspace{1cm}
 - \frac{\sqrt{7} \mathbb{X}_{33}[\mathbb{Q}_{3,0}^{(a,E^{\prime\prime})}] \otimes\mathbb{U}_{6}[\mathbb{T}_{0}^{(u,A_{1}^{\prime})}] \otimes\mathbb{F}_{60}[\mathbb{T}_{2,0}^{(k,E^{\prime\prime})}]}{14} - \frac{\sqrt{7} \mathbb{X}_{33}[\mathbb{Q}_{3,0}^{(a,E^{\prime\prime})}] \otimes\mathbb{U}_{7}[\mathbb{T}_{1}^{(u,A_{2}^{\prime\prime})}] \otimes\mathbb{F}_{56}[\mathbb{T}_{1,0}^{(k,E^{\prime})}]}{14}
 \cr&\hspace{1cm}
 + \frac{\sqrt{7} \mathbb{X}_{34}[\mathbb{Q}_{3,1}^{(a,E^{\prime\prime})}] \otimes\mathbb{U}_{4}[\mathbb{Q}_{0}^{(u,A_{1}^{\prime})}] \otimes\mathbb{F}_{49}[\mathbb{Q}_{2,1}^{(k,E^{\prime\prime})}]}{14} + \frac{\sqrt{7} \mathbb{X}_{34}[\mathbb{Q}_{3,1}^{(a,E^{\prime\prime})}] \otimes\mathbb{U}_{5}[\mathbb{Q}_{1}^{(u,A_{2}^{\prime\prime})}] \otimes\mathbb{F}_{45}[\mathbb{Q}_{1,1}^{(k,E^{\prime})}]}{14}
 \cr&\hspace{1cm}
 - \frac{\sqrt{7} \mathbb{X}_{34}[\mathbb{Q}_{3,1}^{(a,E^{\prime\prime})}] \otimes\mathbb{U}_{6}[\mathbb{T}_{0}^{(u,A_{1}^{\prime})}] \otimes\mathbb{F}_{61}[\mathbb{T}_{2,1}^{(k,E^{\prime\prime})}]}{14} - \frac{\sqrt{7} \mathbb{X}_{34}[\mathbb{Q}_{3,1}^{(a,E^{\prime\prime})}] \otimes\mathbb{U}_{7}[\mathbb{T}_{1}^{(u,A_{2}^{\prime\prime})}] \otimes\mathbb{F}_{57}[\mathbb{T}_{1,1}^{(k,E^{\prime})}]}{14}
\end{align*}
\vspace{4mm}
\noindent \fbox{No. {55}} $\,\,\,\hat{\mathbb{Q}}_{0}^{(A_{1}^{\prime})}$ [M$_{3}$,\,B$_{6}$]
\begin{align*}
\hat{\mathbb{Z}}_{55}=\frac{\sqrt{3} \mathbb{X}_{32}[\mathbb{Q}_{3}^{(a,A_{2}^{\prime})}] \otimes\mathbb{Y}_{48}[\mathbb{Q}_{3}^{(b,A_{2}^{\prime})}]}{3} + \frac{\sqrt{3} \mathbb{X}_{33}[\mathbb{Q}_{3,0}^{(a,E^{\prime\prime})}] \otimes\mathbb{Y}_{49}[\mathbb{Q}_{3,0}^{(b,E^{\prime\prime})}]}{3} + \frac{\sqrt{3} \mathbb{X}_{34}[\mathbb{Q}_{3,1}^{(a,E^{\prime\prime})}] \otimes\mathbb{Y}_{50}[\mathbb{Q}_{3,1}^{(b,E^{\prime\prime})}]}{3}
\end{align*}
\begin{align*}
&
\hat{\mathbb{Z}}_{55}(\bm{k})=\frac{\sqrt{3} \mathbb{X}_{32}[\mathbb{Q}_{3}^{(a,A_{2}^{\prime})}] \otimes\mathbb{U}_{4}[\mathbb{Q}_{0}^{(u,A_{1}^{\prime})}] \otimes\mathbb{F}_{50}[\mathbb{Q}_{3}^{(k,A_{2}^{\prime})}]}{6} + \frac{\sqrt{3} \mathbb{X}_{32}[\mathbb{Q}_{3}^{(a,A_{2}^{\prime})}] \otimes\mathbb{U}_{5}[\mathbb{Q}_{1}^{(u,A_{2}^{\prime\prime})}] \otimes\mathbb{F}_{53}[\mathbb{Q}_{4}^{(k,A_{1}^{\prime\prime})}]}{6}
\cr&\hspace{1cm}
 - \frac{\sqrt{3} \mathbb{X}_{32}[\mathbb{Q}_{3}^{(a,A_{2}^{\prime})}] \otimes\mathbb{U}_{6}[\mathbb{T}_{0}^{(u,A_{1}^{\prime})}] \otimes\mathbb{F}_{62}[\mathbb{T}_{3}^{(k,A_{2}^{\prime})}]}{6} - \frac{\sqrt{3} \mathbb{X}_{32}[\mathbb{Q}_{3}^{(a,A_{2}^{\prime})}] \otimes\mathbb{U}_{7}[\mathbb{T}_{1}^{(u,A_{2}^{\prime\prime})}] \otimes\mathbb{F}_{65}[\mathbb{T}_{4}^{(k,A_{1}^{\prime\prime})}]}{6}
 \cr&\hspace{1cm}
 + \frac{\sqrt{3} \mathbb{X}_{33}[\mathbb{Q}_{3,0}^{(a,E^{\prime\prime})}] \otimes\mathbb{U}_{4}[\mathbb{Q}_{0}^{(u,A_{1}^{\prime})}] \otimes\mathbb{F}_{51}[\mathbb{Q}_{3,0}^{(k,E^{\prime\prime})}]}{6} + \frac{\sqrt{3} \mathbb{X}_{33}[\mathbb{Q}_{3,0}^{(a,E^{\prime\prime})}] \otimes\mathbb{U}_{5}[\mathbb{Q}_{1}^{(u,A_{2}^{\prime\prime})}] \otimes\mathbb{F}_{46}[\mathbb{Q}_{2,0}^{(k,E^{\prime})}]}{6}
 \cr&\hspace{1cm}
 - \frac{\sqrt{3} \mathbb{X}_{33}[\mathbb{Q}_{3,0}^{(a,E^{\prime\prime})}] \otimes\mathbb{U}_{6}[\mathbb{T}_{0}^{(u,A_{1}^{\prime})}] \otimes\mathbb{F}_{63}[\mathbb{T}_{3,0}^{(k,E^{\prime\prime})}]}{6} - \frac{\sqrt{3} \mathbb{X}_{33}[\mathbb{Q}_{3,0}^{(a,E^{\prime\prime})}] \otimes\mathbb{U}_{7}[\mathbb{T}_{1}^{(u,A_{2}^{\prime\prime})}] \otimes\mathbb{F}_{58}[\mathbb{T}_{2,0}^{(k,E^{\prime})}]}{6}
 \cr&\hspace{1cm}
 + \frac{\sqrt{3} \mathbb{X}_{34}[\mathbb{Q}_{3,1}^{(a,E^{\prime\prime})}] \otimes\mathbb{U}_{4}[\mathbb{Q}_{0}^{(u,A_{1}^{\prime})}] \otimes\mathbb{F}_{52}[\mathbb{Q}_{3,1}^{(k,E^{\prime\prime})}]}{6} + \frac{\sqrt{3} \mathbb{X}_{34}[\mathbb{Q}_{3,1}^{(a,E^{\prime\prime})}] \otimes\mathbb{U}_{5}[\mathbb{Q}_{1}^{(u,A_{2}^{\prime\prime})}] \otimes\mathbb{F}_{47}[\mathbb{Q}_{2,1}^{(k,E^{\prime})}]}{6}
 \cr&\hspace{1cm}
 - \frac{\sqrt{3} \mathbb{X}_{34}[\mathbb{Q}_{3,1}^{(a,E^{\prime\prime})}] \otimes\mathbb{U}_{6}[\mathbb{T}_{0}^{(u,A_{1}^{\prime})}] \otimes\mathbb{F}_{64}[\mathbb{T}_{3,1}^{(k,E^{\prime\prime})}]}{6} - \frac{\sqrt{3} \mathbb{X}_{34}[\mathbb{Q}_{3,1}^{(a,E^{\prime\prime})}] \otimes\mathbb{U}_{7}[\mathbb{T}_{1}^{(u,A_{2}^{\prime\prime})}] \otimes\mathbb{F}_{59}[\mathbb{T}_{2,1}^{(k,E^{\prime})}]}{6}
\end{align*}
\vspace{4mm}
\noindent \fbox{No. {56}} $\,\,\,\hat{\mathbb{Q}}_{2}^{(A_{1}^{\prime})}$ [M$_{3}$,\,B$_{6}$]
\begin{align*}
\hat{\mathbb{Z}}_{56}=- \frac{\sqrt{6} \mathbb{X}_{32}[\mathbb{Q}_{3}^{(a,A_{2}^{\prime})}] \otimes\mathbb{Y}_{48}[\mathbb{Q}_{3}^{(b,A_{2}^{\prime})}]}{3} + \frac{\sqrt{6} \mathbb{X}_{33}[\mathbb{Q}_{3,0}^{(a,E^{\prime\prime})}] \otimes\mathbb{Y}_{49}[\mathbb{Q}_{3,0}^{(b,E^{\prime\prime})}]}{6} + \frac{\sqrt{6} \mathbb{X}_{34}[\mathbb{Q}_{3,1}^{(a,E^{\prime\prime})}] \otimes\mathbb{Y}_{50}[\mathbb{Q}_{3,1}^{(b,E^{\prime\prime})}]}{6}
\end{align*}
\begin{align*}
&
\hat{\mathbb{Z}}_{56}(\bm{k})=- \frac{\sqrt{6} \mathbb{X}_{32}[\mathbb{Q}_{3}^{(a,A_{2}^{\prime})}] \otimes\mathbb{U}_{4}[\mathbb{Q}_{0}^{(u,A_{1}^{\prime})}] \otimes\mathbb{F}_{50}[\mathbb{Q}_{3}^{(k,A_{2}^{\prime})}]}{6} - \frac{\sqrt{6} \mathbb{X}_{32}[\mathbb{Q}_{3}^{(a,A_{2}^{\prime})}] \otimes\mathbb{U}_{5}[\mathbb{Q}_{1}^{(u,A_{2}^{\prime\prime})}] \otimes\mathbb{F}_{53}[\mathbb{Q}_{4}^{(k,A_{1}^{\prime\prime})}]}{6}
\cr&\hspace{1cm}
 + \frac{\sqrt{6} \mathbb{X}_{32}[\mathbb{Q}_{3}^{(a,A_{2}^{\prime})}] \otimes\mathbb{U}_{6}[\mathbb{T}_{0}^{(u,A_{1}^{\prime})}] \otimes\mathbb{F}_{62}[\mathbb{T}_{3}^{(k,A_{2}^{\prime})}]}{6} + \frac{\sqrt{6} \mathbb{X}_{32}[\mathbb{Q}_{3}^{(a,A_{2}^{\prime})}] \otimes\mathbb{U}_{7}[\mathbb{T}_{1}^{(u,A_{2}^{\prime\prime})}] \otimes\mathbb{F}_{65}[\mathbb{T}_{4}^{(k,A_{1}^{\prime\prime})}]}{6}
 \cr&\hspace{1cm}
 + \frac{\sqrt{6} \mathbb{X}_{33}[\mathbb{Q}_{3,0}^{(a,E^{\prime\prime})}] \otimes\mathbb{U}_{4}[\mathbb{Q}_{0}^{(u,A_{1}^{\prime})}] \otimes\mathbb{F}_{51}[\mathbb{Q}_{3,0}^{(k,E^{\prime\prime})}]}{12} + \frac{\sqrt{6} \mathbb{X}_{33}[\mathbb{Q}_{3,0}^{(a,E^{\prime\prime})}] \otimes\mathbb{U}_{5}[\mathbb{Q}_{1}^{(u,A_{2}^{\prime\prime})}] \otimes\mathbb{F}_{46}[\mathbb{Q}_{2,0}^{(k,E^{\prime})}]}{12}
 \cr&\hspace{1cm}
 - \frac{\sqrt{6} \mathbb{X}_{33}[\mathbb{Q}_{3,0}^{(a,E^{\prime\prime})}] \otimes\mathbb{U}_{6}[\mathbb{T}_{0}^{(u,A_{1}^{\prime})}] \otimes\mathbb{F}_{63}[\mathbb{T}_{3,0}^{(k,E^{\prime\prime})}]}{12} - \frac{\sqrt{6} \mathbb{X}_{33}[\mathbb{Q}_{3,0}^{(a,E^{\prime\prime})}] \otimes\mathbb{U}_{7}[\mathbb{T}_{1}^{(u,A_{2}^{\prime\prime})}] \otimes\mathbb{F}_{58}[\mathbb{T}_{2,0}^{(k,E^{\prime})}]}{12}
 \cr&\hspace{1cm}
 + \frac{\sqrt{6} \mathbb{X}_{34}[\mathbb{Q}_{3,1}^{(a,E^{\prime\prime})}] \otimes\mathbb{U}_{4}[\mathbb{Q}_{0}^{(u,A_{1}^{\prime})}] \otimes\mathbb{F}_{52}[\mathbb{Q}_{3,1}^{(k,E^{\prime\prime})}]}{12} + \frac{\sqrt{6} \mathbb{X}_{34}[\mathbb{Q}_{3,1}^{(a,E^{\prime\prime})}] \otimes\mathbb{U}_{5}[\mathbb{Q}_{1}^{(u,A_{2}^{\prime\prime})}] \otimes\mathbb{F}_{47}[\mathbb{Q}_{2,1}^{(k,E^{\prime})}]}{12}
 \cr&\hspace{1cm}
 - \frac{\sqrt{6} \mathbb{X}_{34}[\mathbb{Q}_{3,1}^{(a,E^{\prime\prime})}] \otimes\mathbb{U}_{6}[\mathbb{T}_{0}^{(u,A_{1}^{\prime})}] \otimes\mathbb{F}_{64}[\mathbb{T}_{3,1}^{(k,E^{\prime\prime})}]}{12} - \frac{\sqrt{6} \mathbb{X}_{34}[\mathbb{Q}_{3,1}^{(a,E^{\prime\prime})}] \otimes\mathbb{U}_{7}[\mathbb{T}_{1}^{(u,A_{2}^{\prime\prime})}] \otimes\mathbb{F}_{59}[\mathbb{T}_{2,1}^{(k,E^{\prime})}]}{12}
\end{align*}
\vspace{4mm}
\noindent \fbox{No. {57}} $\,\,\,\hat{\mathbb{Q}}_{2}^{(A_{1}^{\prime})}$ [M$_{3}$,\,B$_{6}$]
\begin{align*}
\hat{\mathbb{Z}}_{57}=\frac{\sqrt{2} \mathbb{X}_{29}[\mathbb{M}_{2,0}^{(a,E^{\prime})}] \otimes\mathbb{Y}_{51}[\mathbb{T}_{1,0}^{(b,E^{\prime})}]}{2} + \frac{\sqrt{2} \mathbb{X}_{30}[\mathbb{M}_{2,1}^{(a,E^{\prime})}] \otimes\mathbb{Y}_{52}[\mathbb{T}_{1,1}^{(b,E^{\prime})}]}{2}
\end{align*}
\begin{align*}
&
\hat{\mathbb{Z}}_{57}(\bm{k})=\frac{\sqrt{2} \mathbb{X}_{29}[\mathbb{M}_{2,0}^{(a,E^{\prime})}] \otimes\mathbb{U}_{4}[\mathbb{Q}_{0}^{(u,A_{1}^{\prime})}] \otimes\mathbb{F}_{56}[\mathbb{T}_{1,0}^{(k,E^{\prime})}]}{4} + \frac{\sqrt{2} \mathbb{X}_{29}[\mathbb{M}_{2,0}^{(a,E^{\prime})}] \otimes\mathbb{U}_{5}[\mathbb{Q}_{1}^{(u,A_{2}^{\prime\prime})}] \otimes\mathbb{F}_{60}[\mathbb{T}_{2,0}^{(k,E^{\prime\prime})}]}{4}
\cr&\hspace{1cm}
 + \frac{\sqrt{2} \mathbb{X}_{29}[\mathbb{M}_{2,0}^{(a,E^{\prime})}] \otimes\mathbb{U}_{6}[\mathbb{T}_{0}^{(u,A_{1}^{\prime})}] \otimes\mathbb{F}_{44}[\mathbb{Q}_{1,0}^{(k,E^{\prime})}]}{4} + \frac{\sqrt{2} \mathbb{X}_{29}[\mathbb{M}_{2,0}^{(a,E^{\prime})}] \otimes\mathbb{U}_{7}[\mathbb{T}_{1}^{(u,A_{2}^{\prime\prime})}] \otimes\mathbb{F}_{48}[\mathbb{Q}_{2,0}^{(k,E^{\prime\prime})}]}{4}
 \cr&\hspace{1cm}
 + \frac{\sqrt{2} \mathbb{X}_{30}[\mathbb{M}_{2,1}^{(a,E^{\prime})}] \otimes\mathbb{U}_{4}[\mathbb{Q}_{0}^{(u,A_{1}^{\prime})}] \otimes\mathbb{F}_{57}[\mathbb{T}_{1,1}^{(k,E^{\prime})}]}{4} + \frac{\sqrt{2} \mathbb{X}_{30}[\mathbb{M}_{2,1}^{(a,E^{\prime})}] \otimes\mathbb{U}_{5}[\mathbb{Q}_{1}^{(u,A_{2}^{\prime\prime})}] \otimes\mathbb{F}_{61}[\mathbb{T}_{2,1}^{(k,E^{\prime\prime})}]}{4}
 \cr&\hspace{1cm}
 + \frac{\sqrt{2} \mathbb{X}_{30}[\mathbb{M}_{2,1}^{(a,E^{\prime})}] \otimes\mathbb{U}_{6}[\mathbb{T}_{0}^{(u,A_{1}^{\prime})}] \otimes\mathbb{F}_{45}[\mathbb{Q}_{1,1}^{(k,E^{\prime})}]}{4} + \frac{\sqrt{2} \mathbb{X}_{30}[\mathbb{M}_{2,1}^{(a,E^{\prime})}] \otimes\mathbb{U}_{7}[\mathbb{T}_{1}^{(u,A_{2}^{\prime\prime})}] \otimes\mathbb{F}_{49}[\mathbb{Q}_{2,1}^{(k,E^{\prime\prime})}]}{4}
\end{align*}
\vspace{4mm}
\noindent \fbox{No. {58}} $\,\,\,\hat{\mathbb{Q}}_{3}^{(A_{1}^{\prime})}$ [M$_{3}$,\,B$_{6}$]
\begin{align*}
&
\hat{\mathbb{Z}}_{58}=- \frac{\mathbb{X}_{29}[\mathbb{M}_{2,0}^{(a,E^{\prime})}] \otimes\mathbb{Y}_{53}[\mathbb{T}_{2,0}^{(b,E^{\prime})}]}{2} - \frac{\mathbb{X}_{30}[\mathbb{M}_{2,1}^{(a,E^{\prime})}] \otimes\mathbb{Y}_{54}[\mathbb{T}_{2,1}^{(b,E^{\prime})}]}{2} - \frac{\mathbb{X}_{35}[\mathbb{M}_{2,0}^{(a,E^{\prime\prime})}] \otimes\mathbb{Y}_{55}[\mathbb{T}_{2,0}^{(b,E^{\prime\prime})}]}{2}
\cr&\hspace{1cm}
 - \frac{\mathbb{X}_{36}[\mathbb{M}_{2,1}^{(a,E^{\prime\prime})}] \otimes\mathbb{Y}_{56}[\mathbb{T}_{2,1}^{(b,E^{\prime\prime})}]}{2}
\end{align*}
\begin{align*}
&
\hat{\mathbb{Z}}_{58}(\bm{k})=- \frac{\mathbb{X}_{29}[\mathbb{M}_{2,0}^{(a,E^{\prime})}] \otimes\mathbb{U}_{4}[\mathbb{Q}_{0}^{(u,A_{1}^{\prime})}] \otimes\mathbb{F}_{58}[\mathbb{T}_{2,0}^{(k,E^{\prime})}]}{4} - \frac{\mathbb{X}_{29}[\mathbb{M}_{2,0}^{(a,E^{\prime})}] \otimes\mathbb{U}_{5}[\mathbb{Q}_{1}^{(u,A_{2}^{\prime\prime})}] \otimes\mathbb{F}_{63}[\mathbb{T}_{3,0}^{(k,E^{\prime\prime})}]}{4}
\cr&\hspace{1cm}
 - \frac{\mathbb{X}_{29}[\mathbb{M}_{2,0}^{(a,E^{\prime})}] \otimes\mathbb{U}_{6}[\mathbb{T}_{0}^{(u,A_{1}^{\prime})}] \otimes\mathbb{F}_{46}[\mathbb{Q}_{2,0}^{(k,E^{\prime})}]}{4} - \frac{\mathbb{X}_{29}[\mathbb{M}_{2,0}^{(a,E^{\prime})}] \otimes\mathbb{U}_{7}[\mathbb{T}_{1}^{(u,A_{2}^{\prime\prime})}] \otimes\mathbb{F}_{51}[\mathbb{Q}_{3,0}^{(k,E^{\prime\prime})}]}{4}
 \cr&\hspace{1cm}
 - \frac{\mathbb{X}_{30}[\mathbb{M}_{2,1}^{(a,E^{\prime})}] \otimes\mathbb{U}_{4}[\mathbb{Q}_{0}^{(u,A_{1}^{\prime})}] \otimes\mathbb{F}_{59}[\mathbb{T}_{2,1}^{(k,E^{\prime})}]}{4} - \frac{\mathbb{X}_{30}[\mathbb{M}_{2,1}^{(a,E^{\prime})}] \otimes\mathbb{U}_{5}[\mathbb{Q}_{1}^{(u,A_{2}^{\prime\prime})}] \otimes\mathbb{F}_{64}[\mathbb{T}_{3,1}^{(k,E^{\prime\prime})}]}{4}
 \cr&\hspace{1cm}
 - \frac{\mathbb{X}_{30}[\mathbb{M}_{2,1}^{(a,E^{\prime})}] \otimes\mathbb{U}_{6}[\mathbb{T}_{0}^{(u,A_{1}^{\prime})}] \otimes\mathbb{F}_{47}[\mathbb{Q}_{2,1}^{(k,E^{\prime})}]}{4} - \frac{\mathbb{X}_{30}[\mathbb{M}_{2,1}^{(a,E^{\prime})}] \otimes\mathbb{U}_{7}[\mathbb{T}_{1}^{(u,A_{2}^{\prime\prime})}] \otimes\mathbb{F}_{52}[\mathbb{Q}_{3,1}^{(k,E^{\prime\prime})}]}{4}
 \cr&\hspace{1cm}
 - \frac{\mathbb{X}_{35}[\mathbb{M}_{2,0}^{(a,E^{\prime\prime})}] \otimes\mathbb{U}_{4}[\mathbb{Q}_{0}^{(u,A_{1}^{\prime})}] \otimes\mathbb{F}_{60}[\mathbb{T}_{2,0}^{(k,E^{\prime\prime})}]}{4} - \frac{\mathbb{X}_{35}[\mathbb{M}_{2,0}^{(a,E^{\prime\prime})}] \otimes\mathbb{U}_{5}[\mathbb{Q}_{1}^{(u,A_{2}^{\prime\prime})}] \otimes\mathbb{F}_{56}[\mathbb{T}_{1,0}^{(k,E^{\prime})}]}{4}
 \cr&\hspace{1cm}
 - \frac{\mathbb{X}_{35}[\mathbb{M}_{2,0}^{(a,E^{\prime\prime})}] \otimes\mathbb{U}_{6}[\mathbb{T}_{0}^{(u,A_{1}^{\prime})}] \otimes\mathbb{F}_{48}[\mathbb{Q}_{2,0}^{(k,E^{\prime\prime})}]}{4} - \frac{\mathbb{X}_{35}[\mathbb{M}_{2,0}^{(a,E^{\prime\prime})}] \otimes\mathbb{U}_{7}[\mathbb{T}_{1}^{(u,A_{2}^{\prime\prime})}] \otimes\mathbb{F}_{44}[\mathbb{Q}_{1,0}^{(k,E^{\prime})}]}{4}
 \cr&\hspace{1cm}
 - \frac{\mathbb{X}_{36}[\mathbb{M}_{2,1}^{(a,E^{\prime\prime})}] \otimes\mathbb{U}_{4}[\mathbb{Q}_{0}^{(u,A_{1}^{\prime})}] \otimes\mathbb{F}_{61}[\mathbb{T}_{2,1}^{(k,E^{\prime\prime})}]}{4} - \frac{\mathbb{X}_{36}[\mathbb{M}_{2,1}^{(a,E^{\prime\prime})}] \otimes\mathbb{U}_{5}[\mathbb{Q}_{1}^{(u,A_{2}^{\prime\prime})}] \otimes\mathbb{F}_{57}[\mathbb{T}_{1,1}^{(k,E^{\prime})}]}{4}
 \cr&\hspace{1cm}
 - \frac{\mathbb{X}_{36}[\mathbb{M}_{2,1}^{(a,E^{\prime\prime})}] \otimes\mathbb{U}_{6}[\mathbb{T}_{0}^{(u,A_{1}^{\prime})}] \otimes\mathbb{F}_{49}[\mathbb{Q}_{2,1}^{(k,E^{\prime\prime})}]}{4} - \frac{\mathbb{X}_{36}[\mathbb{M}_{2,1}^{(a,E^{\prime\prime})}] \otimes\mathbb{U}_{7}[\mathbb{T}_{1}^{(u,A_{2}^{\prime\prime})}] \otimes\mathbb{F}_{45}[\mathbb{Q}_{1,1}^{(k,E^{\prime})}]}{4}
\end{align*}
\vspace{4mm}
\noindent \fbox{No. {59}} $\,\,\,\hat{\mathbb{G}}_{4}^{(A_{1}^{\prime})}$ [M$_{3}$,\,B$_{6}$]
\begin{align*}
&
\hat{\mathbb{Z}}_{59}=- \frac{\mathbb{X}_{29}[\mathbb{M}_{2,0}^{(a,E^{\prime})}] \otimes\mathbb{Y}_{53}[\mathbb{T}_{2,0}^{(b,E^{\prime})}]}{2} - \frac{\mathbb{X}_{30}[\mathbb{M}_{2,1}^{(a,E^{\prime})}] \otimes\mathbb{Y}_{54}[\mathbb{T}_{2,1}^{(b,E^{\prime})}]}{2} + \frac{\mathbb{X}_{35}[\mathbb{M}_{2,0}^{(a,E^{\prime\prime})}] \otimes\mathbb{Y}_{55}[\mathbb{T}_{2,0}^{(b,E^{\prime\prime})}]}{2}
\cr&\hspace{1cm}
 + \frac{\mathbb{X}_{36}[\mathbb{M}_{2,1}^{(a,E^{\prime\prime})}] \otimes\mathbb{Y}_{56}[\mathbb{T}_{2,1}^{(b,E^{\prime\prime})}]}{2}
\end{align*}
\begin{align*}
&
\hat{\mathbb{Z}}_{59}(\bm{k})=- \frac{\mathbb{X}_{29}[\mathbb{M}_{2,0}^{(a,E^{\prime})}] \otimes\mathbb{U}_{4}[\mathbb{Q}_{0}^{(u,A_{1}^{\prime})}] \otimes\mathbb{F}_{58}[\mathbb{T}_{2,0}^{(k,E^{\prime})}]}{4} - \frac{\mathbb{X}_{29}[\mathbb{M}_{2,0}^{(a,E^{\prime})}] \otimes\mathbb{U}_{5}[\mathbb{Q}_{1}^{(u,A_{2}^{\prime\prime})}] \otimes\mathbb{F}_{63}[\mathbb{T}_{3,0}^{(k,E^{\prime\prime})}]}{4}
\cr&\hspace{1cm}
 - \frac{\mathbb{X}_{29}[\mathbb{M}_{2,0}^{(a,E^{\prime})}] \otimes\mathbb{U}_{6}[\mathbb{T}_{0}^{(u,A_{1}^{\prime})}] \otimes\mathbb{F}_{46}[\mathbb{Q}_{2,0}^{(k,E^{\prime})}]}{4} - \frac{\mathbb{X}_{29}[\mathbb{M}_{2,0}^{(a,E^{\prime})}] \otimes\mathbb{U}_{7}[\mathbb{T}_{1}^{(u,A_{2}^{\prime\prime})}] \otimes\mathbb{F}_{51}[\mathbb{Q}_{3,0}^{(k,E^{\prime\prime})}]}{4}
 \cr&\hspace{1cm}
 - \frac{\mathbb{X}_{30}[\mathbb{M}_{2,1}^{(a,E^{\prime})}] \otimes\mathbb{U}_{4}[\mathbb{Q}_{0}^{(u,A_{1}^{\prime})}] \otimes\mathbb{F}_{59}[\mathbb{T}_{2,1}^{(k,E^{\prime})}]}{4} - \frac{\mathbb{X}_{30}[\mathbb{M}_{2,1}^{(a,E^{\prime})}] \otimes\mathbb{U}_{5}[\mathbb{Q}_{1}^{(u,A_{2}^{\prime\prime})}] \otimes\mathbb{F}_{64}[\mathbb{T}_{3,1}^{(k,E^{\prime\prime})}]}{4}
 \cr&\hspace{1cm}
 - \frac{\mathbb{X}_{30}[\mathbb{M}_{2,1}^{(a,E^{\prime})}] \otimes\mathbb{U}_{6}[\mathbb{T}_{0}^{(u,A_{1}^{\prime})}] \otimes\mathbb{F}_{47}[\mathbb{Q}_{2,1}^{(k,E^{\prime})}]}{4} - \frac{\mathbb{X}_{30}[\mathbb{M}_{2,1}^{(a,E^{\prime})}] \otimes\mathbb{U}_{7}[\mathbb{T}_{1}^{(u,A_{2}^{\prime\prime})}] \otimes\mathbb{F}_{52}[\mathbb{Q}_{3,1}^{(k,E^{\prime\prime})}]}{4}
 \cr&\hspace{1cm}
 + \frac{\mathbb{X}_{35}[\mathbb{M}_{2,0}^{(a,E^{\prime\prime})}] \otimes\mathbb{U}_{4}[\mathbb{Q}_{0}^{(u,A_{1}^{\prime})}] \otimes\mathbb{F}_{60}[\mathbb{T}_{2,0}^{(k,E^{\prime\prime})}]}{4} + \frac{\mathbb{X}_{35}[\mathbb{M}_{2,0}^{(a,E^{\prime\prime})}] \otimes\mathbb{U}_{5}[\mathbb{Q}_{1}^{(u,A_{2}^{\prime\prime})}] \otimes\mathbb{F}_{56}[\mathbb{T}_{1,0}^{(k,E^{\prime})}]}{4}
 \cr&\hspace{1cm}
 + \frac{\mathbb{X}_{35}[\mathbb{M}_{2,0}^{(a,E^{\prime\prime})}] \otimes\mathbb{U}_{6}[\mathbb{T}_{0}^{(u,A_{1}^{\prime})}] \otimes\mathbb{F}_{48}[\mathbb{Q}_{2,0}^{(k,E^{\prime\prime})}]}{4} + \frac{\mathbb{X}_{35}[\mathbb{M}_{2,0}^{(a,E^{\prime\prime})}] \otimes\mathbb{U}_{7}[\mathbb{T}_{1}^{(u,A_{2}^{\prime\prime})}] \otimes\mathbb{F}_{44}[\mathbb{Q}_{1,0}^{(k,E^{\prime})}]}{4}
 \cr&\hspace{1cm}
 + \frac{\mathbb{X}_{36}[\mathbb{M}_{2,1}^{(a,E^{\prime\prime})}] \otimes\mathbb{U}_{4}[\mathbb{Q}_{0}^{(u,A_{1}^{\prime})}] \otimes\mathbb{F}_{61}[\mathbb{T}_{2,1}^{(k,E^{\prime\prime})}]}{4} + \frac{\mathbb{X}_{36}[\mathbb{M}_{2,1}^{(a,E^{\prime\prime})}] \otimes\mathbb{U}_{5}[\mathbb{Q}_{1}^{(u,A_{2}^{\prime\prime})}] \otimes\mathbb{F}_{57}[\mathbb{T}_{1,1}^{(k,E^{\prime})}]}{4}
 \cr&\hspace{1cm}
 + \frac{\mathbb{X}_{36}[\mathbb{M}_{2,1}^{(a,E^{\prime\prime})}] \otimes\mathbb{U}_{6}[\mathbb{T}_{0}^{(u,A_{1}^{\prime})}] \otimes\mathbb{F}_{49}[\mathbb{Q}_{2,1}^{(k,E^{\prime\prime})}]}{4} + \frac{\mathbb{X}_{36}[\mathbb{M}_{2,1}^{(a,E^{\prime\prime})}] \otimes\mathbb{U}_{7}[\mathbb{T}_{1}^{(u,A_{2}^{\prime\prime})}] \otimes\mathbb{F}_{45}[\mathbb{Q}_{1,1}^{(k,E^{\prime})}]}{4}
\end{align*}
\vspace{4mm}
\noindent \fbox{No. {60}} $\,\,\,\hat{\mathbb{Q}}_{2}^{(A_{1}^{\prime})}$ [M$_{3}$,\,B$_{6}$]
\begin{align*}
\hat{\mathbb{Z}}_{60}=- \frac{\sqrt{2} \mathbb{X}_{35}[\mathbb{M}_{2,0}^{(a,E^{\prime\prime})}] \otimes\mathbb{Y}_{57}[\mathbb{T}_{3,0}^{(b,E^{\prime\prime})}]}{2} - \frac{\sqrt{2} \mathbb{X}_{36}[\mathbb{M}_{2,1}^{(a,E^{\prime\prime})}] \otimes\mathbb{Y}_{58}[\mathbb{T}_{3,1}^{(b,E^{\prime\prime})}]}{2}
\end{align*}
\begin{align*}
&
\hat{\mathbb{Z}}_{60}(\bm{k})=- \frac{\sqrt{2} \mathbb{X}_{35}[\mathbb{M}_{2,0}^{(a,E^{\prime\prime})}] \otimes\mathbb{U}_{4}[\mathbb{Q}_{0}^{(u,A_{1}^{\prime})}] \otimes\mathbb{F}_{63}[\mathbb{T}_{3,0}^{(k,E^{\prime\prime})}]}{4} - \frac{\sqrt{2} \mathbb{X}_{35}[\mathbb{M}_{2,0}^{(a,E^{\prime\prime})}] \otimes\mathbb{U}_{5}[\mathbb{Q}_{1}^{(u,A_{2}^{\prime\prime})}] \otimes\mathbb{F}_{58}[\mathbb{T}_{2,0}^{(k,E^{\prime})}]}{4}
\cr&\hspace{1cm}
 - \frac{\sqrt{2} \mathbb{X}_{35}[\mathbb{M}_{2,0}^{(a,E^{\prime\prime})}] \otimes\mathbb{U}_{6}[\mathbb{T}_{0}^{(u,A_{1}^{\prime})}] \otimes\mathbb{F}_{51}[\mathbb{Q}_{3,0}^{(k,E^{\prime\prime})}]}{4} - \frac{\sqrt{2} \mathbb{X}_{35}[\mathbb{M}_{2,0}^{(a,E^{\prime\prime})}] \otimes\mathbb{U}_{7}[\mathbb{T}_{1}^{(u,A_{2}^{\prime\prime})}] \otimes\mathbb{F}_{46}[\mathbb{Q}_{2,0}^{(k,E^{\prime})}]}{4}
 \cr&\hspace{1cm}
 - \frac{\sqrt{2} \mathbb{X}_{36}[\mathbb{M}_{2,1}^{(a,E^{\prime\prime})}] \otimes\mathbb{U}_{4}[\mathbb{Q}_{0}^{(u,A_{1}^{\prime})}] \otimes\mathbb{F}_{64}[\mathbb{T}_{3,1}^{(k,E^{\prime\prime})}]}{4} - \frac{\sqrt{2} \mathbb{X}_{36}[\mathbb{M}_{2,1}^{(a,E^{\prime\prime})}] \otimes\mathbb{U}_{5}[\mathbb{Q}_{1}^{(u,A_{2}^{\prime\prime})}] \otimes\mathbb{F}_{59}[\mathbb{T}_{2,1}^{(k,E^{\prime})}]}{4}
 \cr&\hspace{1cm}
 - \frac{\sqrt{2} \mathbb{X}_{36}[\mathbb{M}_{2,1}^{(a,E^{\prime\prime})}] \otimes\mathbb{U}_{6}[\mathbb{T}_{0}^{(u,A_{1}^{\prime})}] \otimes\mathbb{F}_{52}[\mathbb{Q}_{3,1}^{(k,E^{\prime\prime})}]}{4} - \frac{\sqrt{2} \mathbb{X}_{36}[\mathbb{M}_{2,1}^{(a,E^{\prime\prime})}] \otimes\mathbb{U}_{7}[\mathbb{T}_{1}^{(u,A_{2}^{\prime\prime})}] \otimes\mathbb{F}_{47}[\mathbb{Q}_{2,1}^{(k,E^{\prime})}]}{4}
\end{align*}
\vspace{4mm}
\noindent \fbox{No. {61}} $\,\,\,\hat{\mathbb{Q}}_{3}^{(A_{1}^{\prime})}$ [M$_{3}$,\,B$_{6}$]
\begin{align*}
\hat{\mathbb{Z}}_{61}=\mathbb{X}_{37}[\mathbb{M}_{2}^{(a,A_{1}^{\prime\prime})}] \otimes\mathbb{Y}_{59}[\mathbb{T}_{4}^{(b,A_{1}^{\prime\prime})}]
\end{align*}
\begin{align*}
&
\hat{\mathbb{Z}}_{61}(\bm{k})=\frac{\mathbb{X}_{37}[\mathbb{M}_{2}^{(a,A_{1}^{\prime\prime})}] \otimes\mathbb{U}_{4}[\mathbb{Q}_{0}^{(u,A_{1}^{\prime})}] \otimes\mathbb{F}_{65}[\mathbb{T}_{4}^{(k,A_{1}^{\prime\prime})}]}{2} + \frac{\mathbb{X}_{37}[\mathbb{M}_{2}^{(a,A_{1}^{\prime\prime})}] \otimes\mathbb{U}_{5}[\mathbb{Q}_{1}^{(u,A_{2}^{\prime\prime})}] \otimes\mathbb{F}_{62}[\mathbb{T}_{3}^{(k,A_{2}^{\prime})}]}{2}
\cr&\hspace{1cm}
 + \frac{\mathbb{X}_{37}[\mathbb{M}_{2}^{(a,A_{1}^{\prime\prime})}] \otimes\mathbb{U}_{6}[\mathbb{T}_{0}^{(u,A_{1}^{\prime})}] \otimes\mathbb{F}_{53}[\mathbb{Q}_{4}^{(k,A_{1}^{\prime\prime})}]}{2} + \frac{\mathbb{X}_{37}[\mathbb{M}_{2}^{(a,A_{1}^{\prime\prime})}] \otimes\mathbb{U}_{7}[\mathbb{T}_{1}^{(u,A_{2}^{\prime\prime})}] \otimes\mathbb{F}_{50}[\mathbb{Q}_{3}^{(k,A_{2}^{\prime})}]}{2}
\end{align*}
\vspace{4mm}
\noindent \fbox{No. {62}} $\,\,\,\hat{\mathbb{Q}}_{0}^{(A_{1}^{\prime})}$ [M$_{2}$,\,B$_{7}$]
\begin{align*}
\hat{\mathbb{Z}}_{62}=\mathbb{X}_{14}[\mathbb{Q}_{0}^{(a,A_{1}^{\prime})}] \otimes\mathbb{Y}_{60}[\mathbb{Q}_{0}^{(b,A_{1}^{\prime})}]
\end{align*}
\begin{align*}
\hat{\mathbb{Z}}_{62}(\bm{k})=\mathbb{X}_{14}[\mathbb{Q}_{0}^{(a,A_{1}^{\prime})}] \otimes\mathbb{U}_{2}[\mathbb{Q}_{0}^{(s,A_{1}^{\prime})}] \otimes\mathbb{F}_{66}[\mathbb{Q}_{0}^{(k,A_{1}^{\prime})}]
\end{align*}
\vspace{4mm}
\noindent \fbox{No. {63}} $\,\,\,\hat{\mathbb{Q}}_{2}^{(A_{1}^{\prime})}$ [M$_{2}$,\,B$_{7}$]
\begin{align*}
\hat{\mathbb{Z}}_{63}=\mathbb{X}_{15}[\mathbb{Q}_{2}^{(a,A_{1}^{\prime})}] \otimes\mathbb{Y}_{60}[\mathbb{Q}_{0}^{(b,A_{1}^{\prime})}]
\end{align*}
\begin{align*}
\hat{\mathbb{Z}}_{63}(\bm{k})=\mathbb{X}_{15}[\mathbb{Q}_{2}^{(a,A_{1}^{\prime})}] \otimes\mathbb{U}_{2}[\mathbb{Q}_{0}^{(s,A_{1}^{\prime})}] \otimes\mathbb{F}_{66}[\mathbb{Q}_{0}^{(k,A_{1}^{\prime})}]
\end{align*}
\vspace{4mm}
\noindent \fbox{No. {64}} $\,\,\,\hat{\mathbb{Q}}_{3}^{(A_{1}^{\prime})}$ [M$_{2}$,\,B$_{7}$]
\begin{align*}
\hat{\mathbb{Z}}_{64}=- \frac{\sqrt{2} \mathbb{X}_{16}[\mathbb{Q}_{2,0}^{(a,E^{\prime})}] \otimes\mathbb{Y}_{61}[\mathbb{Q}_{1,0}^{(b,E^{\prime})}]}{2} - \frac{\sqrt{2} \mathbb{X}_{17}[\mathbb{Q}_{2,1}^{(a,E^{\prime})}] \otimes\mathbb{Y}_{62}[\mathbb{Q}_{1,1}^{(b,E^{\prime})}]}{2}
\end{align*}
\begin{align*}
\hat{\mathbb{Z}}_{64}(\bm{k})=- \frac{\sqrt{2} \mathbb{X}_{16}[\mathbb{Q}_{2,0}^{(a,E^{\prime})}] \otimes\mathbb{U}_{2}[\mathbb{Q}_{0}^{(s,A_{1}^{\prime})}] \otimes\mathbb{F}_{67}[\mathbb{Q}_{1,0}^{(k,E^{\prime})}]}{2} - \frac{\sqrt{2} \mathbb{X}_{17}[\mathbb{Q}_{2,1}^{(a,E^{\prime})}] \otimes\mathbb{U}_{2}[\mathbb{Q}_{0}^{(s,A_{1}^{\prime})}] \otimes\mathbb{F}_{68}[\mathbb{Q}_{1,1}^{(k,E^{\prime})}]}{2}
\end{align*}
\vspace{4mm}
\noindent \fbox{No. {65}} $\,\,\,\hat{\mathbb{Q}}_{0}^{(A_{1}^{\prime})}$ [M$_{2}$,\,B$_{7}$]
\begin{align*}
\hat{\mathbb{Z}}_{65}=\frac{\sqrt{2} \mathbb{X}_{18}[\mathbb{Q}_{2,0}^{(a,E^{\prime\prime})}] \otimes\mathbb{Y}_{63}[\mathbb{Q}_{2,0}^{(b,E^{\prime\prime})}]}{2} + \frac{\sqrt{2} \mathbb{X}_{19}[\mathbb{Q}_{2,1}^{(a,E^{\prime\prime})}] \otimes\mathbb{Y}_{64}[\mathbb{Q}_{2,1}^{(b,E^{\prime\prime})}]}{2}
\end{align*}
\begin{align*}
\hat{\mathbb{Z}}_{65}(\bm{k})=\frac{\sqrt{2} \mathbb{X}_{18}[\mathbb{Q}_{2,0}^{(a,E^{\prime\prime})}] \otimes\mathbb{U}_{3}[\mathbb{Q}_{1}^{(s,A_{2}^{\prime\prime})}] \otimes\mathbb{F}_{67}[\mathbb{Q}_{1,0}^{(k,E^{\prime})}]}{2} + \frac{\sqrt{2} \mathbb{X}_{19}[\mathbb{Q}_{2,1}^{(a,E^{\prime\prime})}] \otimes\mathbb{U}_{3}[\mathbb{Q}_{1}^{(s,A_{2}^{\prime\prime})}] \otimes\mathbb{F}_{68}[\mathbb{Q}_{1,1}^{(k,E^{\prime})}]}{2}
\end{align*}
\vspace{4mm}
\noindent \fbox{No. {66}} $\,\,\,\hat{\mathbb{Q}}_{2}^{(A_{1}^{\prime})}$ [M$_{2}$,\,B$_{7}$]
\begin{align*}
\hat{\mathbb{Z}}_{66}=\frac{\sqrt{2} \mathbb{X}_{21}[\mathbb{M}_{1,0}^{(a,E^{\prime\prime})}] \otimes\mathbb{Y}_{65}[\mathbb{T}_{2,0}^{(b,E^{\prime\prime})}]}{2} + \frac{\sqrt{2} \mathbb{X}_{22}[\mathbb{M}_{1,1}^{(a,E^{\prime\prime})}] \otimes\mathbb{Y}_{66}[\mathbb{T}_{2,1}^{(b,E^{\prime\prime})}]}{2}
\end{align*}
\begin{align*}
\hat{\mathbb{Z}}_{66}(\bm{k})=\frac{\sqrt{2} \mathbb{X}_{21}[\mathbb{M}_{1,0}^{(a,E^{\prime\prime})}] \otimes\mathbb{U}_{3}[\mathbb{Q}_{1}^{(s,A_{2}^{\prime\prime})}] \otimes\mathbb{F}_{69}[\mathbb{T}_{1,0}^{(k,E^{\prime})}]}{2} + \frac{\sqrt{2} \mathbb{X}_{22}[\mathbb{M}_{1,1}^{(a,E^{\prime\prime})}] \otimes\mathbb{U}_{3}[\mathbb{Q}_{1}^{(s,A_{2}^{\prime\prime})}] \otimes\mathbb{F}_{70}[\mathbb{T}_{1,1}^{(k,E^{\prime})}]}{2}
\end{align*}
\vspace{4mm}
\noindent \fbox{No. {67}} $\,\,\,\hat{\mathbb{Q}}_{3}^{(A_{1}^{\prime})}$ [M$_{2}$,\,B$_{7}$]
\begin{align*}
\hat{\mathbb{Z}}_{67}=\mathbb{X}_{20}[\mathbb{M}_{1}^{(a,A_{2}^{\prime})}] \otimes\mathbb{Y}_{67}[\mathbb{T}_{3}^{(b,A_{2}^{\prime})}]
\end{align*}
\begin{align*}
\hat{\mathbb{Z}}_{67}(\bm{k})=\mathbb{X}_{20}[\mathbb{M}_{1}^{(a,A_{2}^{\prime})}] \otimes\mathbb{U}_{2}[\mathbb{Q}_{0}^{(s,A_{1}^{\prime})}] \otimes\mathbb{F}_{71}[\mathbb{T}_{3}^{(k,A_{2}^{\prime})}]
\end{align*}
\vspace{4mm}
\noindent \fbox{No. {68}} $\,\,\,\hat{\mathbb{Q}}_{0}^{(A_{1}^{\prime})}$ [M$_{2}$,\,B$_{8}$]
\begin{align*}
\hat{\mathbb{Z}}_{68}=\mathbb{X}_{14}[\mathbb{Q}_{0}^{(a,A_{1}^{\prime})}] \otimes\mathbb{Y}_{68}[\mathbb{Q}_{0}^{(b,A_{1}^{\prime})}]
\end{align*}
\begin{align*}
\hat{\mathbb{Z}}_{68}(\bm{k})=\frac{\sqrt{2} \mathbb{X}_{14}[\mathbb{Q}_{0}^{(a,A_{1}^{\prime})}] \otimes\mathbb{U}_{8}[\mathbb{Q}_{0}^{(u,A_{1}^{\prime})}] \otimes\mathbb{F}_{72}[\mathbb{Q}_{0}^{(k,A_{1}^{\prime})}]}{2} - \frac{\sqrt{2} \mathbb{X}_{14}[\mathbb{Q}_{0}^{(a,A_{1}^{\prime})}] \otimes\mathbb{U}_{9}[\mathbb{T}_{1}^{(u,A_{2}^{\prime\prime})}] \otimes\mathbb{F}_{73}[\mathbb{T}_{1}^{(k,A_{2}^{\prime\prime})}]}{2}
\end{align*}
\vspace{4mm}
\noindent \fbox{No. {69}} $\,\,\,\hat{\mathbb{Q}}_{2}^{(A_{1}^{\prime})}$ [M$_{2}$,\,B$_{8}$]
\begin{align*}
\hat{\mathbb{Z}}_{69}=\mathbb{X}_{15}[\mathbb{Q}_{2}^{(a,A_{1}^{\prime})}] \otimes\mathbb{Y}_{68}[\mathbb{Q}_{0}^{(b,A_{1}^{\prime})}]
\end{align*}
\begin{align*}
\hat{\mathbb{Z}}_{69}(\bm{k})=\frac{\sqrt{2} \mathbb{X}_{15}[\mathbb{Q}_{2}^{(a,A_{1}^{\prime})}] \otimes\mathbb{U}_{8}[\mathbb{Q}_{0}^{(u,A_{1}^{\prime})}] \otimes\mathbb{F}_{72}[\mathbb{Q}_{0}^{(k,A_{1}^{\prime})}]}{2} - \frac{\sqrt{2} \mathbb{X}_{15}[\mathbb{Q}_{2}^{(a,A_{1}^{\prime})}] \otimes\mathbb{U}_{9}[\mathbb{T}_{1}^{(u,A_{2}^{\prime\prime})}] \otimes\mathbb{F}_{73}[\mathbb{T}_{1}^{(k,A_{2}^{\prime\prime})}]}{2}
\end{align*}
\vspace{4mm}
\noindent \fbox{No. {70}} $\,\,\,\hat{\mathbb{Q}}_{0}^{(A_{1}^{\prime})}$ [M$_{2}$,\,B$_{9}$]
\begin{align*}
\hat{\mathbb{Z}}_{70}=\mathbb{X}_{14}[\mathbb{Q}_{0}^{(a,A_{1}^{\prime})}] \otimes\mathbb{Y}_{69}[\mathbb{Q}_{0}^{(b,A_{1}^{\prime})}]
\end{align*}
\begin{align*}
\hat{\mathbb{Z}}_{70}(\bm{k})=\mathbb{X}_{14}[\mathbb{Q}_{0}^{(a,A_{1}^{\prime})}] \otimes\mathbb{U}_{2}[\mathbb{Q}_{0}^{(s,A_{1}^{\prime})}] \otimes\mathbb{F}_{74}[\mathbb{Q}_{0}^{(k,A_{1}^{\prime})}]
\end{align*}
\vspace{4mm}
\noindent \fbox{No. {71}} $\,\,\,\hat{\mathbb{Q}}_{2}^{(A_{1}^{\prime})}$ [M$_{2}$,\,B$_{9}$]
\begin{align*}
\hat{\mathbb{Z}}_{71}=\mathbb{X}_{15}[\mathbb{Q}_{2}^{(a,A_{1}^{\prime})}] \otimes\mathbb{Y}_{69}[\mathbb{Q}_{0}^{(b,A_{1}^{\prime})}]
\end{align*}
\begin{align*}
\hat{\mathbb{Z}}_{71}(\bm{k})=\mathbb{X}_{15}[\mathbb{Q}_{2}^{(a,A_{1}^{\prime})}] \otimes\mathbb{U}_{2}[\mathbb{Q}_{0}^{(s,A_{1}^{\prime})}] \otimes\mathbb{F}_{74}[\mathbb{Q}_{0}^{(k,A_{1}^{\prime})}]
\end{align*}
\vspace{4mm}
\noindent \fbox{No. {72}} $\,\,\,\hat{\mathbb{Q}}_{3}^{(A_{1}^{\prime})}$ [M$_{2}$,\,B$_{9}$]
\begin{align*}
\hat{\mathbb{Z}}_{72}=- \frac{\sqrt{2} \mathbb{X}_{16}[\mathbb{Q}_{2,0}^{(a,E^{\prime})}] \otimes\mathbb{Y}_{70}[\mathbb{Q}_{1,0}^{(b,E^{\prime})}]}{2} - \frac{\sqrt{2} \mathbb{X}_{17}[\mathbb{Q}_{2,1}^{(a,E^{\prime})}] \otimes\mathbb{Y}_{71}[\mathbb{Q}_{1,1}^{(b,E^{\prime})}]}{2}
\end{align*}
\begin{align*}
\hat{\mathbb{Z}}_{72}(\bm{k})=- \frac{\sqrt{2} \mathbb{X}_{16}[\mathbb{Q}_{2,0}^{(a,E^{\prime})}] \otimes\mathbb{U}_{2}[\mathbb{Q}_{0}^{(s,A_{1}^{\prime})}] \otimes\mathbb{F}_{75}[\mathbb{Q}_{1,0}^{(k,E^{\prime})}]}{2} - \frac{\sqrt{2} \mathbb{X}_{17}[\mathbb{Q}_{2,1}^{(a,E^{\prime})}] \otimes\mathbb{U}_{2}[\mathbb{Q}_{0}^{(s,A_{1}^{\prime})}] \otimes\mathbb{F}_{76}[\mathbb{Q}_{1,1}^{(k,E^{\prime})}]}{2}
\end{align*}
\vspace{4mm}
\noindent \fbox{No. {73}} $\,\,\,\hat{\mathbb{Q}}_{0}^{(A_{1}^{\prime})}$ [M$_{2}$,\,B$_{9}$]
\begin{align*}
\hat{\mathbb{Z}}_{73}=\frac{\sqrt{2} \mathbb{X}_{18}[\mathbb{Q}_{2,0}^{(a,E^{\prime\prime})}] \otimes\mathbb{Y}_{72}[\mathbb{Q}_{2,0}^{(b,E^{\prime\prime})}]}{2} + \frac{\sqrt{2} \mathbb{X}_{19}[\mathbb{Q}_{2,1}^{(a,E^{\prime\prime})}] \otimes\mathbb{Y}_{73}[\mathbb{Q}_{2,1}^{(b,E^{\prime\prime})}]}{2}
\end{align*}
\begin{align*}
\hat{\mathbb{Z}}_{73}(\bm{k})=\frac{\sqrt{2} \mathbb{X}_{18}[\mathbb{Q}_{2,0}^{(a,E^{\prime\prime})}] \otimes\mathbb{U}_{3}[\mathbb{Q}_{1}^{(s,A_{2}^{\prime\prime})}] \otimes\mathbb{F}_{75}[\mathbb{Q}_{1,0}^{(k,E^{\prime})}]}{2} + \frac{\sqrt{2} \mathbb{X}_{19}[\mathbb{Q}_{2,1}^{(a,E^{\prime\prime})}] \otimes\mathbb{U}_{3}[\mathbb{Q}_{1}^{(s,A_{2}^{\prime\prime})}] \otimes\mathbb{F}_{76}[\mathbb{Q}_{1,1}^{(k,E^{\prime})}]}{2}
\end{align*}
\vspace{4mm}
\noindent \fbox{No. {74}} $\,\,\,\hat{\mathbb{Q}}_{2}^{(A_{1}^{\prime})}$ [M$_{2}$,\,B$_{9}$]
\begin{align*}
\hat{\mathbb{Z}}_{74}=\frac{\sqrt{2} \mathbb{X}_{21}[\mathbb{M}_{1,0}^{(a,E^{\prime\prime})}] \otimes\mathbb{Y}_{74}[\mathbb{T}_{2,0}^{(b,E^{\prime\prime})}]}{2} + \frac{\sqrt{2} \mathbb{X}_{22}[\mathbb{M}_{1,1}^{(a,E^{\prime\prime})}] \otimes\mathbb{Y}_{75}[\mathbb{T}_{2,1}^{(b,E^{\prime\prime})}]}{2}
\end{align*}
\begin{align*}
\hat{\mathbb{Z}}_{74}(\bm{k})=\frac{\sqrt{2} \mathbb{X}_{21}[\mathbb{M}_{1,0}^{(a,E^{\prime\prime})}] \otimes\mathbb{U}_{3}[\mathbb{Q}_{1}^{(s,A_{2}^{\prime\prime})}] \otimes\mathbb{F}_{77}[\mathbb{T}_{1,0}^{(k,E^{\prime})}]}{2} + \frac{\sqrt{2} \mathbb{X}_{22}[\mathbb{M}_{1,1}^{(a,E^{\prime\prime})}] \otimes\mathbb{U}_{3}[\mathbb{Q}_{1}^{(s,A_{2}^{\prime\prime})}] \otimes\mathbb{F}_{78}[\mathbb{T}_{1,1}^{(k,E^{\prime})}]}{2}
\end{align*}

\item The atomic SAMBs are given in Table XII.

\begin{center}
\renewcommand{\arraystretch}{1.3}
\begin{longtable}{c|c|c|c}
\caption{Atomic SAMB.}
 \\
 \hline \hline
symbol & type & group & form \\ \hline \endfirsthead

\multicolumn{3}{l}{\tablename\ \thetable{}} \\
 \hline \hline
symbol & type & group & form \\ \hline \endhead

 \hline \hline
\multicolumn{3}{r}{\footnotesize\it continued ...} \\ \endfoot

 \hline \hline
\multicolumn{3}{r}{} \\ \endlastfoot

$ \mathbb{X}_{1} $ & $\mathbb{Q}_{0}^{(a,A_{1}^{\prime})}$ & M$_{1}$ & $\begin{pmatrix} \frac{\sqrt{5}}{5} & 0 & 0 & 0 & 0 \\ 0 & \frac{\sqrt{5}}{5} & 0 & 0 & 0 \\ 0 & 0 & \frac{\sqrt{5}}{5} & 0 & 0 \\ 0 & 0 & 0 & \frac{\sqrt{5}}{5} & 0 \\ 0 & 0 & 0 & 0 & \frac{\sqrt{5}}{5} \end{pmatrix}$ \\
$ \mathbb{X}_{2} $ & $\mathbb{Q}_{2}^{(a,A_{1}^{\prime})}$ & M$_{1}$ & $\begin{pmatrix} \frac{\sqrt{14}}{7} & 0 & 0 & 0 & 0 \\ 0 & - \frac{\sqrt{14}}{7} & 0 & 0 & 0 \\ 0 & 0 & \frac{\sqrt{14}}{14} & 0 & 0 \\ 0 & 0 & 0 & \frac{\sqrt{14}}{14} & 0 \\ 0 & 0 & 0 & 0 & - \frac{\sqrt{14}}{7} \end{pmatrix}$ \\
$ \mathbb{X}_{3} $ & $\mathbb{Q}_{4}^{(a,A_{1}^{\prime})}$ & M$_{1}$ & $\begin{pmatrix} \frac{3 \sqrt{70}}{35} & 0 & 0 & 0 & 0 \\ 0 & \frac{\sqrt{70}}{70} & 0 & 0 & 0 \\ 0 & 0 & - \frac{2 \sqrt{70}}{35} & 0 & 0 \\ 0 & 0 & 0 & - \frac{2 \sqrt{70}}{35} & 0 \\ 0 & 0 & 0 & 0 & \frac{\sqrt{70}}{70} \end{pmatrix}$ \\
$ \mathbb{X}_{4} $ & $\mathbb{Q}_{2,0}^{(a,E^{\prime})}$ & M$_{1}$ & $\begin{pmatrix} 0 & 0 & 0 & 0 & - \frac{\sqrt{14}}{7} \\ 0 & 0 & 0 & 0 & 0 \\ 0 & 0 & 0 & - \frac{\sqrt{42}}{14} & 0 \\ 0 & 0 & - \frac{\sqrt{42}}{14} & 0 & 0 \\ - \frac{\sqrt{14}}{7} & 0 & 0 & 0 & 0 \end{pmatrix}$ \\
$ \mathbb{X}_{5} $ & $\mathbb{Q}_{2,1}^{(a,E^{\prime})}$ & M$_{1}$ & $\begin{pmatrix} 0 & \frac{\sqrt{14}}{7} & 0 & 0 & 0 \\ \frac{\sqrt{14}}{7} & 0 & 0 & 0 & 0 \\ 0 & 0 & \frac{\sqrt{42}}{14} & 0 & 0 \\ 0 & 0 & 0 & - \frac{\sqrt{42}}{14} & 0 \\ 0 & 0 & 0 & 0 & 0 \end{pmatrix}$ \\
$ \mathbb{X}_{6} $ & $\mathbb{Q}_{4,0}^{(a,E^{\prime},1)}$ & M$_{1}$ & $\begin{pmatrix} 0 & 0 & 0 & 0 & 0 \\ 0 & 0 & 0 & 0 & - \frac{\sqrt{2}}{2} \\ 0 & 0 & 0 & 0 & 0 \\ 0 & 0 & 0 & 0 & 0 \\ 0 & - \frac{\sqrt{2}}{2} & 0 & 0 & 0 \end{pmatrix}$ \\
$ \mathbb{X}_{7} $ & $\mathbb{Q}_{4,1}^{(a,E^{\prime},1)}$ & M$_{1}$ & $\begin{pmatrix} 0 & 0 & 0 & 0 & 0 \\ 0 & - \frac{\sqrt{2}}{2} & 0 & 0 & 0 \\ 0 & 0 & 0 & 0 & 0 \\ 0 & 0 & 0 & 0 & 0 \\ 0 & 0 & 0 & 0 & \frac{\sqrt{2}}{2} \end{pmatrix}$ \\
$ \mathbb{X}_{8} $ & $\mathbb{Q}_{4,0}^{(a,E^{\prime},2)}$ & M$_{1}$ & $\begin{pmatrix} 0 & 0 & 0 & 0 & \frac{\sqrt{42}}{14} \\ 0 & 0 & 0 & 0 & 0 \\ 0 & 0 & 0 & - \frac{\sqrt{14}}{7} & 0 \\ 0 & 0 & - \frac{\sqrt{14}}{7} & 0 & 0 \\ \frac{\sqrt{42}}{14} & 0 & 0 & 0 & 0 \end{pmatrix}$ \\
$ \mathbb{X}_{9} $ & $\mathbb{Q}_{4,1}^{(a,E^{\prime},2)}$ & M$_{1}$ & $\begin{pmatrix} 0 & - \frac{\sqrt{42}}{14} & 0 & 0 & 0 \\ - \frac{\sqrt{42}}{14} & 0 & 0 & 0 & 0 \\ 0 & 0 & \frac{\sqrt{14}}{7} & 0 & 0 \\ 0 & 0 & 0 & - \frac{\sqrt{14}}{7} & 0 \\ 0 & 0 & 0 & 0 & 0 \end{pmatrix}$ \\
$ \mathbb{X}_{10} $ & $\mathbb{M}_{1}^{(a,A_{2}^{\prime})}$ & M$_{1}$ & $\begin{pmatrix} 0 & 0 & 0 & 0 & 0 \\ 0 & 0 & 0 & 0 & \frac{\sqrt{10} i}{5} \\ 0 & 0 & 0 & \frac{\sqrt{10} i}{10} & 0 \\ 0 & 0 & - \frac{\sqrt{10} i}{10} & 0 & 0 \\ 0 & - \frac{\sqrt{10} i}{5} & 0 & 0 & 0 \end{pmatrix}$ \\
$ \mathbb{X}_{11} $ & $\mathbb{M}_{3}^{(a,A_{2}^{\prime})}$ & M$_{1}$ & $\begin{pmatrix} 0 & 0 & 0 & 0 & 0 \\ 0 & 0 & 0 & 0 & - \frac{\sqrt{10} i}{10} \\ 0 & 0 & 0 & \frac{\sqrt{10} i}{5} & 0 \\ 0 & 0 & - \frac{\sqrt{10} i}{5} & 0 & 0 \\ 0 & \frac{\sqrt{10} i}{10} & 0 & 0 & 0 \end{pmatrix}$ \\
$ \mathbb{X}_{12} $ & $\mathbb{M}_{3,0}^{(a,E^{\prime})}$ & M$_{1}$ & $\begin{pmatrix} 0 & 0 & 0 & 0 & - \frac{\sqrt{2} i}{2} \\ 0 & 0 & 0 & 0 & 0 \\ 0 & 0 & 0 & 0 & 0 \\ 0 & 0 & 0 & 0 & 0 \\ \frac{\sqrt{2} i}{2} & 0 & 0 & 0 & 0 \end{pmatrix}$ \\
$ \mathbb{X}_{13} $ & $\mathbb{M}_{3,1}^{(a,E^{\prime})}$ & M$_{1}$ & $\begin{pmatrix} 0 & \frac{\sqrt{2} i}{2} & 0 & 0 & 0 \\ - \frac{\sqrt{2} i}{2} & 0 & 0 & 0 & 0 \\ 0 & 0 & 0 & 0 & 0 \\ 0 & 0 & 0 & 0 & 0 \\ 0 & 0 & 0 & 0 & 0 \end{pmatrix}$ \\ \hline
$ \mathbb{X}_{14} $ & $\mathbb{Q}_{0}^{(a,A_{1}^{\prime})}$ & M$_{2}$ & $\begin{pmatrix} \frac{\sqrt{3}}{3} & 0 & 0 \\ 0 & \frac{\sqrt{3}}{3} & 0 \\ 0 & 0 & \frac{\sqrt{3}}{3} \end{pmatrix}$ \\
$ \mathbb{X}_{15} $ & $\mathbb{Q}_{2}^{(a,A_{1}^{\prime})}$ & M$_{2}$ & $\begin{pmatrix} - \frac{\sqrt{6}}{6} & 0 & 0 \\ 0 & - \frac{\sqrt{6}}{6} & 0 \\ 0 & 0 & \frac{\sqrt{6}}{3} \end{pmatrix}$ \\
$ \mathbb{X}_{16} $ & $\mathbb{Q}_{2,0}^{(a,E^{\prime})}$ & M$_{2}$ & $\begin{pmatrix} 0 & - \frac{\sqrt{2}}{2} & 0 \\ - \frac{\sqrt{2}}{2} & 0 & 0 \\ 0 & 0 & 0 \end{pmatrix}$ \\
$ \mathbb{X}_{17} $ & $\mathbb{Q}_{2,1}^{(a,E^{\prime})}$ & M$_{2}$ & $\begin{pmatrix} - \frac{\sqrt{2}}{2} & 0 & 0 \\ 0 & \frac{\sqrt{2}}{2} & 0 \\ 0 & 0 & 0 \end{pmatrix}$ \\
$ \mathbb{X}_{18} $ & $\mathbb{Q}_{2,0}^{(a,E^{\prime\prime})}$ & M$_{2}$ & $\begin{pmatrix} 0 & 0 & \frac{\sqrt{2}}{2} \\ 0 & 0 & 0 \\ \frac{\sqrt{2}}{2} & 0 & 0 \end{pmatrix}$ \\
$ \mathbb{X}_{19} $ & $\mathbb{Q}_{2,1}^{(a,E^{\prime\prime})}$ & M$_{2}$ & $\begin{pmatrix} 0 & 0 & 0 \\ 0 & 0 & \frac{\sqrt{2}}{2} \\ 0 & \frac{\sqrt{2}}{2} & 0 \end{pmatrix}$ \\
$ \mathbb{X}_{20} $ & $\mathbb{M}_{1}^{(a,A_{2}^{\prime})}$ & M$_{2}$ & $\begin{pmatrix} 0 & - \frac{\sqrt{2} i}{2} & 0 \\ \frac{\sqrt{2} i}{2} & 0 & 0 \\ 0 & 0 & 0 \end{pmatrix}$ \\
$ \mathbb{X}_{21} $ & $\mathbb{M}_{1,0}^{(a,E^{\prime\prime})}$ & M$_{2}$ & $\begin{pmatrix} 0 & 0 & - \frac{\sqrt{2} i}{2} \\ 0 & 0 & 0 \\ \frac{\sqrt{2} i}{2} & 0 & 0 \end{pmatrix}$ \\
$ \mathbb{X}_{22} $ & $\mathbb{M}_{1,1}^{(a,E^{\prime\prime})}$ & M$_{2}$ & $\begin{pmatrix} 0 & 0 & 0 \\ 0 & 0 & - \frac{\sqrt{2} i}{2} \\ 0 & \frac{\sqrt{2} i}{2} & 0 \end{pmatrix}$ \\ \hline
$ \mathbb{X}_{23} $ & $\mathbb{Q}_{1}^{(a,A_{2}^{\prime\prime})}$ & M$_{3}$ & $\begin{pmatrix} 0 & 0 & \frac{\sqrt{10}}{5} \\ 0 & 0 & 0 \\ 0 & \frac{\sqrt{30}}{10} & 0 \\ \frac{\sqrt{30}}{10} & 0 & 0 \\ 0 & 0 & 0 \end{pmatrix}$ \\
$ \mathbb{X}_{24} $ & $\mathbb{Q}_{3}^{(a,A_{2}^{\prime\prime})}$ & M$_{3}$ & $\begin{pmatrix} 0 & 0 & \frac{\sqrt{15}}{5} \\ 0 & 0 & 0 \\ 0 & - \frac{\sqrt{5}}{5} & 0 \\ - \frac{\sqrt{5}}{5} & 0 & 0 \\ 0 & 0 & 0 \end{pmatrix}$ \\
$ \mathbb{X}_{25} $ & $\mathbb{Q}_{1,0}^{(a,E^{\prime})}$ & M$_{3}$ & $\begin{pmatrix} - \frac{\sqrt{10}}{10} & 0 & 0 \\ \frac{\sqrt{30}}{10} & 0 & 0 \\ 0 & 0 & 0 \\ 0 & 0 & \frac{\sqrt{30}}{10} \\ 0 & - \frac{\sqrt{30}}{10} & 0 \end{pmatrix}$ \\
$ \mathbb{X}_{26} $ & $\mathbb{Q}_{1,1}^{(a,E^{\prime})}$ & M$_{3}$ & $\begin{pmatrix} 0 & - \frac{\sqrt{10}}{10} & 0 \\ 0 & - \frac{\sqrt{30}}{10} & 0 \\ 0 & 0 & \frac{\sqrt{30}}{10} \\ 0 & 0 & 0 \\ - \frac{\sqrt{30}}{10} & 0 & 0 \end{pmatrix}$ \\
$ \mathbb{X}_{27} $ & $\mathbb{Q}_{3,0}^{(a,E^{\prime})}$ & M$_{3}$ & $\begin{pmatrix} \frac{\sqrt{10}}{5} & 0 & 0 \\ - \frac{\sqrt{30}}{30} & 0 & 0 \\ 0 & 0 & 0 \\ 0 & 0 & \frac{2 \sqrt{30}}{15} \\ 0 & \frac{\sqrt{30}}{30} & 0 \end{pmatrix}$ \\
$ \mathbb{X}_{28} $ & $\mathbb{Q}_{3,1}^{(a,E^{\prime})}$ & M$_{3}$ & $\begin{pmatrix} 0 & \frac{\sqrt{10}}{5} & 0 \\ 0 & \frac{\sqrt{30}}{30} & 0 \\ 0 & 0 & \frac{2 \sqrt{30}}{15} \\ 0 & 0 & 0 \\ \frac{\sqrt{30}}{30} & 0 & 0 \end{pmatrix}$ \\
$ \mathbb{X}_{29} $ & $\mathbb{M}_{2,0}^{(a,E^{\prime})}$ & M$_{3}$ & $\begin{pmatrix} \frac{\sqrt{2} i}{2} & 0 & 0 \\ \frac{\sqrt{6} i}{6} & 0 & 0 \\ 0 & 0 & 0 \\ 0 & 0 & - \frac{\sqrt{6} i}{6} \\ 0 & - \frac{\sqrt{6} i}{6} & 0 \end{pmatrix}$ \\
$ \mathbb{X}_{30} $ & $\mathbb{M}_{2,1}^{(a,E^{\prime})}$ & M$_{3}$ & $\begin{pmatrix} 0 & \frac{\sqrt{2} i}{2} & 0 \\ 0 & - \frac{\sqrt{6} i}{6} & 0 \\ 0 & 0 & - \frac{\sqrt{6} i}{6} \\ 0 & 0 & 0 \\ - \frac{\sqrt{6} i}{6} & 0 & 0 \end{pmatrix}$ \\
$ \mathbb{X}_{31} $ & $\mathbb{Q}_{3}^{(a,A_{1}^{\prime})}$ & M$_{3}$ & $\begin{pmatrix} 0 & 0 & 0 \\ 0 & \frac{\sqrt{2}}{2} & 0 \\ 0 & 0 & 0 \\ 0 & 0 & 0 \\ - \frac{\sqrt{2}}{2} & 0 & 0 \end{pmatrix}$ \\
$ \mathbb{X}_{32} $ & $\mathbb{Q}_{3}^{(a,A_{2}^{\prime})}$ & M$_{3}$ & $\begin{pmatrix} 0 & 0 & 0 \\ \frac{\sqrt{2}}{2} & 0 & 0 \\ 0 & 0 & 0 \\ 0 & 0 & 0 \\ 0 & \frac{\sqrt{2}}{2} & 0 \end{pmatrix}$ \\
$ \mathbb{X}_{33} $ & $\mathbb{Q}_{3,0}^{(a,E^{\prime\prime})}$ & M$_{3}$ & $\begin{pmatrix} 0 & 0 & 0 \\ 0 & 0 & 0 \\ - \frac{\sqrt{3}}{3} & 0 & 0 \\ 0 & - \frac{\sqrt{3}}{3} & 0 \\ 0 & 0 & \frac{\sqrt{3}}{3} \end{pmatrix}$ \\
$ \mathbb{X}_{34} $ & $\mathbb{Q}_{3,1}^{(a,E^{\prime\prime})}$ & M$_{3}$ & $\begin{pmatrix} 0 & 0 & 0 \\ 0 & 0 & - \frac{\sqrt{3}}{3} \\ 0 & \frac{\sqrt{3}}{3} & 0 \\ - \frac{\sqrt{3}}{3} & 0 & 0 \\ 0 & 0 & 0 \end{pmatrix}$ \\
$ \mathbb{X}_{35} $ & $\mathbb{M}_{2,0}^{(a,E^{\prime\prime})}$ & M$_{3}$ & $\begin{pmatrix} 0 & 0 & 0 \\ 0 & 0 & 0 \\ \frac{\sqrt{6} i}{6} & 0 & 0 \\ 0 & \frac{\sqrt{6} i}{6} & 0 \\ 0 & 0 & \frac{\sqrt{6} i}{3} \end{pmatrix}$ \\
$ \mathbb{X}_{36} $ & $\mathbb{M}_{2,1}^{(a,E^{\prime\prime})}$ & M$_{3}$ & $\begin{pmatrix} 0 & 0 & 0 \\ 0 & 0 & - \frac{\sqrt{6} i}{3} \\ 0 & - \frac{\sqrt{6} i}{6} & 0 \\ \frac{\sqrt{6} i}{6} & 0 & 0 \\ 0 & 0 & 0 \end{pmatrix}$ \\
$ \mathbb{X}_{37} $ & $\mathbb{M}_{2}^{(a,A_{1}^{\prime\prime})}$ & M$_{3}$ & $\begin{pmatrix} 0 & 0 & 0 \\ 0 & 0 & 0 \\ \frac{\sqrt{2} i}{2} & 0 & 0 \\ 0 & - \frac{\sqrt{2} i}{2} & 0 \\ 0 & 0 & 0 \end{pmatrix}$ \\
\end{longtable}
\end{center}

\item The site/bond cluster SAMBs are given in Table XIII.

\begin{center}
\renewcommand{\arraystretch}{1.3}
\begin{longtable}{c|c|c|c}
\caption{Cluster SAMB.}
 \\
 \hline \hline
symbol & type & cluster & form \\ \hline \endfirsthead

\multicolumn{3}{l}{\tablename\ \thetable{}} \\
 \hline \hline
symbol & type & cluster & form \\ \hline \endhead

 \hline \hline
\multicolumn{3}{r}{\footnotesize\it continued ...} \\ \endfoot

 \hline \hline
\multicolumn{3}{r}{} \\ \endlastfoot

$ \mathbb{Y}_{1} $ & $\mathbb{Q}_{0}^{(s,A_{1}^{\prime})}$ & S$_{1}$ & $\begin{pmatrix} 1 \end{pmatrix}$ \\ \hline
$ \mathbb{Y}_{2} $ & $\mathbb{Q}_{0}^{(s,A_{1}^{\prime})}$ & S$_{2}$ & $\begin{pmatrix} \frac{\sqrt{2}}{2} & \frac{\sqrt{2}}{2} \end{pmatrix}$ \\ \hline
$ \mathbb{Y}_{3} $ & $\mathbb{Q}_{0}^{(b,A_{1}^{\prime})}$ & B$_{1}$ & $\begin{pmatrix} \frac{\sqrt{3}}{3} & \frac{\sqrt{3}}{3} & \frac{\sqrt{3}}{3} \end{pmatrix}$ \\
$ \mathbb{Y}_{4} $ & $\mathbb{Q}_{1,0}^{(b,E^{\prime})}$ & B$_{1}$ & $\begin{pmatrix} - \frac{\sqrt{2}}{2} & \frac{\sqrt{2}}{2} & 0 \end{pmatrix}$ \\
$ \mathbb{Y}_{5} $ & $\mathbb{Q}_{1,1}^{(b,E^{\prime})}$ & B$_{1}$ & $\begin{pmatrix} - \frac{\sqrt{6}}{6} & - \frac{\sqrt{6}}{6} & \frac{\sqrt{6}}{3} \end{pmatrix}$ \\
$ \mathbb{Y}_{6} $ & $\mathbb{T}_{1,0}^{(b,E^{\prime})}$ & B$_{1}$ & $\begin{pmatrix} \frac{\sqrt{6} i}{6} & - \frac{\sqrt{6} i}{6} & - \frac{\sqrt{6} i}{3} \end{pmatrix}$ \\
$ \mathbb{Y}_{7} $ & $\mathbb{T}_{1,1}^{(b,E^{\prime})}$ & B$_{1}$ & $\begin{pmatrix} - \frac{\sqrt{2} i}{2} & - \frac{\sqrt{2} i}{2} & 0 \end{pmatrix}$ \\
$ \mathbb{Y}_{8} $ & $\mathbb{T}_{3}^{(b,A_{2}^{\prime})}$ & B$_{1}$ & $\begin{pmatrix} \frac{\sqrt{3} i}{3} & - \frac{\sqrt{3} i}{3} & \frac{\sqrt{3} i}{3} \end{pmatrix}$ \\ \hline
$ \mathbb{Y}_{9} $ & $\mathbb{Q}_{0}^{(b,A_{1}^{\prime})}$ & B$_{2}$ & $\begin{pmatrix} \frac{\sqrt{3}}{3} & \frac{\sqrt{3}}{3} & \frac{\sqrt{3}}{3} \end{pmatrix}$ \\
$ \mathbb{Y}_{10} $ & $\mathbb{Q}_{1,0}^{(b,E^{\prime})}$ & B$_{2}$ & $\begin{pmatrix} - \frac{\sqrt{2}}{2} & \frac{\sqrt{2}}{2} & 0 \end{pmatrix}$ \\
$ \mathbb{Y}_{11} $ & $\mathbb{Q}_{1,1}^{(b,E^{\prime})}$ & B$_{2}$ & $\begin{pmatrix} - \frac{\sqrt{6}}{6} & - \frac{\sqrt{6}}{6} & \frac{\sqrt{6}}{3} \end{pmatrix}$ \\
$ \mathbb{Y}_{12} $ & $\mathbb{T}_{1,0}^{(b,E^{\prime})}$ & B$_{2}$ & $\begin{pmatrix} - \frac{\sqrt{2} i}{2} & \frac{\sqrt{2} i}{2} & 0 \end{pmatrix}$ \\
$ \mathbb{Y}_{13} $ & $\mathbb{T}_{1,1}^{(b,E^{\prime})}$ & B$_{2}$ & $\begin{pmatrix} - \frac{\sqrt{6} i}{6} & - \frac{\sqrt{6} i}{6} & \frac{\sqrt{6} i}{3} \end{pmatrix}$ \\ \hline
$ \mathbb{Y}_{14} $ & $\mathbb{Q}_{0}^{(b,A_{1}^{\prime})}$ & B$_{3}$ & $\begin{pmatrix} \frac{\sqrt{3}}{3} & \frac{\sqrt{3}}{3} & \frac{\sqrt{3}}{3} \end{pmatrix}$ \\
$ \mathbb{Y}_{15} $ & $\mathbb{Q}_{2,0}^{(b,E^{\prime})}$ & B$_{3}$ & $\begin{pmatrix} - \frac{\sqrt{2}}{2} & \frac{\sqrt{2}}{2} & 0 \end{pmatrix}$ \\
$ \mathbb{Y}_{16} $ & $\mathbb{Q}_{2,1}^{(b,E^{\prime})}$ & B$_{3}$ & $\begin{pmatrix} \frac{\sqrt{6}}{6} & \frac{\sqrt{6}}{6} & - \frac{\sqrt{6}}{3} \end{pmatrix}$ \\
$ \mathbb{Y}_{17} $ & $\mathbb{T}_{1,0}^{(b,E^{\prime})}$ & B$_{3}$ & $\begin{pmatrix} - \frac{\sqrt{6} i}{6} & \frac{\sqrt{6} i}{6} & - \frac{\sqrt{6} i}{3} \end{pmatrix}$ \\
$ \mathbb{Y}_{18} $ & $\mathbb{T}_{1,1}^{(b,E^{\prime})}$ & B$_{3}$ & $\begin{pmatrix} - \frac{\sqrt{2} i}{2} & - \frac{\sqrt{2} i}{2} & 0 \end{pmatrix}$ \\
$ \mathbb{Y}_{19} $ & $\mathbb{T}_{3}^{(b,A_{2}^{\prime})}$ & B$_{3}$ & $\begin{pmatrix} \frac{\sqrt{3} i}{3} & - \frac{\sqrt{3} i}{3} & - \frac{\sqrt{3} i}{3} \end{pmatrix}$ \\ \hline
$ \mathbb{Y}_{20} $ & $\mathbb{Q}_{0}^{(b,A_{1}^{\prime})}$ & B$_{4}$ & $\begin{pmatrix} \frac{\sqrt{6}}{6} & \frac{\sqrt{6}}{6} & \frac{\sqrt{6}}{6} & \frac{\sqrt{6}}{6} & \frac{\sqrt{6}}{6} & \frac{\sqrt{6}}{6} \end{pmatrix}$ \\
$ \mathbb{Y}_{21} $ & $\mathbb{Q}_{1}^{(b,A_{2}^{\prime\prime})}$ & B$_{4}$ & $\begin{pmatrix} \frac{\sqrt{6}}{6} & - \frac{\sqrt{6}}{6} & - \frac{\sqrt{6}}{6} & - \frac{\sqrt{6}}{6} & \frac{\sqrt{6}}{6} & \frac{\sqrt{6}}{6} \end{pmatrix}$ \\
$ \mathbb{Y}_{22} $ & $\mathbb{Q}_{1,0}^{(b,E^{\prime})}$ & B$_{4}$ & $\begin{pmatrix} - \frac{1}{2} & \frac{1}{2} & - \frac{1}{2} & 0 & \frac{1}{2} & 0 \end{pmatrix}$ \\
$ \mathbb{Y}_{23} $ & $\mathbb{Q}_{1,1}^{(b,E^{\prime})}$ & B$_{4}$ & $\begin{pmatrix} - \frac{\sqrt{3}}{6} & - \frac{\sqrt{3}}{6} & - \frac{\sqrt{3}}{6} & \frac{\sqrt{3}}{3} & - \frac{\sqrt{3}}{6} & \frac{\sqrt{3}}{3} \end{pmatrix}$ \\
$ \mathbb{Y}_{24} $ & $\mathbb{Q}_{2,0}^{(b,E^{\prime\prime})}$ & B$_{4}$ & $\begin{pmatrix} - \frac{1}{2} & - \frac{1}{2} & \frac{1}{2} & 0 & \frac{1}{2} & 0 \end{pmatrix}$ \\
$ \mathbb{Y}_{25} $ & $\mathbb{Q}_{2,1}^{(b,E^{\prime\prime})}$ & B$_{4}$ & $\begin{pmatrix} - \frac{\sqrt{3}}{6} & \frac{\sqrt{3}}{6} & \frac{\sqrt{3}}{6} & - \frac{\sqrt{3}}{3} & - \frac{\sqrt{3}}{6} & \frac{\sqrt{3}}{3} \end{pmatrix}$ \\
$ \mathbb{Y}_{26} $ & $\mathbb{T}_{1,0}^{(b,E^{\prime})}$ & B$_{4}$ & $\begin{pmatrix} - \frac{i}{2} & \frac{i}{2} & - \frac{i}{2} & 0 & \frac{i}{2} & 0 \end{pmatrix}$ \\
$ \mathbb{Y}_{27} $ & $\mathbb{T}_{1,1}^{(b,E^{\prime})}$ & B$_{4}$ & $\begin{pmatrix} - \frac{\sqrt{3} i}{6} & - \frac{\sqrt{3} i}{6} & - \frac{\sqrt{3} i}{6} & \frac{\sqrt{3} i}{3} & - \frac{\sqrt{3} i}{6} & \frac{\sqrt{3} i}{3} \end{pmatrix}$ \\
$ \mathbb{Y}_{28} $ & $\mathbb{T}_{2,0}^{(b,E^{\prime\prime})}$ & B$_{4}$ & $\begin{pmatrix} - \frac{i}{2} & - \frac{i}{2} & \frac{i}{2} & 0 & \frac{i}{2} & 0 \end{pmatrix}$ \\
$ \mathbb{Y}_{29} $ & $\mathbb{T}_{2,1}^{(b,E^{\prime\prime})}$ & B$_{4}$ & $\begin{pmatrix} - \frac{\sqrt{3} i}{6} & \frac{\sqrt{3} i}{6} & \frac{\sqrt{3} i}{6} & - \frac{\sqrt{3} i}{3} & - \frac{\sqrt{3} i}{6} & \frac{\sqrt{3} i}{3} \end{pmatrix}$ \\ \hline
$ \mathbb{Y}_{30} $ & $\mathbb{Q}_{0}^{(b,A_{1}^{\prime})}$ & B$_{5}$ & $\begin{pmatrix} \frac{\sqrt{6}}{6} & \frac{\sqrt{6}}{6} & \frac{\sqrt{6}}{6} & \frac{\sqrt{6}}{6} & \frac{\sqrt{6}}{6} & \frac{\sqrt{6}}{6} \end{pmatrix}$ \\
$ \mathbb{Y}_{31} $ & $\mathbb{Q}_{1}^{(b,A_{2}^{\prime\prime})}$ & B$_{5}$ & $\begin{pmatrix} \frac{\sqrt{6}}{6} & - \frac{\sqrt{6}}{6} & - \frac{\sqrt{6}}{6} & - \frac{\sqrt{6}}{6} & \frac{\sqrt{6}}{6} & \frac{\sqrt{6}}{6} \end{pmatrix}$ \\
$ \mathbb{Y}_{32} $ & $\mathbb{Q}_{1,0}^{(b,E^{\prime})}$ & B$_{5}$ & $\begin{pmatrix} 0 & 0 & - \frac{1}{2} & \frac{1}{2} & \frac{1}{2} & - \frac{1}{2} \end{pmatrix}$ \\
$ \mathbb{Y}_{33} $ & $\mathbb{Q}_{1,1}^{(b,E^{\prime})}$ & B$_{5}$ & $\begin{pmatrix} - \frac{\sqrt{3}}{3} & - \frac{\sqrt{3}}{3} & \frac{\sqrt{3}}{6} & \frac{\sqrt{3}}{6} & \frac{\sqrt{3}}{6} & \frac{\sqrt{3}}{6} \end{pmatrix}$ \\
$ \mathbb{Y}_{34} $ & $\mathbb{Q}_{2,0}^{(b,E^{\prime\prime})}$ & B$_{5}$ & $\begin{pmatrix} 0 & 0 & \frac{1}{2} & - \frac{1}{2} & \frac{1}{2} & - \frac{1}{2} \end{pmatrix}$ \\
$ \mathbb{Y}_{35} $ & $\mathbb{Q}_{2,1}^{(b,E^{\prime\prime})}$ & B$_{5}$ & $\begin{pmatrix} - \frac{\sqrt{3}}{3} & \frac{\sqrt{3}}{3} & - \frac{\sqrt{3}}{6} & - \frac{\sqrt{3}}{6} & \frac{\sqrt{3}}{6} & \frac{\sqrt{3}}{6} \end{pmatrix}$ \\
$ \mathbb{Y}_{36} $ & $\mathbb{T}_{1,0}^{(b,E^{\prime})}$ & B$_{5}$ & $\begin{pmatrix} 0 & 0 & - \frac{i}{2} & \frac{i}{2} & \frac{i}{2} & - \frac{i}{2} \end{pmatrix}$ \\
$ \mathbb{Y}_{37} $ & $\mathbb{T}_{1,1}^{(b,E^{\prime})}$ & B$_{5}$ & $\begin{pmatrix} - \frac{\sqrt{3} i}{3} & - \frac{\sqrt{3} i}{3} & \frac{\sqrt{3} i}{6} & \frac{\sqrt{3} i}{6} & \frac{\sqrt{3} i}{6} & \frac{\sqrt{3} i}{6} \end{pmatrix}$ \\
$ \mathbb{Y}_{38} $ & $\mathbb{T}_{2,0}^{(b,E^{\prime\prime})}$ & B$_{5}$ & $\begin{pmatrix} 0 & 0 & \frac{i}{2} & - \frac{i}{2} & \frac{i}{2} & - \frac{i}{2} \end{pmatrix}$ \\
$ \mathbb{Y}_{39} $ & $\mathbb{T}_{2,1}^{(b,E^{\prime\prime})}$ & B$_{5}$ & $\begin{pmatrix} - \frac{\sqrt{3} i}{3} & \frac{\sqrt{3} i}{3} & - \frac{\sqrt{3} i}{6} & - \frac{\sqrt{3} i}{6} & \frac{\sqrt{3} i}{6} & \frac{\sqrt{3} i}{6} \end{pmatrix}$ \\ \hline
$ \mathbb{Y}_{40} $ & $\mathbb{Q}_{0}^{(b,A_{1}^{\prime})}$ & B$_{6}$ & $\begin{pmatrix} \frac{\sqrt{3}}{6} & \frac{\sqrt{3}}{6} & \frac{\sqrt{3}}{6} & \frac{\sqrt{3}}{6} & \frac{\sqrt{3}}{6} & \frac{\sqrt{3}}{6} & \frac{\sqrt{3}}{6} & \frac{\sqrt{3}}{6} & \frac{\sqrt{3}}{6} & \frac{\sqrt{3}}{6} & \frac{\sqrt{3}}{6} & \frac{\sqrt{3}}{6} \end{pmatrix}$ \\
$ \mathbb{Y}_{41} $ & $\mathbb{Q}_{1}^{(b,A_{2}^{\prime\prime})}$ & B$_{6}$ & $\begin{pmatrix} \frac{\sqrt{3}}{6} & - \frac{\sqrt{3}}{6} & - \frac{\sqrt{3}}{6} & - \frac{\sqrt{3}}{6} & \frac{\sqrt{3}}{6} & \frac{\sqrt{3}}{6} & \frac{\sqrt{3}}{6} & \frac{\sqrt{3}}{6} & \frac{\sqrt{3}}{6} & - \frac{\sqrt{3}}{6} & - \frac{\sqrt{3}}{6} & - \frac{\sqrt{3}}{6} \end{pmatrix}$ \\
$ \mathbb{Y}_{42} $ & $\mathbb{Q}_{1,0}^{(b,E^{\prime})}$ & B$_{6}$ & $\begin{pmatrix} - \frac{\sqrt{6}}{12} & \frac{\sqrt{6}}{12} & - \frac{\sqrt{6}}{6} & \frac{\sqrt{6}}{12} & \frac{\sqrt{6}}{6} & - \frac{\sqrt{6}}{12} & \frac{\sqrt{6}}{12} & - \frac{\sqrt{6}}{6} & \frac{\sqrt{6}}{12} & - \frac{\sqrt{6}}{12} & - \frac{\sqrt{6}}{12} & \frac{\sqrt{6}}{6} \end{pmatrix}$ \\
$ \mathbb{Y}_{43} $ & $\mathbb{Q}_{1,1}^{(b,E^{\prime})}$ & B$_{6}$ & $\begin{pmatrix} - \frac{\sqrt{2}}{4} & - \frac{\sqrt{2}}{4} & 0 & \frac{\sqrt{2}}{4} & 0 & \frac{\sqrt{2}}{4} & - \frac{\sqrt{2}}{4} & 0 & \frac{\sqrt{2}}{4} & - \frac{\sqrt{2}}{4} & \frac{\sqrt{2}}{4} & 0 \end{pmatrix}$ \\
$ \mathbb{Y}_{44} $ & $\mathbb{Q}_{2,0}^{(b,E^{\prime})}$ & B$_{6}$ & $\begin{pmatrix} - \frac{\sqrt{2}}{4} & \frac{\sqrt{2}}{4} & 0 & - \frac{\sqrt{2}}{4} & 0 & \frac{\sqrt{2}}{4} & \frac{\sqrt{2}}{4} & 0 & - \frac{\sqrt{2}}{4} & - \frac{\sqrt{2}}{4} & \frac{\sqrt{2}}{4} & 0 \end{pmatrix}$ \\
$ \mathbb{Y}_{45} $ & $\mathbb{Q}_{2,1}^{(b,E^{\prime})}$ & B$_{6}$ & $\begin{pmatrix} \frac{\sqrt{6}}{12} & \frac{\sqrt{6}}{12} & - \frac{\sqrt{6}}{6} & \frac{\sqrt{6}}{12} & - \frac{\sqrt{6}}{6} & \frac{\sqrt{6}}{12} & \frac{\sqrt{6}}{12} & - \frac{\sqrt{6}}{6} & \frac{\sqrt{6}}{12} & \frac{\sqrt{6}}{12} & \frac{\sqrt{6}}{12} & - \frac{\sqrt{6}}{6} \end{pmatrix}$ \\
$ \mathbb{Y}_{46} $ & $\mathbb{Q}_{2,0}^{(b,E^{\prime\prime})}$ & B$_{6}$ & $\begin{pmatrix} - \frac{\sqrt{6}}{12} & - \frac{\sqrt{6}}{12} & \frac{\sqrt{6}}{6} & - \frac{\sqrt{6}}{12} & \frac{\sqrt{6}}{6} & - \frac{\sqrt{6}}{12} & \frac{\sqrt{6}}{12} & - \frac{\sqrt{6}}{6} & \frac{\sqrt{6}}{12} & \frac{\sqrt{6}}{12} & \frac{\sqrt{6}}{12} & - \frac{\sqrt{6}}{6} \end{pmatrix}$ \\
$ \mathbb{Y}_{47} $ & $\mathbb{Q}_{2,1}^{(b,E^{\prime\prime})}$ & B$_{6}$ & $\begin{pmatrix} - \frac{\sqrt{2}}{4} & \frac{\sqrt{2}}{4} & 0 & - \frac{\sqrt{2}}{4} & 0 & \frac{\sqrt{2}}{4} & - \frac{\sqrt{2}}{4} & 0 & \frac{\sqrt{2}}{4} & \frac{\sqrt{2}}{4} & - \frac{\sqrt{2}}{4} & 0 \end{pmatrix}$ \\
$ \mathbb{Y}_{48} $ & $\mathbb{Q}_{3}^{(b,A_{2}^{\prime})}$ & B$_{6}$ & $\begin{pmatrix} \frac{\sqrt{3}}{6} & - \frac{\sqrt{3}}{6} & - \frac{\sqrt{3}}{6} & - \frac{\sqrt{3}}{6} & \frac{\sqrt{3}}{6} & \frac{\sqrt{3}}{6} & - \frac{\sqrt{3}}{6} & - \frac{\sqrt{3}}{6} & - \frac{\sqrt{3}}{6} & \frac{\sqrt{3}}{6} & \frac{\sqrt{3}}{6} & \frac{\sqrt{3}}{6} \end{pmatrix}$ \\
$ \mathbb{Y}_{49} $ & $\mathbb{Q}_{3,0}^{(b,E^{\prime\prime})}$ & B$_{6}$ & $\begin{pmatrix} - \frac{\sqrt{2}}{4} & - \frac{\sqrt{2}}{4} & 0 & \frac{\sqrt{2}}{4} & 0 & \frac{\sqrt{2}}{4} & \frac{\sqrt{2}}{4} & 0 & - \frac{\sqrt{2}}{4} & \frac{\sqrt{2}}{4} & - \frac{\sqrt{2}}{4} & 0 \end{pmatrix}$ \\
$ \mathbb{Y}_{50} $ & $\mathbb{Q}_{3,1}^{(b,E^{\prime\prime})}$ & B$_{6}$ & $\begin{pmatrix} \frac{\sqrt{6}}{12} & - \frac{\sqrt{6}}{12} & \frac{\sqrt{6}}{6} & - \frac{\sqrt{6}}{12} & - \frac{\sqrt{6}}{6} & \frac{\sqrt{6}}{12} & \frac{\sqrt{6}}{12} & - \frac{\sqrt{6}}{6} & \frac{\sqrt{6}}{12} & - \frac{\sqrt{6}}{12} & - \frac{\sqrt{6}}{12} & \frac{\sqrt{6}}{6} \end{pmatrix}$ \\
$ \mathbb{Y}_{51} $ & $\mathbb{T}_{1,0}^{(b,E^{\prime})}$ & B$_{6}$ & $\begin{pmatrix} - \frac{\sqrt{6} i}{12} & \frac{\sqrt{6} i}{12} & - \frac{\sqrt{6} i}{6} & \frac{\sqrt{6} i}{12} & \frac{\sqrt{6} i}{6} & - \frac{\sqrt{6} i}{12} & \frac{\sqrt{6} i}{12} & - \frac{\sqrt{6} i}{6} & \frac{\sqrt{6} i}{12} & - \frac{\sqrt{6} i}{12} & - \frac{\sqrt{6} i}{12} & \frac{\sqrt{6} i}{6} \end{pmatrix}$ \\
$ \mathbb{Y}_{52} $ & $\mathbb{T}_{1,1}^{(b,E^{\prime})}$ & B$_{6}$ & $\begin{pmatrix} - \frac{\sqrt{2} i}{4} & - \frac{\sqrt{2} i}{4} & 0 & \frac{\sqrt{2} i}{4} & 0 & \frac{\sqrt{2} i}{4} & - \frac{\sqrt{2} i}{4} & 0 & \frac{\sqrt{2} i}{4} & - \frac{\sqrt{2} i}{4} & \frac{\sqrt{2} i}{4} & 0 \end{pmatrix}$ \\
$ \mathbb{Y}_{53} $ & $\mathbb{T}_{2,0}^{(b,E^{\prime})}$ & B$_{6}$ & $\begin{pmatrix} - \frac{\sqrt{2} i}{4} & \frac{\sqrt{2} i}{4} & 0 & - \frac{\sqrt{2} i}{4} & 0 & \frac{\sqrt{2} i}{4} & \frac{\sqrt{2} i}{4} & 0 & - \frac{\sqrt{2} i}{4} & - \frac{\sqrt{2} i}{4} & \frac{\sqrt{2} i}{4} & 0 \end{pmatrix}$ \\
$ \mathbb{Y}_{54} $ & $\mathbb{T}_{2,1}^{(b,E^{\prime})}$ & B$_{6}$ & $\begin{pmatrix} \frac{\sqrt{6} i}{12} & \frac{\sqrt{6} i}{12} & - \frac{\sqrt{6} i}{6} & \frac{\sqrt{6} i}{12} & - \frac{\sqrt{6} i}{6} & \frac{\sqrt{6} i}{12} & \frac{\sqrt{6} i}{12} & - \frac{\sqrt{6} i}{6} & \frac{\sqrt{6} i}{12} & \frac{\sqrt{6} i}{12} & \frac{\sqrt{6} i}{12} & - \frac{\sqrt{6} i}{6} \end{pmatrix}$ \\
$ \mathbb{Y}_{55} $ & $\mathbb{T}_{2,0}^{(b,E^{\prime\prime})}$ & B$_{6}$ & $\begin{pmatrix} - \frac{\sqrt{6} i}{12} & - \frac{\sqrt{6} i}{12} & \frac{\sqrt{6} i}{6} & - \frac{\sqrt{6} i}{12} & \frac{\sqrt{6} i}{6} & - \frac{\sqrt{6} i}{12} & \frac{\sqrt{6} i}{12} & - \frac{\sqrt{6} i}{6} & \frac{\sqrt{6} i}{12} & \frac{\sqrt{6} i}{12} & \frac{\sqrt{6} i}{12} & - \frac{\sqrt{6} i}{6} \end{pmatrix}$ \\
$ \mathbb{Y}_{56} $ & $\mathbb{T}_{2,1}^{(b,E^{\prime\prime})}$ & B$_{6}$ & $\begin{pmatrix} - \frac{\sqrt{2} i}{4} & \frac{\sqrt{2} i}{4} & 0 & - \frac{\sqrt{2} i}{4} & 0 & \frac{\sqrt{2} i}{4} & - \frac{\sqrt{2} i}{4} & 0 & \frac{\sqrt{2} i}{4} & \frac{\sqrt{2} i}{4} & - \frac{\sqrt{2} i}{4} & 0 \end{pmatrix}$ \\
$ \mathbb{Y}_{57} $ & $\mathbb{T}_{3,0}^{(b,E^{\prime\prime})}$ & B$_{6}$ & $\begin{pmatrix} - \frac{\sqrt{2} i}{4} & - \frac{\sqrt{2} i}{4} & 0 & \frac{\sqrt{2} i}{4} & 0 & \frac{\sqrt{2} i}{4} & \frac{\sqrt{2} i}{4} & 0 & - \frac{\sqrt{2} i}{4} & \frac{\sqrt{2} i}{4} & - \frac{\sqrt{2} i}{4} & 0 \end{pmatrix}$ \\
$ \mathbb{Y}_{58} $ & $\mathbb{T}_{3,1}^{(b,E^{\prime\prime})}$ & B$_{6}$ & $\begin{pmatrix} \frac{\sqrt{6} i}{12} & - \frac{\sqrt{6} i}{12} & \frac{\sqrt{6} i}{6} & - \frac{\sqrt{6} i}{12} & - \frac{\sqrt{6} i}{6} & \frac{\sqrt{6} i}{12} & \frac{\sqrt{6} i}{12} & - \frac{\sqrt{6} i}{6} & \frac{\sqrt{6} i}{12} & - \frac{\sqrt{6} i}{12} & - \frac{\sqrt{6} i}{12} & \frac{\sqrt{6} i}{6} \end{pmatrix}$ \\
$ \mathbb{Y}_{59} $ & $\mathbb{T}_{4}^{(b,A_{1}^{\prime\prime})}$ & B$_{6}$ & $\begin{pmatrix} \frac{\sqrt{3} i}{6} & \frac{\sqrt{3} i}{6} & \frac{\sqrt{3} i}{6} & \frac{\sqrt{3} i}{6} & \frac{\sqrt{3} i}{6} & \frac{\sqrt{3} i}{6} & - \frac{\sqrt{3} i}{6} & - \frac{\sqrt{3} i}{6} & - \frac{\sqrt{3} i}{6} & - \frac{\sqrt{3} i}{6} & - \frac{\sqrt{3} i}{6} & - \frac{\sqrt{3} i}{6} \end{pmatrix}$ \\ \hline
$ \mathbb{Y}_{60} $ & $\mathbb{Q}_{0}^{(b,A_{1}^{\prime})}$ & B$_{7}$ & $\begin{pmatrix} \frac{\sqrt{6}}{6} & \frac{\sqrt{6}}{6} & \frac{\sqrt{6}}{6} & \frac{\sqrt{6}}{6} & \frac{\sqrt{6}}{6} & \frac{\sqrt{6}}{6} \end{pmatrix}$ \\
$ \mathbb{Y}_{61} $ & $\mathbb{Q}_{1,0}^{(b,E^{\prime})}$ & B$_{7}$ & $\begin{pmatrix} 0 & 0 & - \frac{1}{2} & \frac{1}{2} & \frac{1}{2} & - \frac{1}{2} \end{pmatrix}$ \\
$ \mathbb{Y}_{62} $ & $\mathbb{Q}_{1,1}^{(b,E^{\prime})}$ & B$_{7}$ & $\begin{pmatrix} - \frac{\sqrt{3}}{3} & - \frac{\sqrt{3}}{3} & \frac{\sqrt{3}}{6} & \frac{\sqrt{3}}{6} & \frac{\sqrt{3}}{6} & \frac{\sqrt{3}}{6} \end{pmatrix}$ \\
$ \mathbb{Y}_{63} $ & $\mathbb{Q}_{2,0}^{(b,E^{\prime\prime})}$ & B$_{7}$ & $\begin{pmatrix} 0 & 0 & \frac{1}{2} & - \frac{1}{2} & \frac{1}{2} & - \frac{1}{2} \end{pmatrix}$ \\
$ \mathbb{Y}_{64} $ & $\mathbb{Q}_{2,1}^{(b,E^{\prime\prime})}$ & B$_{7}$ & $\begin{pmatrix} - \frac{\sqrt{3}}{3} & \frac{\sqrt{3}}{3} & - \frac{\sqrt{3}}{6} & - \frac{\sqrt{3}}{6} & \frac{\sqrt{3}}{6} & \frac{\sqrt{3}}{6} \end{pmatrix}$ \\
$ \mathbb{Y}_{65} $ & $\mathbb{T}_{2,0}^{(b,E^{\prime\prime})}$ & B$_{7}$ & $\begin{pmatrix} - \frac{\sqrt{3} i}{3} & \frac{\sqrt{3} i}{3} & \frac{\sqrt{3} i}{6} & - \frac{\sqrt{3} i}{6} & \frac{\sqrt{3} i}{6} & - \frac{\sqrt{3} i}{6} \end{pmatrix}$ \\
$ \mathbb{Y}_{66} $ & $\mathbb{T}_{2,1}^{(b,E^{\prime\prime})}$ & B$_{7}$ & $\begin{pmatrix} 0 & 0 & \frac{i}{2} & \frac{i}{2} & - \frac{i}{2} & - \frac{i}{2} \end{pmatrix}$ \\
$ \mathbb{Y}_{67} $ & $\mathbb{T}_{3}^{(b,A_{2}^{\prime})}$ & B$_{7}$ & $\begin{pmatrix} \frac{\sqrt{6} i}{6} & \frac{\sqrt{6} i}{6} & - \frac{\sqrt{6} i}{6} & \frac{\sqrt{6} i}{6} & \frac{\sqrt{6} i}{6} & - \frac{\sqrt{6} i}{6} \end{pmatrix}$ \\ \hline
$ \mathbb{Y}_{68} $ & $\mathbb{Q}_{0}^{(b,A_{1}^{\prime})}$ & B$_{8}$ & $\begin{pmatrix} 1 \end{pmatrix}$ \\ \hline
$ \mathbb{Y}_{69} $ & $\mathbb{Q}_{0}^{(b,A_{1}^{\prime})}$ & B$_{9}$ & $\begin{pmatrix} \frac{\sqrt{6}}{6} & \frac{\sqrt{6}}{6} & \frac{\sqrt{6}}{6} & \frac{\sqrt{6}}{6} & \frac{\sqrt{6}}{6} & \frac{\sqrt{6}}{6} \end{pmatrix}$ \\
$ \mathbb{Y}_{70} $ & $\mathbb{Q}_{1,0}^{(b,E^{\prime})}$ & B$_{9}$ & $\begin{pmatrix} 0 & 0 & - \frac{1}{2} & \frac{1}{2} & \frac{1}{2} & - \frac{1}{2} \end{pmatrix}$ \\
$ \mathbb{Y}_{71} $ & $\mathbb{Q}_{1,1}^{(b,E^{\prime})}$ & B$_{9}$ & $\begin{pmatrix} - \frac{\sqrt{3}}{3} & - \frac{\sqrt{3}}{3} & \frac{\sqrt{3}}{6} & \frac{\sqrt{3}}{6} & \frac{\sqrt{3}}{6} & \frac{\sqrt{3}}{6} \end{pmatrix}$ \\
$ \mathbb{Y}_{72} $ & $\mathbb{Q}_{2,0}^{(b,E^{\prime\prime})}$ & B$_{9}$ & $\begin{pmatrix} 0 & 0 & \frac{1}{2} & - \frac{1}{2} & \frac{1}{2} & - \frac{1}{2} \end{pmatrix}$ \\
$ \mathbb{Y}_{73} $ & $\mathbb{Q}_{2,1}^{(b,E^{\prime\prime})}$ & B$_{9}$ & $\begin{pmatrix} - \frac{\sqrt{3}}{3} & \frac{\sqrt{3}}{3} & - \frac{\sqrt{3}}{6} & - \frac{\sqrt{3}}{6} & \frac{\sqrt{3}}{6} & \frac{\sqrt{3}}{6} \end{pmatrix}$ \\
$ \mathbb{Y}_{74} $ & $\mathbb{T}_{2,0}^{(b,E^{\prime\prime})}$ & B$_{9}$ & $\begin{pmatrix} 0 & 0 & \frac{i}{2} & - \frac{i}{2} & \frac{i}{2} & - \frac{i}{2} \end{pmatrix}$ \\
$ \mathbb{Y}_{75} $ & $\mathbb{T}_{2,1}^{(b,E^{\prime\prime})}$ & B$_{9}$ & $\begin{pmatrix} - \frac{\sqrt{3} i}{3} & \frac{\sqrt{3} i}{3} & - \frac{\sqrt{3} i}{6} & - \frac{\sqrt{3} i}{6} & \frac{\sqrt{3} i}{6} & \frac{\sqrt{3} i}{6} \end{pmatrix}$ \\
\end{longtable}
\end{center}

\item The relevant polar and axial harmonics are summarized in Tables XIV and XV.

\begin{center}
\renewcommand{\arraystretch}{1.3}
\begin{longtable}{ccccccc}
\caption{Polar harmonics.}
 \\
 \hline \hline
No. & symbol & rank & irrep. & mul. & comp. & form \\ \hline \endfirsthead

\multicolumn{6}{l}{\tablename\ \thetable{}} \\
 \hline \hline
No. & symbol & rank & irrep. & mul. & comp. & form \\ \hline \endhead

 \hline \hline
\multicolumn{6}{r}{\footnotesize\it continued ...} \\ \endfoot

 \hline \hline
\multicolumn{6}{r}{} \\ \endlastfoot

$ 1 $ & $ \mathbb{Q}_{0}^{(A_{1}^{\prime})} $ & $ 0 $ & $ A_{1}^{\prime} $ & $ - $ & $ - $ & $ 1 $ \\ \hline
$ 2 $ & $ \mathbb{Q}_{1}^{(A_{2}^{\prime\prime})} $ & $ 1 $ & $ A_{2}^{\prime\prime} $ & $ - $ & $ - $ & $ z $ \\
$ 3 $ & $ \mathbb{Q}_{1,0}^{(E^{\prime})} $ & $ 1 $ & $ E^{\prime} $ & $ - $ & $ 0 $ & $ x $ \\
$ 4 $ & $ \mathbb{Q}_{1,1}^{(E^{\prime})} $ & $ 1 $ & $ E^{\prime} $ & $ - $ & $ 1 $ & $ y $ \\ \hline
$ 5 $ & $ \mathbb{Q}_{2}^{(A_{1}^{\prime})} $ & $ 2 $ & $ A_{1}^{\prime} $ & $ - $ & $ - $ & $ - \frac{x^{2}}{2} - \frac{y^{2}}{2} + z^{2} $ \\
$ 6 $ & $ \mathbb{Q}_{2,0}^{(E^{\prime\prime})} $ & $ 2 $ & $ E^{\prime\prime} $ & $ - $ & $ 0 $ & $ \sqrt{3} x z $ \\
$ 7 $ & $ \mathbb{Q}_{2,1}^{(E^{\prime\prime})} $ & $ 2 $ & $ E^{\prime\prime} $ & $ - $ & $ 1 $ & $ \sqrt{3} y z $ \\
$ 8 $ & $ \mathbb{Q}_{2,0}^{(E^{\prime})} $ & $ 2 $ & $ E^{\prime} $ & $ - $ & $ 0 $ & $ - \sqrt{3} x y $ \\
$ 9 $ & $ \mathbb{Q}_{2,1}^{(E^{\prime})} $ & $ 2 $ & $ E^{\prime} $ & $ - $ & $ 1 $ & $ - \frac{\sqrt{3} \left(x - y\right) \left(x + y\right)}{2} $ \\ \hline
$ 10 $ & $ \mathbb{Q}_{3}^{(A_{1}^{\prime})} $ & $ 3 $ & $ A_{1}^{\prime} $ & $ - $ & $ - $ & $ \frac{\sqrt{10} y \left(3 x^{2} - y^{2}\right)}{4} $ \\
$ 11 $ & $ \mathbb{Q}_{3}^{(A_{2}^{\prime\prime})} $ & $ 3 $ & $ A_{2}^{\prime\prime} $ & $ - $ & $ - $ & $ - \frac{z \left(3 x^{2} + 3 y^{2} - 2 z^{2}\right)}{2} $ \\
$ 12 $ & $ \mathbb{Q}_{3}^{(A_{2}^{\prime})} $ & $ 3 $ & $ A_{2}^{\prime} $ & $ - $ & $ - $ & $ \frac{\sqrt{10} x \left(x^{2} - 3 y^{2}\right)}{4} $ \\
$ 13 $ & $ \mathbb{Q}_{3,0}^{(E^{\prime\prime})} $ & $ 3 $ & $ E^{\prime\prime} $ & $ - $ & $ 0 $ & $ - \sqrt{15} x y z $ \\
$ 14 $ & $ \mathbb{Q}_{3,1}^{(E^{\prime\prime})} $ & $ 3 $ & $ E^{\prime\prime} $ & $ - $ & $ 1 $ & $ - \frac{\sqrt{15} z \left(x - y\right) \left(x + y\right)}{2} $ \\
$ 15 $ & $ \mathbb{Q}_{3,0}^{(E^{\prime})} $ & $ 3 $ & $ E^{\prime} $ & $ - $ & $ 0 $ & $ - \frac{\sqrt{6} x \left(x^{2} + y^{2} - 4 z^{2}\right)}{4} $ \\
$ 16 $ & $ \mathbb{Q}_{3,1}^{(E^{\prime})} $ & $ 3 $ & $ E^{\prime} $ & $ - $ & $ 1 $ & $ - \frac{\sqrt{6} y \left(x^{2} + y^{2} - 4 z^{2}\right)}{4} $ \\ \hline
$ 17 $ & $ \mathbb{Q}_{4}^{(A_{1}^{\prime\prime})} $ & $ 4 $ & $ A_{1}^{\prime\prime} $ & $ - $ & $ - $ & $ \frac{\sqrt{70} x z \left(x^{2} - 3 y^{2}\right)}{4} $ \\
$ 18 $ & $ \mathbb{Q}_{4}^{(A_{1}^{\prime})} $ & $ 4 $ & $ A_{1}^{\prime} $ & $ - $ & $ - $ & $ \frac{3 x^{4}}{8} + \frac{3 x^{2} y^{2}}{4} - 3 x^{2} z^{2} + \frac{3 y^{4}}{8} - 3 y^{2} z^{2} + z^{4} $ \\
$ 19 $ & $ \mathbb{Q}_{4,0}^{(E^{\prime},1)} $ & $ 4 $ & $ E^{\prime} $ & $ 1 $ & $ 0 $ & $ \frac{\sqrt{35} x y \left(x - y\right) \left(x + y\right)}{2} $ \\
$ 20 $ & $ \mathbb{Q}_{4,1}^{(E^{\prime},1)} $ & $ 4 $ & $ E^{\prime} $ & $ 1 $ & $ 1 $ & $ - \frac{\sqrt{35} \left(x^{2} - 2 x y - y^{2}\right) \left(x^{2} + 2 x y - y^{2}\right)}{8} $ \\
$ 21 $ & $ \mathbb{Q}_{4,0}^{(E^{\prime},2)} $ & $ 4 $ & $ E^{\prime} $ & $ 2 $ & $ 0 $ & $ \frac{\sqrt{5} x y \left(x^{2} + y^{2} - 6 z^{2}\right)}{2} $ \\
$ 22 $ & $ \mathbb{Q}_{4,1}^{(E^{\prime},2)} $ & $ 4 $ & $ E^{\prime} $ & $ 2 $ & $ 1 $ & $ \frac{\sqrt{5} \left(x - y\right) \left(x + y\right) \left(x^{2} + y^{2} - 6 z^{2}\right)}{4} $ \\
\end{longtable}
\end{center}
\begin{center}
\renewcommand{\arraystretch}{1.3}
\begin{longtable}{ccccccc}
\caption{Axial harmonics.}
 \\
 \hline \hline
No. & symbol & rank & irrep. & mul. & comp. & form \\ \hline \endfirsthead

\multicolumn{6}{l}{\tablename\ \thetable{}} \\
 \hline \hline
No. & symbol & rank & irrep. & mul. & comp. & form \\ \hline \endhead

 \hline \hline
\multicolumn{6}{r}{\footnotesize\it continued ...} \\ \endfoot

 \hline \hline
\multicolumn{6}{r}{} \\ \endlastfoot

$ 1 $ & $ \mathbb{G}_{1}^{(A_{2}^{\prime})} $ & $ 1 $ & $ A_{2}^{\prime} $ & $ - $ & $ - $ & $ Z $ \\
$ 2 $ & $ \mathbb{G}_{1,0}^{(E^{\prime\prime})} $ & $ 1 $ & $ E^{\prime\prime} $ & $ - $ & $ 0 $ & $ - Y $ \\
$ 3 $ & $ \mathbb{G}_{1,1}^{(E^{\prime\prime})} $ & $ 1 $ & $ E^{\prime\prime} $ & $ - $ & $ 1 $ & $ X $ \\ \hline
$ 4 $ & $ \mathbb{G}_{2}^{(A_{1}^{\prime\prime})} $ & $ 2 $ & $ A_{1}^{\prime\prime} $ & $ - $ & $ - $ & $ - \frac{X^{2}}{2} - \frac{Y^{2}}{2} + Z^{2} $ \\
$ 5 $ & $ \mathbb{G}_{2,0}^{(E^{\prime\prime})} $ & $ 2 $ & $ E^{\prime\prime} $ & $ - $ & $ 0 $ & $ \frac{\sqrt{3} \left(X - Y\right) \left(X + Y\right)}{2} $ \\
$ 6 $ & $ \mathbb{G}_{2,1}^{(E^{\prime\prime})} $ & $ 2 $ & $ E^{\prime\prime} $ & $ - $ & $ 1 $ & $ - \sqrt{3} X Y $ \\
$ 7 $ & $ \mathbb{G}_{2,0}^{(E^{\prime})} $ & $ 2 $ & $ E^{\prime} $ & $ - $ & $ 0 $ & $ - \sqrt{3} Y Z $ \\
$ 8 $ & $ \mathbb{G}_{2,1}^{(E^{\prime})} $ & $ 2 $ & $ E^{\prime} $ & $ - $ & $ 1 $ & $ \sqrt{3} X Z $ \\ \hline
$ 9 $ & $ \mathbb{G}_{3}^{(A_{2}^{\prime})} $ & $ 3 $ & $ A_{2}^{\prime} $ & $ - $ & $ - $ & $ - \frac{Z \left(3 X^{2} + 3 Y^{2} - 2 Z^{2}\right)}{2} $ \\
$ 10 $ & $ \mathbb{G}_{3,0}^{(E^{\prime})} $ & $ 3 $ & $ E^{\prime} $ & $ - $ & $ 0 $ & $ \frac{\sqrt{15} Z \left(X - Y\right) \left(X + Y\right)}{2} $ \\
$ 11 $ & $ \mathbb{G}_{3,1}^{(E^{\prime})} $ & $ 3 $ & $ E^{\prime} $ & $ - $ & $ 1 $ & $ - \sqrt{15} X Y Z $ \\
\end{longtable}
\end{center}

\end{itemize}

\subsection{Parameter optimization}
\label{sec_usage_ex_mos2_fit}

In this section, we show the results of the parameter optimization.
We chose the high symmetry lines $\Gamma-$K$-$M$-\Gamma$, and $50$ $\bm{k}$ points in each line are used to evaluate the loss function.
The maximum number of iterations is $N_{\rm iter} = 500$, the learning rate is $\alpha = 0.01$, and $N_{\rm h} = 3$ hidden layers are used.
The total number of the optimization parameters including the hyper-parameters in the hidden layers is about 1,200,000.
We also perform 50 optimizations with different random initial parameters.

\begin{figure}[htbp]
   \begin{center}
      \includegraphics[width=16.5cm]{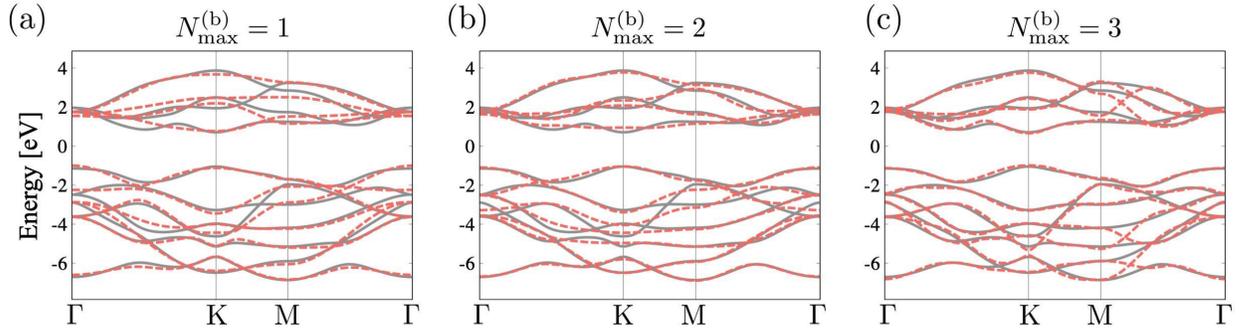}
        \caption{
         The comparisons of the band dispersion between the Wannier TB model (solid gray lines) and our TB model (dashed red lines).
         (a)-(c) The optimized results obtained by using the maximum number of neighboring bonds $N_{\rm max}^{\rm (b)} = $1-3.
         The Fermi energy is taken as the origin.
        }
        \label{fig_mos2_opt_band_comp}
   \end{center}
\end{figure}

The results of the optimized dispersions are shown in Figs.~\ref{fig_mos2_opt_band_comp} (a)-(c).
As shown in Figs.~\ref{fig_mos2_opt_band_comp} (a)-(c), the quality of the optimization are improved gradually by increasing $N_{\rm max}^{\rm (b)}$.
In particular, when $N_{\rm max}^{\rm (b)} = 3$, the obtained TB model reproduces the DF Wannier band dispersions with high accuracy.
The optimized model parameters $z_{j}$ [eV] of up to nearest-neighbor hopping are given by
\begin{align}
&
z_{1}=2.170,\,\,\,z_{2}=-1.653,\,\,\,z_{3}=-2.842,\,\,\,z_{4}=3.683,\,\,\,
z_{5}=-1.219,\,\,\,z_{6}=-1.298,\,\,\,z_{7}=0.121,\,\,\,z_{8}=-0.018,\cr&
z_{9}=0.166,\,\,\,z_{10}=-0.500,\,\,\,z_{11}=0.851,\,\,\,z_{12}=-1.450,\,\,\,
z_{13}=0.489,\,\,\,z_{14}=-0.462,\,\,\,z_{15}=-0.106,\,\,\,z_{16}=-0.439,\cr&
z_{17}=0.007,\,\,\,z_{18}=0.047,\,\,\,z_{19}=-0.539,\,\,\,z_{20}=-0.385,\,\,\,
z_{21}=-1.127,\,\,\,z_{22}=0.172,\,\,\,z_{23}=-0.018,\,\,\,z_{24}=-0.055,\cr&
z_{25}=-0.136,\,\,\,z_{26}=-0.682,\,\,\,z_{27}=-0.122,\,\,\,z_{28}=0.443.
\end{align}

\begin{figure}[htbp]
   \begin{center}
   \includegraphics[width=8cm]{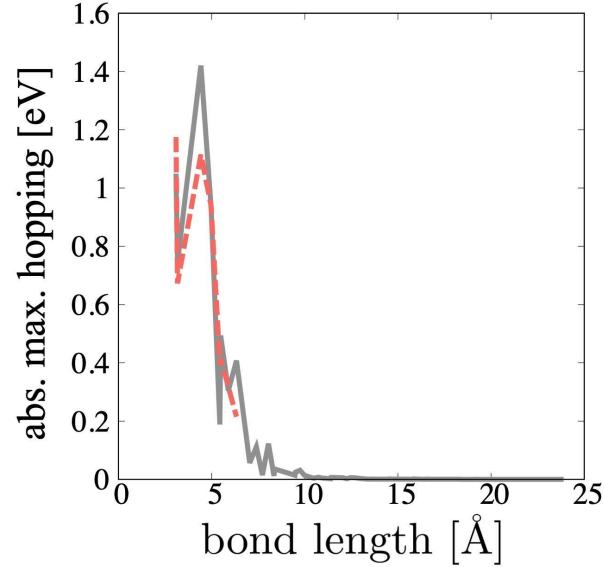}
      \caption{
         Bond length dependence of the absolute maximum value of the hopping parameters ($6 \leq j \leq 74$) in eV units between the Wannier TB model (solid gray lines) and our TB model (dashed red lines).
      }
      \label{fig_mos2_bond_dep_Nh3_Nb3}
   \end{center}
\end{figure}

As shown in Fig.~\ref{fig_mos2_bond_dep_Nh3_Nb3}, the magnitude of the hopping parameters of our TB model decreases for further neighbor hoppings, and much less number of parameters are required as compared with the Wannier TB model.

\end{widetext}

\end{document}